\documentclass[emulateapj,twocolumn]{aastex631}

\usepackage{graphicx}
\usepackage{natbib}
\usepackage{enumitem}
\usepackage{subfigure}
\usepackage{epstopdf}
\usepackage{epsfig}
\usepackage{amsmath}
\usepackage{longtable}
\usepackage{rotating}
\usepackage{parskip}
\usepackage{color}
\usepackage[normalem]{ulem}

\definecolor{orange}{rgb}{1,0.5,0}
\definecolor{purple}{rgb}{0.5,0.1,0.7} 
\definecolor{light-gray}{gray}{0.5}

\newcommand{\etal}{{et al.\ }}
\newcommand{\eg}{{e.g.,}}

\graphicspath{{./}{figures/}}

\shorttitle{The Nature of LoBAL QSOs: II. Morphologies Dominated by Mergers}
\shortauthors{Lazarova et al.}

\received{2022 July 28}
\accepted{2023 March 15}

\begin{document}

\title{The Nature of LoBAL QSOs: II. HST/WFC3 Observations Reveal Host Galaxies Dominated by Mergers }

\correspondingauthor{Mariana Lazarova}
\email{mariana.lazarova@unco.edu}

\author[0000-0003-3818-6691]{Mariana S. Lazarova}
\affiliation{Department of Physics and Astronomy, University of Northern Colorado, \\
Greeley, CO 80639, USA}

\author[0000-0003-4693-6157]{Gabriela Canalizo}
\affiliation{Department of Physics and Astronomy, University of California, \\
Riverside, CA 92521, USA}

\author[0000-0002-3032-1783]{Mark Lacy}
\affiliation{National ALMA Science Center, National Radio Astronomy Observatory,\\
 520 Edgemont Rd., Charlottesville, VA 22903, USA}

\author{Wyatt Behn}
\affiliation{Department of Physics, University of Wisconsin-Madison, \\
Madison, WI 53706 USA}
\affiliation{previously at Department of Physics and Astronomy, University of Nebraska-Kearney,\\
 Kearney, NE 68849, USA}

\author{Kaitlyn Raub}
\affiliation{Department of Physics and Astronomy, University of Northern Colorado, \\
Greeley, CO 80639, USA}

\author[0000-0003-2064-0518]{Vardha N. Bennert}
\affiliation{Physics Department California Polytechnic State University,\\
 1 Grand Avenue, San Luis Obispo, CA 93407, USA}

\author[0000-0003-1748-2010]{Duncan Farrah}
\affiliation{Department of Physics and Astronomy, University of Hawai'i at M\~{a}noa, \\2505 Correa Road, Honolulu, HI 96822, USA}

\begin{abstract}

Low-ionization Broad Absorption Line QSOs (LoBALs) are suspected to be merging systems in which extreme, AGN-driven outflows have been triggered.  Whether or not LoBALs are uniquely associated with mergers, however, has yet to be established. To characterize the morphologies of LoBALs, we present the first high-resolution morphological analysis of a volume-limited sample of 22 SDSS-selected LoBALs at $0.5 < z < 0.6$ from {Hubble Space Telescope} Wide Field Camera 3 observations. Host galaxies are resolved in 86\% of the systems in F125W, which is sensitive to old stellar populations, while only 18\% are detected in F475W, which traces young, unobscured stellar populations.  Signs of recent or ongoing tidal interaction are present in 45$-$64\% of the hosts, including double nuclei, tidal tails, bridges, plumes, shells, and extended debris.  Ongoing interaction with a companion is apparent in 27$-$41\% of the LoBALs, with as much as 1/3 of the sample representing late-stage mergers at projected nuclear separations $<$10 kpc. Detailed surface brightness modeling indicates that 41\% of the hosts are bulge-dominated while only 18\% are disks.  We  discuss trends in various properties as a function of merger stage and parametric morphology. Notably, mergers are associated with slower, dustier winds than those seen in undisturbed/unresolved hosts.  Our results favor an evolutionary scenario in which quasar-level accretion during various merger stages is associated with the observed outflows in low-z LoBALs. We discuss differences between LoBALs and FeLoBALs and show that selection via the traditional Balnicity index would have excluded all but one of the mergers.\\

\end{abstract}

\keywords{galaxies: active -- galaxies: interactions --- galaxies: evolution --- quasars: absorption lines --- quasars: general}

\section{Introduction}

Since observations established the existence of supermassive black holes (SMBHs) in the centers of galaxies with bulges, as well as correlations between their masses and properties of the host galaxies \citep[for a review, see][]{Kormendy2013}, much effort has been made in understanding the underlying connection and possible interaction between these two physically distinct scales - the black hole's direct sub-parsec influence and the galaxy's kpc size. The extent and importance of various mechanisms via which the galaxy and black hole regulate each other's growth are still debated \citep[e.g.,][]{Harrison2017}, as are the dominant triggers of accretion onto galactic SMBHs, hence, the onset of active galactic nucleus (AGN) activity, even for the most luminous AGN found in quasi-stellar objects (QSOs) \citep[\eg][]{Sharma2021}. Outflows at different spatial scales have been discovered in quasars\footnote{We use the terms 'quasar' and 'QSO' interchangeably to refer to the general population of quasi-stellar objects (QSOs), regardless of identification with radio emission, which was the original definition of 'quasar' by \citet{Chiu1964}.} in molecular \citep[\eg][]{Vayner2021,Alatalo2015,Feruglio2015,Spoon2013}, atomic \citep[\eg][]{Morganti2018}, and ionized gas \citep[\eg][]{Veilleux2021,Rupke2013}, with evidence that some are powerful enough to be the main AGN feedback agent \citep[\eg][]{Miller2020,Nardini2015,Harrison2014}. Yet, strong observational evidence is missing of the role mergers play in contributing to the emergence of such outflows in quasars, and the extent to which they affect the environment on galactic scale.

The extreme outflows observed in Broad Absorption Line (BAL) QSOs are of particular interest in our search for the predicted, but observationally elusive, AGN feedback, a key ingredient in galaxy evolution recipes. LoBALs are part of the larger class of BAL QSOs, objects characterized by broad absorption troughs of UV resonance lines, blue-shifted relative to the QSO's rest frame, indicating gas outflows with speeds of up to 0.2$c$ \citep{Foltz1983}. Based on the ions producing BAL troughs, there are three main subclasses of BAL QSOs.  The high-ionization BAL QSOs (HiBALs) are identified via the broad absorption from \ion{C}{4} $\lambda$1549 and other high-ionization species such as Ly$\alpha$, \ion{N}{5} $\lambda$1240, and \ion{Si}{4} $\lambda$1394 \citep{Hall2002}. HiBALs are found in 10$-$30\% of optically-selected QSOs, and make up the majority ($\sim$90\%) of the BAL QSOs. The low-ionization BAL QSOs (LoBALs), in addition to the lines present in HiBALs, feature absorption lines from \ion{Mg}{2} $\lambda$$\lambda$ 2796,2803, \ion{Al}{3}, and \ion{Al}{2}. Only 10\% of the BAL QSOs are LoBALs. A very small fraction of LoBALs, which represents $\sim$3\% of all BAL QSOs, called FeLoBALs, show absorption in their rest-frame UV spectra from metastable, excited states of \ion{Fe}{2} \citep{Hazard1987}.  Selection criteria for BAL QSOs have changed over the years. BALs were defined by \citet{Weymann1991} via a Balnicity Index (BI$>$0) to include only objects with broad absorption lines wider than 2000 km s$^{-1}$, blue-shifted past the first 3000 km s$^{-1}$. In their BAL QSO catalogues, \citet{Trump2006} use a more inclusive equivalent width called the Absorption Index (AI$>$0), which relaxes the criterion of the minimum absorption width to 1000 km s$^{-1}$, starting from zero velocity, while \citet{Gibson2009} employ a modified BI integrating the troughs from zero velocity. 

Low-ionization BAL QSOs (LoBALs), in particular, are suspected to be young, recently fueled QSOs showing disturbed host morphologies \citep{Canalizo2002}. While some simulations predict that the energy released during the luminous quasar phase can generate large-scale galactic outflows and have global impact on the host galaxy by quenching the star formation \citep{DiMatteo2005,Hopkins2008,Wagner2012}, others suggest mergers do not lead to such galactic-scale winds \citep[\eg][]{Debuhr2010}, or that fast outflows arise but have little effect on galactic disks \citep{Gabor2014}.   Yet, most theoretical galaxy evolution models support the co-evolution of galaxies and their SMBHs \citep[for review, see][]{Somerville2015} by invoking galaxy mergers and ensuing AGN-driven feedback to account for key observables in galactic properties and galaxy populations.  Observations support the hierarchical model for galaxy formation and evolution, in which larger structures form through successive gravitational merging of low-mass dark matter halos, and ordinary matter follows the dynamics of dark matter \citep{White1978,Cole2000}. The history of the cosmic star formation follows that of central black hole accretion, offering further evidence for the co-evolution of black holes and their host galaxies \citep[\eg][]{Madau2014}.  The cosmic star formation rate density and the AGN luminosity function do roughly trace each other, but there is new evidence that black hole growth might be outpacing the build up of stellar mass at high redshift \citep[\eg][]{Runburg2022}. Mergers play an inextricable part of the cosmic history of the universe, yet with somewhat uncertain role.

Major mergers of gas-rich galaxies can offer both the ingredients for growth (ample gas supply) and feedback mechanisms to curtail it (outflows, shock heating, ionization). Mergers can catalyze galaxy growth through bursts of star formation and fuel episodic accretion onto the SMBHs \citep[\eg][]{Sanders1988,Silk1998,Fabian2012,Heckman2014,Hickox2018}.  Mergers can be associated with episodes of AGN-driven feedback, as inflowing gas fuels black hole accretion, which gives rise to radiative-driven outflows and, sometimes, jets. But observations in support of a dominant role for gas-rich mergers in fueling AGN, even at the highest luminosities, are still scarce and debated \citep{Sharma2021}. One of the biggest obstacles to studying the host galaxy morphologies for signs of tidal interaction in quasars, in particular, is the high contrast between nuclear and host emission, and, hence, the requirement of deep observations to uncover low surface brightness tidal features indicative of past merger activity. The host galaxies of QSOs, once thought to be mostly normal, quiescent ellipticals \citep[e.g.,][]{Dunlop2003,Floyd2004}, have been shown to have unambiguous relics of past tidal interaction through the presence of faint tidal features in deep HST images \citep{Canalizo2007, Bennert2008}. Currently, there is no consensus on the issue, with observations of samples across various redshifts and luminosities supporting two competing claims. On one hand, many studies find high fraction of mergers in their samples and argue for a key role played by mergers \citep[\eg][]{Guyon2006,Bennert2008,Urrutia2008,Veilleux2009,Koss2010,Koss2011,Treister2012,Glikman2012,Satyapal2014,Glikman2015,Kocevski2015,Hong2015,Fan2016,Goulding2018,Pfeifle2019,Gao2020}. Parallel to those are studies that conclude AGN are not predominantly triggered by major mergers \citep[\eg][]{Grogin2005,Gabor2009,Cisternas2011a,Cisternas2011b,Bohm2013,Kocevski2012,Schawinski2012,Villforth2014,Mechtley2016,Villforth2017,Chang2017,Marian2019,Villforth2019,Marian2020,Sharma2021}. The various studied samples are not matched in redshift and luminosity, making it hard to compare directly. Even if secular processes, such as bar instabilities and cooling flows \citep[\eg][]{Nulsen2000,Kormendy2004,Buta2013}, are the main driver of galaxy evolution and AGN activity, it is still important to understand if a specific population of objects are associated with mergers and why.

The BAL outflows seen in QSOs have been of great interest in the search for the elusive feedback stage, as they might represent a short-lived episode of rapid black hole growth launching fast accretion disk winds capable of reaching the host \citep[\eg][]{Hopkins2008a}. There is still much uncertainty about the locations of these outflows, with some studies estimating pc-scale \citep[\eg][]{Hamann2019,Leighly2018,Chamberlain2015,deKool2002}, while other work finding galaxy-scale kpc flows \citep[\eg][]{FaucherGiguere2012,Bautista2010,Dunn2010,Moe2009}. Most interesting have been recent spectral-synthesis analysis results from ionic column densities in BAL QSOs: \citet{Leighly2018} and \citet{Hamann2019} report kinetic luminosities greater than 5\% of the bolometric luminosity, enough to curtail galaxy growth, as predicted by co-evolution models \citep{Hopkins2010}. Hence, the kinetic power of the winds is sufficient to unbind large mass of gas on galactic scales \citep[see for a review][]{Fabian2012}. But linking outflows to definitive evidence for quenching of star formation has been challenging.

The relative rarity of LoBAL is possibly a key piece to the puzzle, and has lead to the two competing interpretations of the BAL phenomena: orientation or evolution. Although hydrodynamic models show that AGNs are capable of launching the high-velocity winds seen in BAL QSOs \citep{Murray1995, Proga2000, Gallagher2012}, in optically-selected QSOs, only 10\%$-$30\% have BALs \citep{Tolea2002, Hewett2003,Reichard2003a,Trump2006, Gibson2009}, and only about one tenth of these are LoBALs \citep{Reichard2003a}. The remarkable similarities in the broadband SEDs of BAL and non-BAL QSOs suggest they are from the same parent population \citep{Weymann1991, Gallagher2007,Lazarova2012}, hence the efforts to explain the low occurrence of BAL QSOs within the framework of the AGN unification model \citep{Antonucci1993,Urry1995,Netzer2015} as an orientation effect: BALs would be seen in classical QSOs only when viewed along a narrow range of lines of sight passing through the accretion disk wind \citep[\eg][]{Elvis2000}.  Low occurrence would mean LoBALs, in particularly, are observed only at a small range of viewing angles. But radio observation have observed BALs at a wide range of inclinations \citep{Becker2000, Gregg2000, Brotherton2006, Montenegro-Montes2008, DiPompeo2010} suggesting that the occurrence of BALs is not a simple orientation effect \citep[e.g.,][]{DiPompeo2012,Allen2011}. 

Due to the much redder continua of LoBALs \citep[\eg][]{Dunn2015}, optical identification possibly omits a large fraction of them.  Infrared-selected QSO samples find much higher number of LoBALs \citep{Urrutia2009, Dai2010}, and \citet{Allen2011} estimate that the intrinsic fraction of BAL QSOs in the SDSS can be as high as $\sim$40\% when the spectroscopic incompleteness and bias are taken into account. The association of BAL QSOs with obscuration leads to an alternative evolutionary explanation, in which they are young QSOs caught during a short-lived phase when powerful QSO-driven winds are blowing away a dusty obscuring cocoon \citep[\eg][]{Hazard1984, Voit1993, Hall2002}. This model appears to be particularly applicable to LoBALs since these objects are suspected to be young or recently refueled QSOs \citep{Boroson1992, Lipari1994} and might be exclusively associated with mergers \citep{Canalizo2001}. Observations by \citet{Canalizo2002} of the only four known LoBALs at $z < 0.4$ at the time show infrared luminosities equivalent to ULIRGs, strongly disturbed host morphologies as a result of major mergers, and unambiguous interaction-induced star formation with post-starburst ages $\leq$ 100 Myr. At higher redshifts, \citet{Wethers2020} find an enhancement of the star formation rates and the FIR emission in LoBALs compared to HiBALs and non-BALs. While recent work by \citet{Chen2022} suggests an evolutionary sequence from LoBALs to HiBALs to non-BALs, and find evidence for the suppression of star formation in HiBALs due to the negative feedback from the outflows and subsequent rebound in non-BAL quasars.  In addition, FeLoBALs at low \citep{Farrah2005} and high redshifts \citep{Farrah2007, Farrah2010} are associated with extremely star-forming ULIRGs and show direct evidence for negative impact on star formation in their host galaxies \citep{Farrah2012}. Yet the morphologies of FeLoBALs at z$\sim$0.9 show mostly disk profiles and dispute strong association with major mergers \citep{Villforth2019}. 

With only a few hosts imaged to date, currently little is know about the host galaxies of LoBALs, which is the focus of this work. We take a detailed look at the morphologies of a sample of low-redshift LoBALs with the goal of distinguishing between the two competing explanations for them. Finding a large fraction of mergers would strongly favor the evolutionary explanation since the AGN unification model may not apply to merging systems and is likely restricted to secularly evolving galaxies \citep{Netzer2015}.

In \citet{Lazarova2012}, we compared the optical-to-FIR SEDs of this sample of LoBALs to non-BAL QSOs and found no statistically significant differences. In this second paper of the series, we present results from the $HST$ WFC3/IR and UVIS imaging campaign of the 22 optically-selected LoBAL QSOs introduced in \citet{Lazarova2012}. A detailed high-resolution study of their morphologies should reveal unambiguous signs of tidal interaction if LoBALs are indeed young transition objects resulting from mergers. The sample is introduced in $\S$~\ref{sample}. Details on the observations and data reduction are explained in $\S$~\ref{observations}. We present the analysis in $\S$~\ref{analysis} and results in $\S$~\ref{results}. Discussion and conclusions are given in $\S$~\ref{discussion} and $\S$~\ref{conclusion}. We assume a flat universe cosmology with $H_0$ = 71 km s$^{-1}$ Mpc$^{-1}$, $\Omega_M$ = 0.27, and $\Omega_{\Lambda}$ = 0.73. All luminosities in units of the bolometric solar luminosity were calculated using $L_{\odot}$ = 3.839 $\times$ 10$^{33}$ erg s$^{-1}$.

\section{Sample selection}\label{sample}

Our sample was selected to include all LoBALs within the redshift range 0.5~$<z<$~0.6 from the \citet{Trump2006} catalog of BAL quasars, drawn from Data Release 3 of the Sloan Digital Sky Survey (SDSS) quasar catalog by \citet{Schneider2005}. 
The lower redshift limit ($z>$ 0.5) is set by requiring the blue-shifted broad absorption of \ion{Mg}{2} $\lambda$2800 - which defines LoBALs - to be redshifted to the optical wavelength range covered by SDSS; the upper redshift limit ($z<$ 0.6) was chosen to make it feasible to resolve the host galaxy morphologies despite of the bright nuclear emission, while ensuring sufficient number of objects in the sample ($>$20) for statistical analysis. In the \citet{Trump2006} catalog, QSOs with regions of flux at least 10\% below the continuum, spanning over a velocity range of at least 1000 km s$^{-1}$ blue-ward of Mg~{\sc II} $\lambda$2800, were identified as LoBALs. They found 457 LoBALs in the redshift range 0.5 $< z <$ 2.15. Of those, only 22 fall within 0.5 $< z <$ 0.6, when we exclude one object which is classified as a narrow-line LoBAL and one identified as an uncertain FeLoBAL. The volume-limited sample of 22 low-redshift LoBALs is listed in Table~\ref{table:observations}.  Note that some of the objects in our sample are not identified as LoBALs in the catalog by \citet{Gibson2009} since they introduce a new balnicity index, integrating the troughs starting from zero velocity shifts, which is different from the absorption index of \citet{Trump2006}.  Yet, we note that the total number of LoBALs within that redshift range in both catalogs is still 22, with 16 objects in each sample shared between them.  For more details on the sample selection, see \citet{Lazarova2012}.

We emphasize that these LoBALs were drawn from a sample of optically selected type-1 BAL QSOs \citep{Trump2006}, chosen simply based on their Mg{II} BAL moving into the SDSS spectral range at this redshift. While we make the distinction between the properties of FeLoBALs and LoBALs (\S~\ref{comparison}), we do not have UV data to rule out FeLoBAL interlopers in our sample. Most of the studies trying to understand BAL QSOs either compile heterogenous samples of HiBAL, LoBALs and FeLoBALs, thus mostly uncover properties of HiBALs due to their relative abudance, or focus on the more extreme subpopulation of FeLoBALs, due to their much redder continua and strange spectral properties. The work we present here aim to bridge our gap in understanding the LoBAL population, excluding FeLoBALs, when possible.

The optical selection should offer less bias towards mergers compared to an infrared-selected sample, which would preferentially pick obscured systems possibly associated with mergers by the selection criteria. Considering that the majority of the AGN population is obscured \citep[for a review, see][]{Hickox2018} and optically-selected quasars comprise only a fraction of the total quasar population \citep[\eg][]{Lacy2004,Lacy2007,Martinez-Sansigre2005,Stern2005,Donley2012,Lacy2015}, it would be interesting to investigate in the future whether or not the results from this optically-selected sample would apply to LoBALs selected differently.

\section{Observations and Data Reduction}\label{observations}

Imaging data were obtained with the Hubble Space telescope using the infrared (IR) and the ultraviolet-visible (UVIS) channels of the Wide Field Camera 3 (WFC3) between August 6, 2009, and April 29, 2011, as part of our Cycle 17 program GO-11557 (PI: Canalizo). All the {\it HST} data used in this paper can be found in MAST: \dataset[10.17909/6bj2-c984]{http://dx.doi.org/10.17909/6bj2-c984}. This program was granted 23 orbits to observe the entire sample of 22 LoBALs and a PSF star in the IR channel using the wide $J$-band F125W filter ($\lambda_{pivot}$ = 1249 nm; $\Delta\lambda$=302 nm; detector pixel scale $\sim$ 0{\farcs}13 pixel$^{-1}$) and the UVIS channel using the SDSS-g'-band F475W filter ($\lambda_{pivot}$ = 478 nm; $\Delta\lambda$=149 nm; detector pixel scale $\sim$ 0{\farcs}04 pixel$^{-1}$). One full orbit was used to acquire both the IR and UVIS observations for each target.  An additional orbit was granted to repeat and replace the observations for SDSS J1614+5238, as guide-star acquisition failure caused drift and inaccurate pointing in the initial F475W observations for that target. The WFC3/IR  observations were obtained in the MULTIACCUM mode, with sampling sequence SPARS50 and NSAMP of 10, 11, or 12. The WFC3/UVIS observations were obtained in the ACCUM mode, and whenever time allocation allowed, we used a CR-SPLIT. A two-point dither pattern, with a spacing of 0{\farcs}636 for the IR and 0{\farcs}145 for the UVIS, was used to ensure better sampling of the PSF and to help with the removal of data artifacts, such as cosmic rays and hot and dead pixels. The resulting total exposure times for each object ranged within 806$-$1006 s for the IR and 1436$-$1748 s for the UVIS exposures; the exposure times for each object are listed in Table 1.  In this paper, individual objects are referred to with a shorthand of their SDSS object identification, e.g., SDSS J023102.49$-$083141.2 would be called as J0231$-$0831.  

Starting with the bias-subtracted and flat-field corrected data products from the HST pipeline (i.e., the flt.fits files), the data were further processed with the Multidrizzle software package (Koekemoer et al. 2002) to clean the cosmic rays, apply the geometric corrections using the latest calibration files, and combine the individual dithered exposures. The images were drizzled to pixel scale of 0{\farcs}035 pixel$^{-1}$ for the UVIS/F475W and 0{\farcs}07 pixel$^{-1}$ for the IR/F125W using a pixel droplet fraction (pixfrac) of 1.0. The sky pedestal was not subtracted for the images to be analyzed with GALFIT.

The final spatial resolution of the images in F125W is PSF FWHM = 0{\farcs}16 (1.0 kpc for median z = 0.55 at 6.4 kpc/") and in F475W PSF FWHM = 0{\farcs}12 (0.77 kpc for z=0.55). The 1$\sigma$ fluctuations in the surface brightness, $\mu$ , distribution of the images corresponds to an object with surface brightness of 24.7$^{+0.2}_{-0.3}$ mag arcsec$^{-2}$ in the F125W and 25.6$\pm$0.5 mag arcsec$^{-2}$ in the F475W. The 3$\sigma$ surface brightness limits are estimated at $\mu_{F125W}\sim$ 23.5$^{+0.2}_{-0.3}$ mag arcsec$^{-2}$ and $\mu_{F475W}\sim$  24.4$\pm$0.5 mag arcsec$^{-2}$.

\section{Analysis}\label{analysis}

The two-dimensional image-fitting program GALFIT \citep{Peng2002, Peng2010} was used for each object to model the central point source and the surface brightness profile of the host galaxy. GALFIT is a least-squares, $\chi^2$, minimization algorithm which fits any number of light profiles to an input image, allowing for decomposition of the different contributions. In our investigation of LoBALs, we are interested in determining whether these galaxies have been involved in tidal interaction. Hence, we are not aiming in deriving models that fit every surface brightness feature, producing residuals close to the noise level. Instead, our goal is to fit classical galaxy components, such as a disk and/or a bulge, and use the residuals to reveal any possible past or present interactions in the host galaxy.

Hence, we fit the LoBALs with a PSF component, to account for the bright QSO point source, and one or two S\'{e}rsic profiles convolved with the PSF. The radial surface brightness S\'{e}rsic profile, one of the the most used functions to study galaxy morphology, has the form:

\begin{equation*}
\Sigma ( r)= \Sigma_{e} \exp\Bigg[ -\kappa \Bigg(\left(\frac{r}{r_e} \right) ^{1/n} -1\Bigg)\Bigg],
\end{equation*}

\noindent
where $r_e$ is the effective radius such that half of the total flux is within $r_e$, $\Sigma_{e}$ is the pixel surface brightness at the effective radius $r_e$, $n$ is the concentration parameter, more commonly known as the S\'{e}rsic index, and $\kappa$ is a variable coupled to $n$.  When $n$ is large, the central profile is steeper and the wings are extended. When $n$ is small, the central peak is flatter and the profile is truncated sharply at smaller radii. A special case of the S\'{e}rsic profile  when $n = 4$ is the de Vaucouleurs profile, which describes galaxy bulges. Another special case when $n = 1$ is the exponential profile, which describes galaxy disks.

We proceeded by fitting the two-dimensional surface brightness distribution of each object with five different galaxy models: single disk ($n=1$), single bulge ($n=4$), single variable S\'{e}rsic profile ($n=free$), disk + bulge ($n=1$ and $n=4$), and two variable S\'{e}rsic profiles ($n=free$ and $n=free$). Bright sources and faint extended structure within the image frames were masked out during the GALFIT modeling. The sky level was estimated separately using $IRAF$ and was held fixed in the GALFIT modeling to prevent possible incorrect GALFIT sky estimation from affecting the S\'{e}rsic component fit (as it is known that the sky anti-correlates with the S\'{e}rsic index \citep{Peng2010}). The PSF and S\'{e}rsic component centroids were allowed to vary. Any sources or bright debris in the vicinity of the objects was masked to avoid contamination, or in few cases was fit with an additional S\'{e}rsic component. The PSF used in the GALFIT modeling was observed during a dedicated HST orbit.  

\subsection{One-S\'{e}rsic Component Models} \label{onemodel}

First, we fit the surface brightness distribution of each LoBAL with a single S\'{e}rsic component and a PSF component to simulate the host galaxy and the unresolved QSO emission.  The S\'{e}rsic component is convolved with the PSF.

Each object was fit by three different one-S\'{e}rsic component models: a S\'{e}rsic profile with unconstrained S\'{e}rsic index (variable n = free), a de Vaucouleurs profile (fixed n = 4), and an exponential disk profile (fixed n = 1). In all cases, the centroids, the effective radius (r$_e$), the axis ratio (b/a), the position angle (PA), and the magnitude were left unconstrained.  Our initial visual assessment of the images determined that most of the systems have complex morphology that would not be well fit by classical bulge or disk models. However, this simple one-galaxy model analysis allows us to determine whether the system is bulge-dominated, disk-dominated, or exhibits an intermediate morphology. It also allows us to expose morphological disturbances, if any, from the residual maps produced by subtracting the model from the data.

\subsection{Two-S\'{e}rsic Component Models }
A more realistic galaxy model would include two S\'{e}rsic profiles. Any significant remaining residuals from the one-S\'{e}rsic component models may be merger-induced morphological anomalies, or may, in principle, indicate the presence of a second low-surface brightness galaxy component. To account for the possibility of a second galaxy component, in addition to the one-S\'{e}rsic models, we fit each object with models that included two S\'{e}rsic components and a PSF: one made of a classical disk and a bulge profile (n=1 + n=4), and another one in which both S\'{e}rsic indices were left unconstrained and free to vary (n-free + n-free).  The centroids, r$_e$, b/a, PA, and magnitude were left unconstrained. In many of the cases, the $\chi^2$ statistics indicated a better fit, which is not surprising given the large number of free parameters. However, a close examination of the final models for each object revealed that at least one of the fitted components was not physically meaningful. Whenever the model parameter converged to extreme value, the models were considered unphysical. For instance, a large S\'{e}rsic index, producing a profile with very steep core and extended wings, compensates for a mismatch between the PSF model and the nucleus emission, extended faint surface brightness, or sky mismatch. Very large effective radius attempts to fit extended low-brightness structure or indicates a mismatch with the sky value. Upon close inspection, none of the second S\'{e}rsic components were deemed physical and we conclude that the quality of the data does not allow us to determine whether a second low surface brightness or very compact galaxy component is present. The failure of the two-S\'{e}rsic models may be due to the large fraction of objects with apparent merger signatures in the images even prior to PSF subtraction.  

In all cases, the morphological analysis with GALFIT was independently performed by two members of the group to verify the validity of the results: MSL and WB modeled the F475W images and MSL and KR modeled the F125W ones.  The results of the surface brightness modeling below discuss only the one-S\'{e}rsic component models.

\section{Results}\label{results}

\begin{figure}
\centering
\begin{tabular}{cc}
\epsfig{file=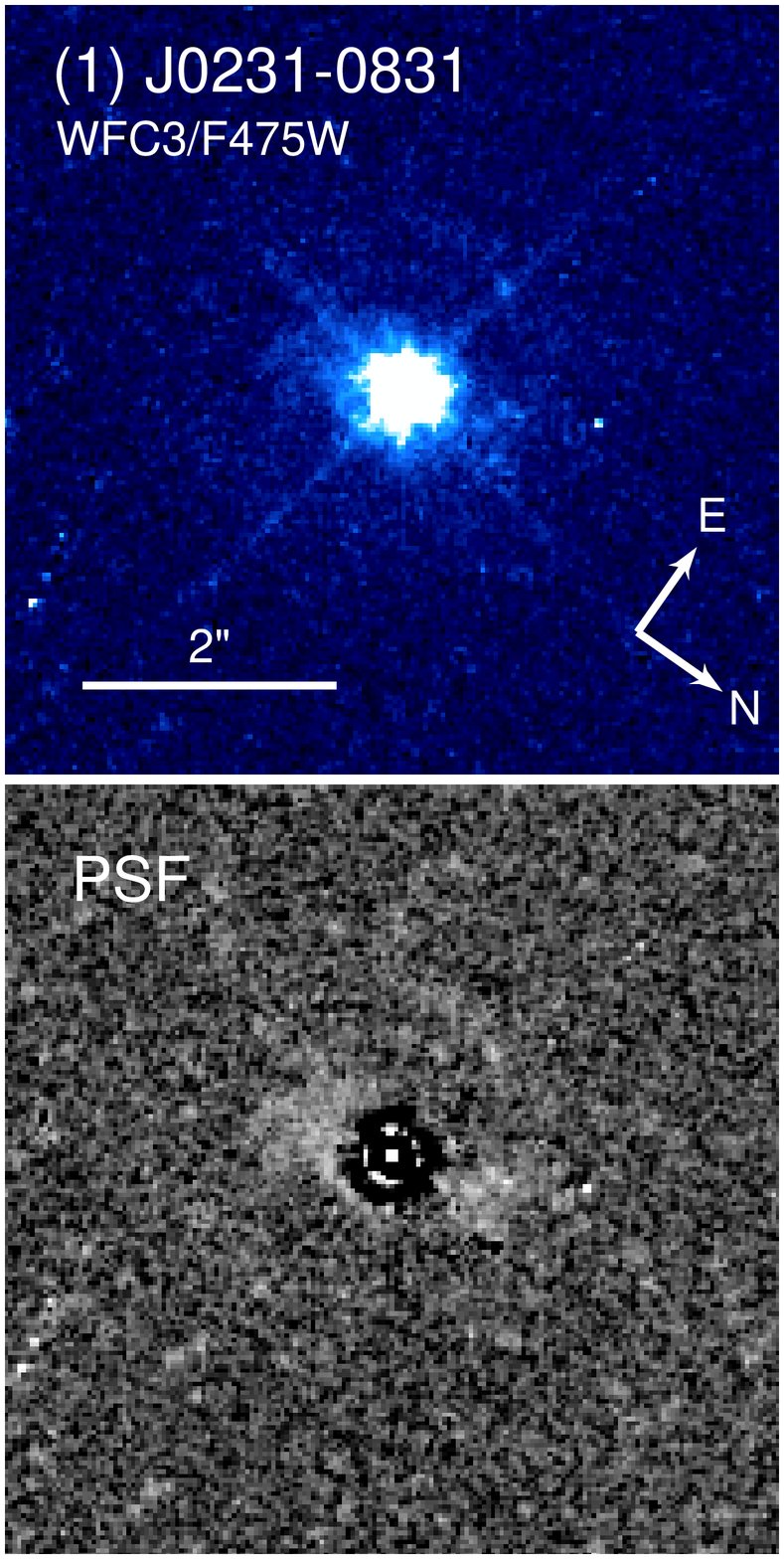,width=0.233\linewidth,clip=} & \epsfig{file=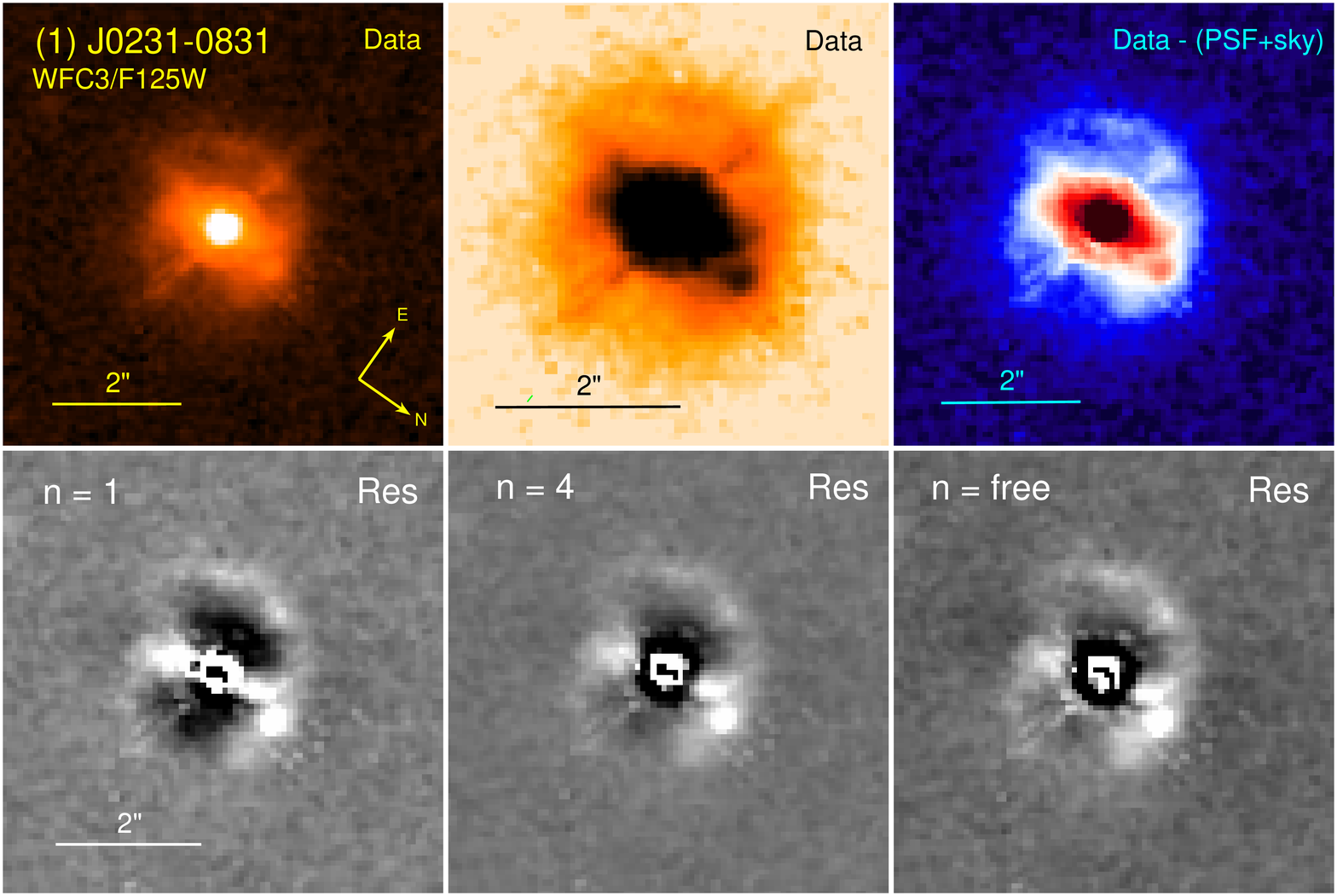,width=0.70\linewidth,clip=} \\
\epsfig{file=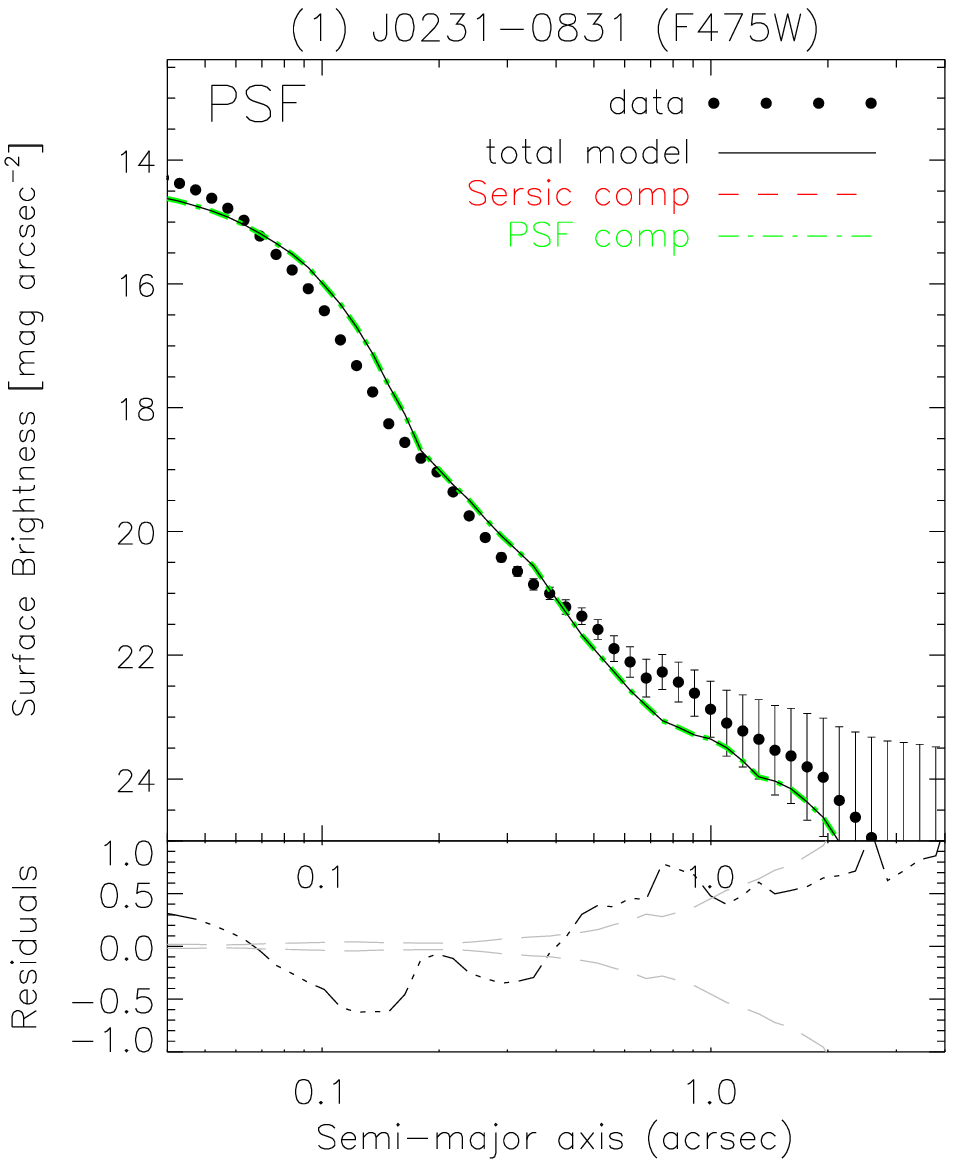,width=0.265\linewidth,clip=} &\epsfig{file=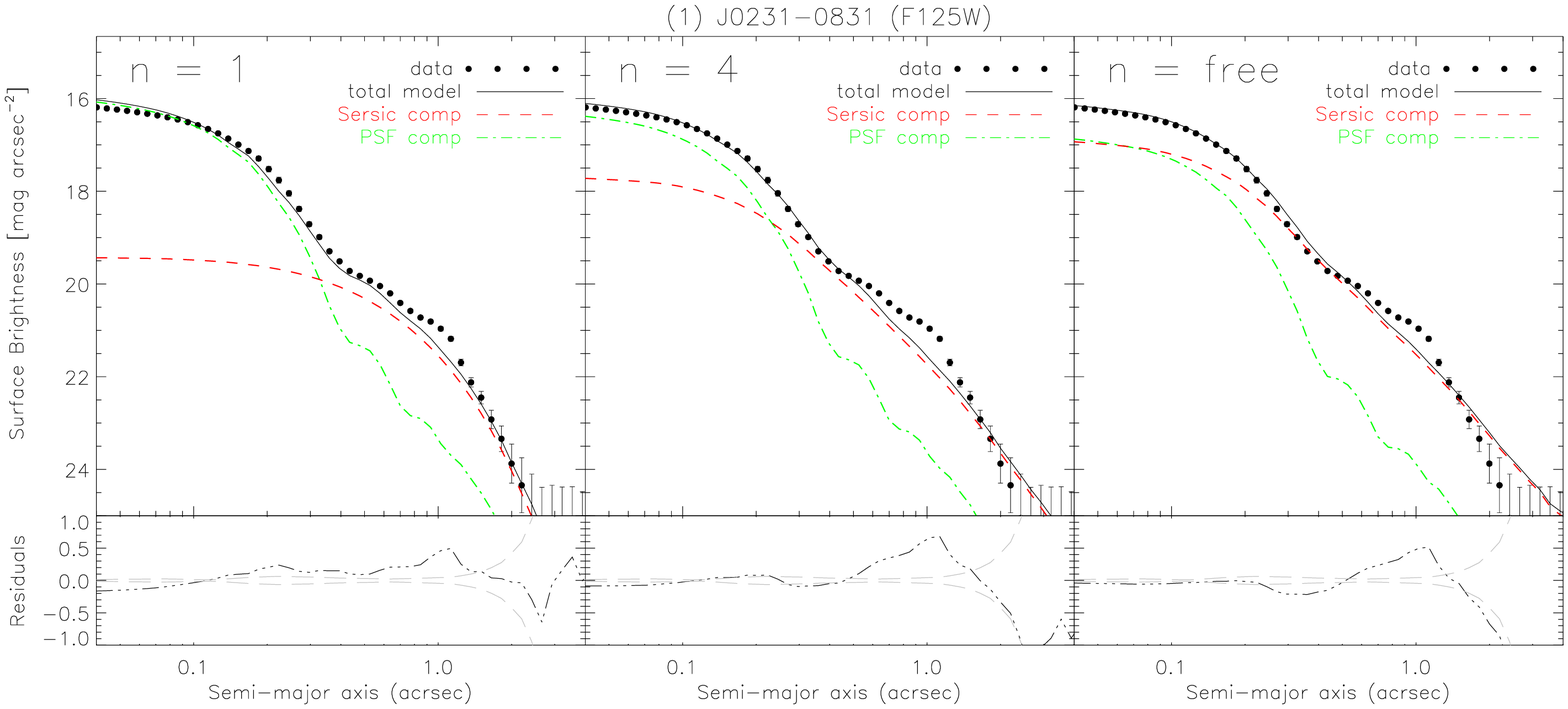,width=0.715\linewidth,clip=}
\end{tabular}
\caption{Object SDSS J0231-0831: HST/WFC3 images and residuals from the GALFIT modeling. Top row, in color, from left to right: F475W image prior to model subtraction (in blue, top left); F125W images prior to model subtraction (in heat, top second and third panel); F125W image after subtracting only the PSF+sky model to highlight the underlying host light (in blue-red, top fourth panel). Second row, in gray, shows the residual images after subtracting the best fit for the GALFIT model indicated in the upper left corner in white text: PSF indicates a PSF + sky model; n = 1 is for a fixed exponential disk + PSF + sky; n = 4 is for a fixed de Vaucouleur profile + PSF + sky; and n = free is for unconstrained S\'{e}rsic index profile + PSF + sky. The second row images have the same scale as the top row image with the object name for each filter. Third row shows radial surface brightness profiles created with the {\it IRAF} task {\it ellipse} of each model; the model is indicated  in the top left corner of each plot: the data is plotted with black dots, the total model with a black solid line, as well as individual model components: the PSF component with a green dash-dot line and the S\'{e}rsic profile with a red dashed line. The residuals plot on the bottom is the difference between the data and the total model as a function of semi-major axis.}
\label{fig:images1}
\end{figure}

\begin{figure}
\centering
\begin{tabular}{cc}
\epsfig{file=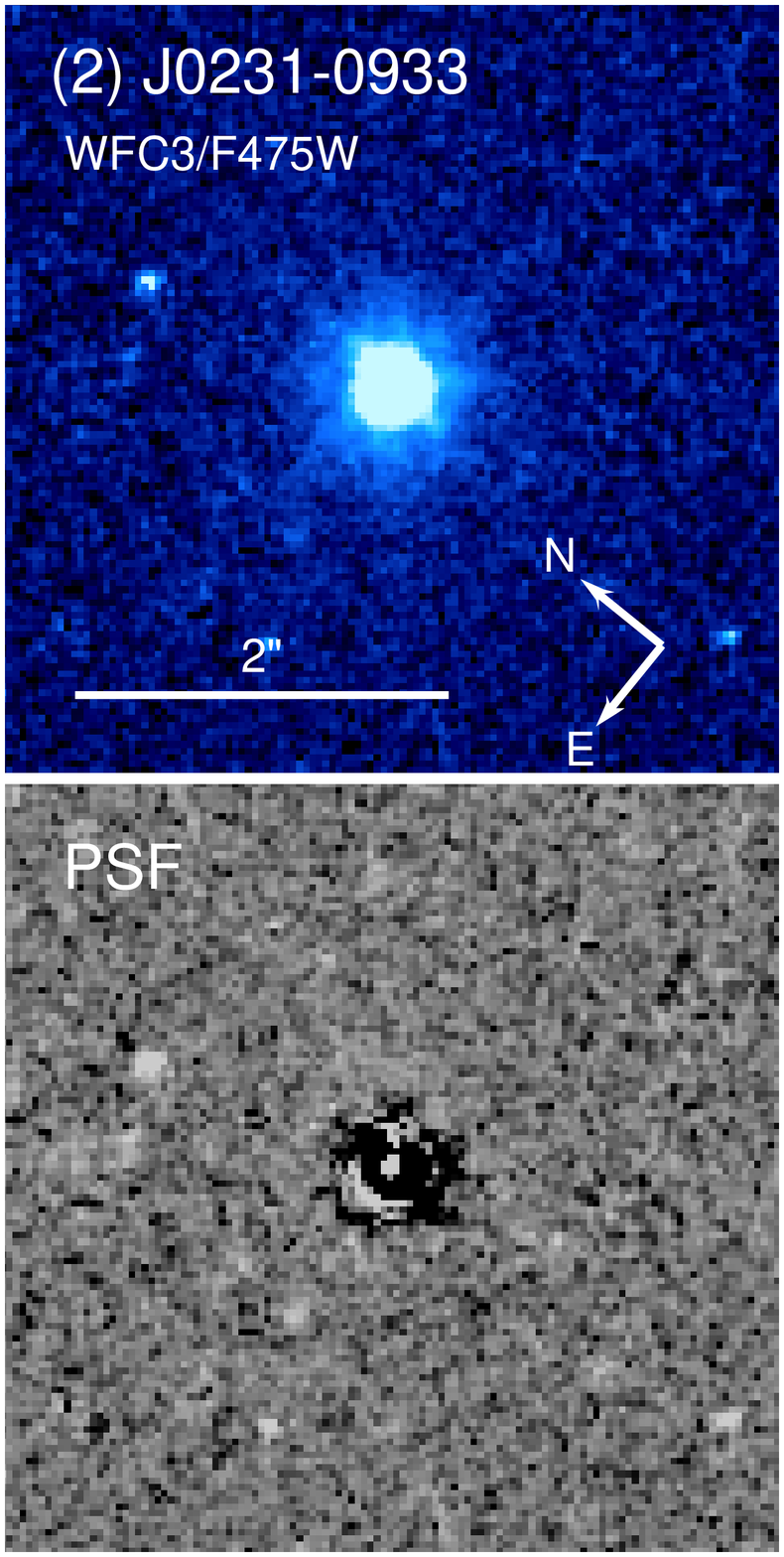,width=0.233\linewidth,clip=} & \epsfig{file=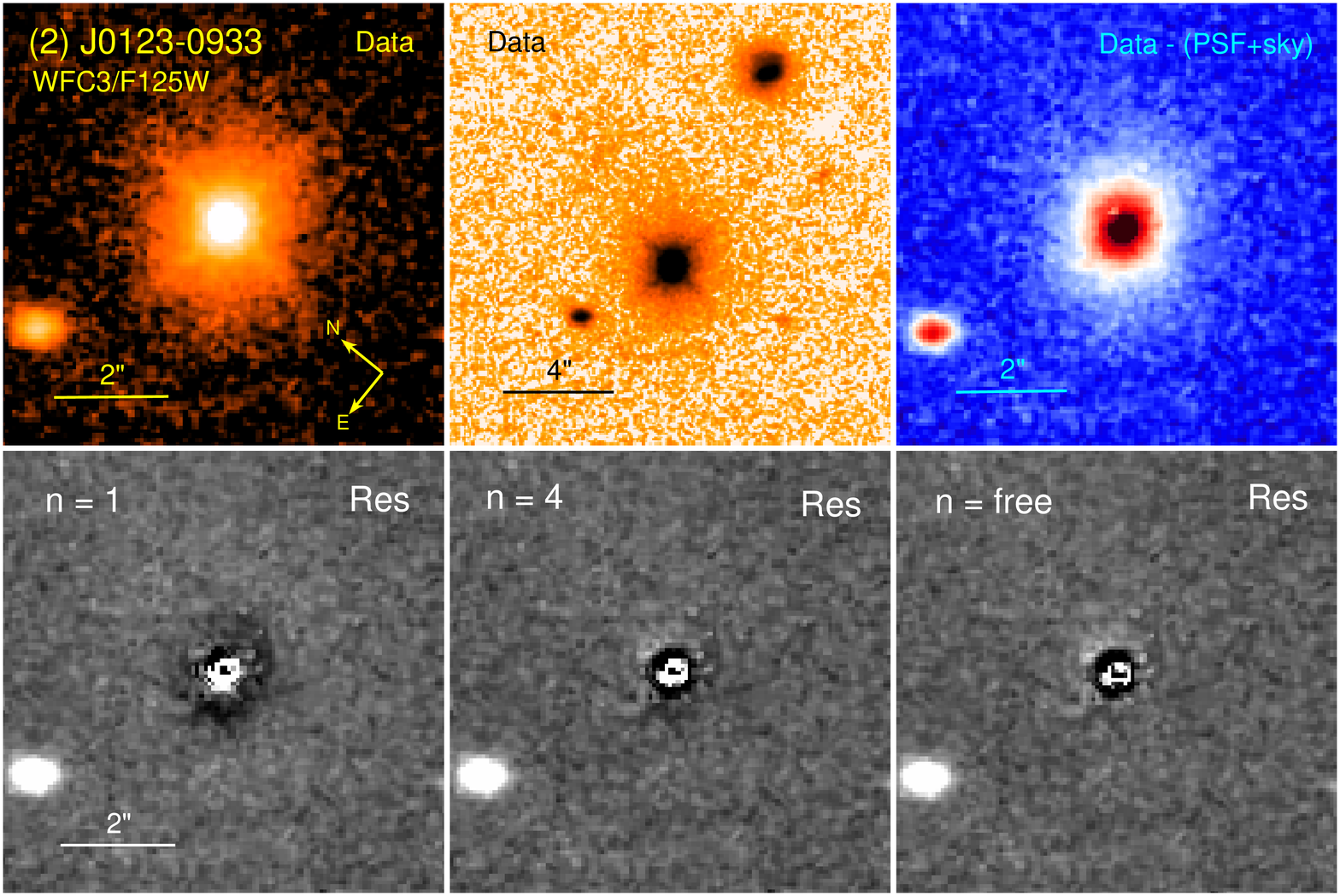,width=0.70\linewidth,clip=} \\
\epsfig{file=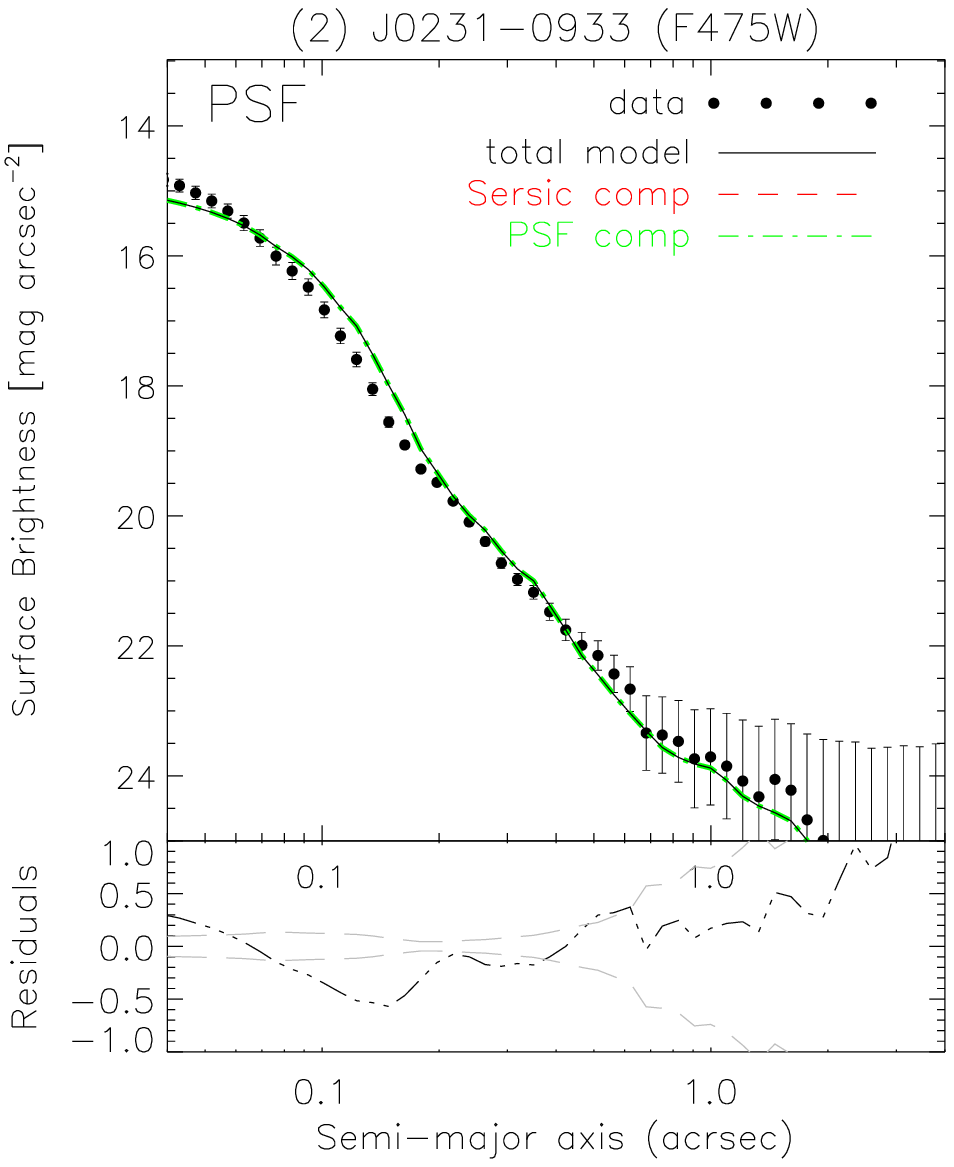,width=0.265\linewidth,clip=} &\epsfig{file=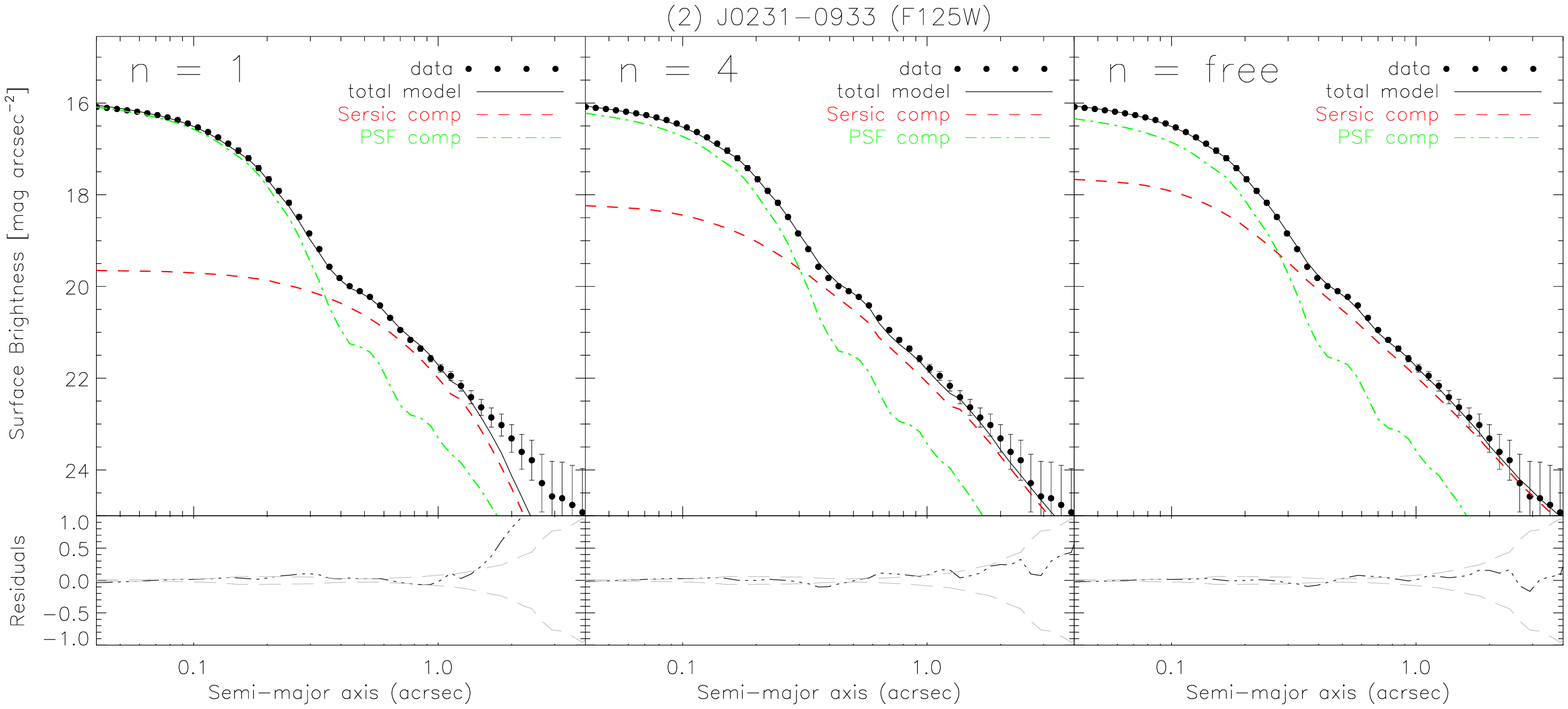,width=0.715\linewidth,clip=}
\end{tabular}
\caption{Object SDSS J0231-0933: Caption, as in Fig. \ref{fig:images1}.}
\label{fig:images2}
\end{figure}

\begin{figure}
\centering
\begin{tabular}{cc}
\epsfig{file=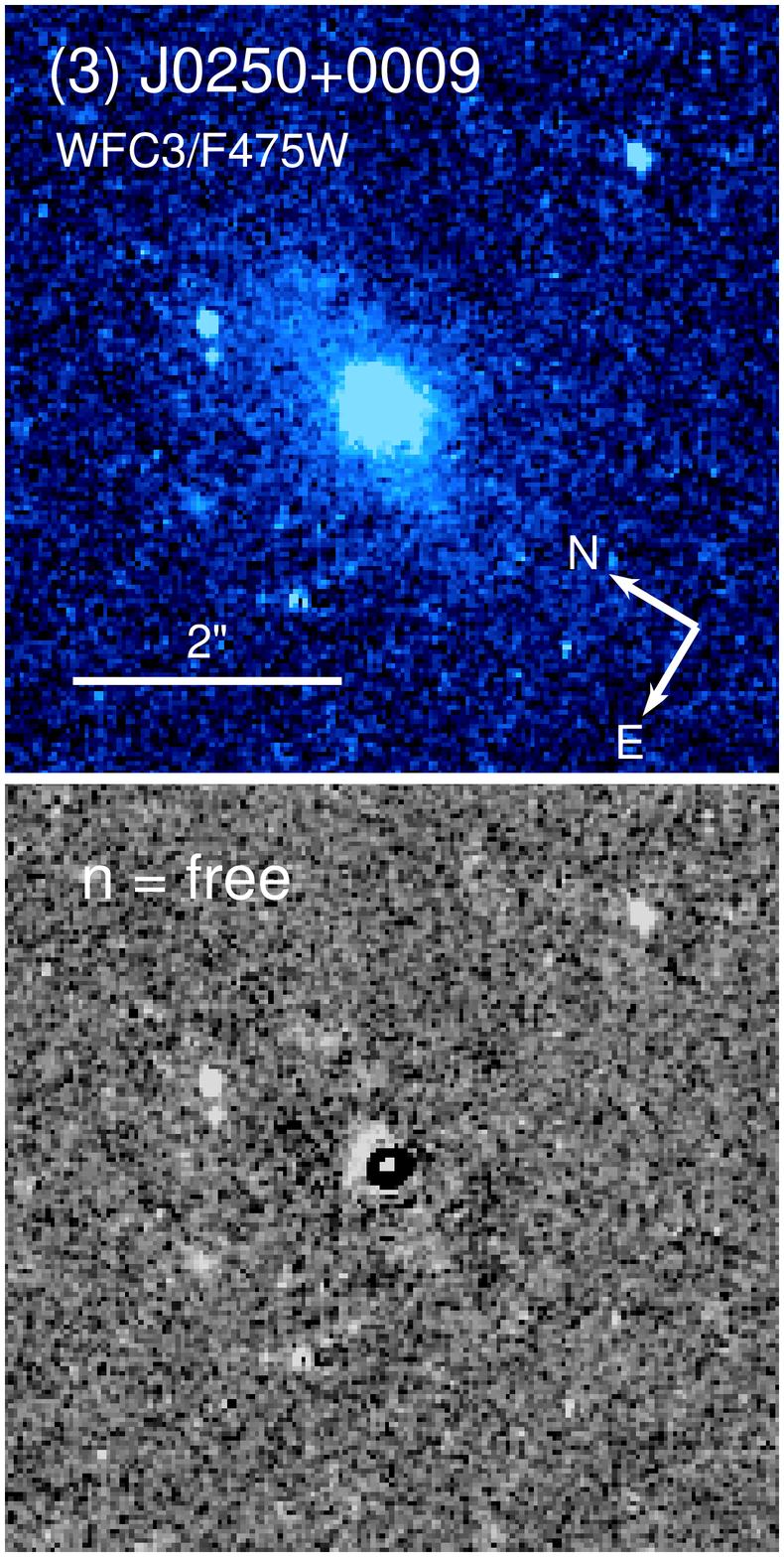,width=0.233\linewidth,clip=} & \epsfig{file=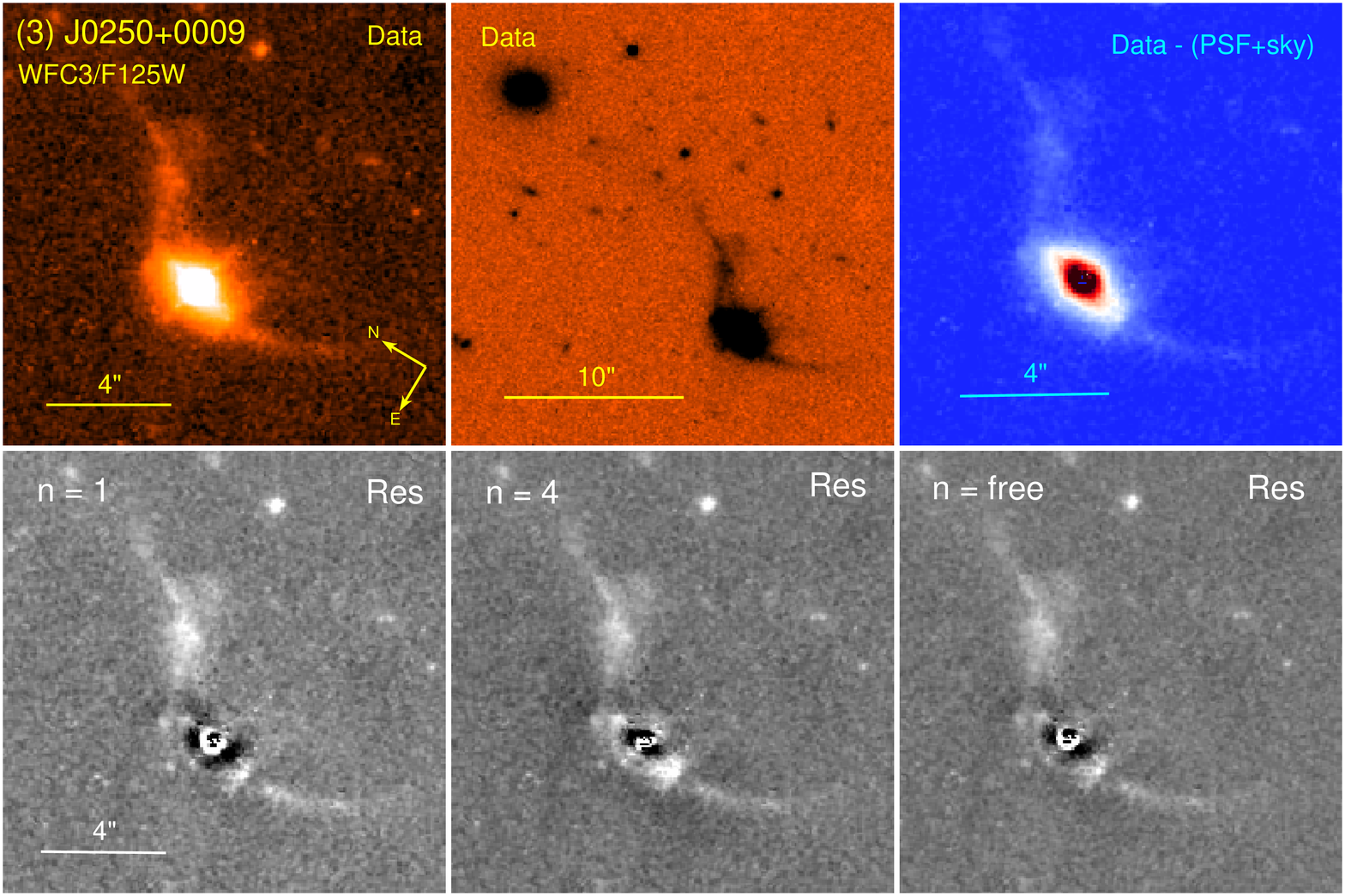,width=0.70\linewidth,clip=} \\
\epsfig{file=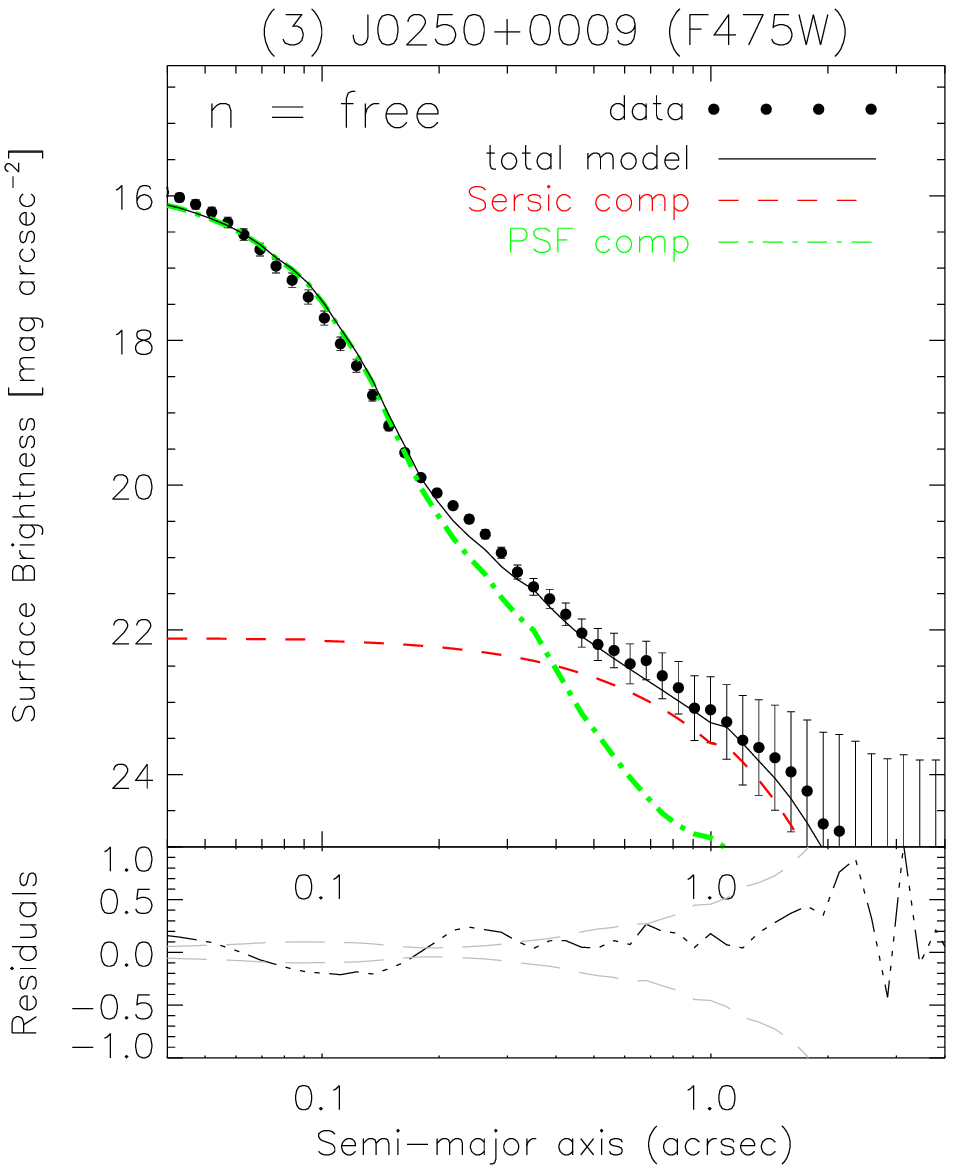,width=0.265\linewidth,clip=} &\epsfig{file=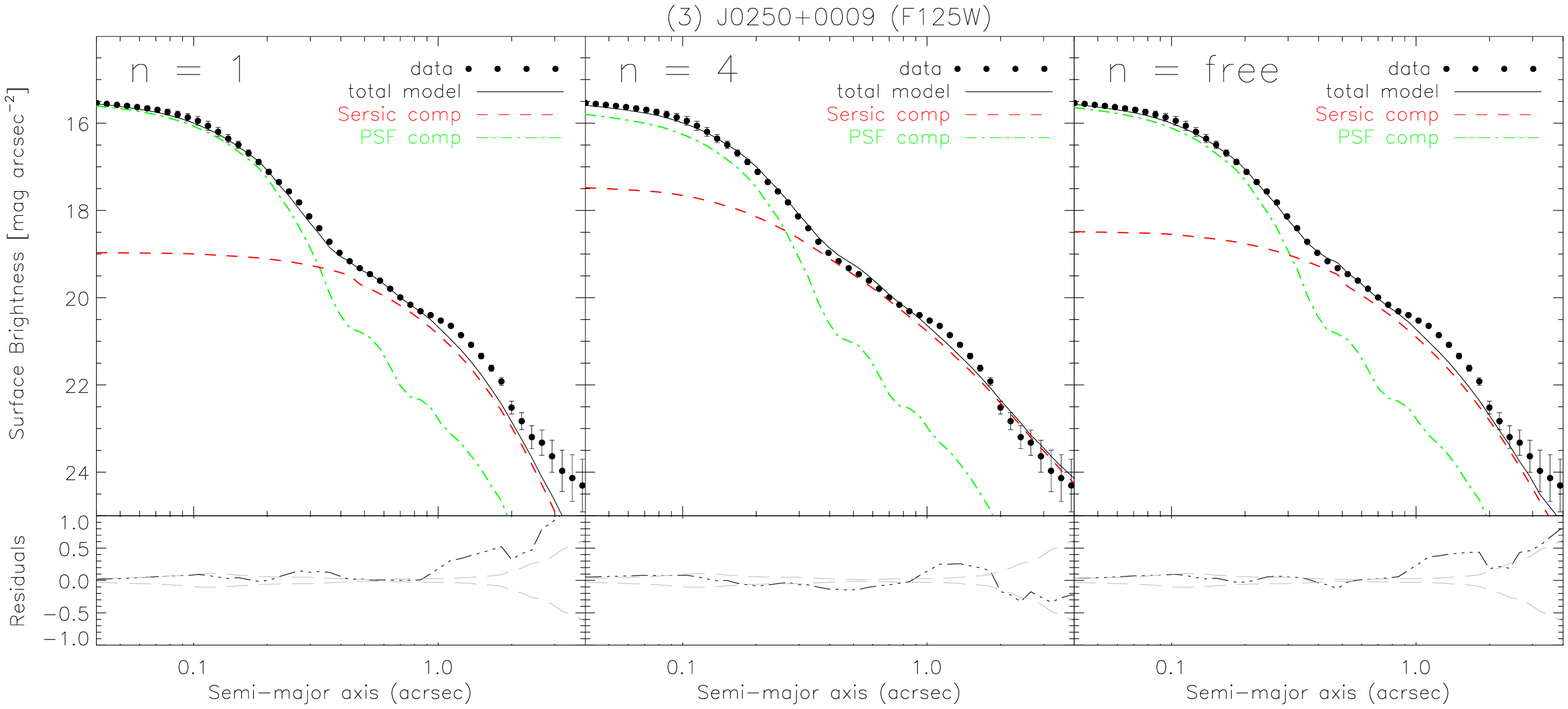,width=0.715\linewidth,clip=}
\end{tabular}
\caption{Object SDSS J0250+0009. Caption, as in Fig. \ref{fig:images1}.}
\label{fig:images3}
\end{figure}

\begin{figure}
\centering
\begin{tabular}{cc}
\epsfig{file=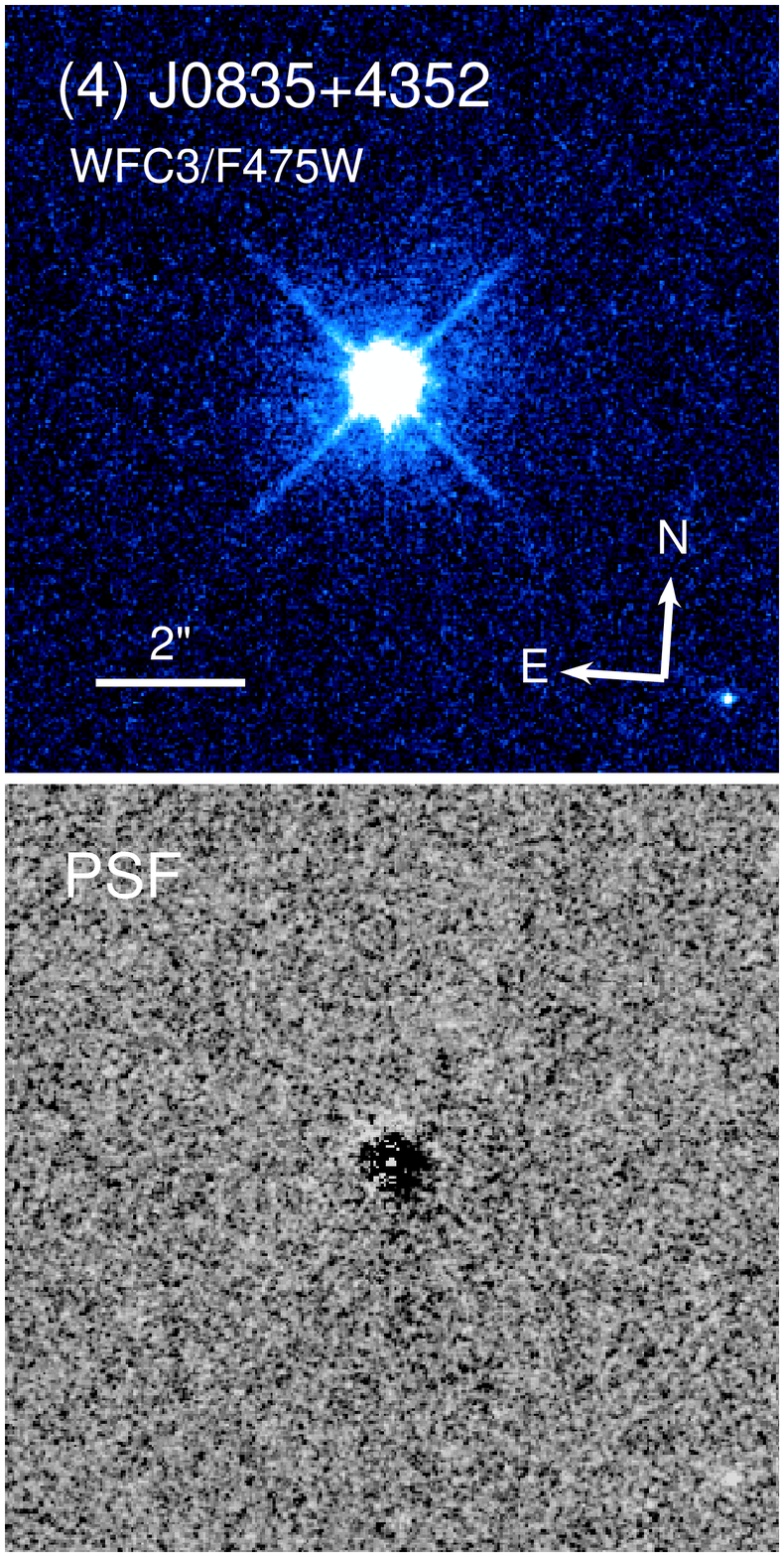,width=0.233\linewidth,clip=} & \epsfig{file=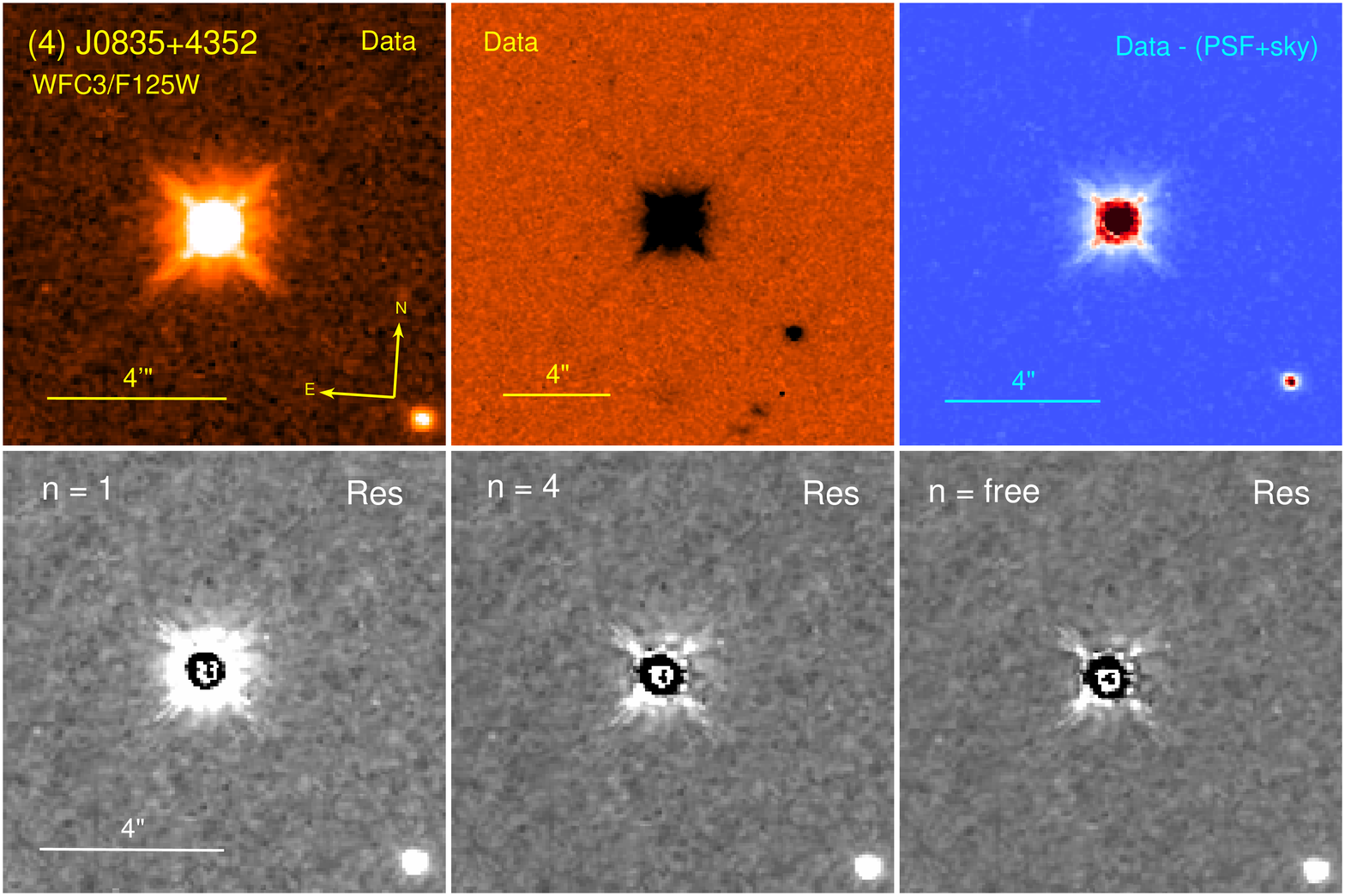,width=0.70\linewidth,clip=} \\
\epsfig{file=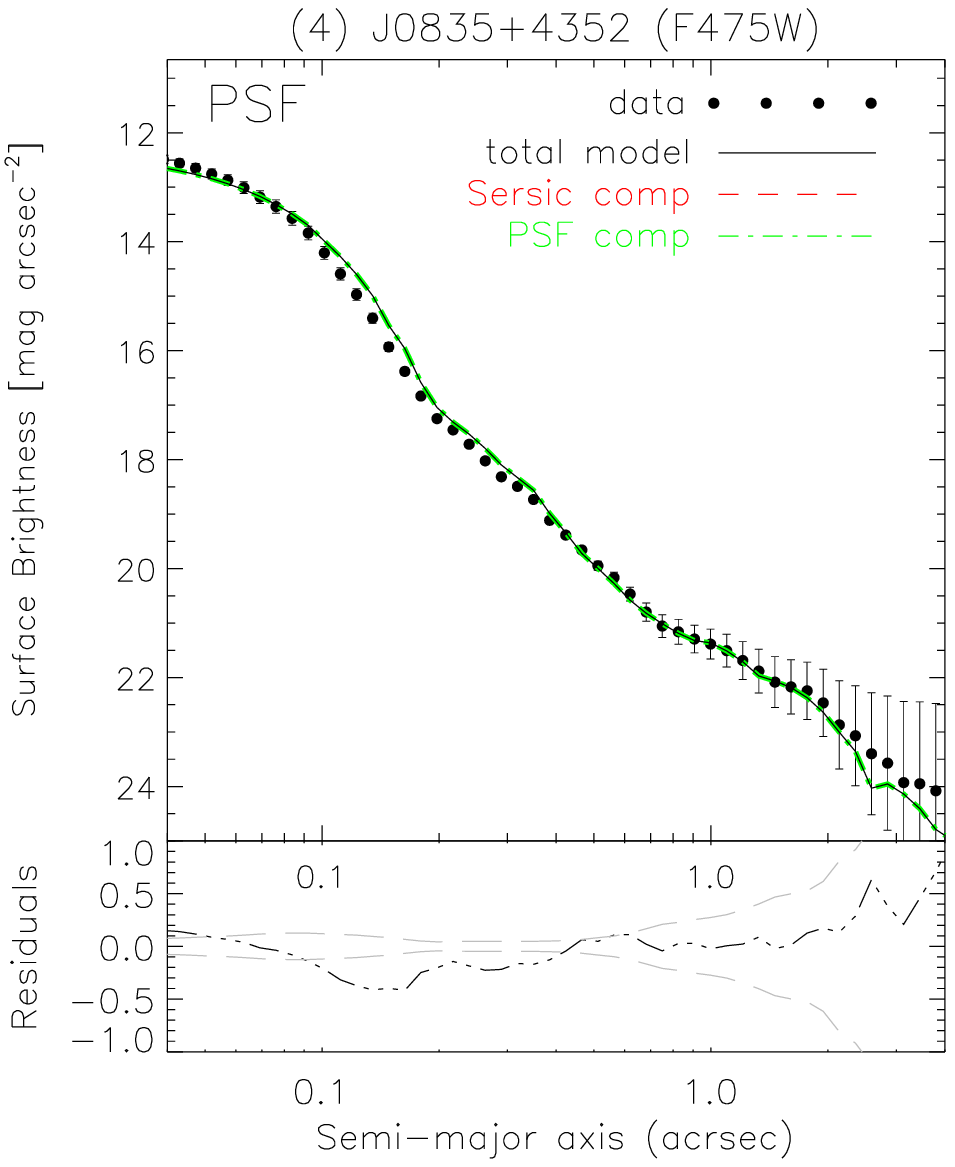,width=0.265\linewidth,clip=} &\epsfig{file=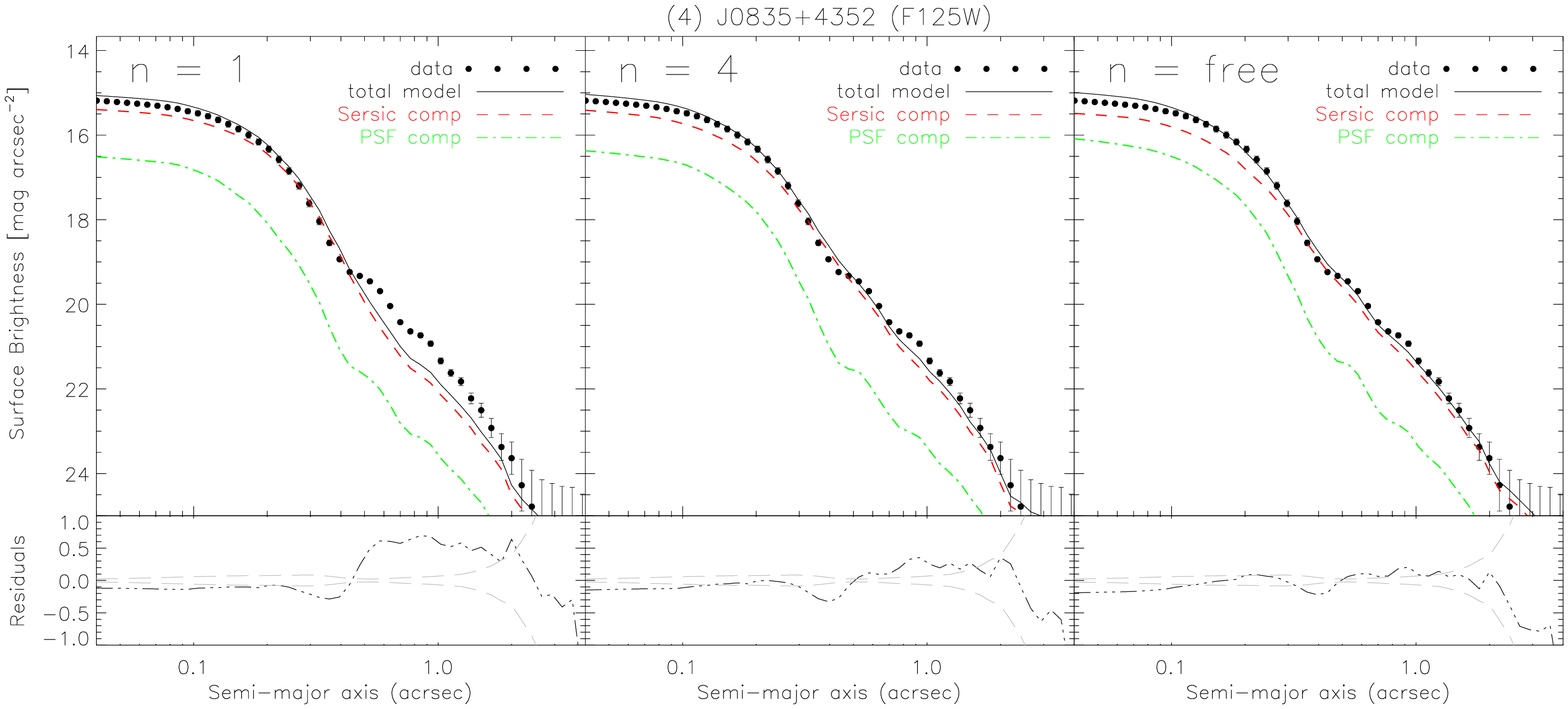,width=0.715\linewidth,clip=}
\end{tabular}
\caption{Object SDSS J0835+4352. Caption, as in Fig. \ref{fig:images1}.}
\label{fig:images4}
\end{figure}

\begin{figure}
\centering
\begin{tabular}{cc}
\epsfig{file=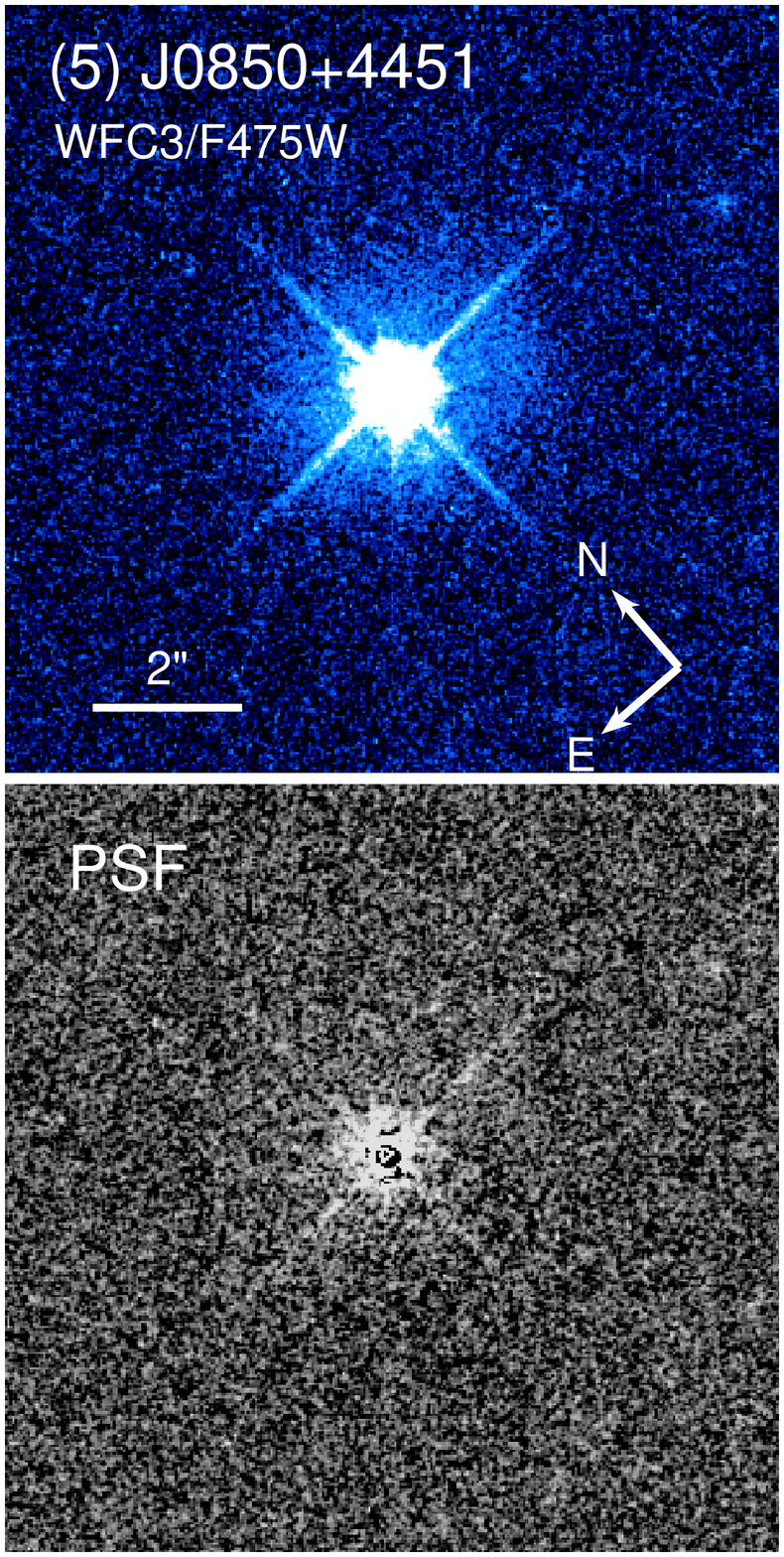,width=0.233\linewidth,clip=} & \epsfig{file=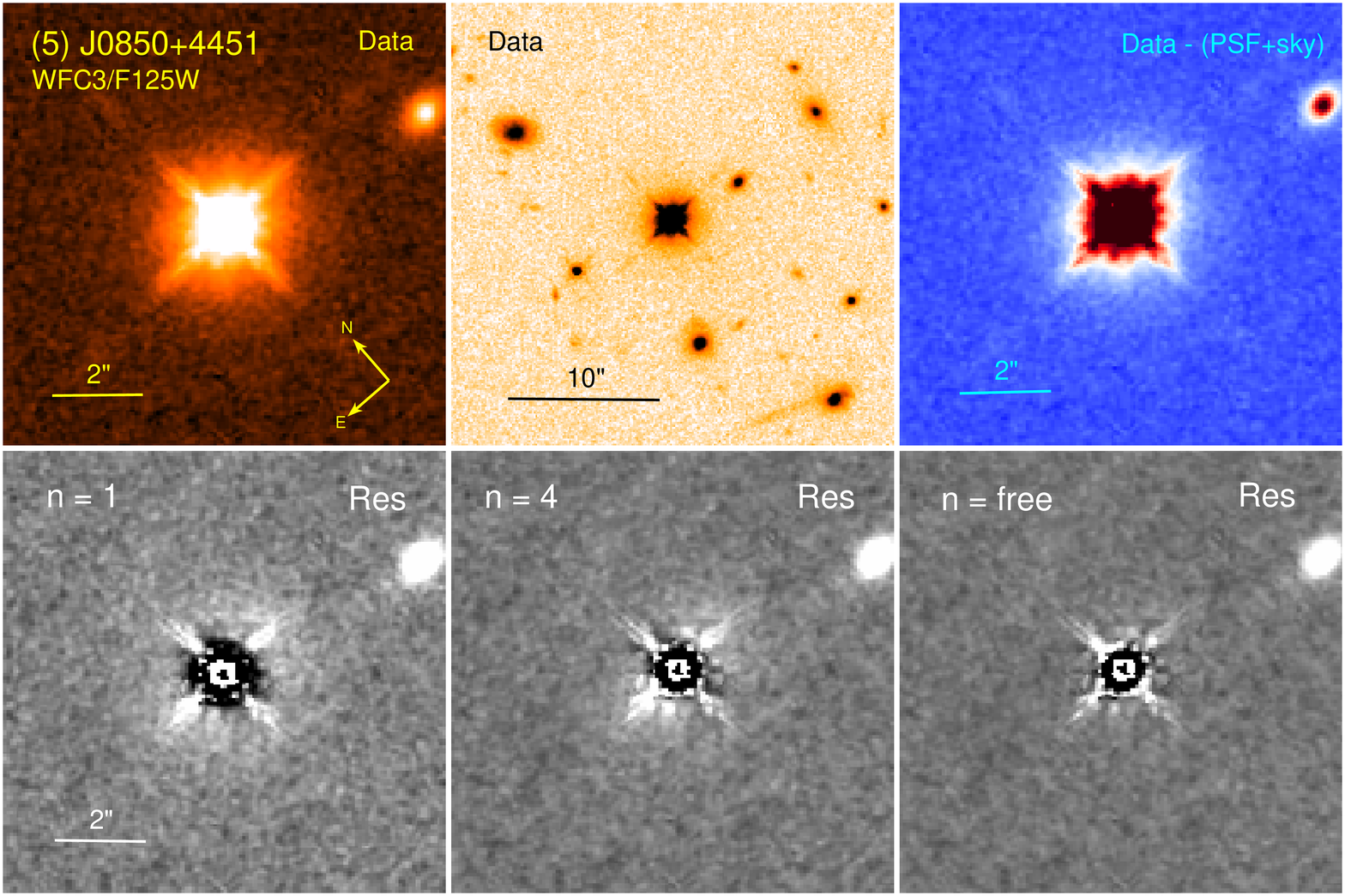,width=0.70\linewidth,clip=} \\
\epsfig{file=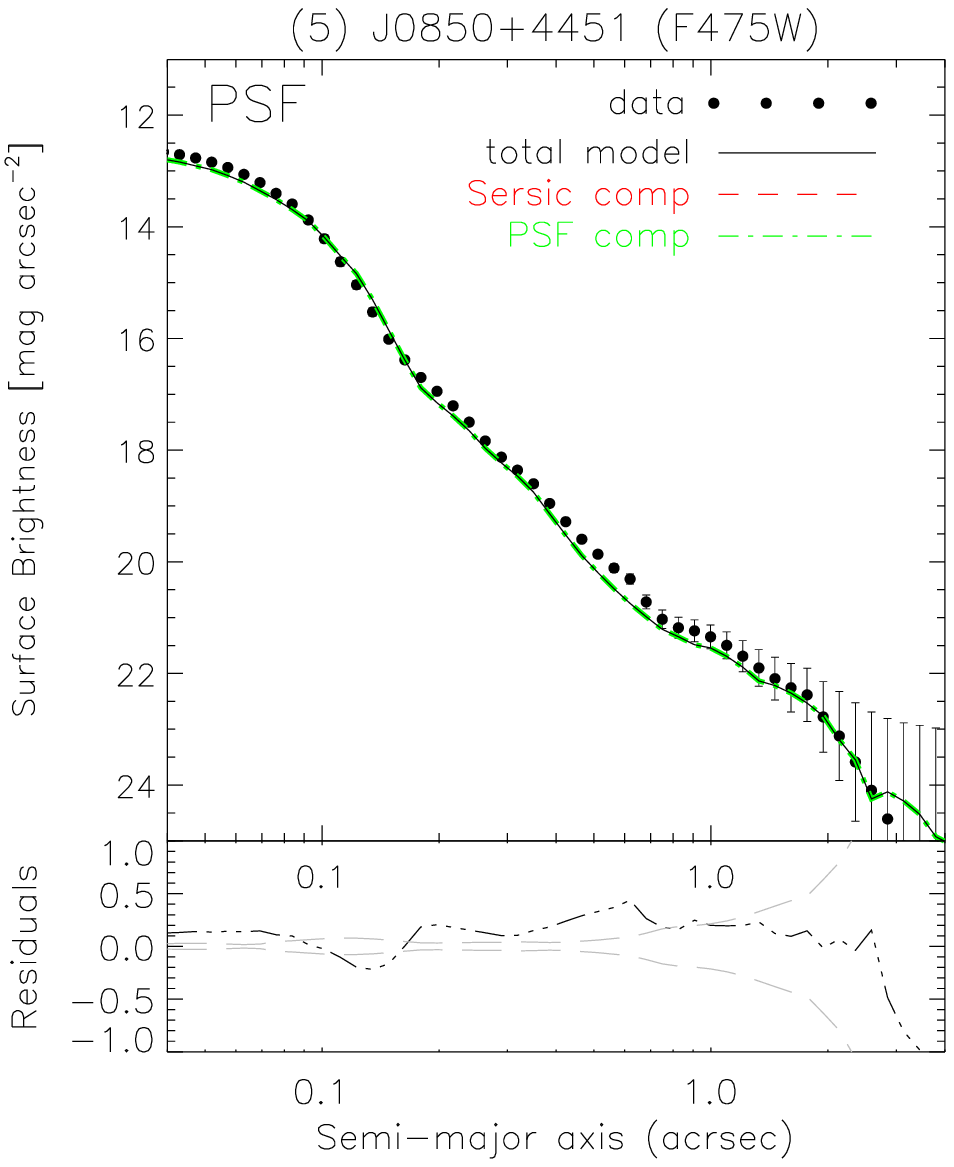,width=0.265\linewidth,clip=} &\epsfig{file=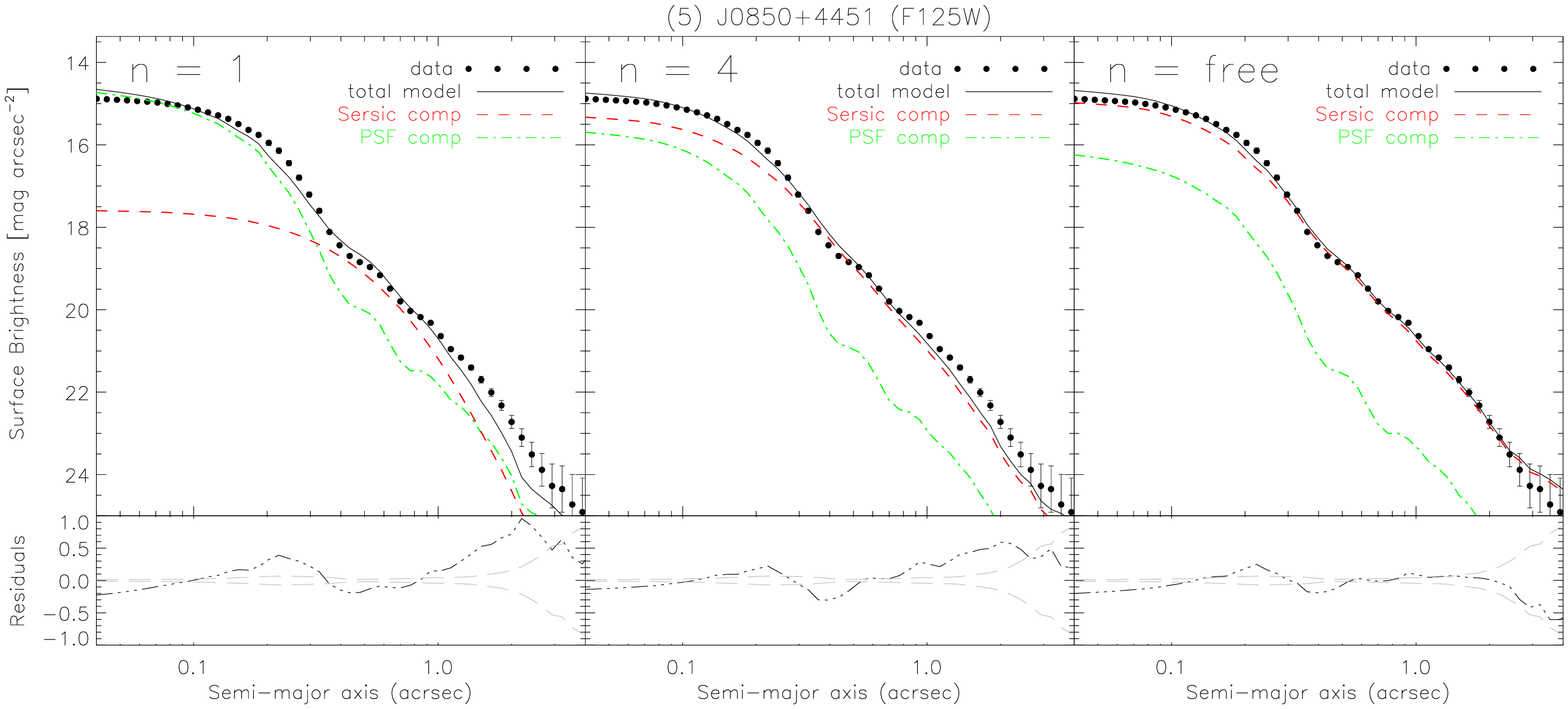,width=0.715\linewidth,clip=}
\end{tabular}
\caption{Object SDSS J0850+4451. Caption, as in Fig. \ref{fig:images1}. The distorted galaxy $\sim$8$\farcs$4SE of the LoBAL is located at the same photometric redshift within the uncertainties (z$\sim$0.55$\pm$0.15).} 
\label{fig:images5}
\end{figure}

\begin{figure}
\centering
\begin{tabular}{cc}
\epsfig{file=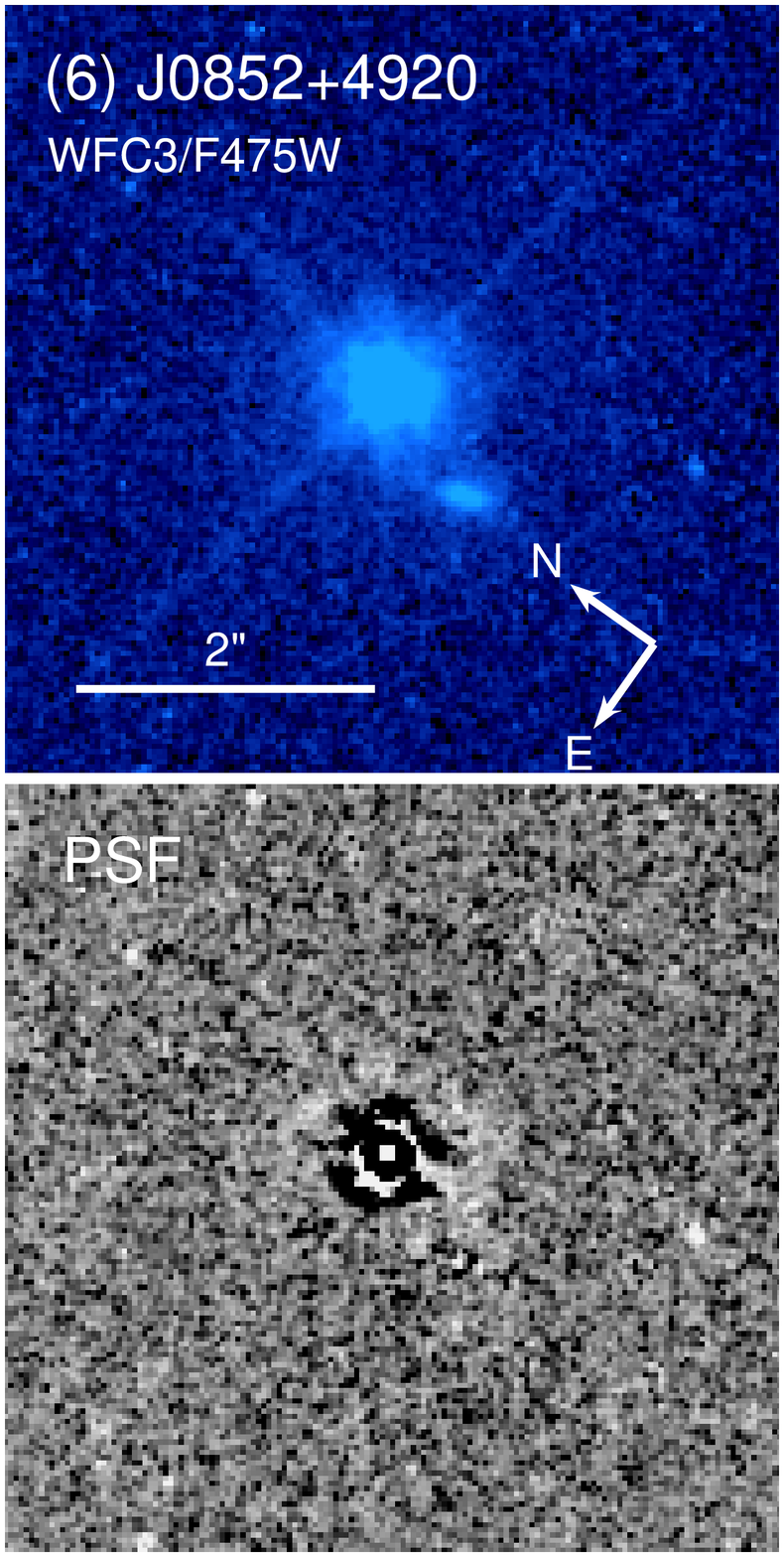,width=0.233\linewidth,clip=} & \epsfig{file=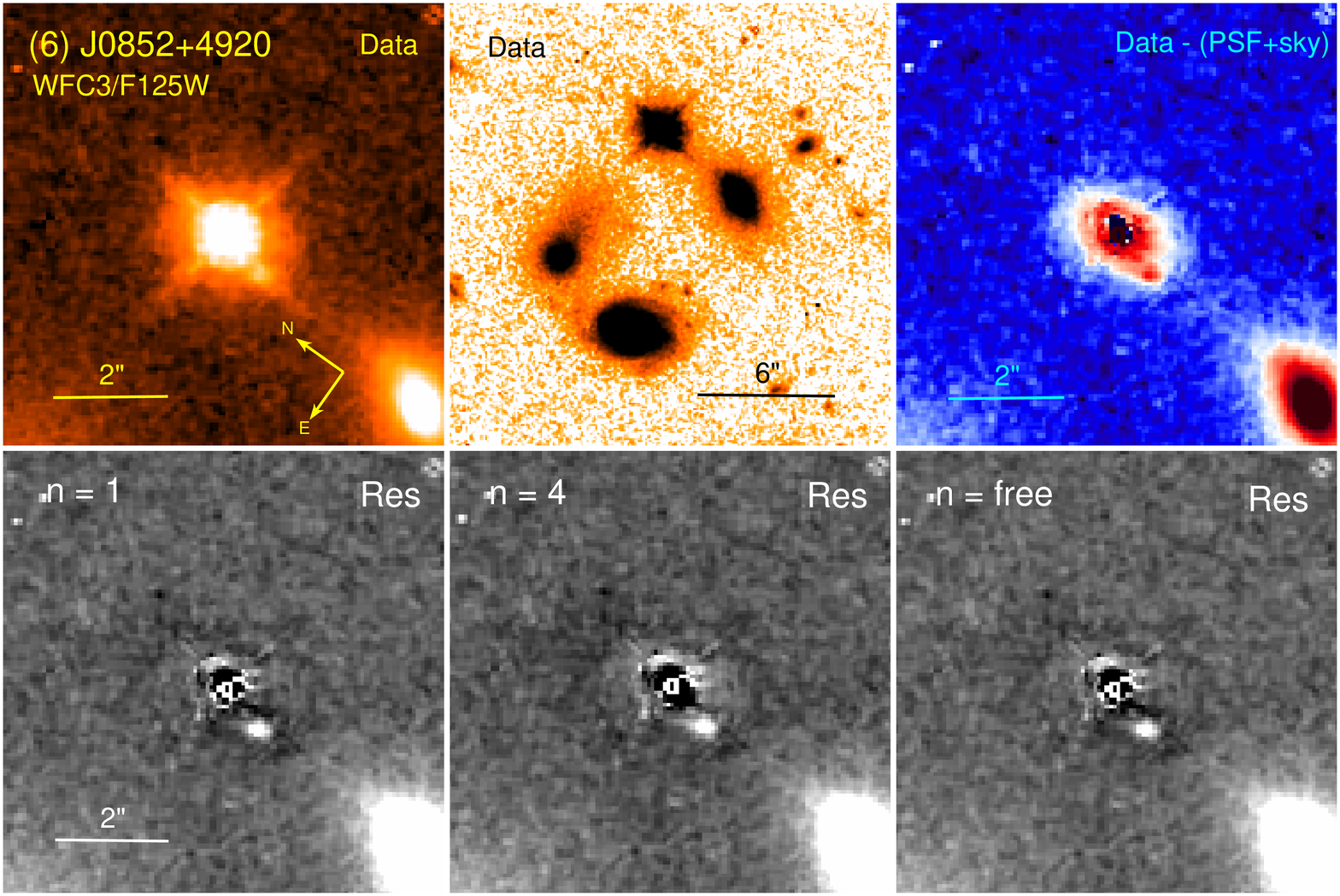,width=0.70\linewidth,clip=} \\
\epsfig{file=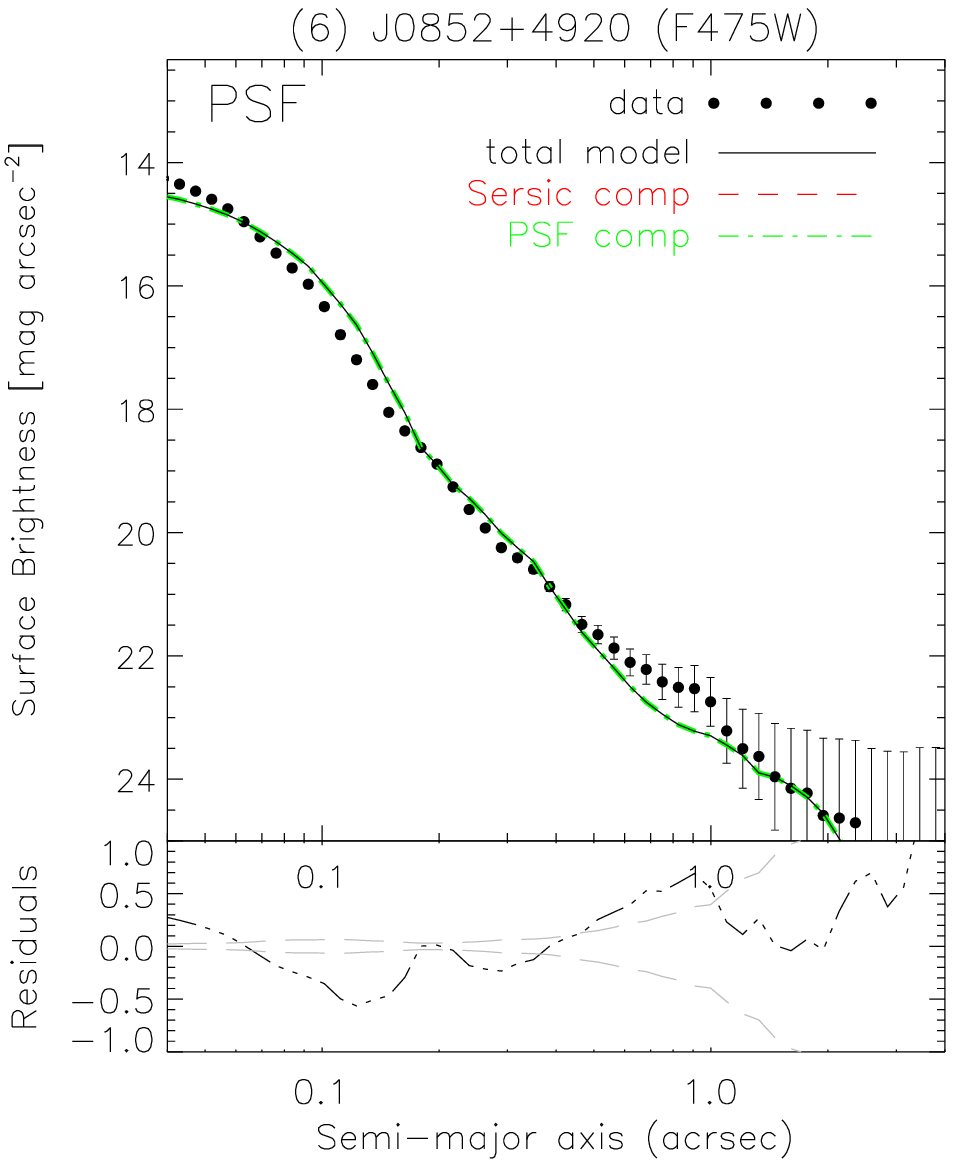,width=0.265\linewidth,clip=} &\epsfig{file=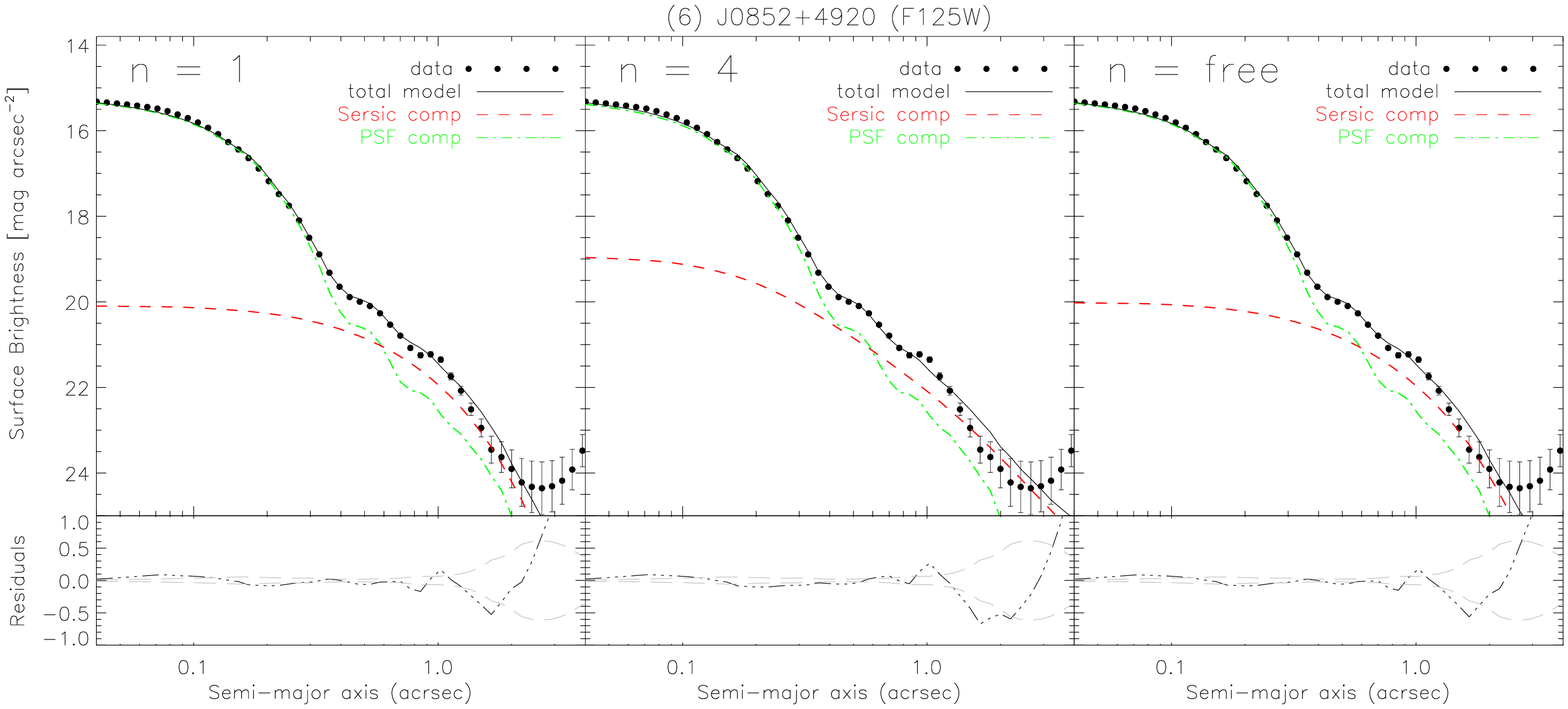,width=0.715\linewidth,clip=}
\end{tabular}
\caption{Object SDSS J0852+4920. Caption, as in Fig. \ref{fig:images1}. There is a red galaxy $\sim$4$\farcs$4S of the LoBAL at the same photometric redshift (z$\sim$0.58$\pm$0.04), with a bridge toward the QSO.} 
\label{fig:images6}
\end{figure}

\begin{figure}
\centering
\begin{tabular}{cc}
\epsfig{file=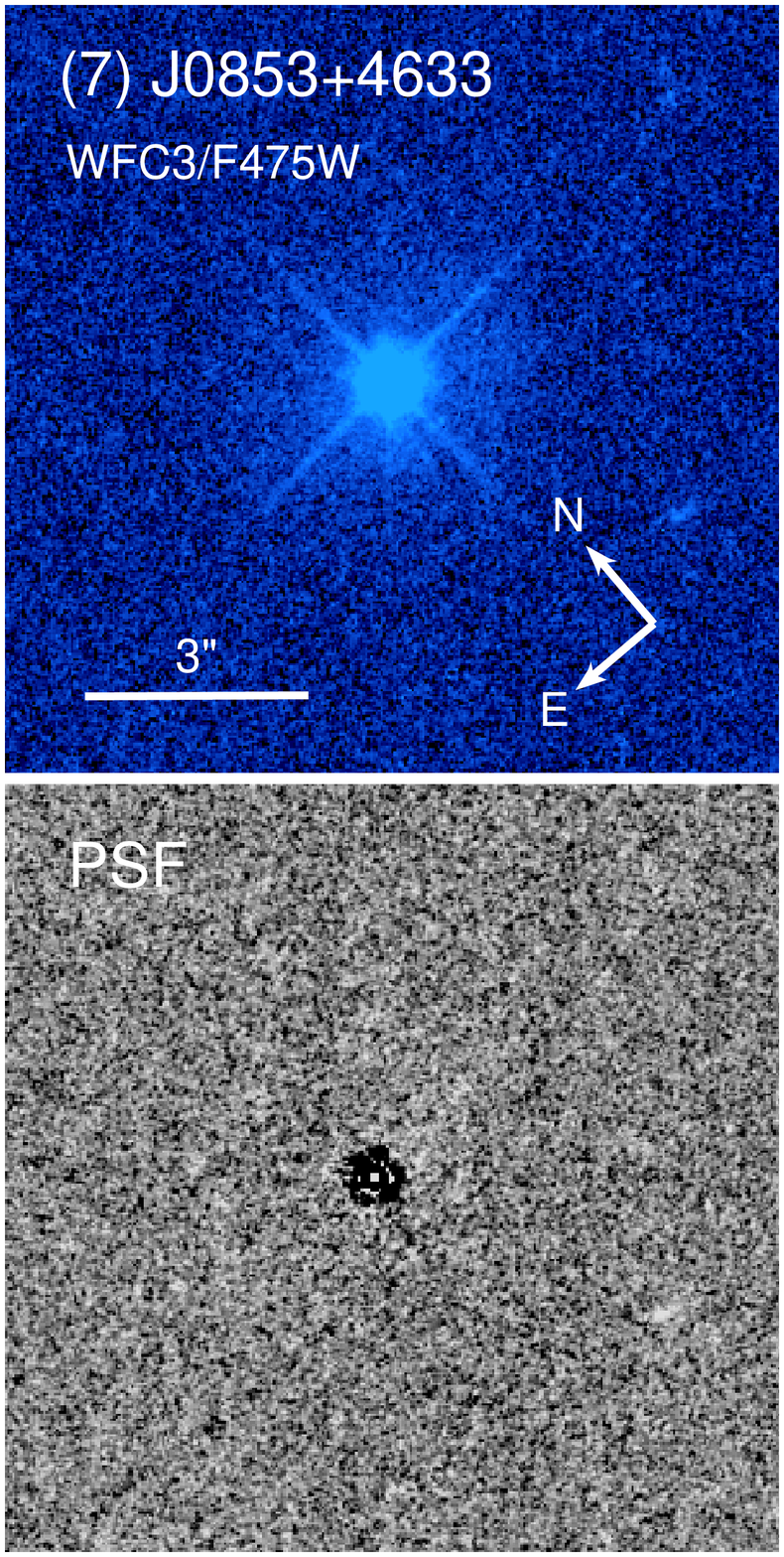,width=0.233\linewidth,clip=} & \epsfig{file=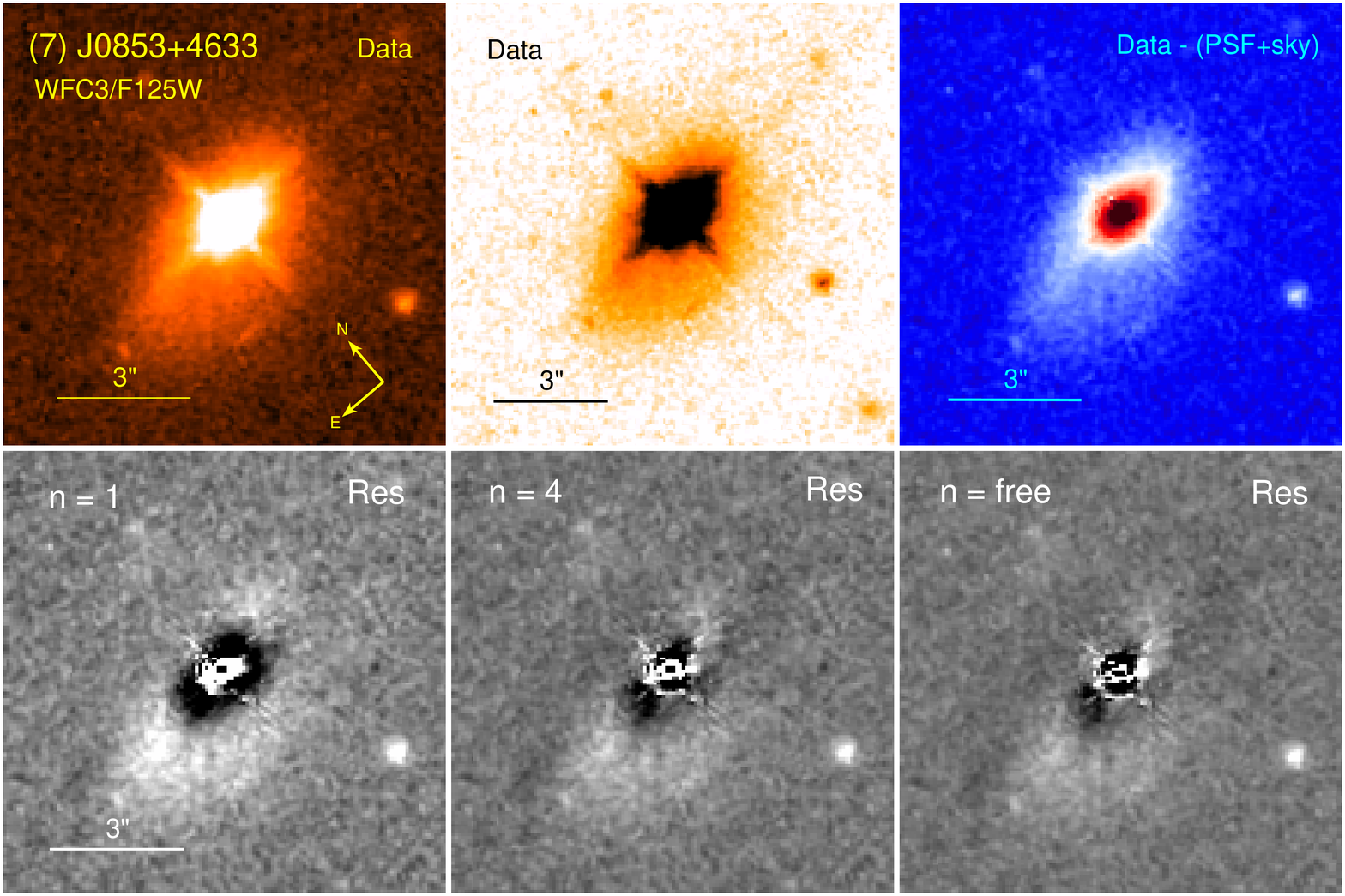,width=0.70\linewidth,clip=} \\
\epsfig{file=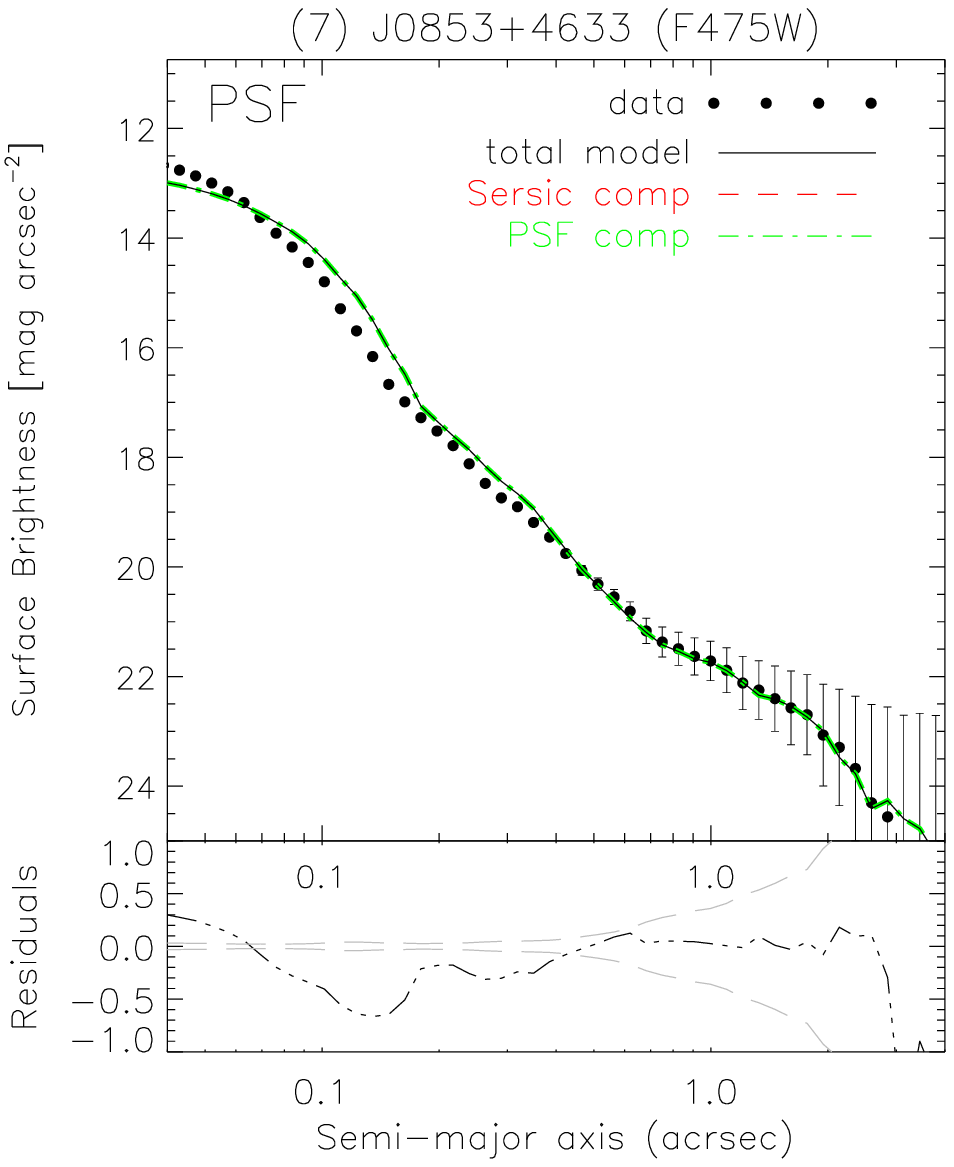,width=0.265\linewidth,clip=} &\epsfig{file=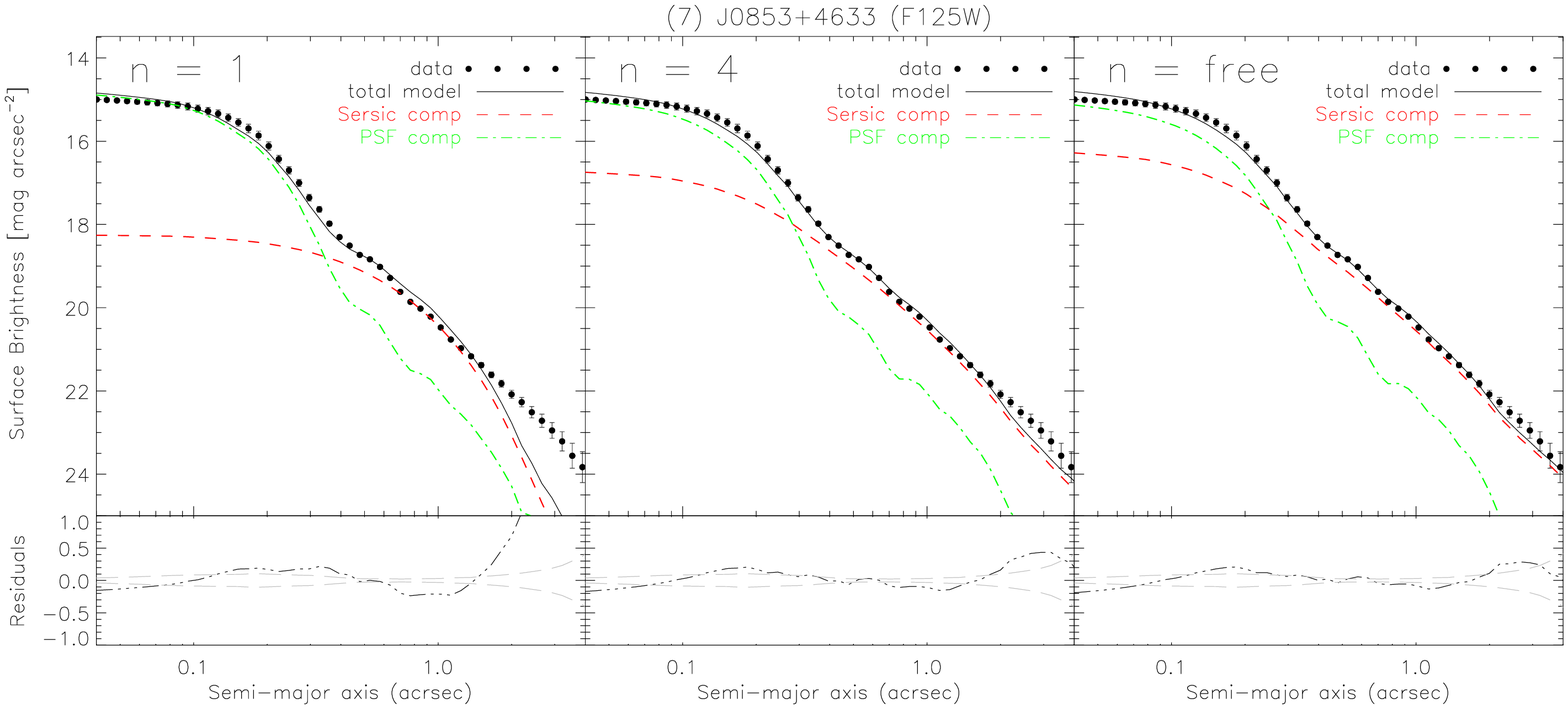,width=0.715\linewidth,clip=}
\end{tabular}
\caption{Object SDSS J0853+4633. Caption, as in Fig. \ref{fig:images1}.}
\label{fig:images7}
\end{figure}

\begin{figure}
\centering
\begin{tabular}{cc}
\epsfig{file=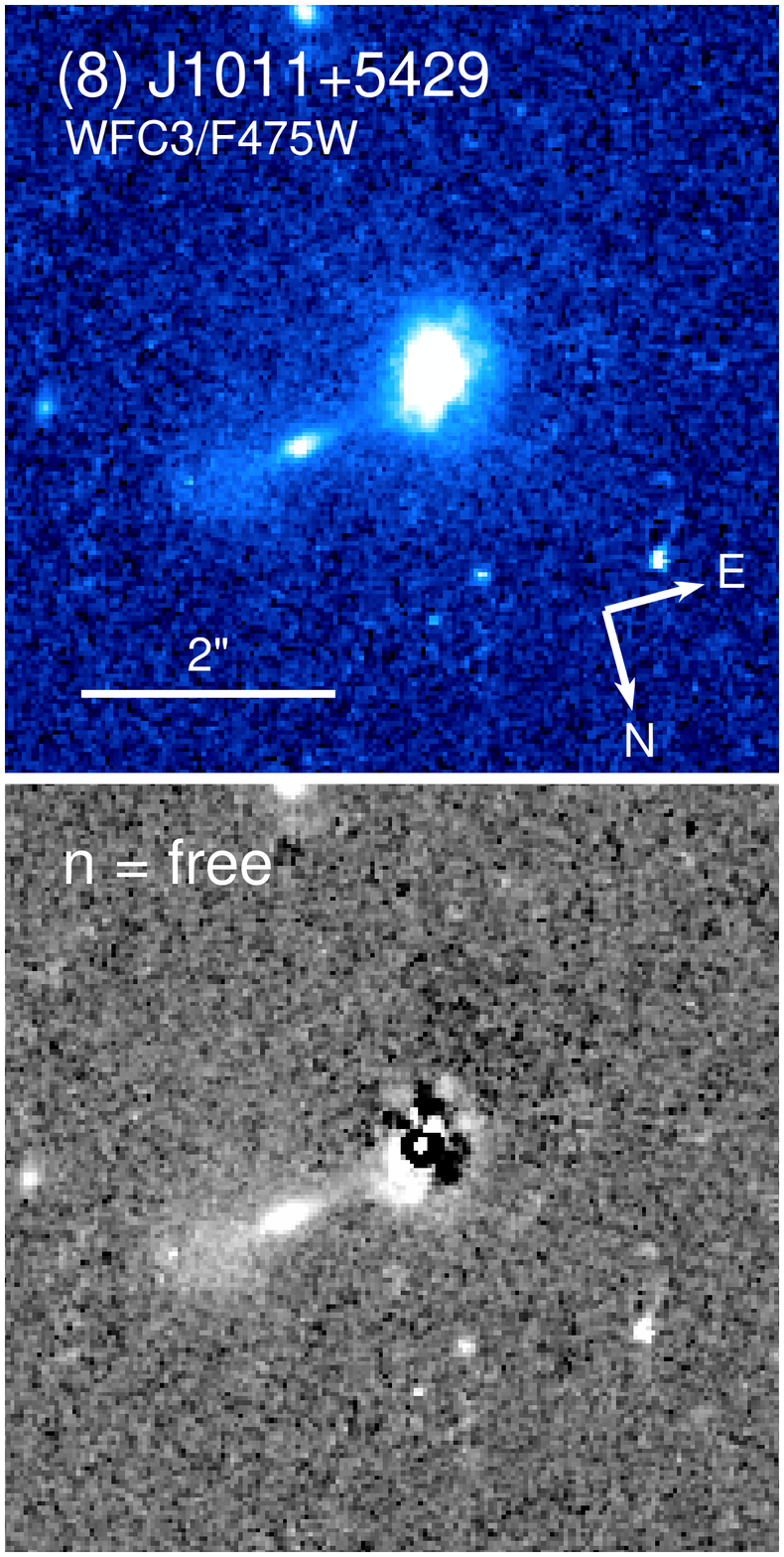,width=0.233\linewidth,clip=} & \epsfig{file=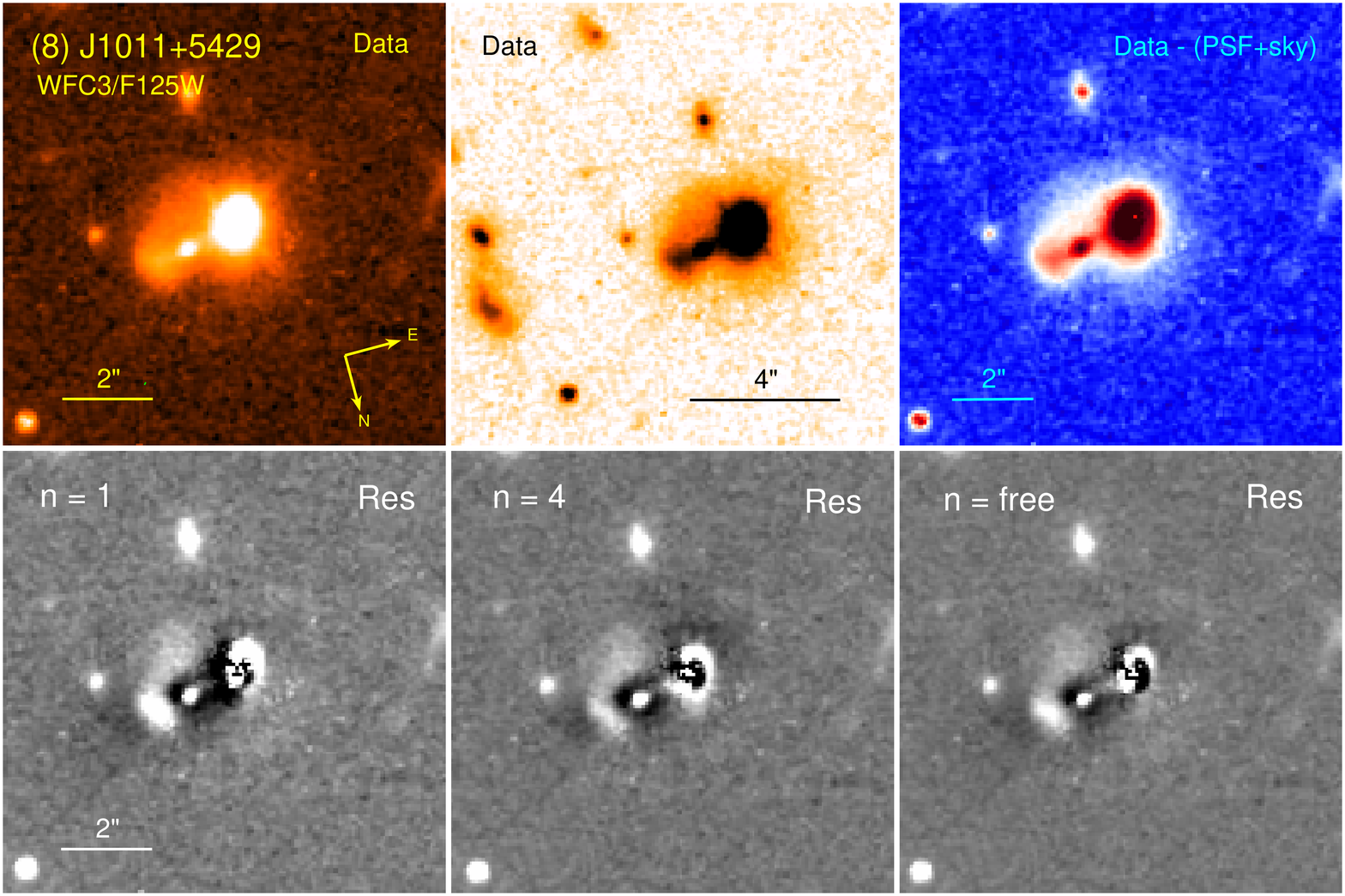,width=0.70\linewidth,clip=} \\
\epsfig{file=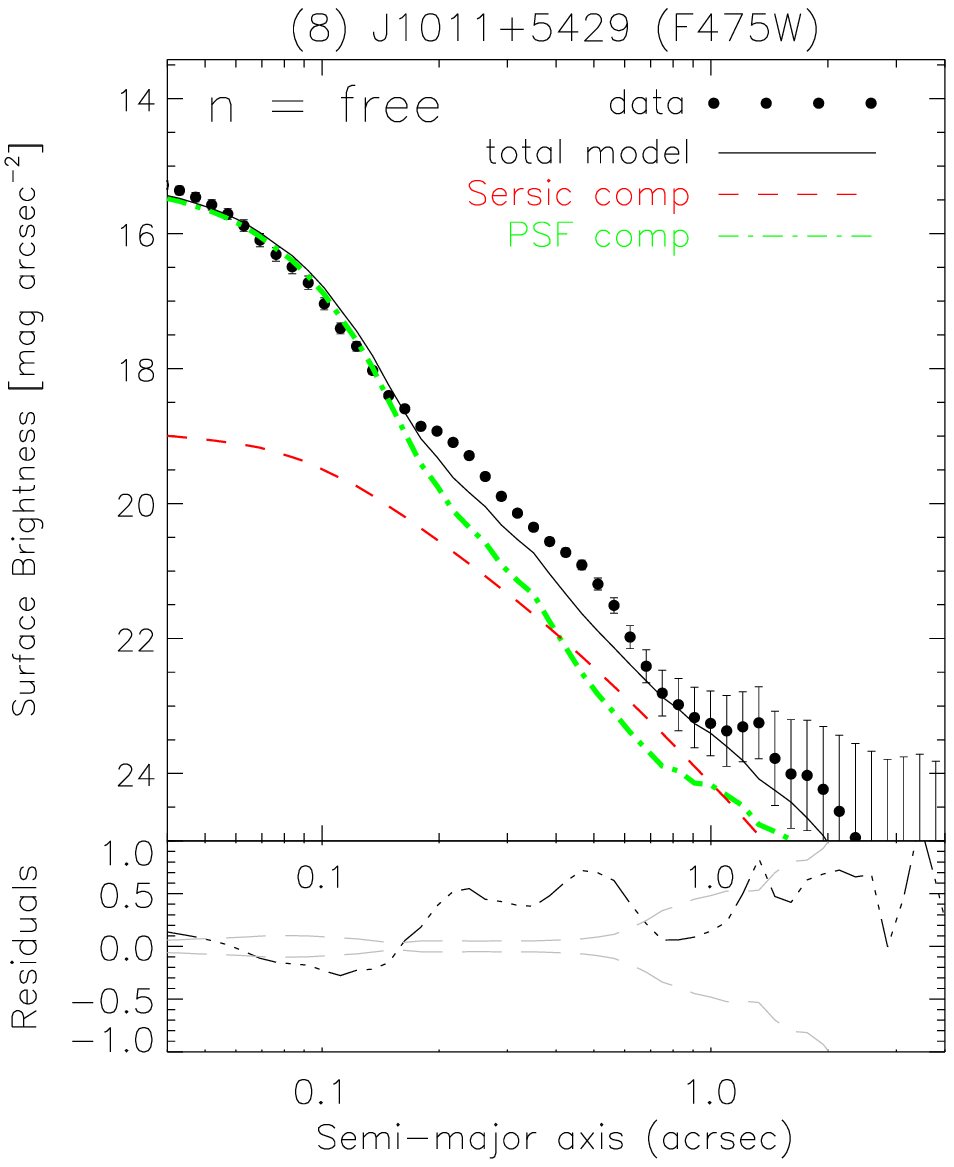,width=0.265\linewidth,clip=} &\epsfig{file=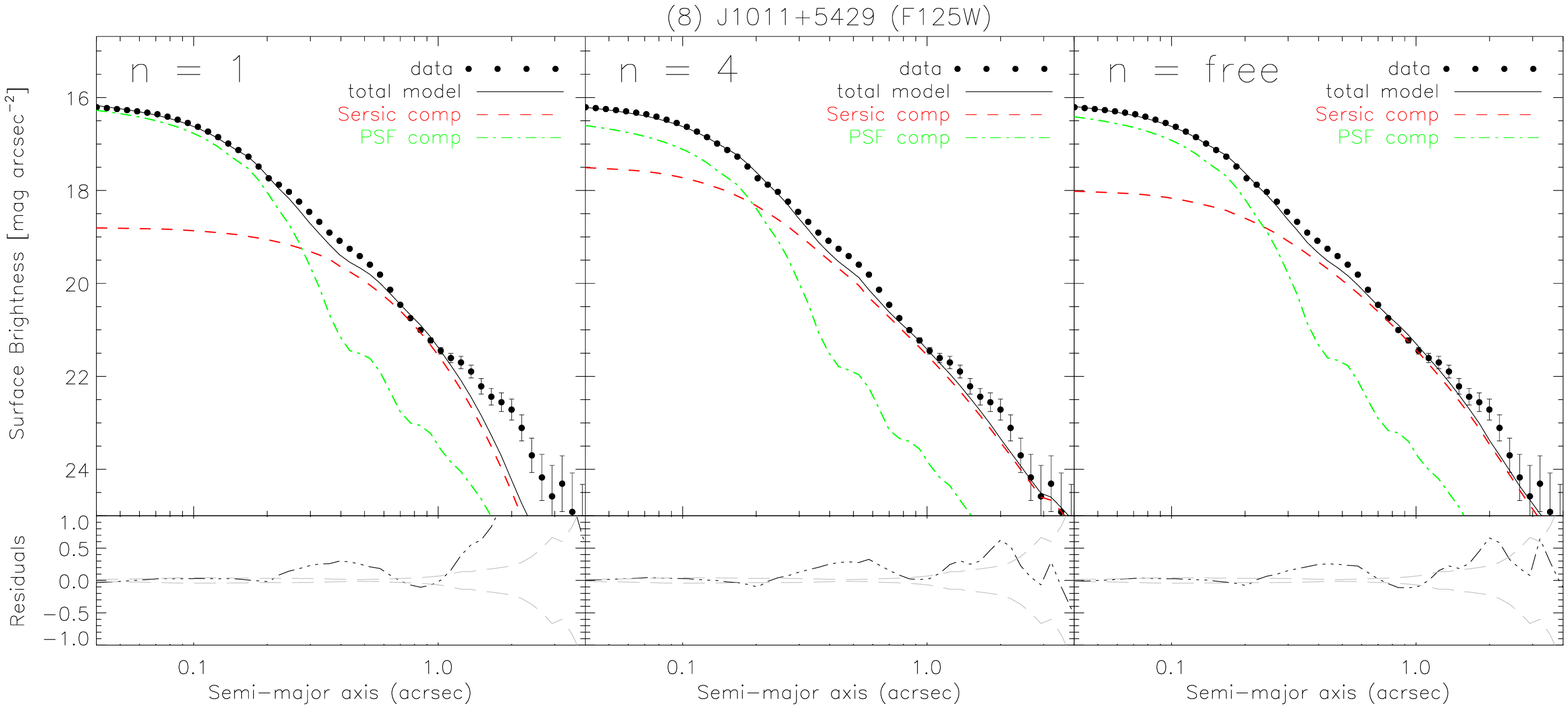,width=0.715\linewidth,clip=}
\end{tabular}
\caption{Object SDSS J1011+5429. Caption, as in Fig. \ref{fig:images1}.}
\label{fig:images8}
\end{figure}

\begin{figure}
\centering
\begin{tabular}{cc}
\epsfig{file=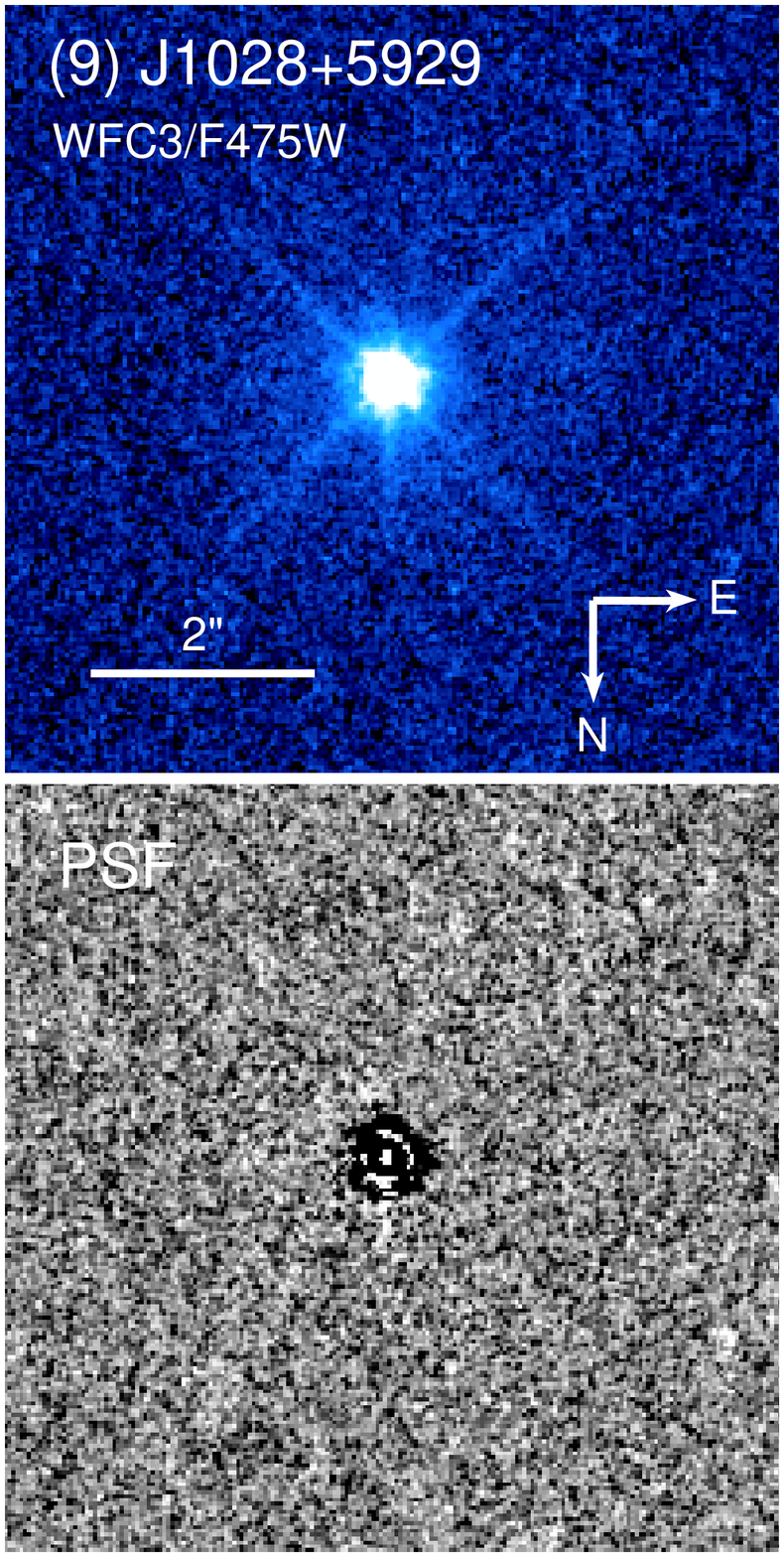,width=0.233\linewidth,clip=} & \epsfig{file=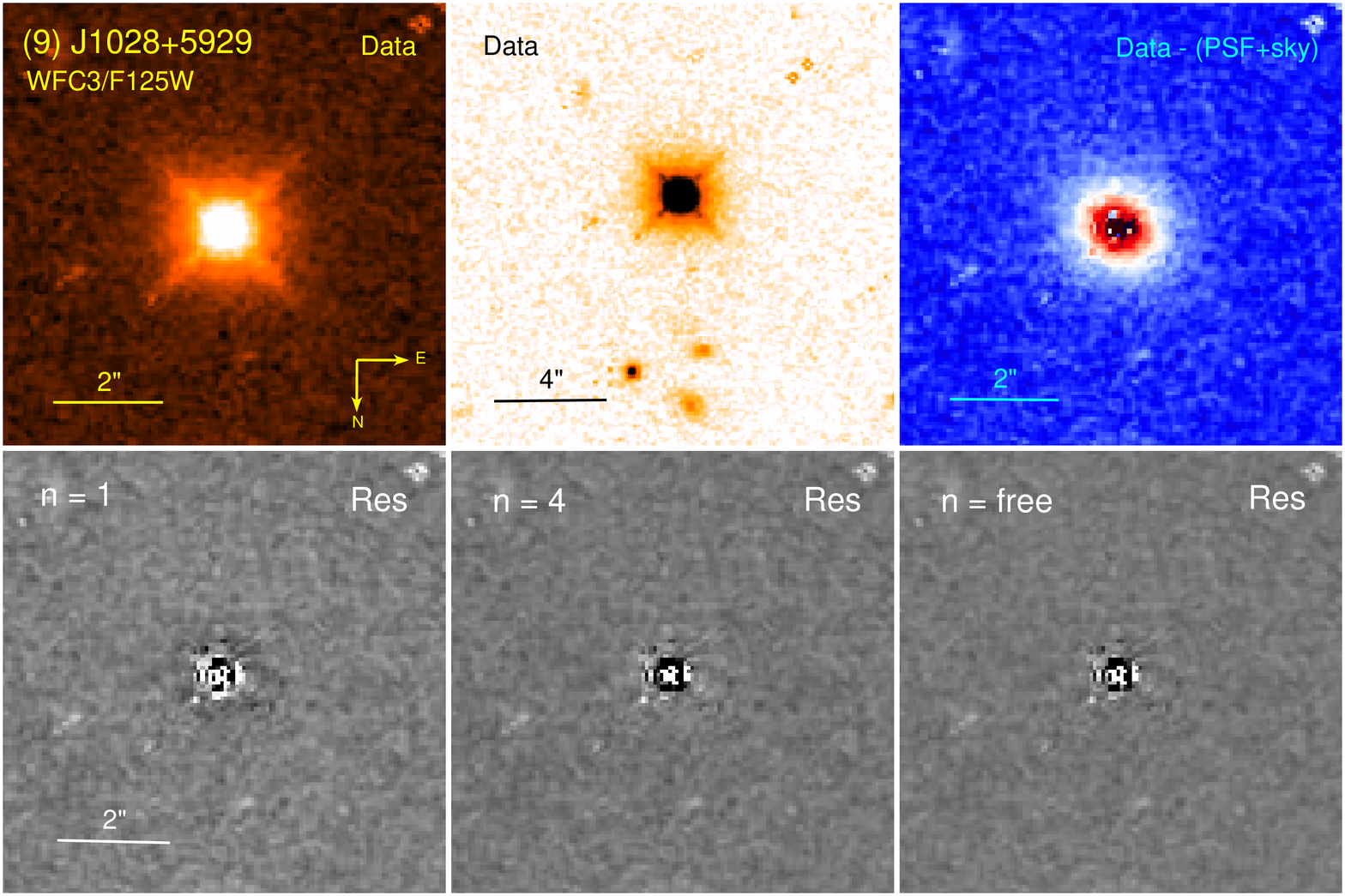,width=0.70\linewidth,clip=} \\
\epsfig{file=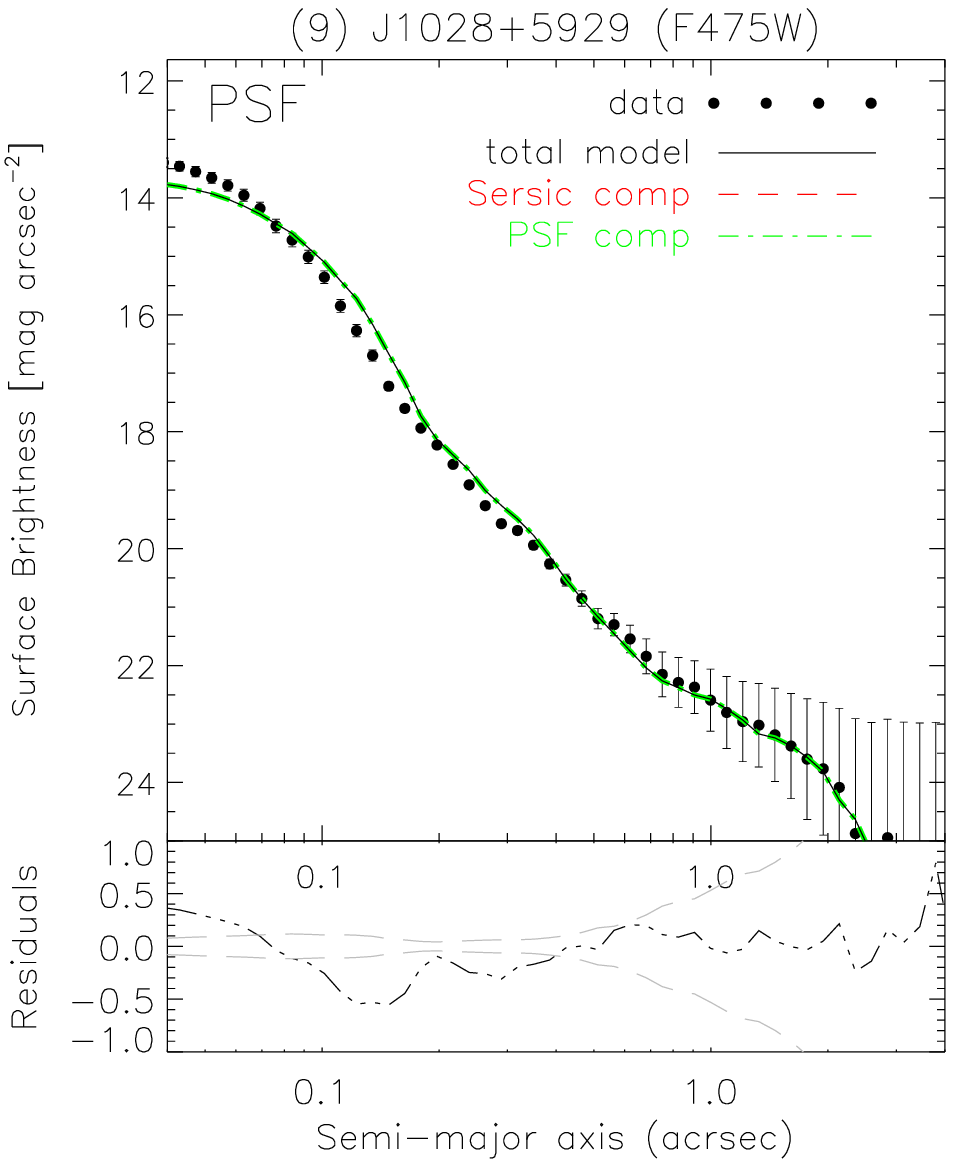,width=0.265\linewidth,clip=} &\epsfig{file=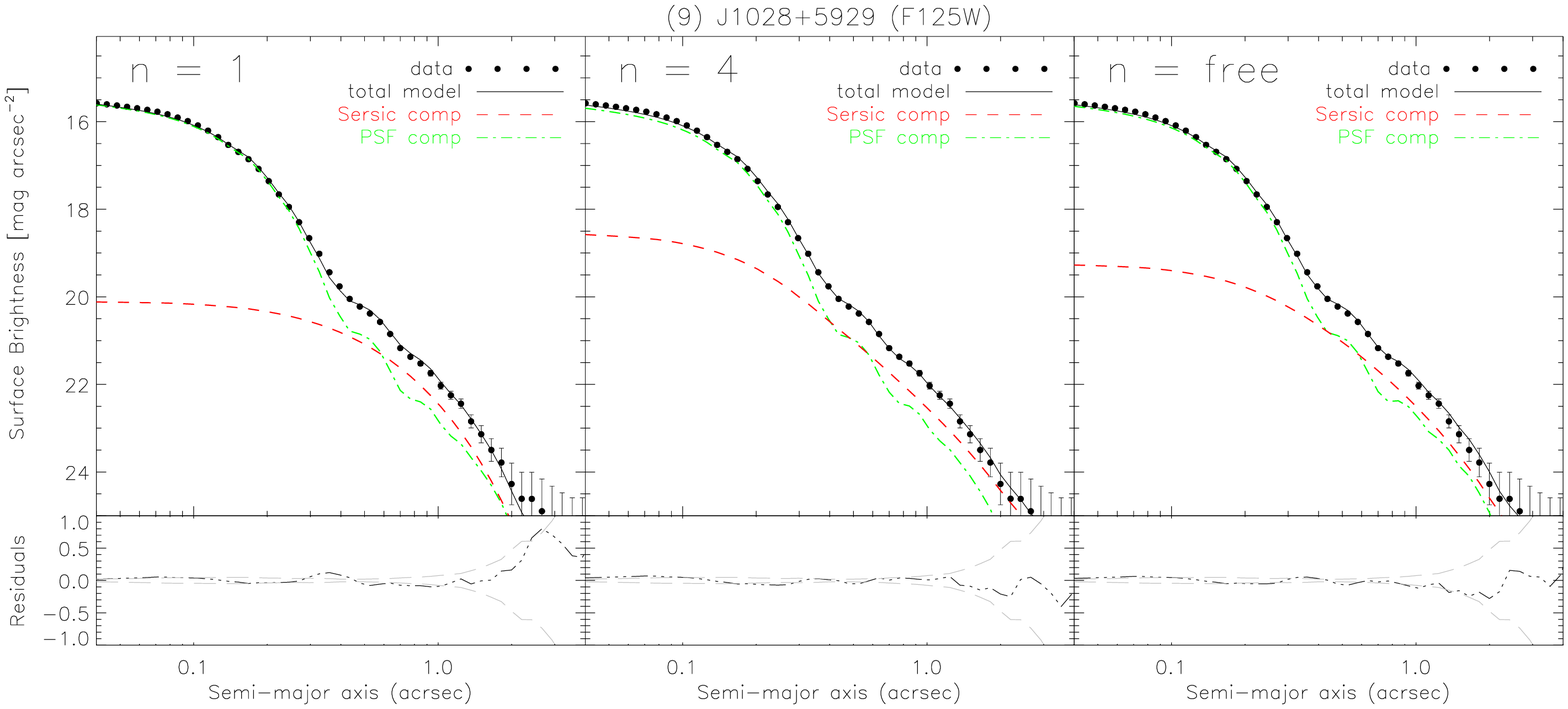,width=0.715\linewidth,clip=}
\end{tabular}
\caption{Object SDSS J1028+5929. Caption, as in Fig. \ref{fig:images1}.}
\label{fig:images9}
\end{figure}

\begin{figure}
\centering
\begin{tabular}{cc}
\epsfig{file=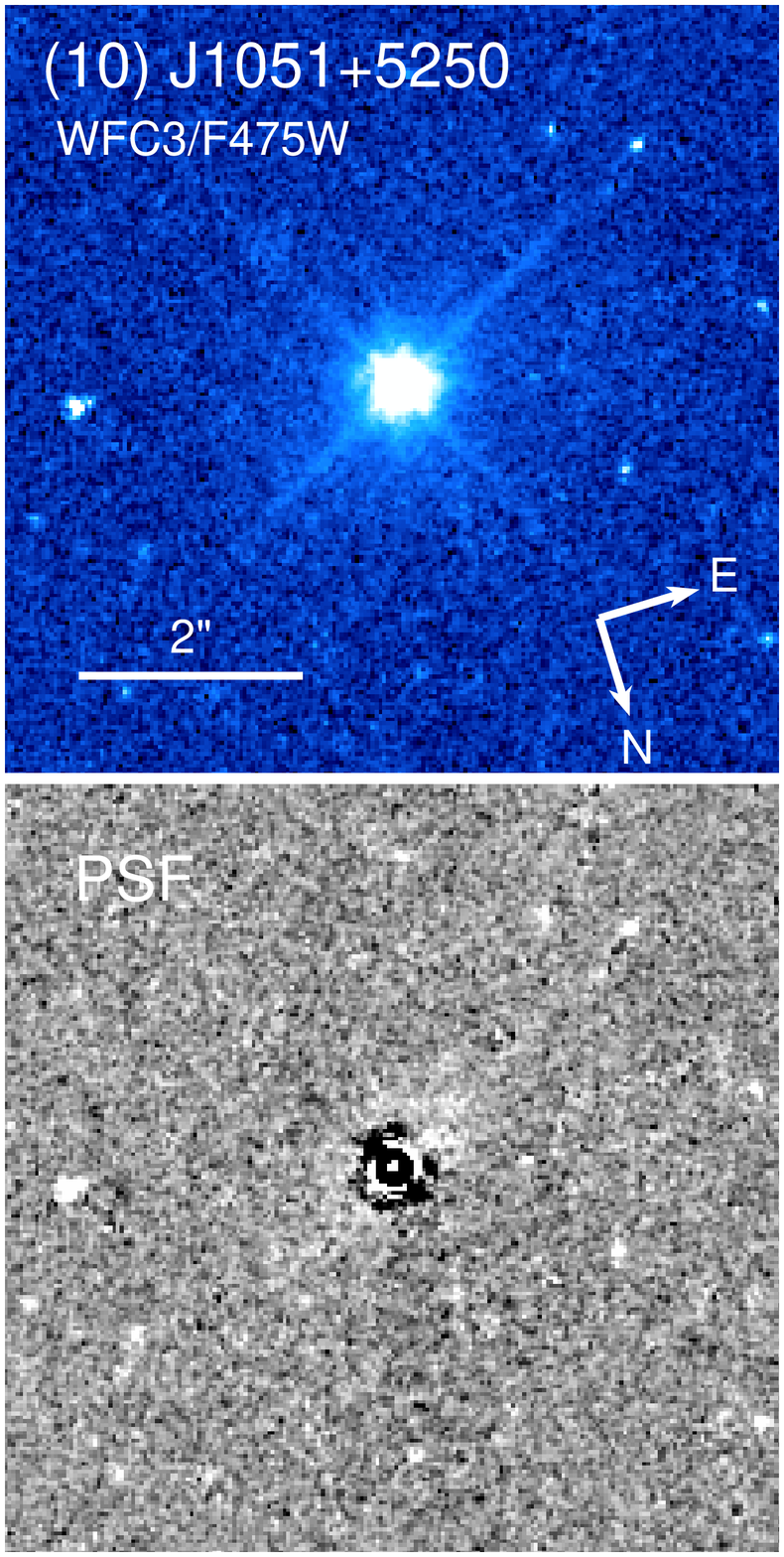,width=0.233\linewidth,clip=} & \epsfig{file=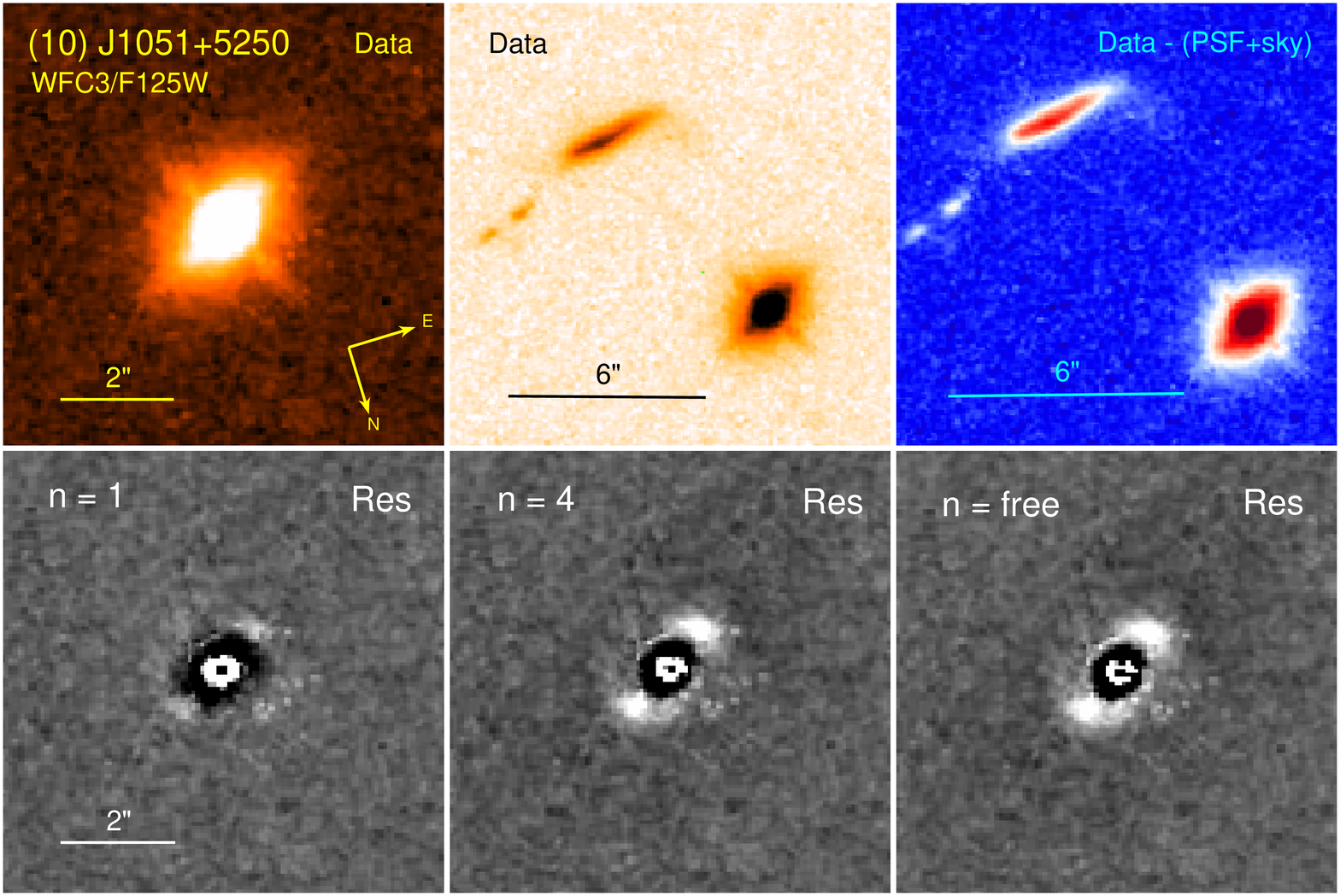,width=0.70\linewidth,clip=} \\
\epsfig{file=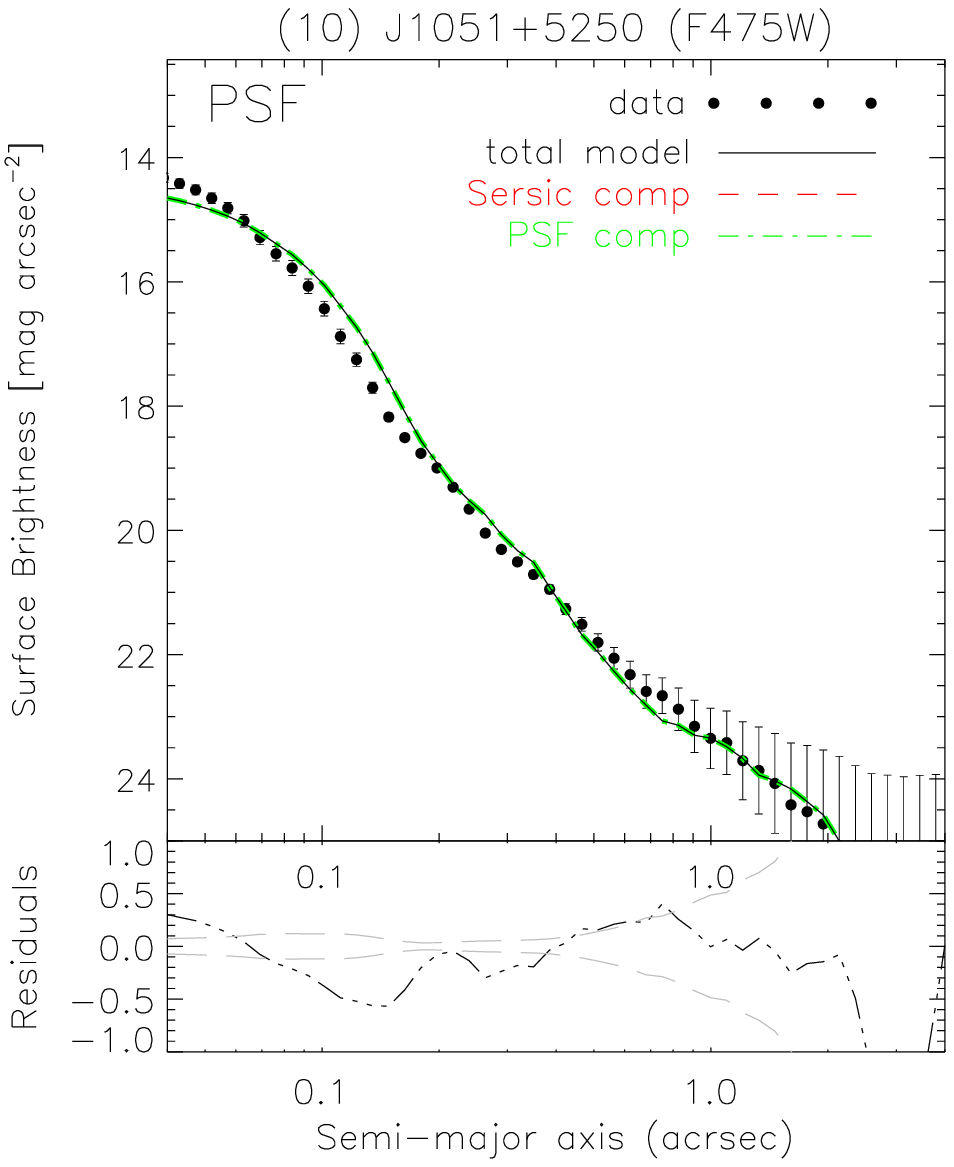,width=0.265\linewidth,clip=} &\epsfig{file=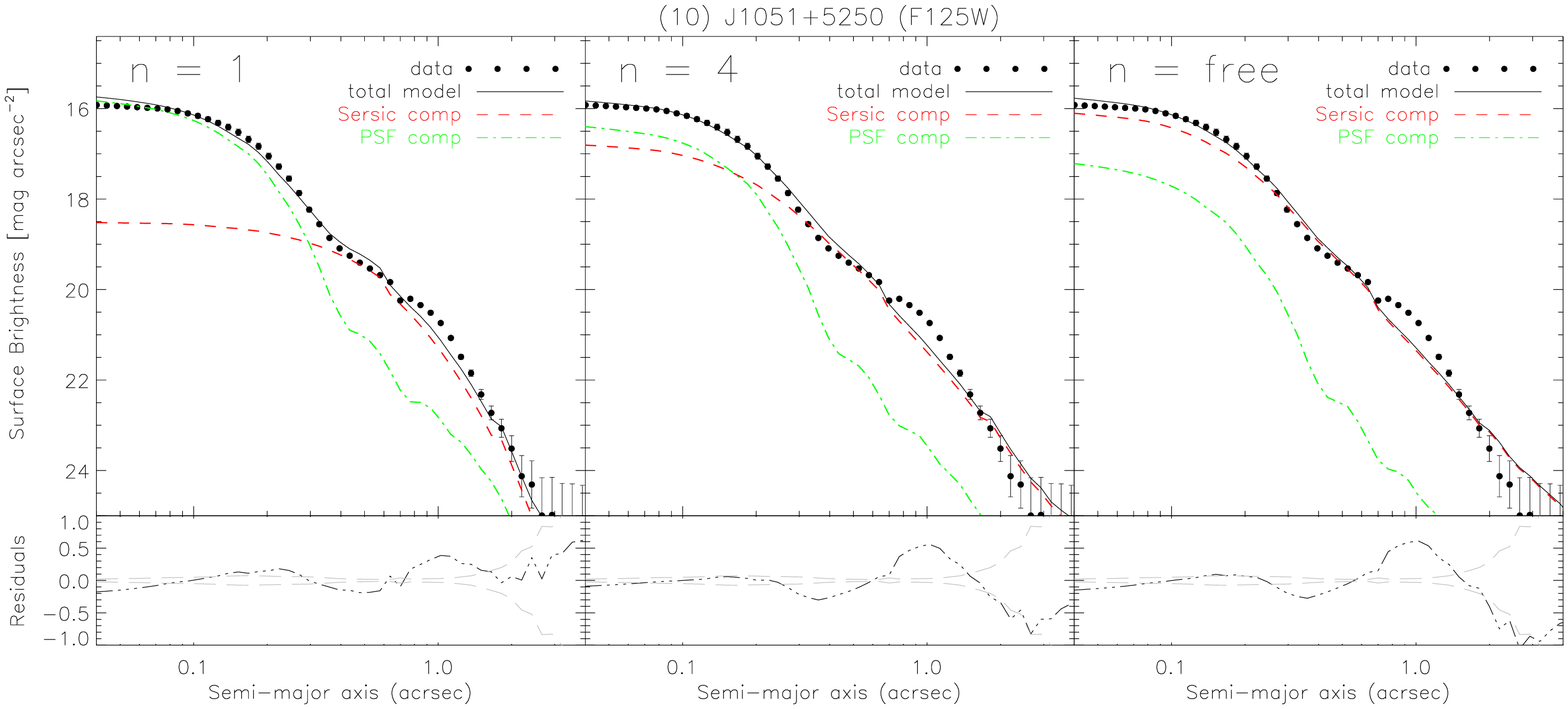,width=0.715\linewidth,clip=}
\end{tabular}
\caption{Object SDSS J1051+5250. Caption, as in Fig. \ref{fig:images1}.}
\label{fig:images10}
\end{figure}

\begin{figure}
\centering
\begin{tabular}{cc}
\epsfig{file=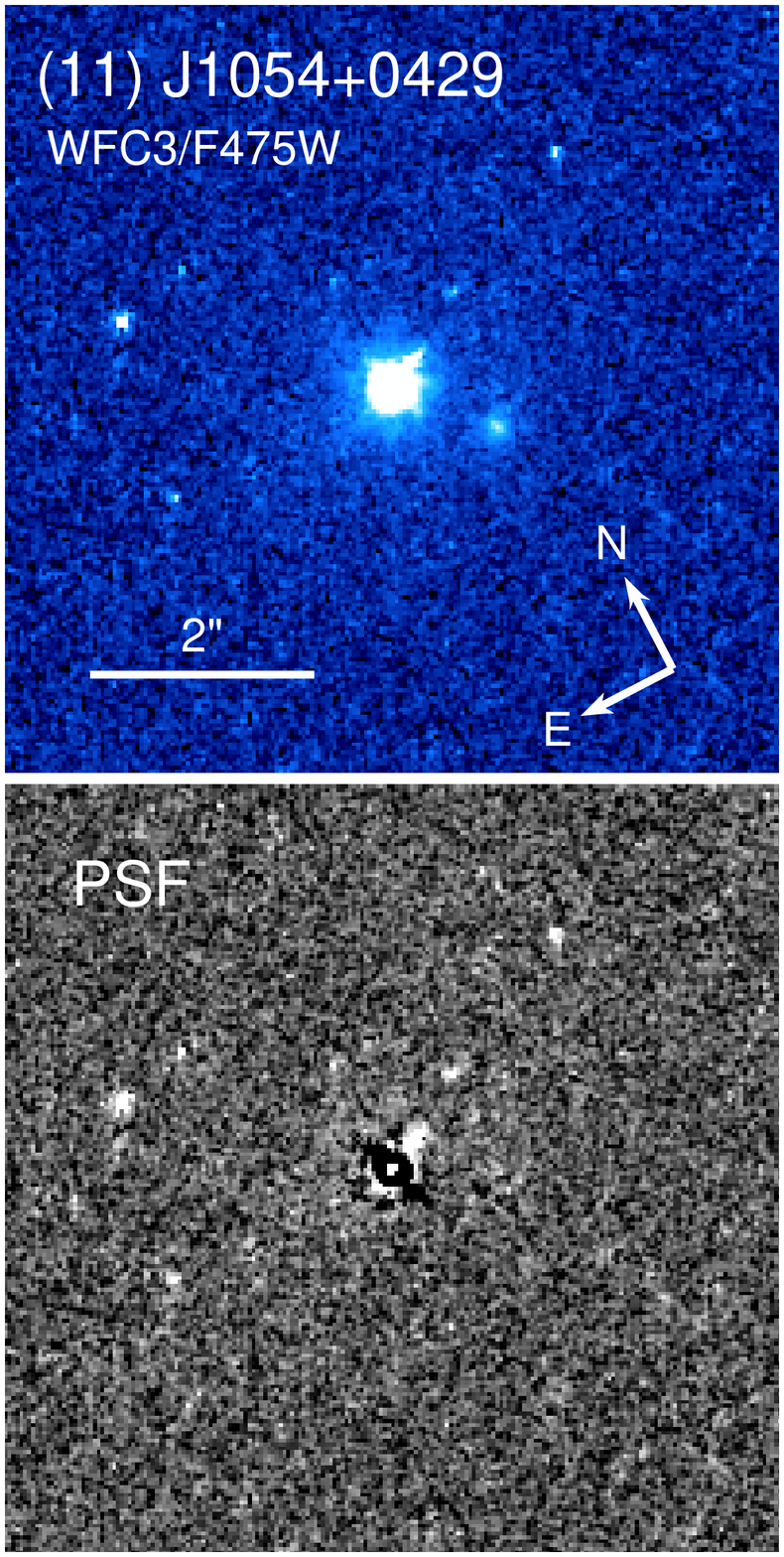,width=0.233\linewidth,clip=} & \epsfig{file=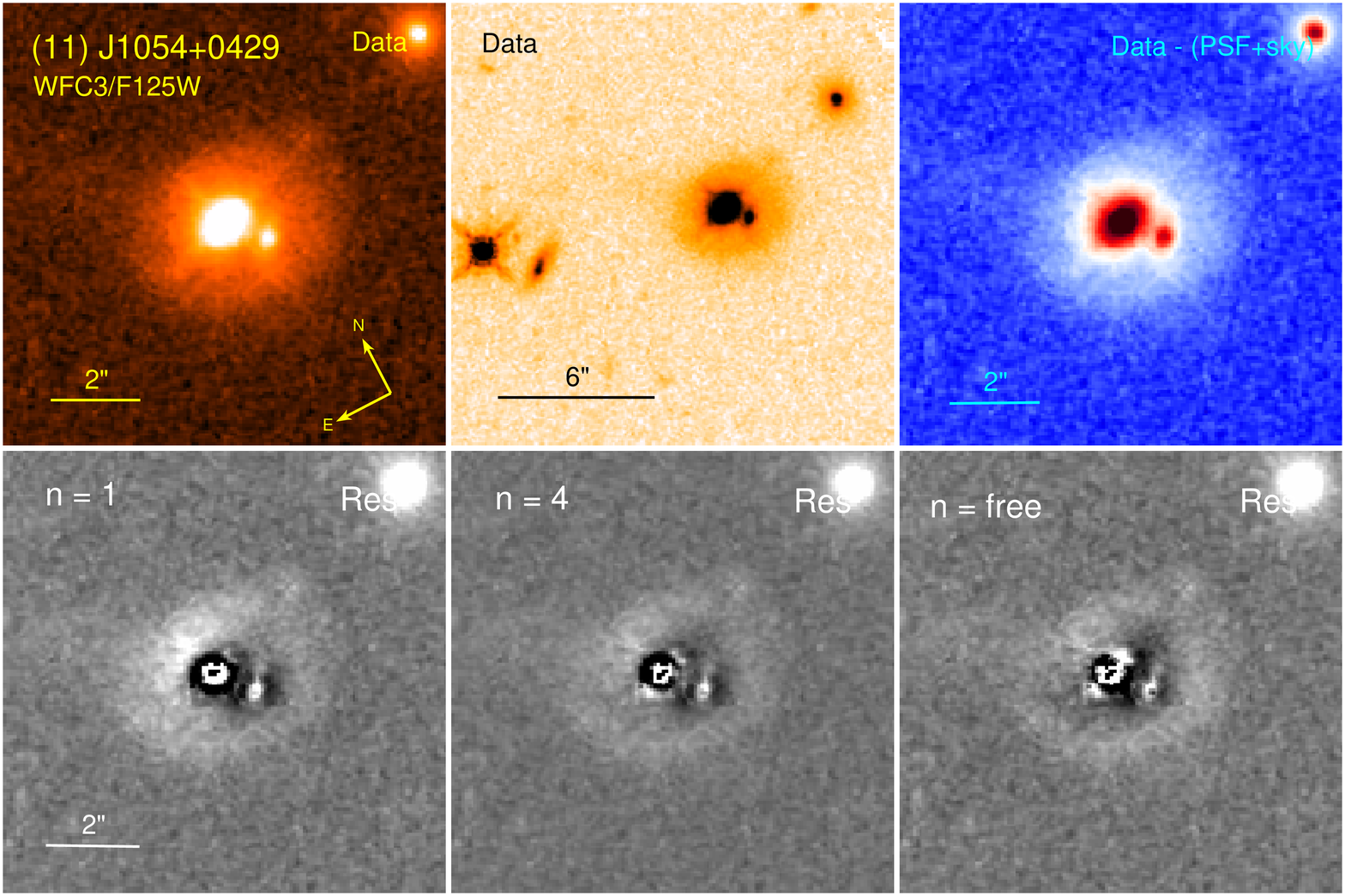,width=0.70\linewidth,clip=} \\
\epsfig{file=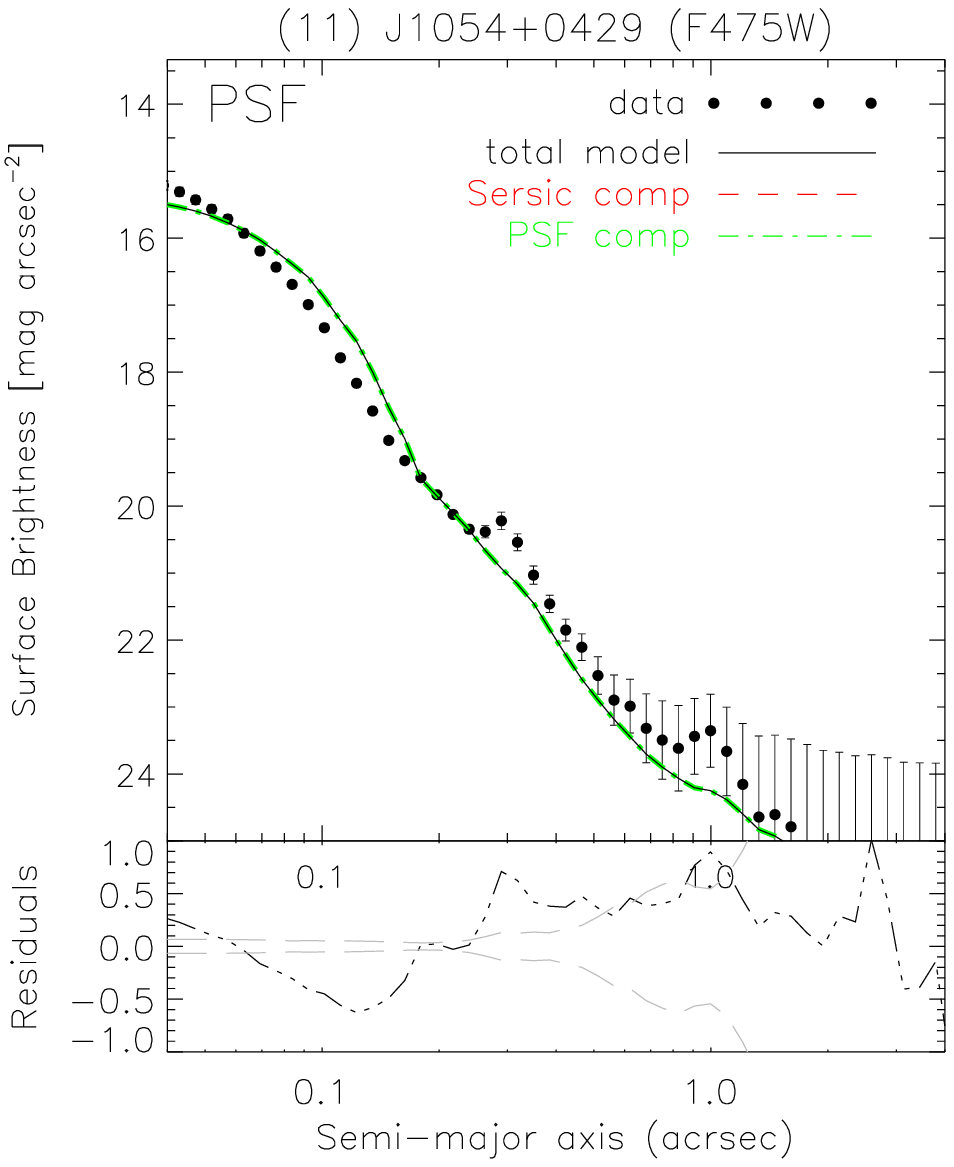,width=0.265\linewidth,clip=} &\epsfig{file=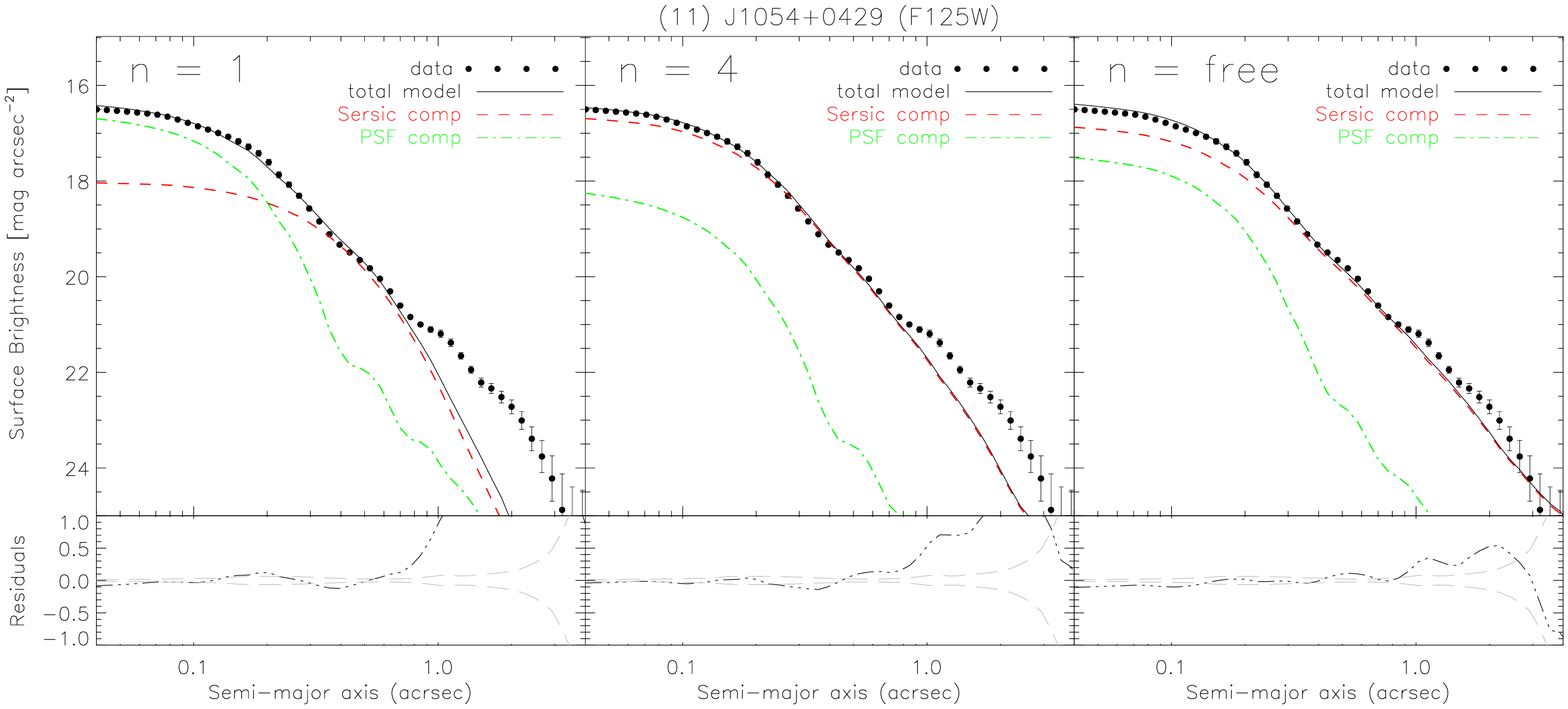,width=0.715\linewidth,clip=}
\end{tabular}
\caption{Object SDSS J1054+0429. Caption, as in Fig. \ref{fig:images1}.}
\label{fig:images11}
\end{figure}

\begin{figure}
\centering
\begin{tabular}{cc}
\epsfig{file=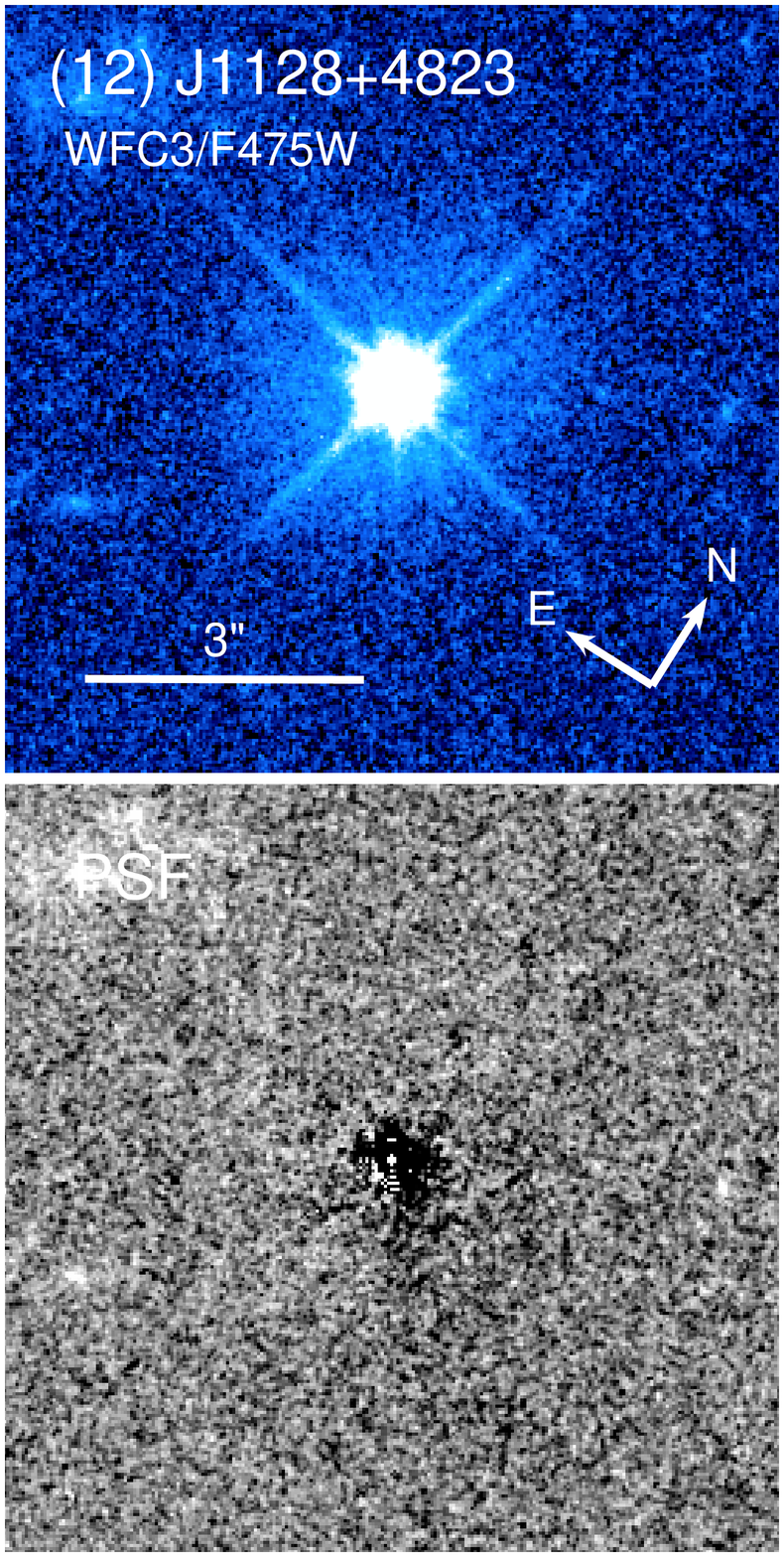,width=0.233\linewidth,clip=} & \epsfig{file=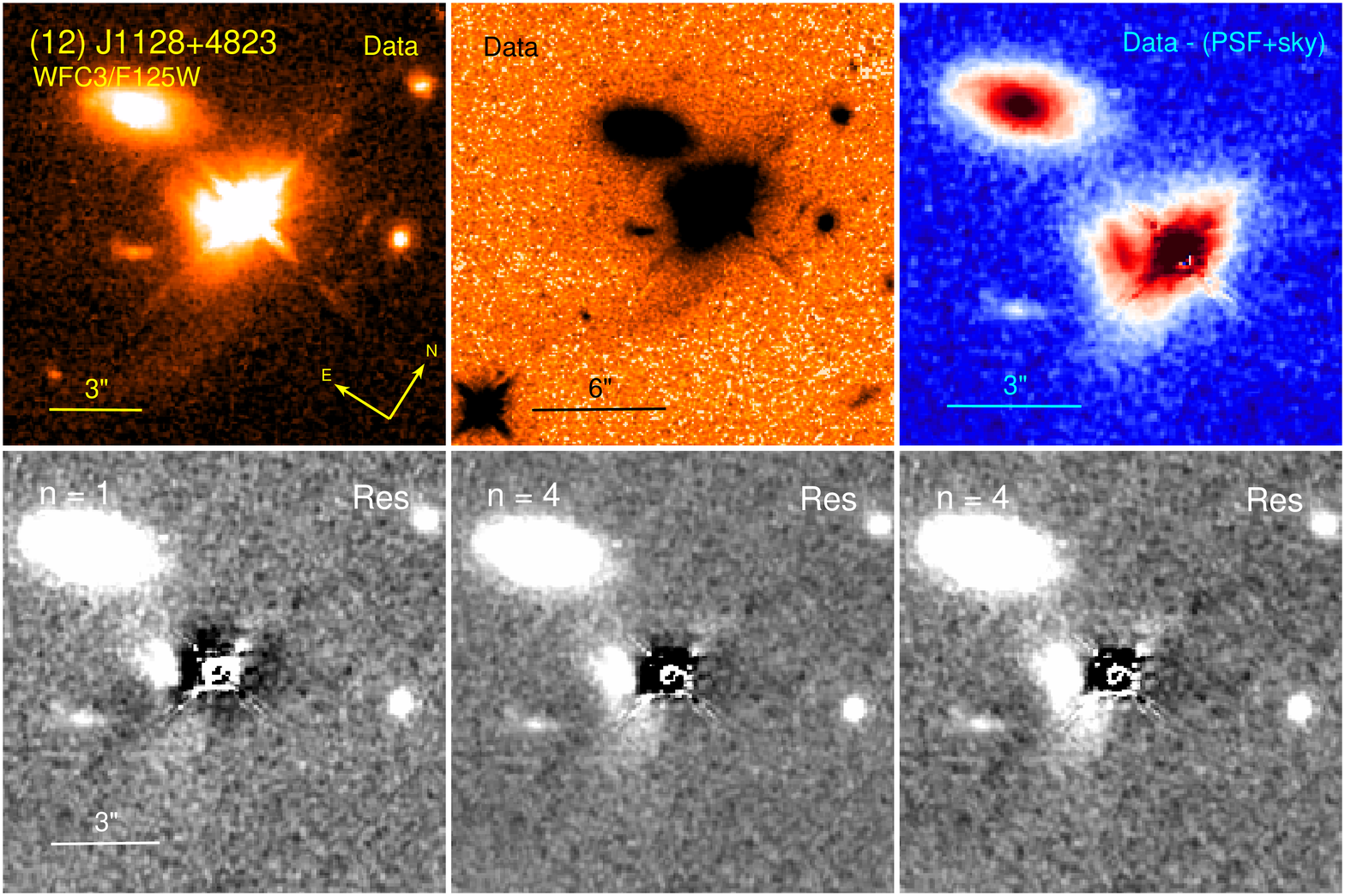,width=0.70\linewidth,clip=} \\
\epsfig{file=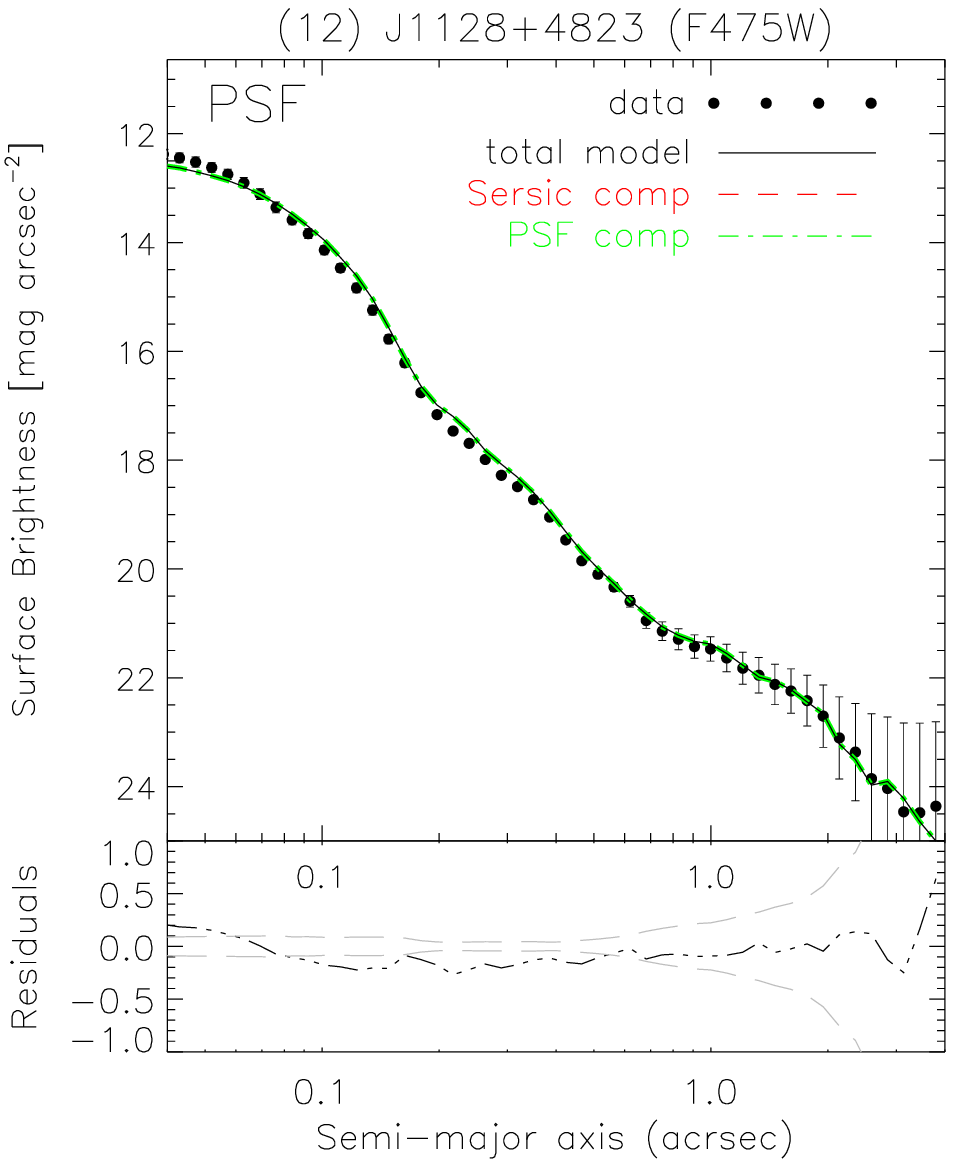,width=0.265\linewidth,clip=} &\epsfig{file=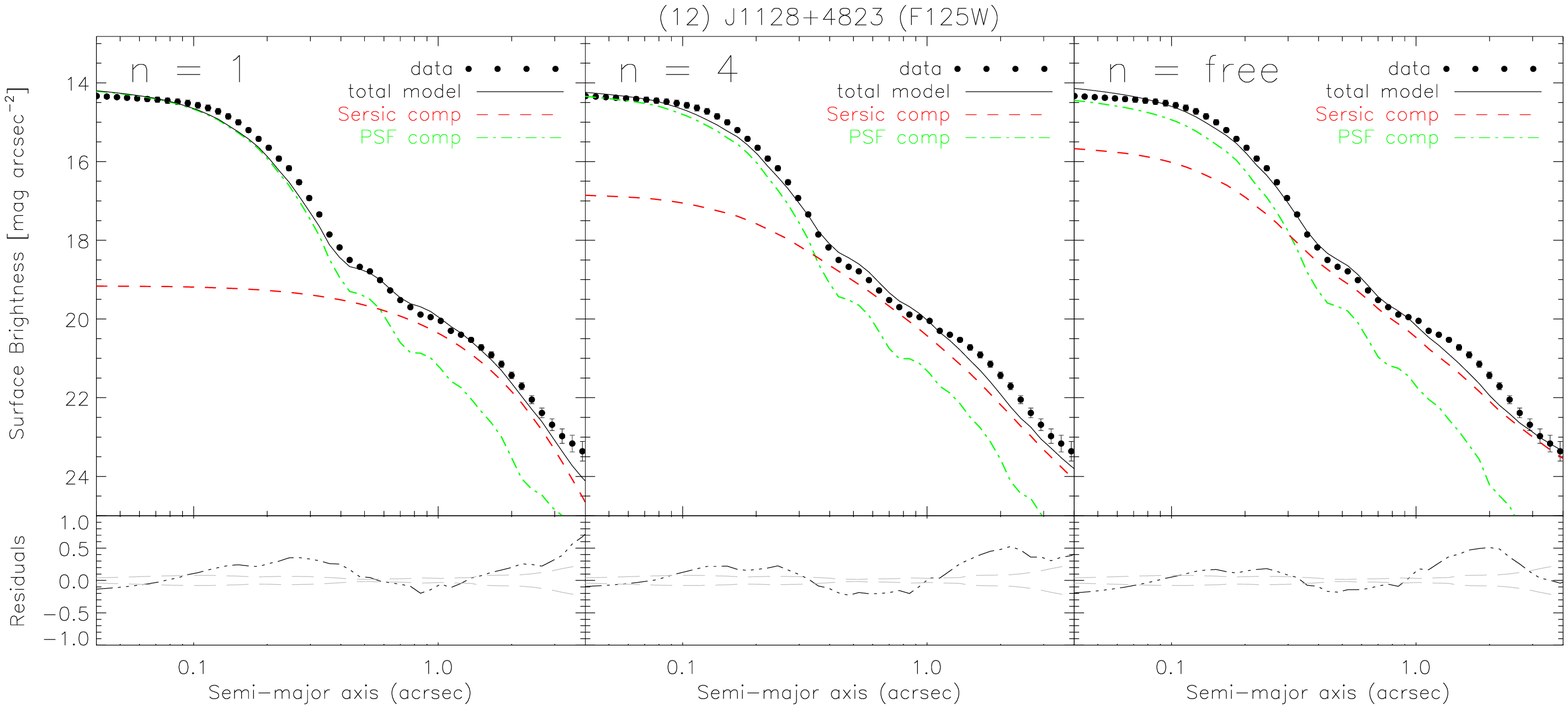,width=0.715\linewidth,clip=}
\end{tabular}
\caption{Object SDSS J1128+4823. Caption, as in Fig. \ref{fig:images1}. The galaxy $\sim$4$\farcs$8E from the LoBAL is at the same photometric redshift (z$\sim$0.52$\pm$0.04). } 
\label{fig:images12}
\end{figure}

\begin{figure}
\centering
\begin{tabular}{cc}
\epsfig{file=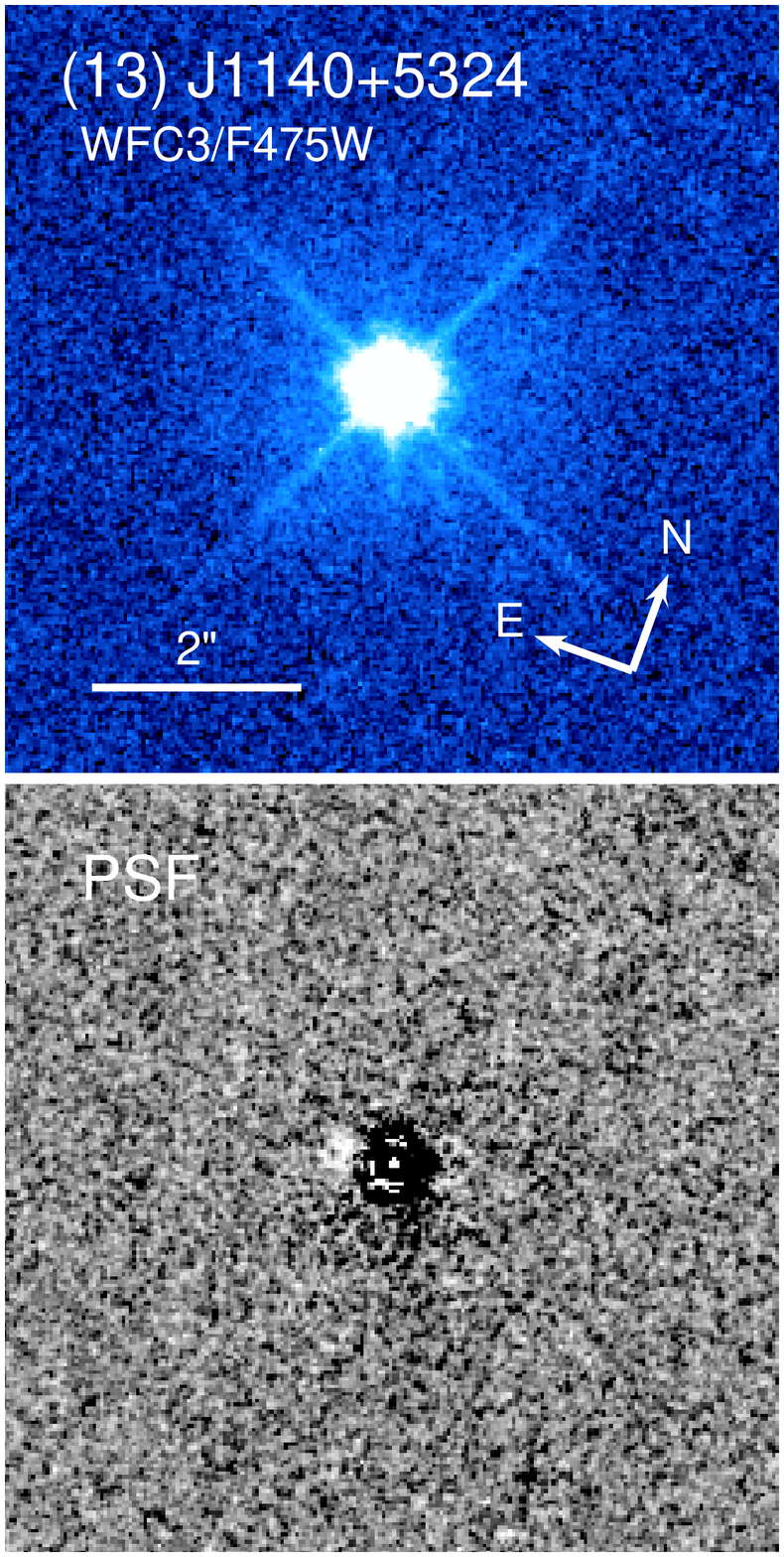,width=0.233\linewidth,clip=} & \epsfig{file=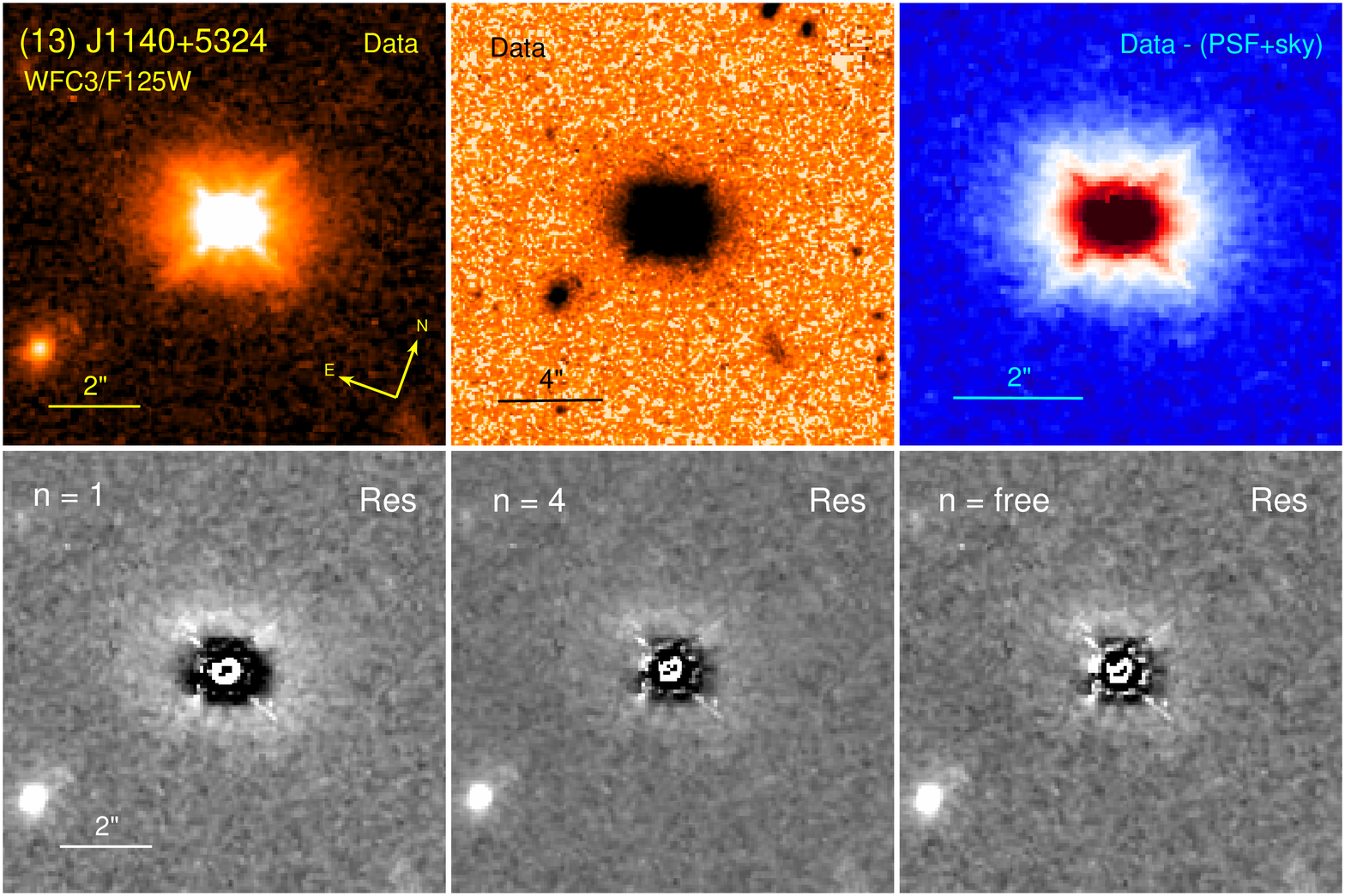,width=0.70\linewidth,clip=} \\
\epsfig{file=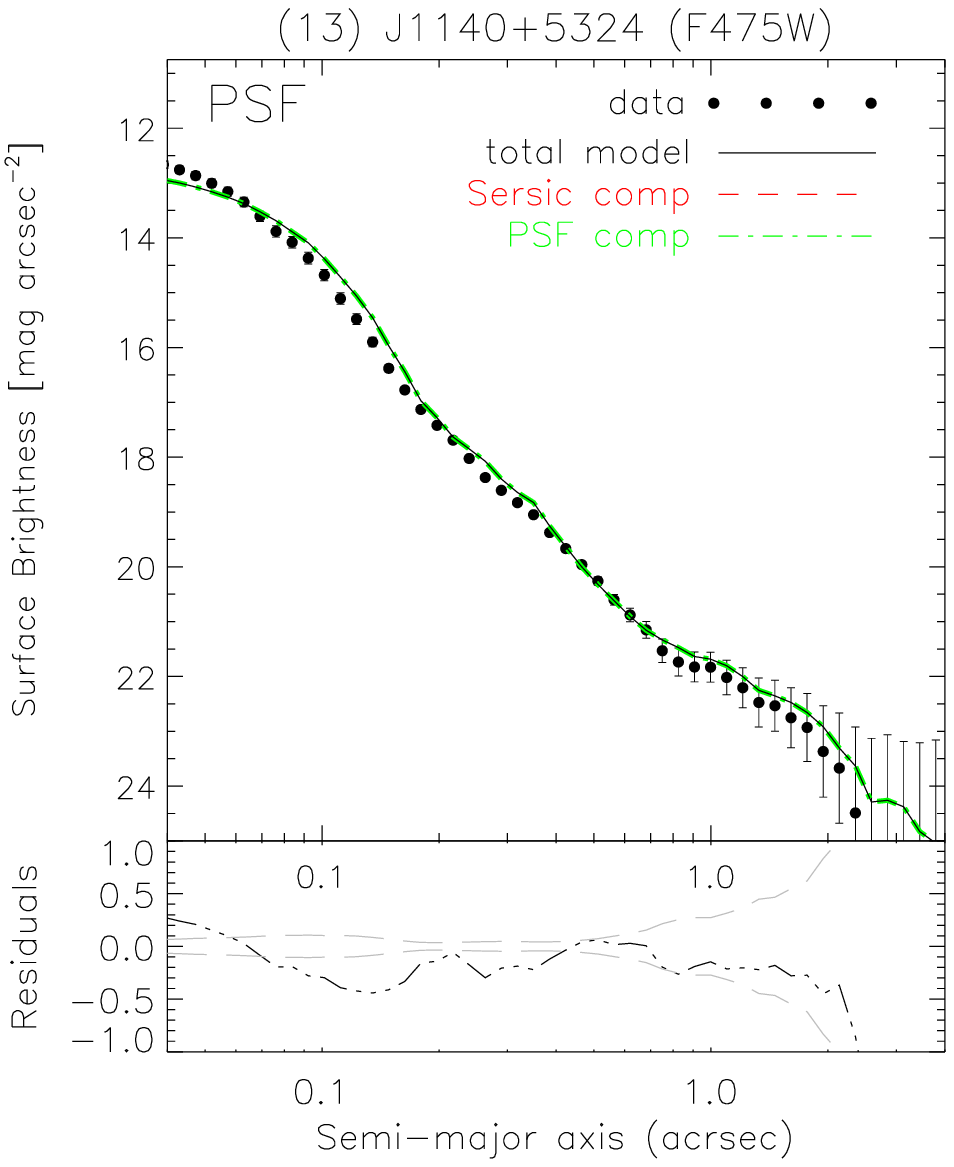,width=0.265\linewidth,clip=} &\epsfig{file=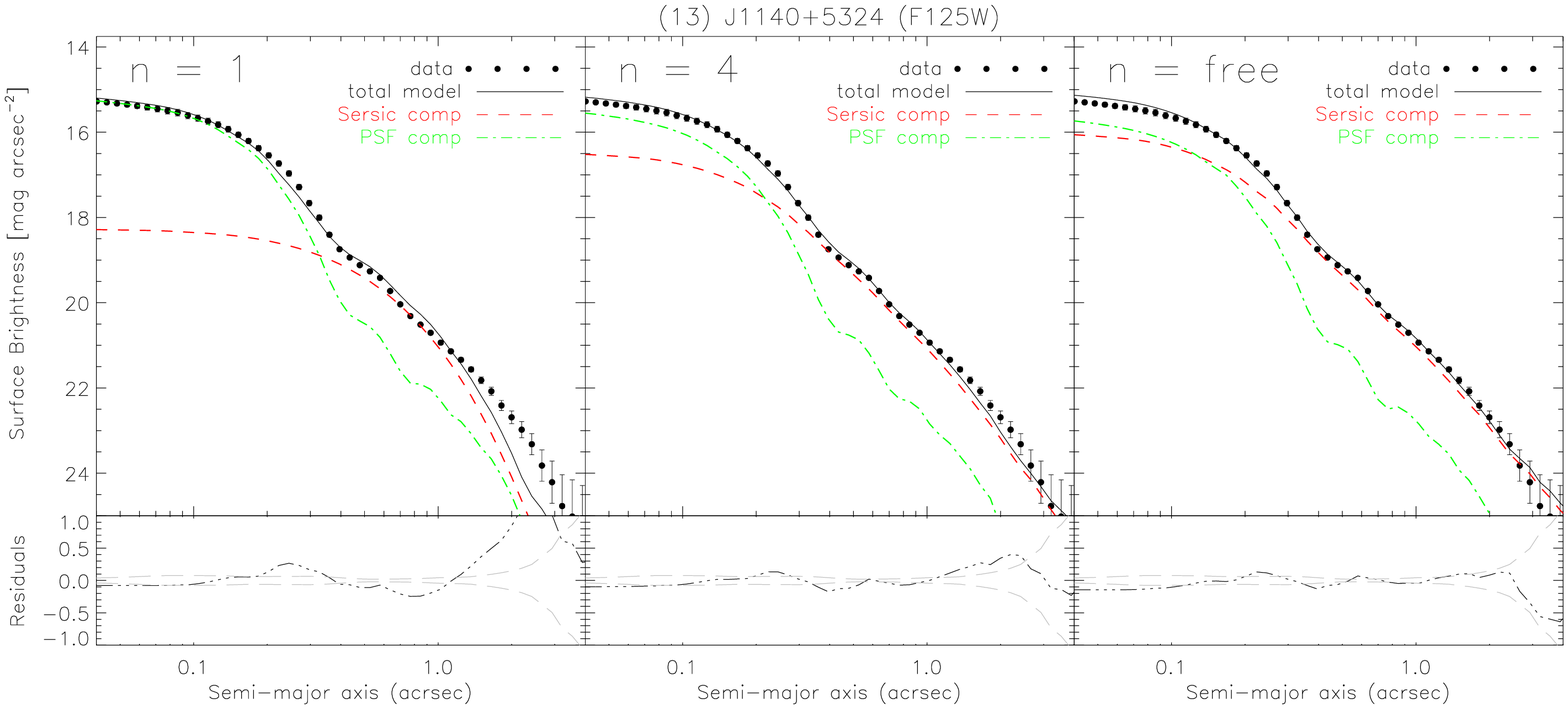,width=0.715\linewidth,clip=}
\end{tabular}
\caption{Object SDSS J1140+5324. Caption, as in Fig. \ref{fig:images1}.}
\label{fig:images13}
\end{figure}

\begin{figure}
\centering
\begin{tabular}{cc}
\epsfig{file=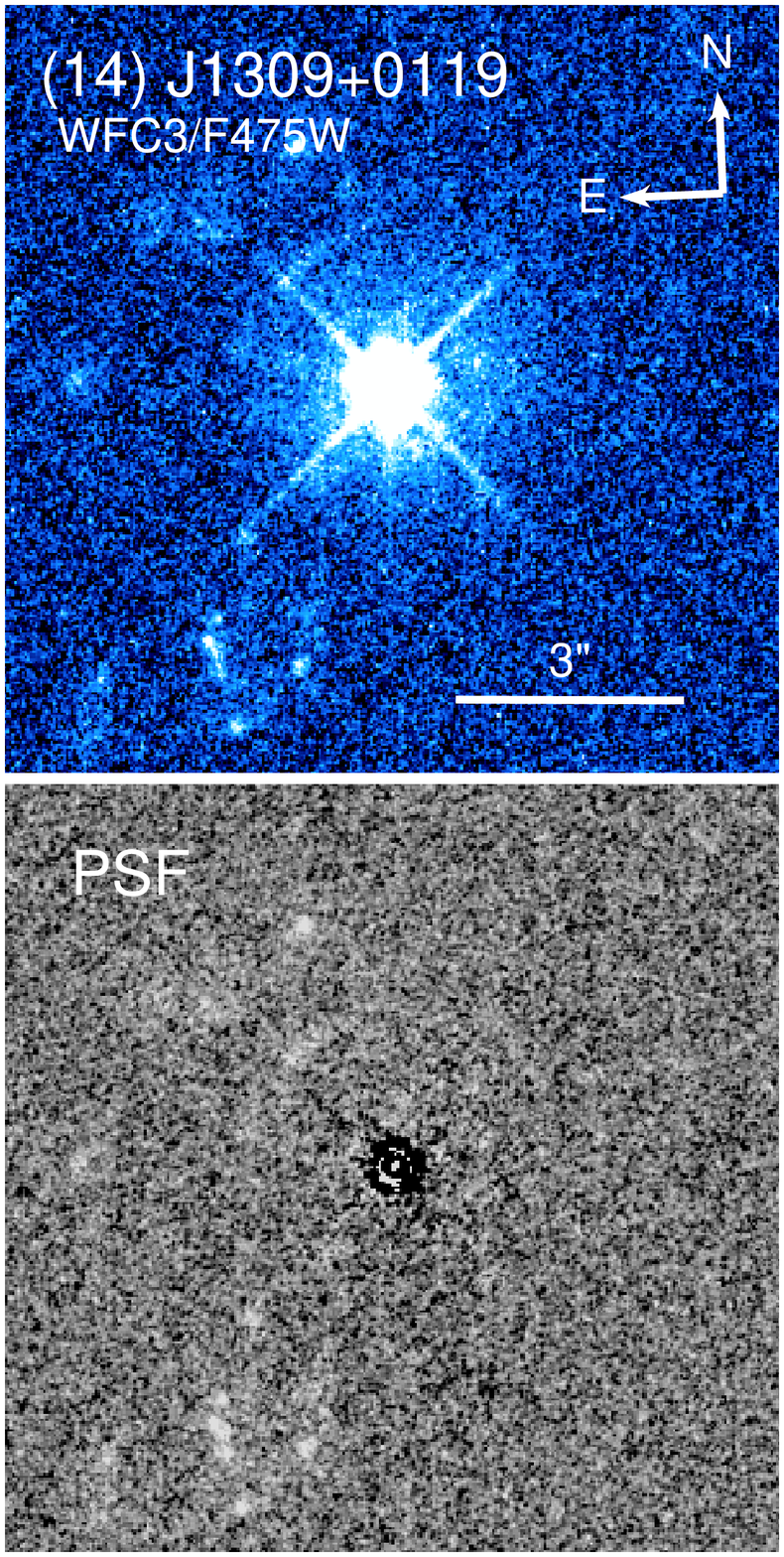,width=0.233\linewidth,clip=} & \epsfig{file=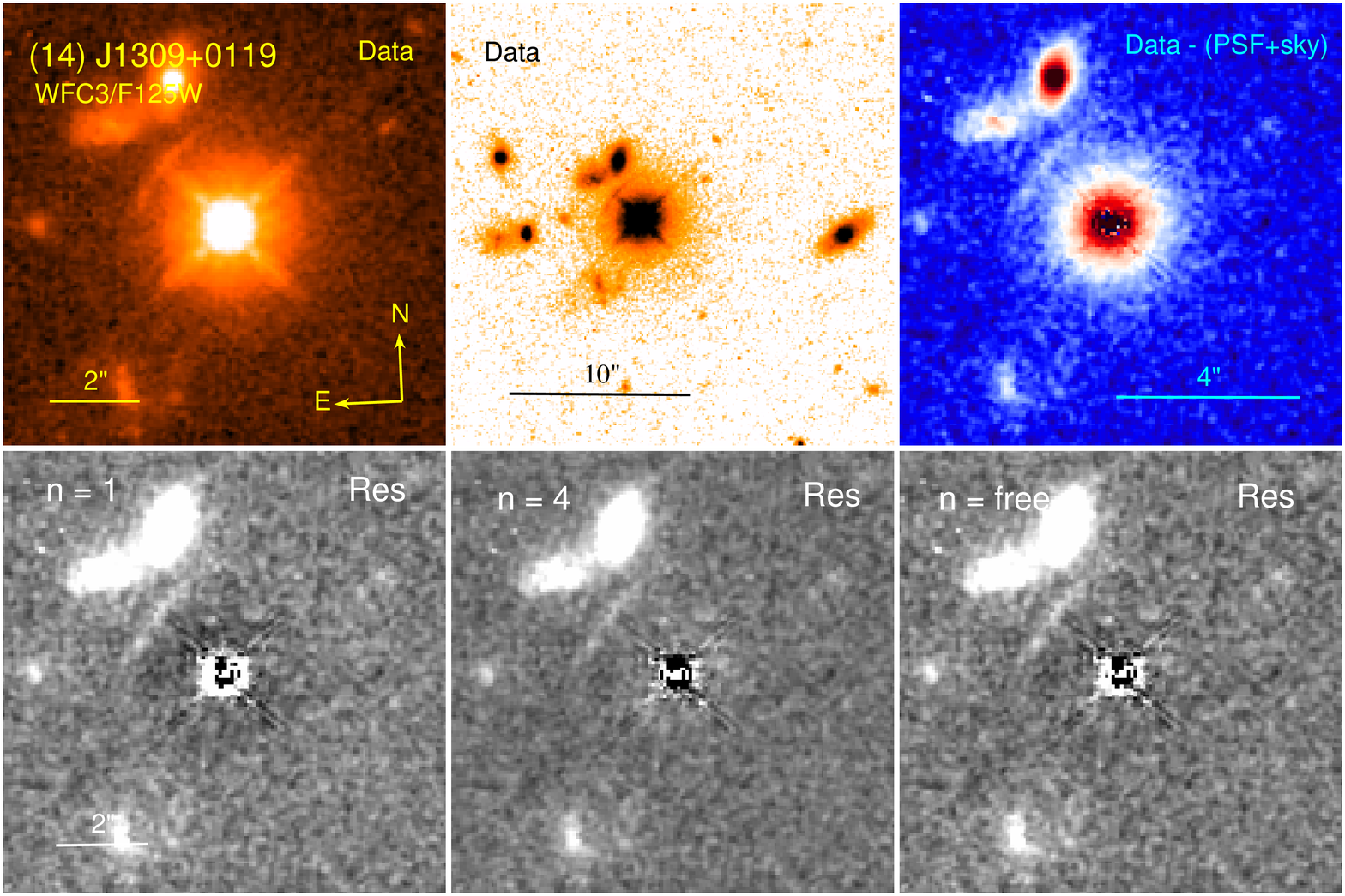,width=0.70\linewidth,clip=} \\
\epsfig{file=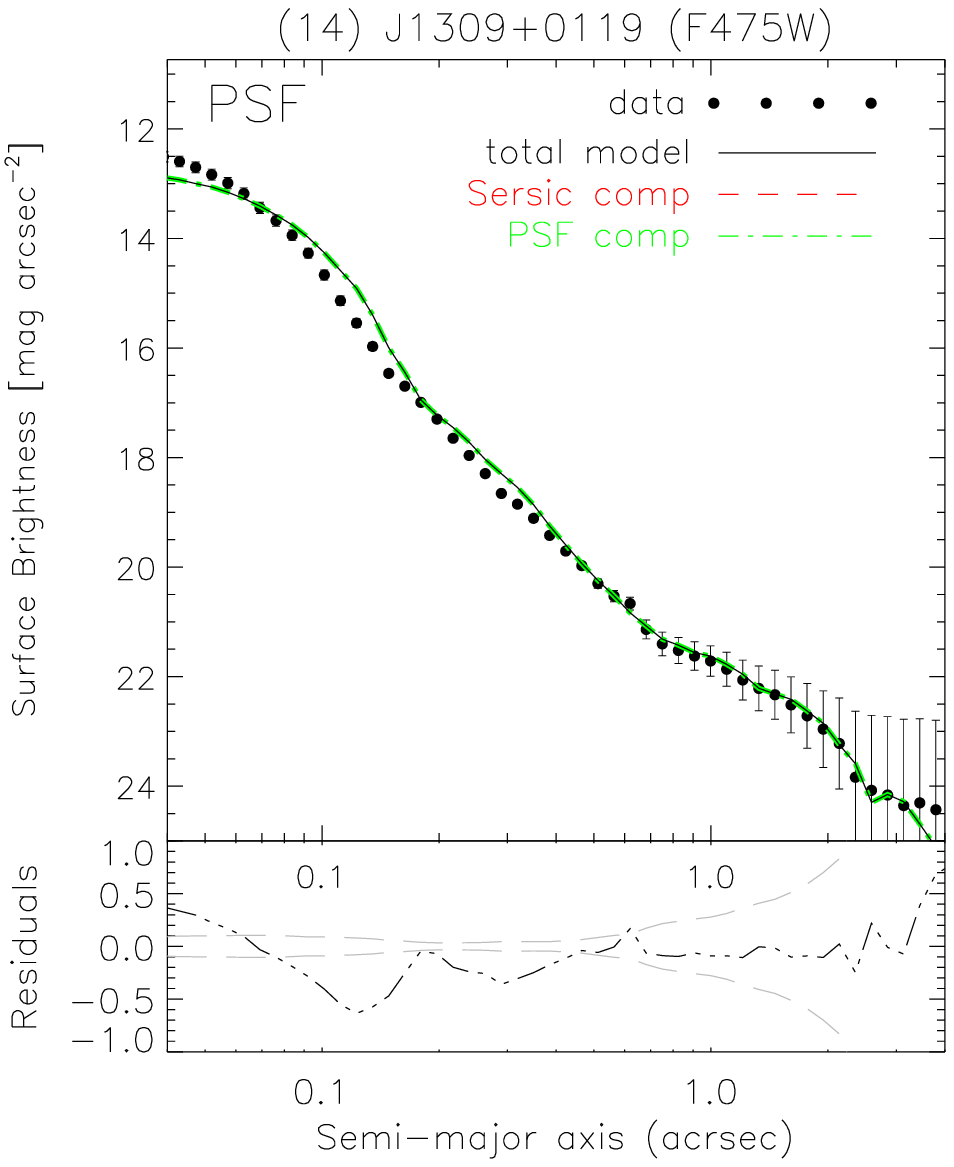,width=0.265\linewidth,clip=} &\epsfig{file=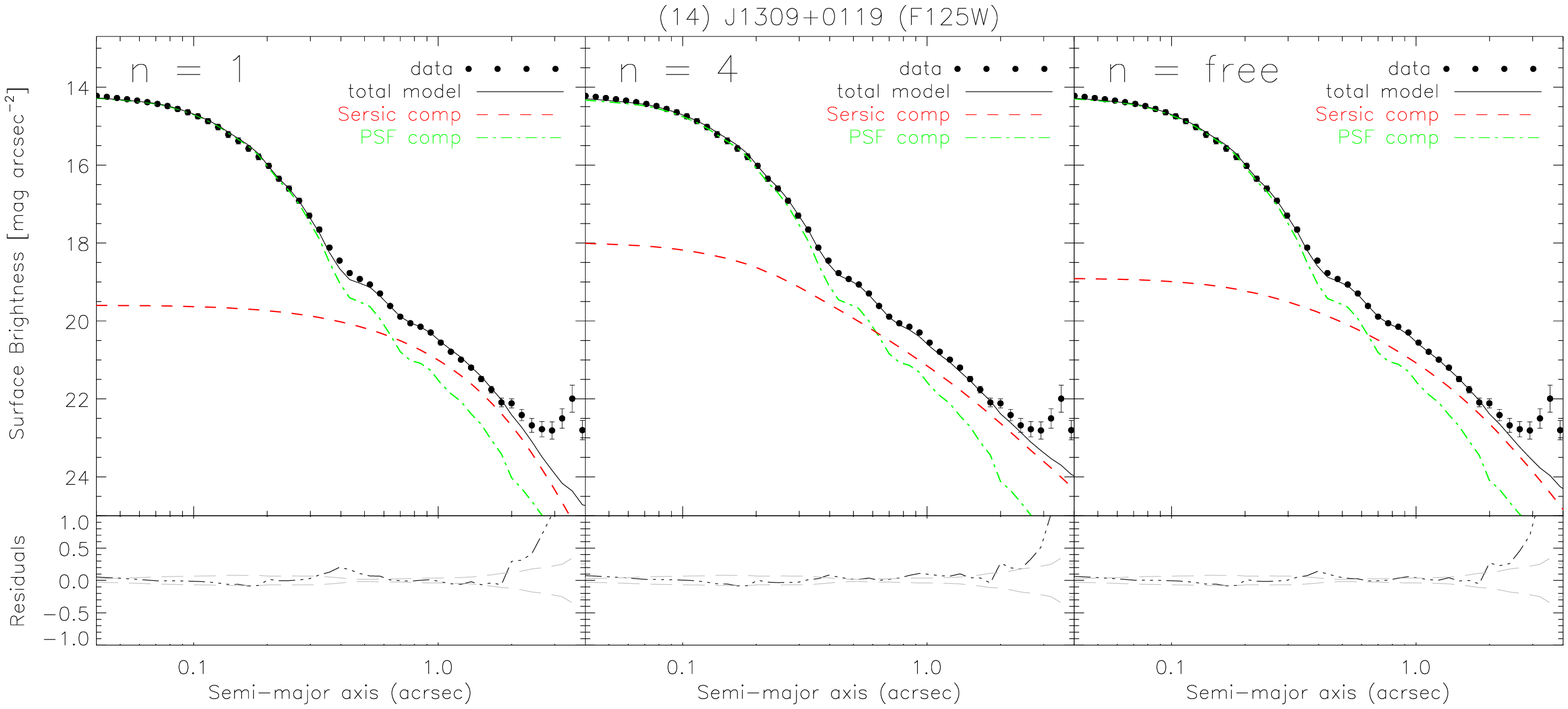,width=0.715\linewidth,clip=}
\end{tabular}
\caption{Object SDSS J1309+0119. Caption, as in Fig. \ref{fig:images1}. The galaxy $\sim$11$\farcs$3W of the LoBAL is at a similar photometric redshift (z$\sim$0.60$\pm$0.12) within the uncertainties.} 
\label{fig:images14}
\end{figure}

\begin{figure}
\centering
\begin{tabular}{cc}
\epsfig{file=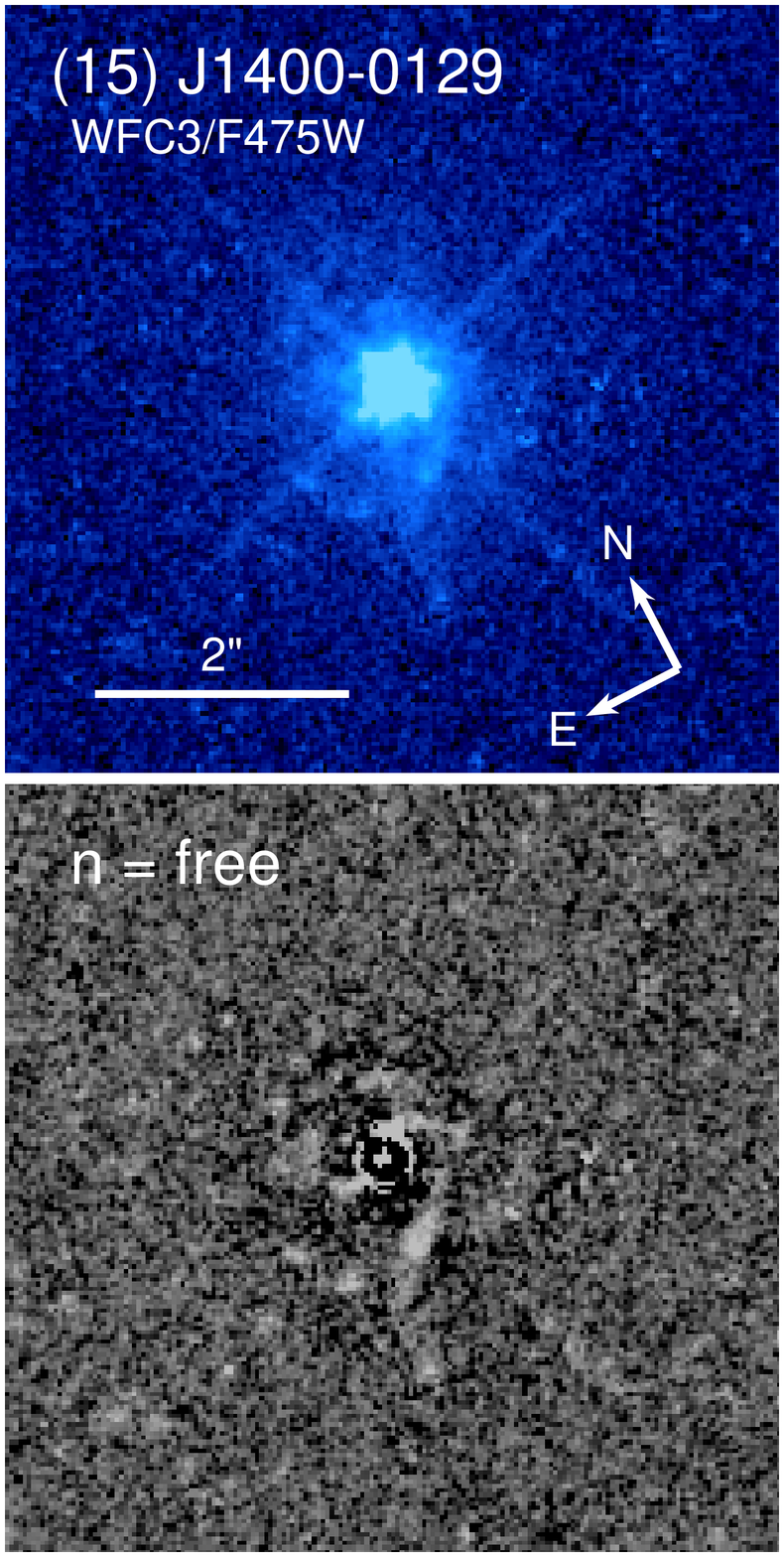,width=0.233\linewidth,clip=} & \epsfig{file=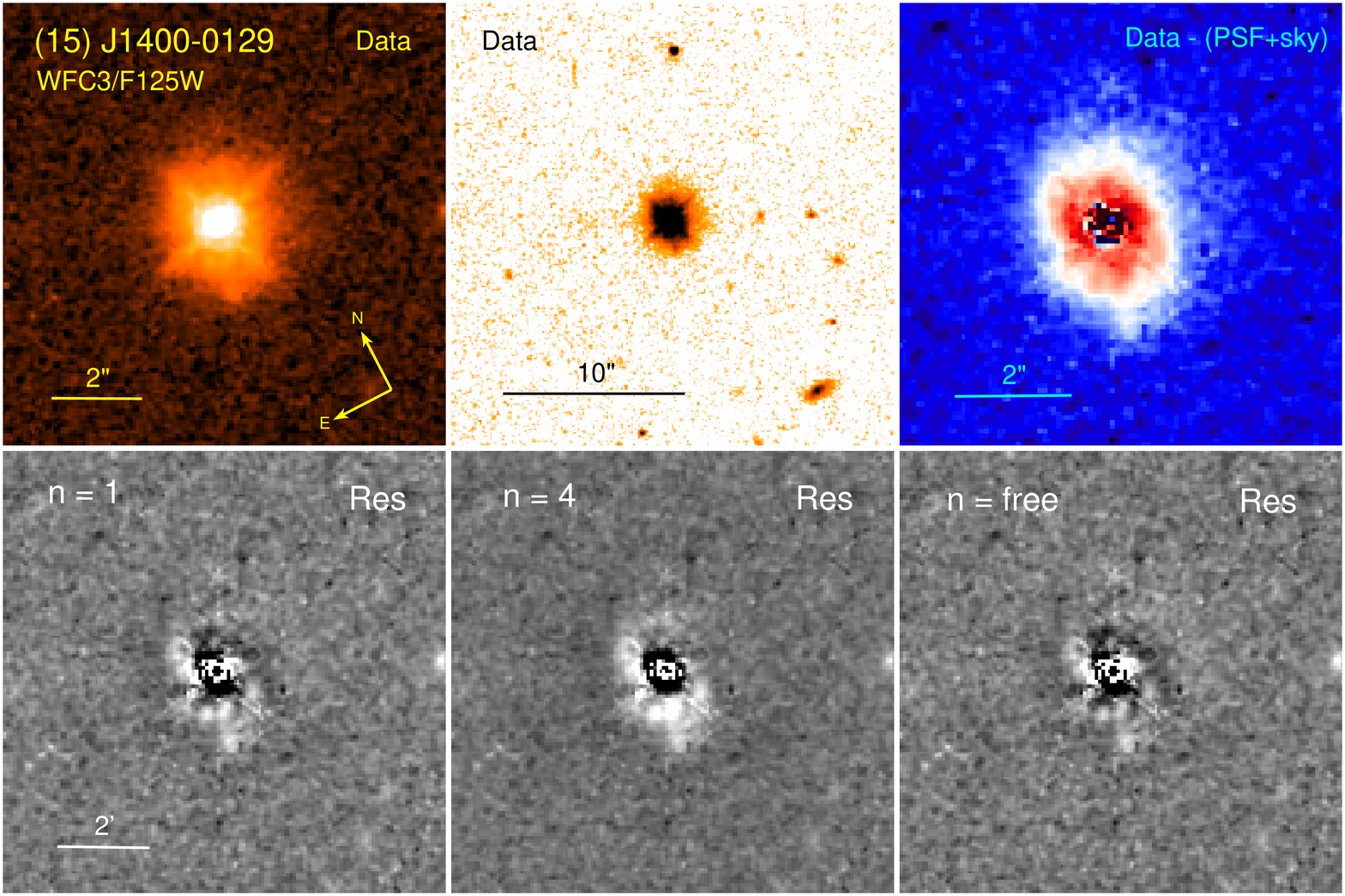,width=0.70\linewidth,clip=} \\
\epsfig{file=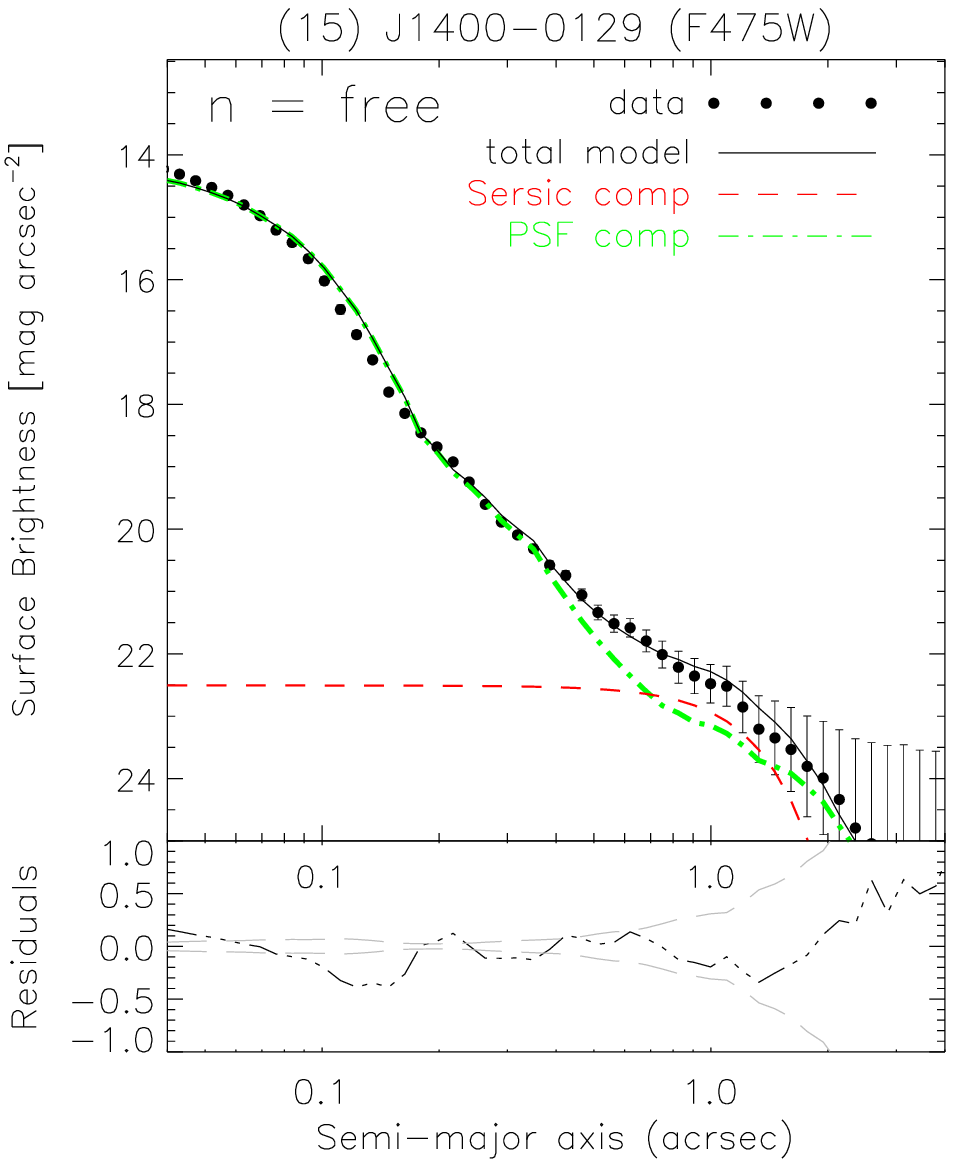,width=0.265\linewidth,clip=} &\epsfig{file=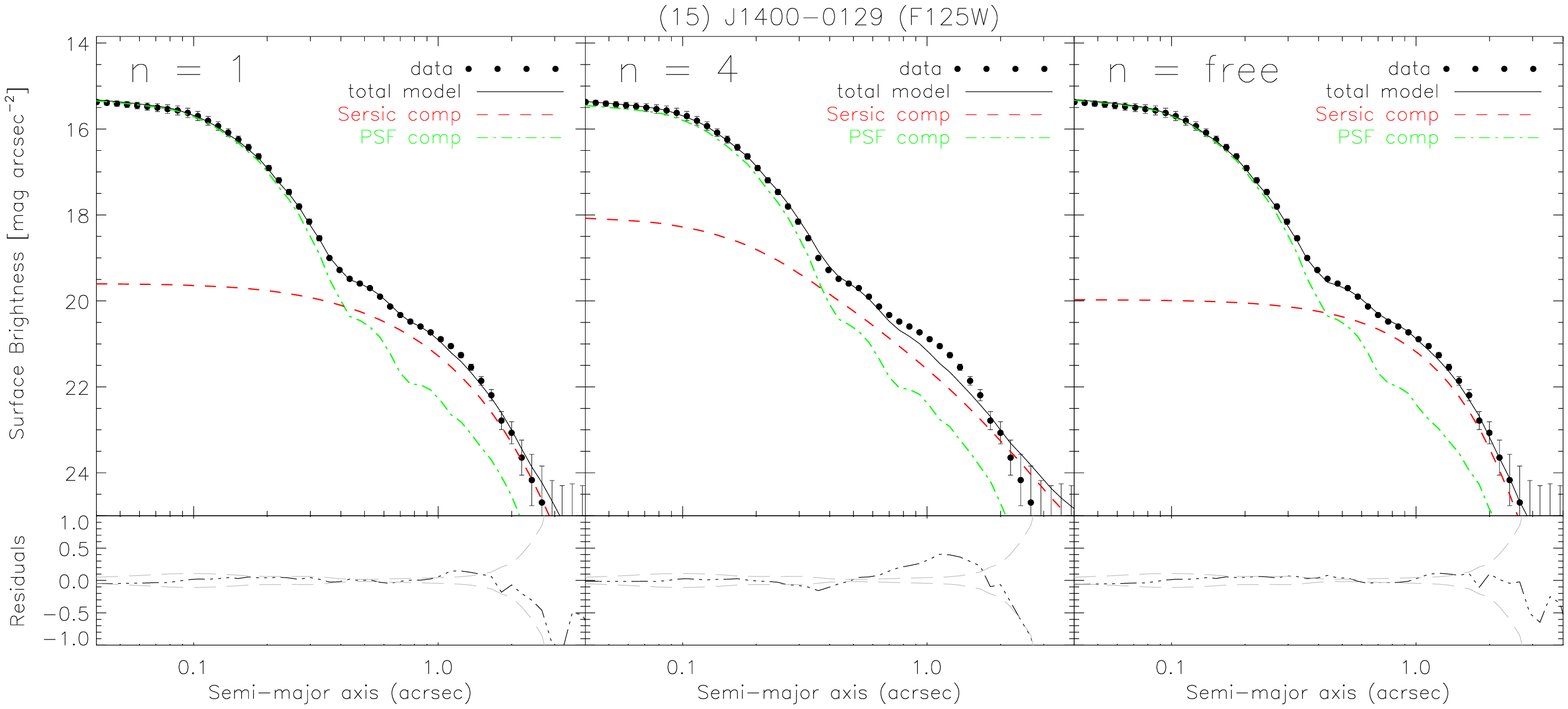,width=0.715\linewidth,clip=}
\end{tabular}
\caption{Object SDSS J1400-0129. Caption, as in Fig. \ref{fig:images1}.}
\label{fig:images15}
\end{figure}

\begin{figure}
\centering
\begin{tabular}{cc}
\epsfig{file=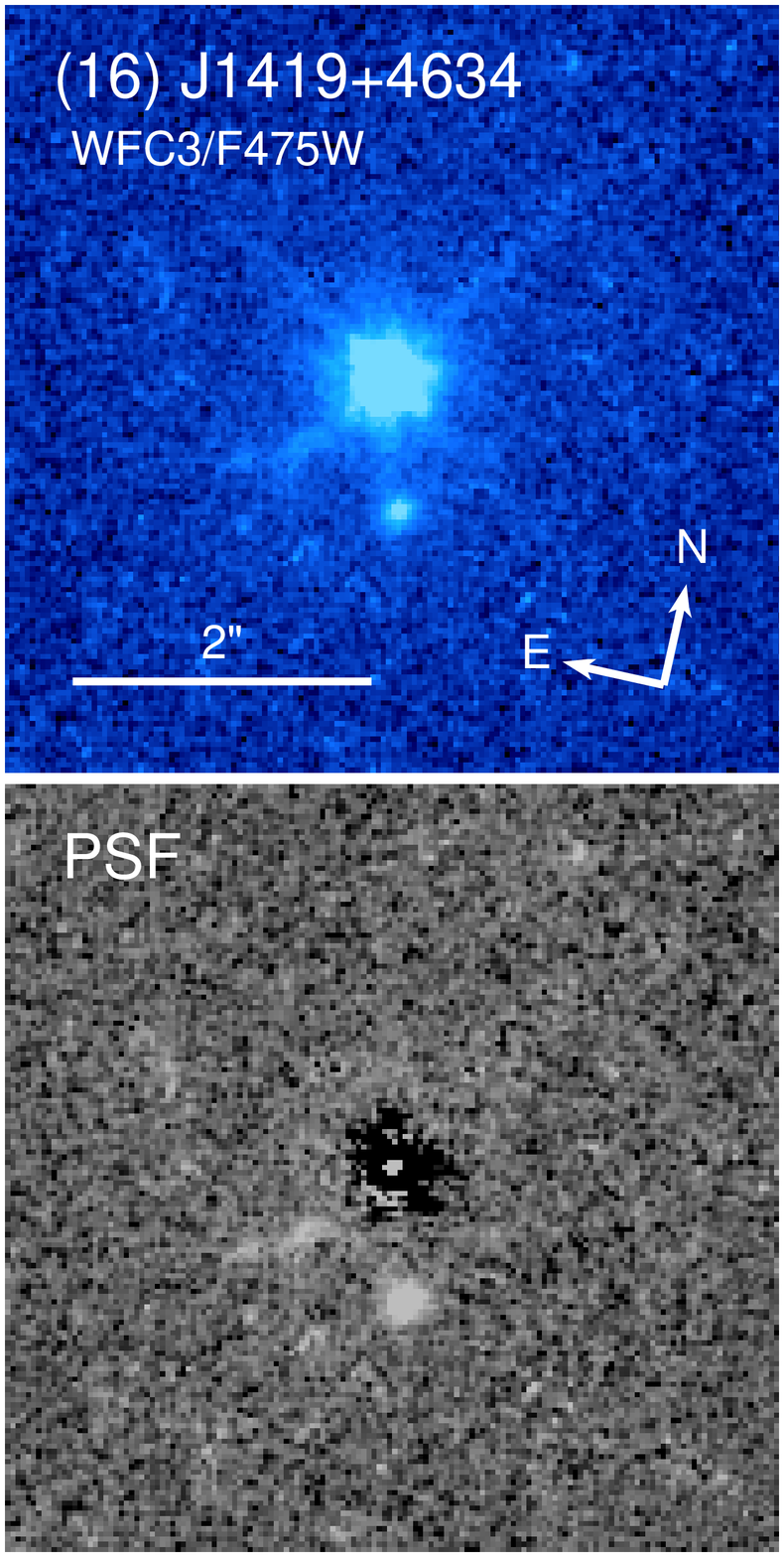,width=0.233\linewidth,clip=} & \epsfig{file=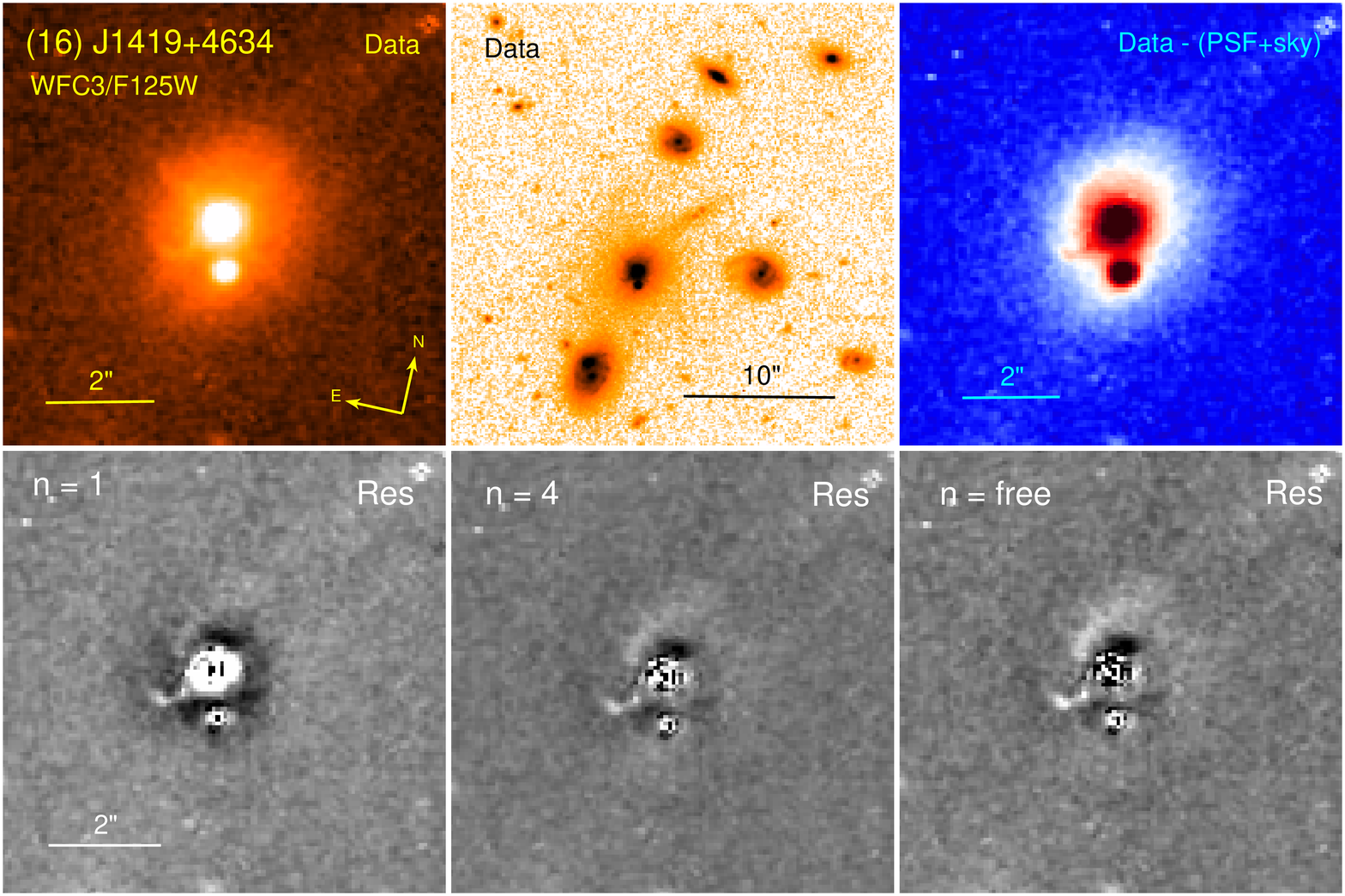,width=0.70\linewidth,clip=} \\
\epsfig{file=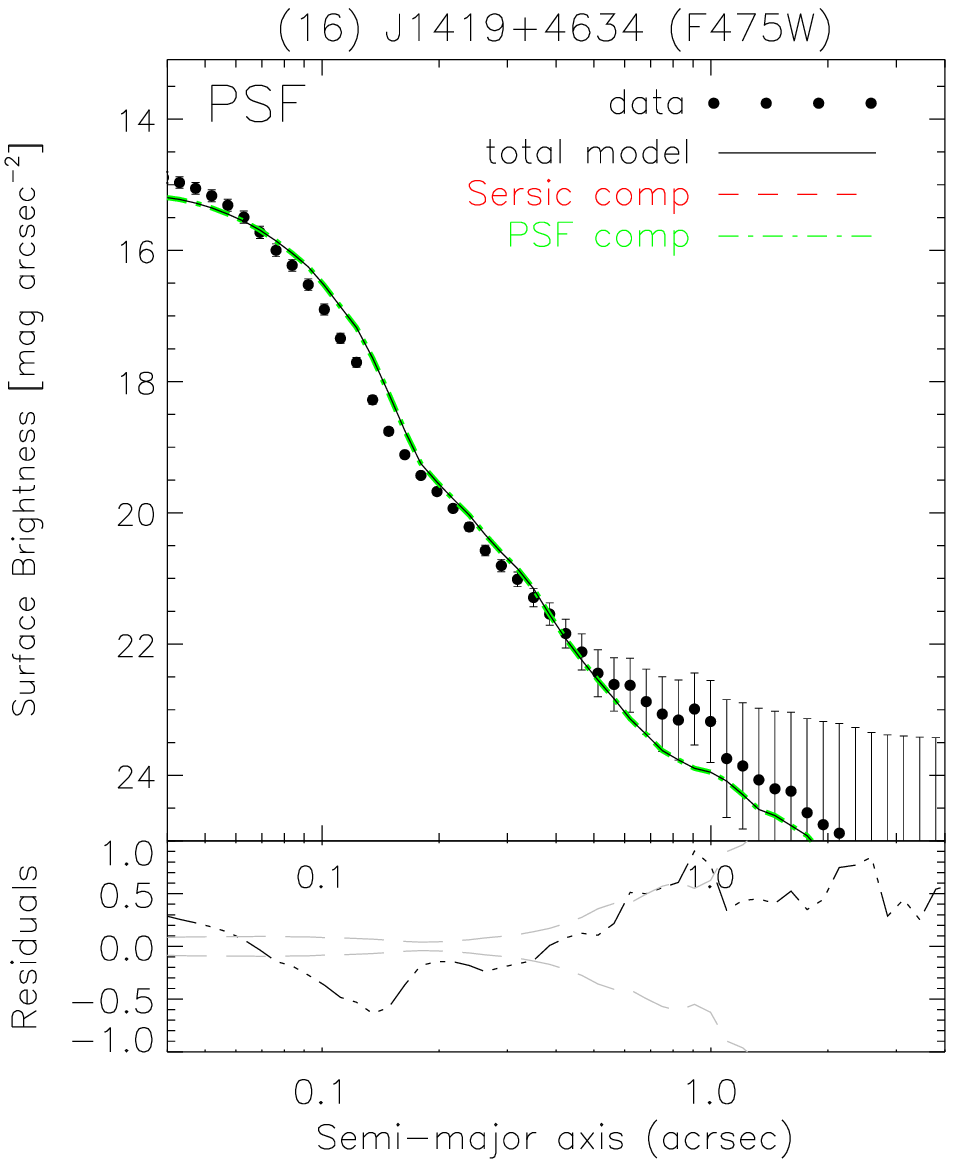,width=0.265\linewidth,clip=} &\epsfig{file=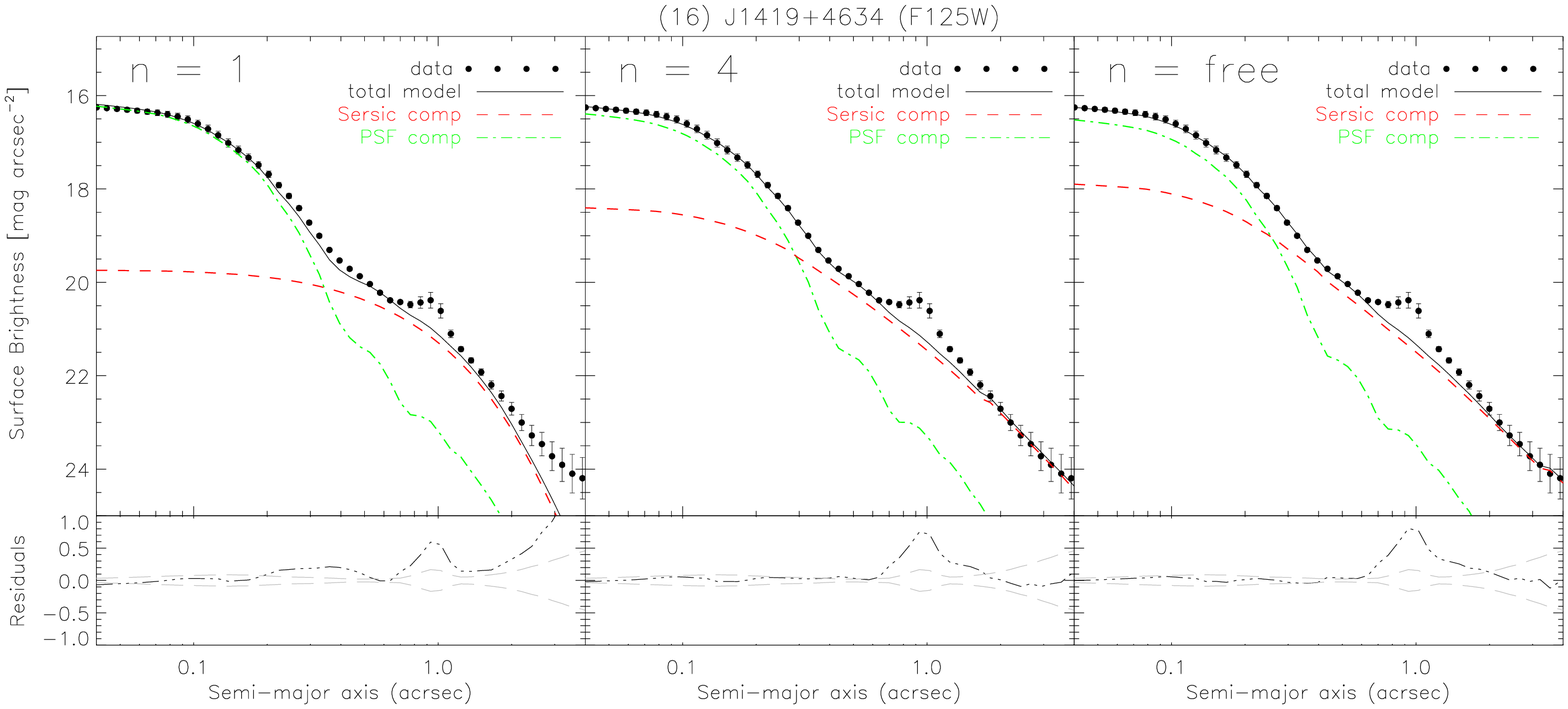,width=0.715\linewidth,clip=}
\end{tabular}
\caption{Object SDSS J1419+4634. Caption, as in Fig. \ref{fig:images1}. Four of the nearby galaxies seen in the middle panel are at a similar photometric redshift: the pair of merging galaxies at $\sim$6$\farcs$8S (z$\sim$0.55$\pm$0.06), the distorted spiral galaxy at $\sim$8$\farcs$3W (z$\sim$0.38$\pm$0.16), the compact galaxy at $\sim$19$\farcs$3NNW (z$\sim$0.49$\pm$0.06) and the galaxy at $\sim$14$\arcsec$N (z$\sim$0.60$\pm$0.13).} 
\label{fig:images16}
\end{figure}

\begin{figure}
\centering
\begin{tabular}{cc}
\epsfig{file=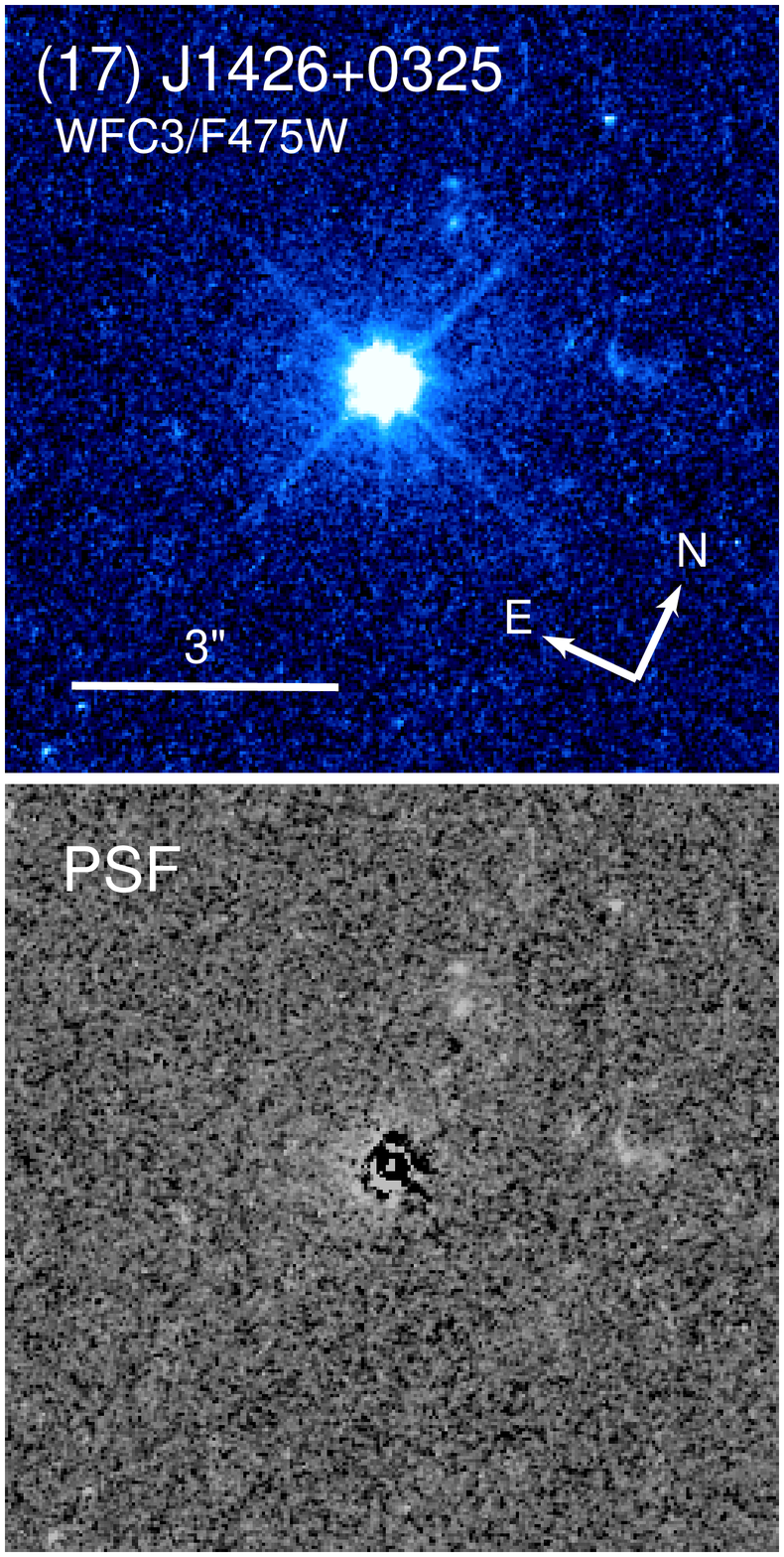,width=0.233\linewidth,clip=} & \epsfig{file=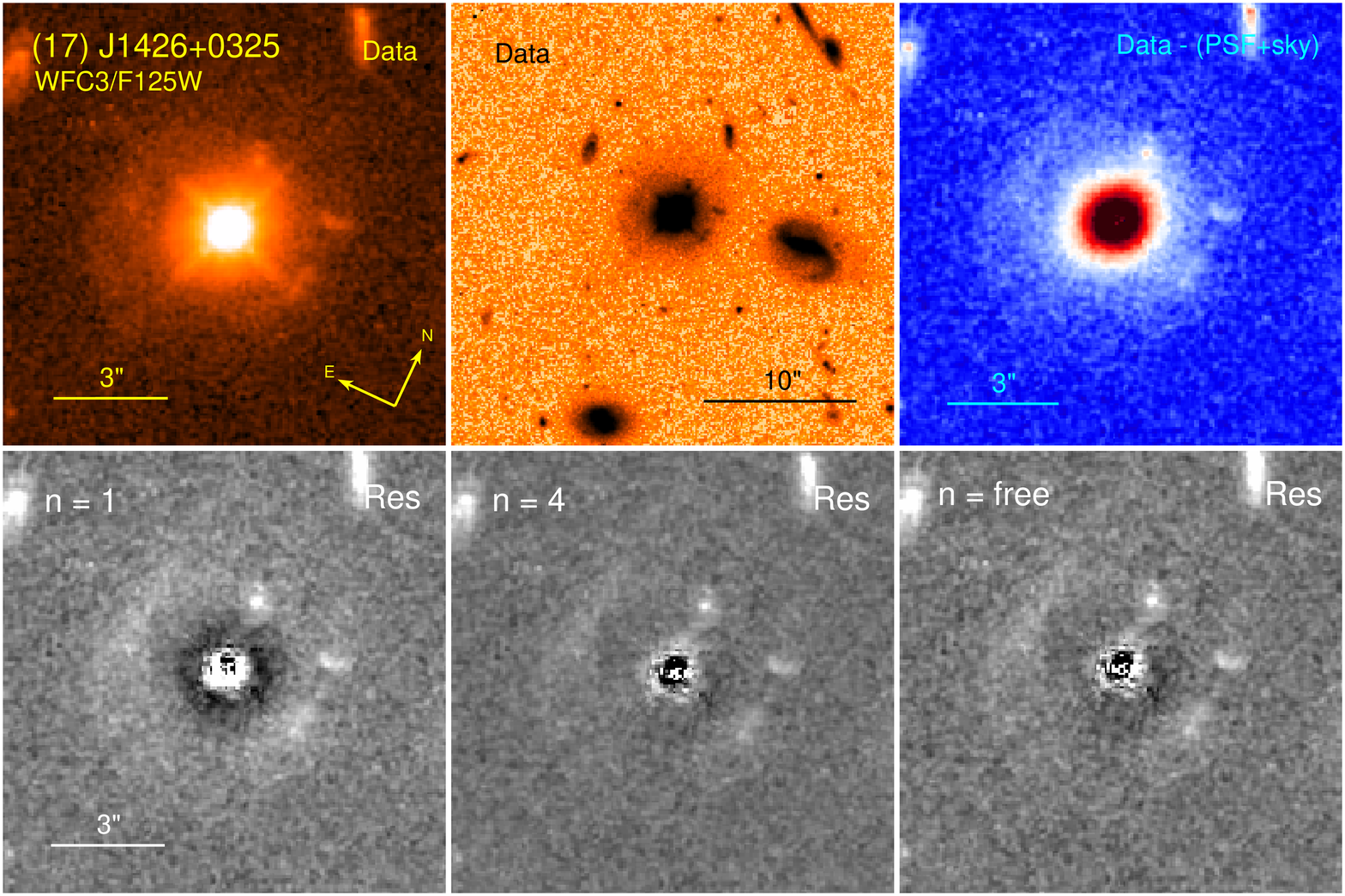,width=0.70\linewidth,clip=} \\
\epsfig{file=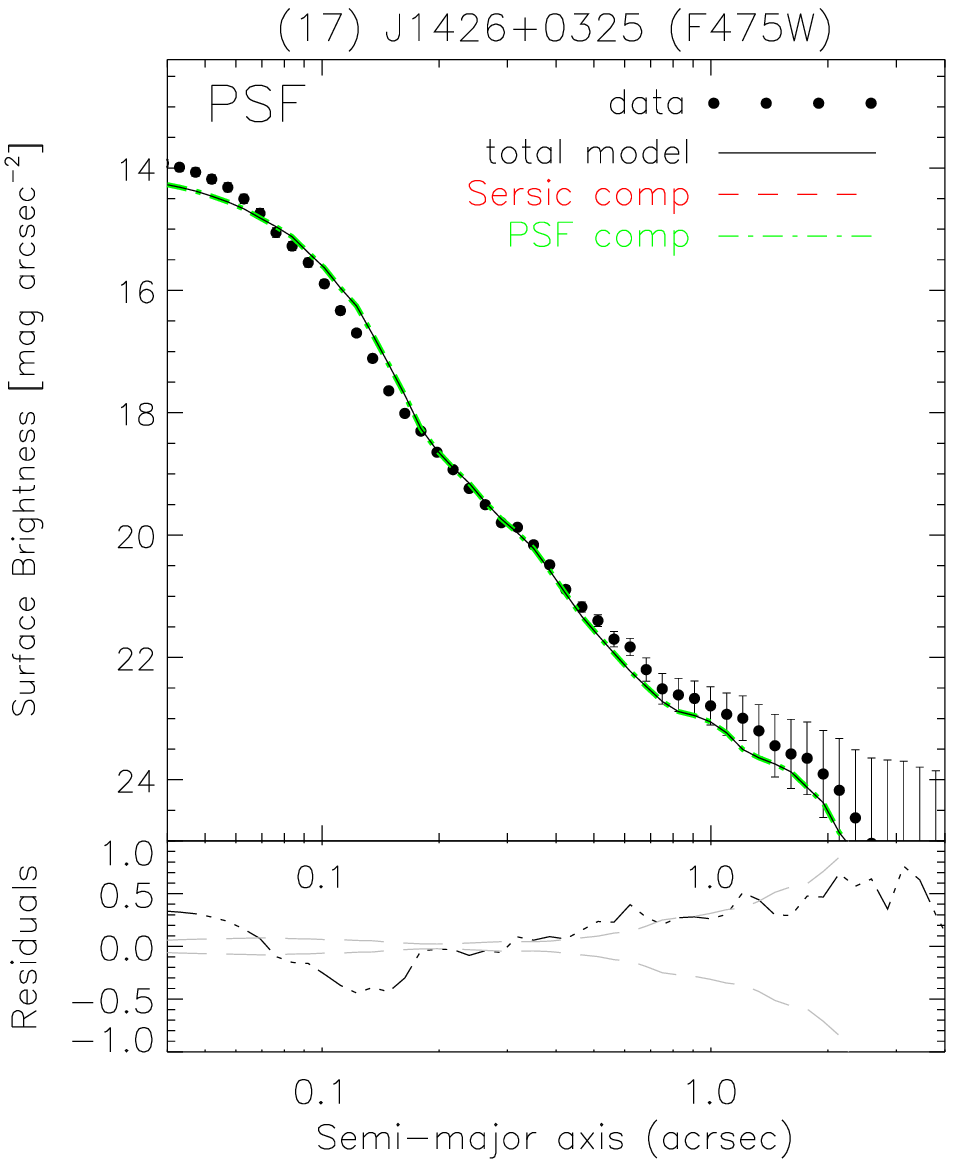,width=0.265\linewidth,clip=} &\epsfig{file=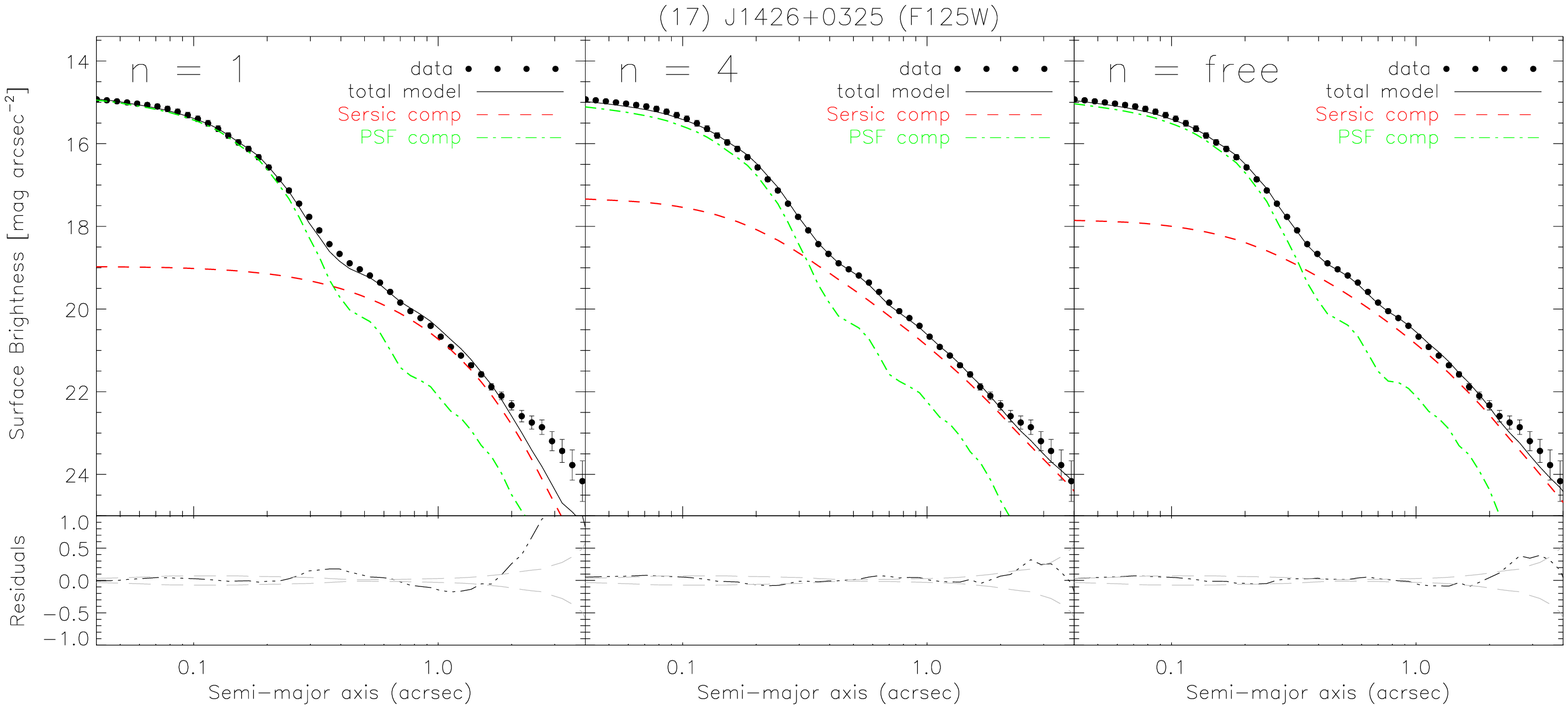,width=0.715\linewidth,clip=}
\end{tabular}
\caption{Object SDSS J1426+0325. Caption, as in Fig. \ref{fig:images1}.}
\label{fig:images17}
\end{figure}

\begin{figure}
\centering
\begin{tabular}{cc}
\epsfig{file=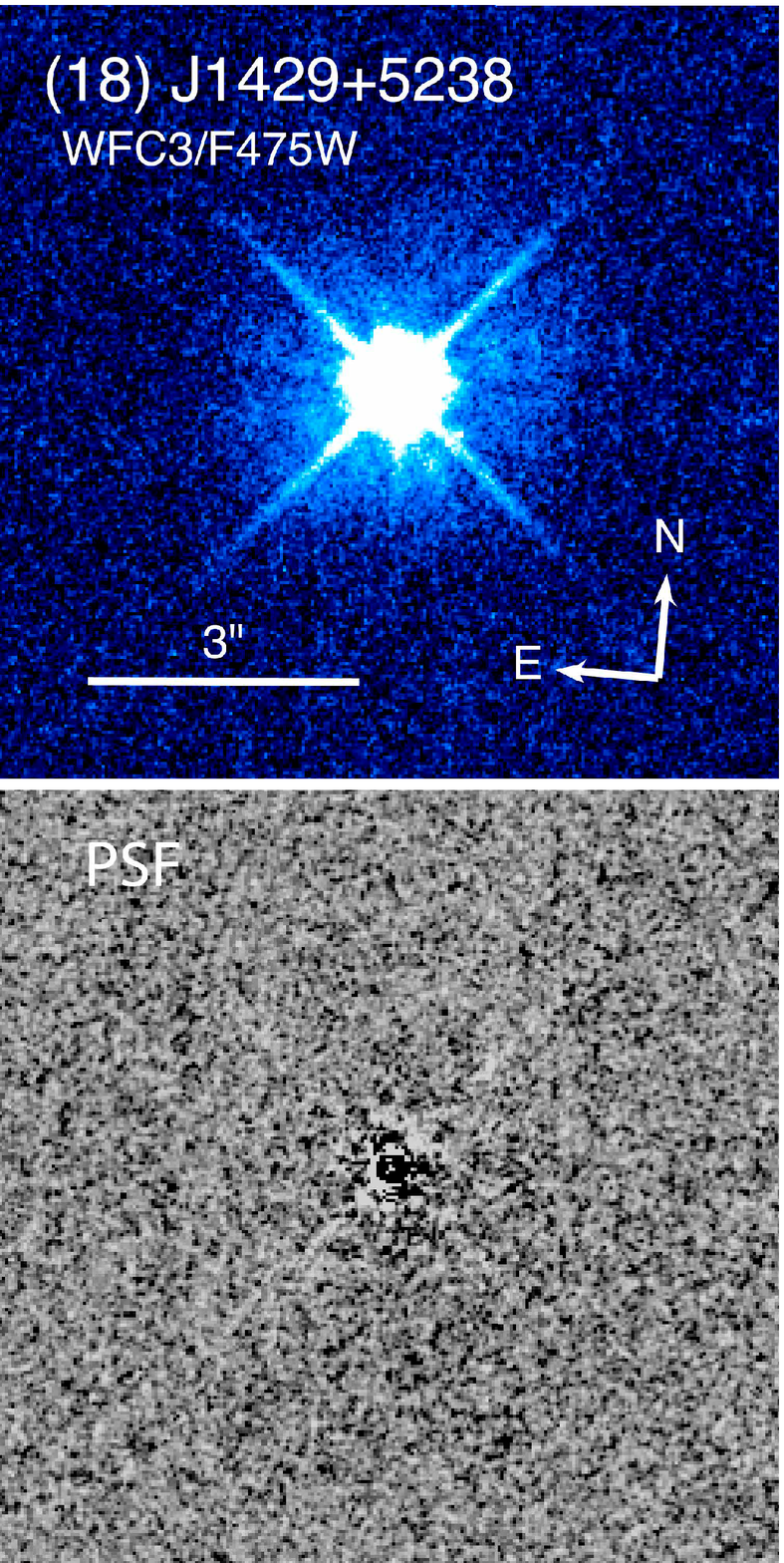,width=0.233\linewidth,clip=} & \epsfig{file=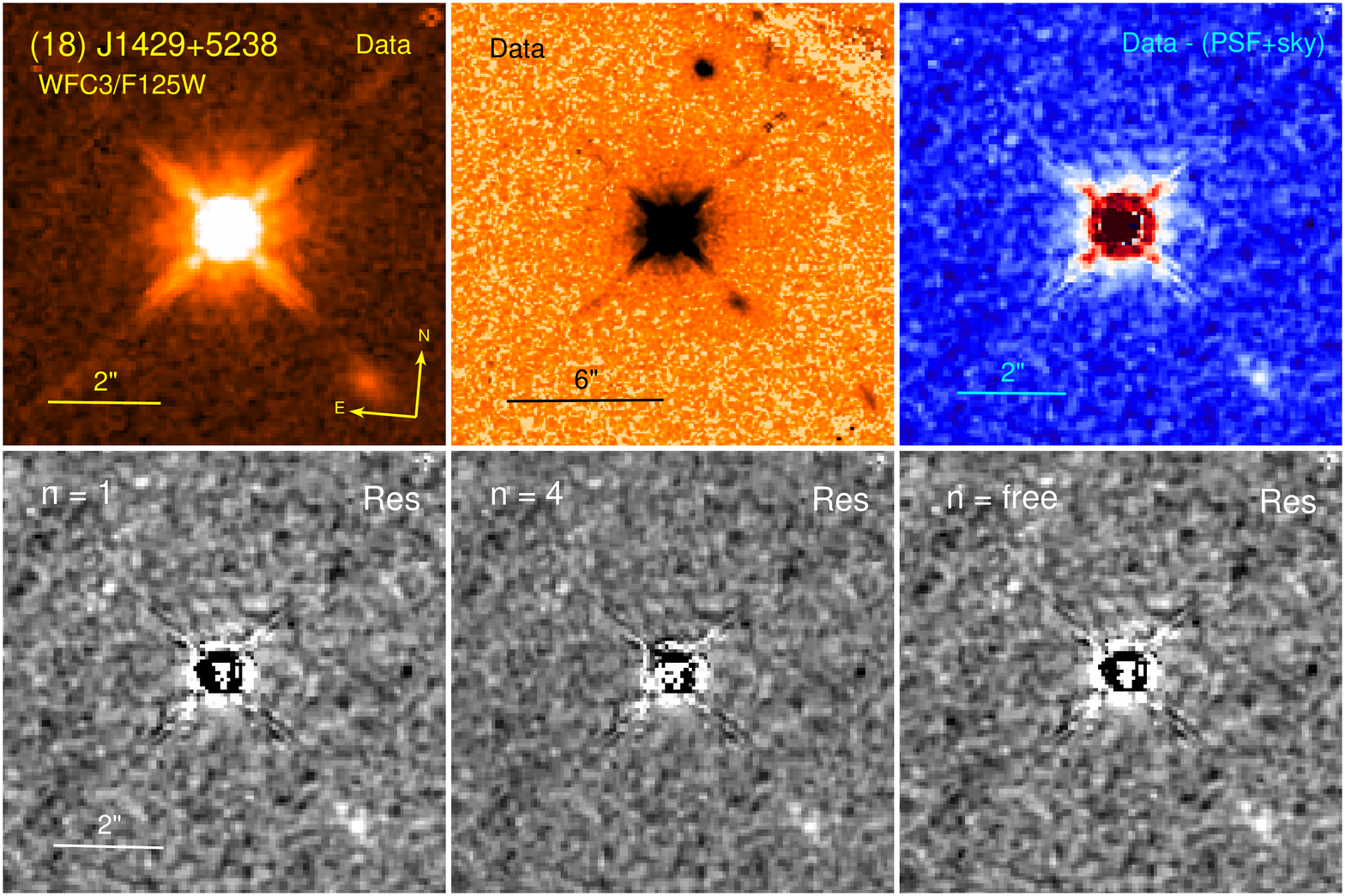,width=0.70\linewidth,clip=} \\
\epsfig{file=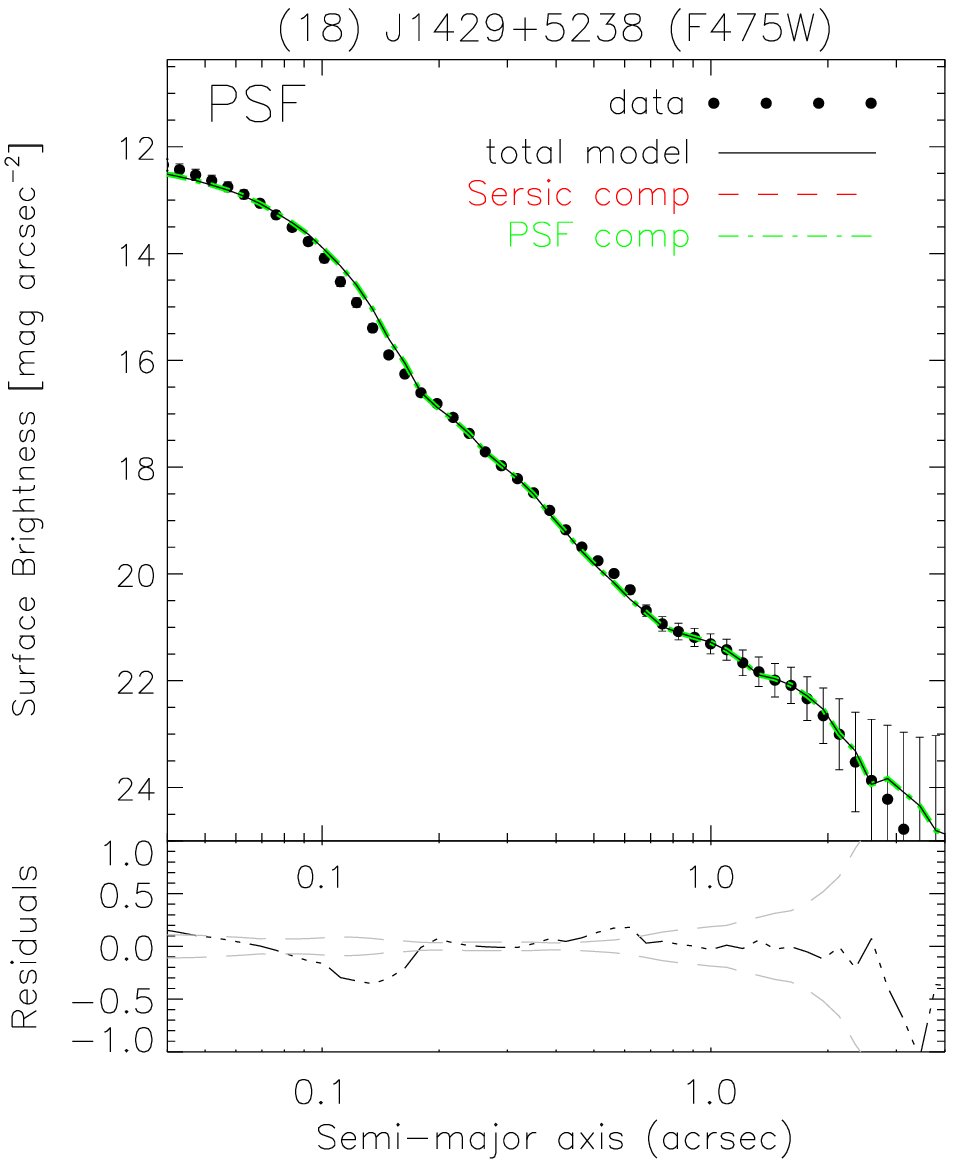,width=0.265\linewidth,clip=} &\epsfig{file=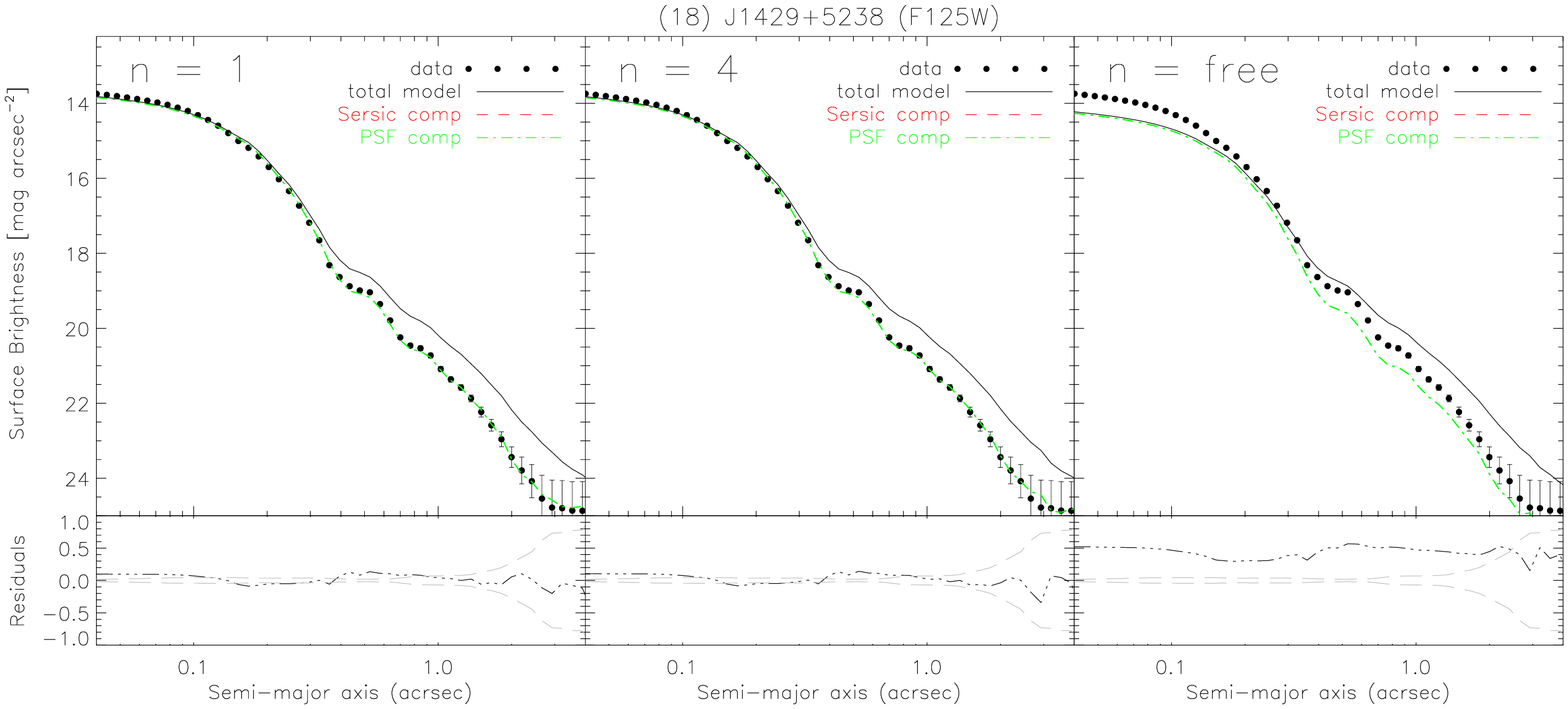,width=0.715\linewidth,clip=}
\end{tabular}
\caption{Object SDSS J1429+5238. Caption, as in Fig. \ref{fig:images1}.}
\label{fig:images18}
\end{figure}

\begin{figure}
\centering
\begin{tabular}{cc}
\epsfig{file=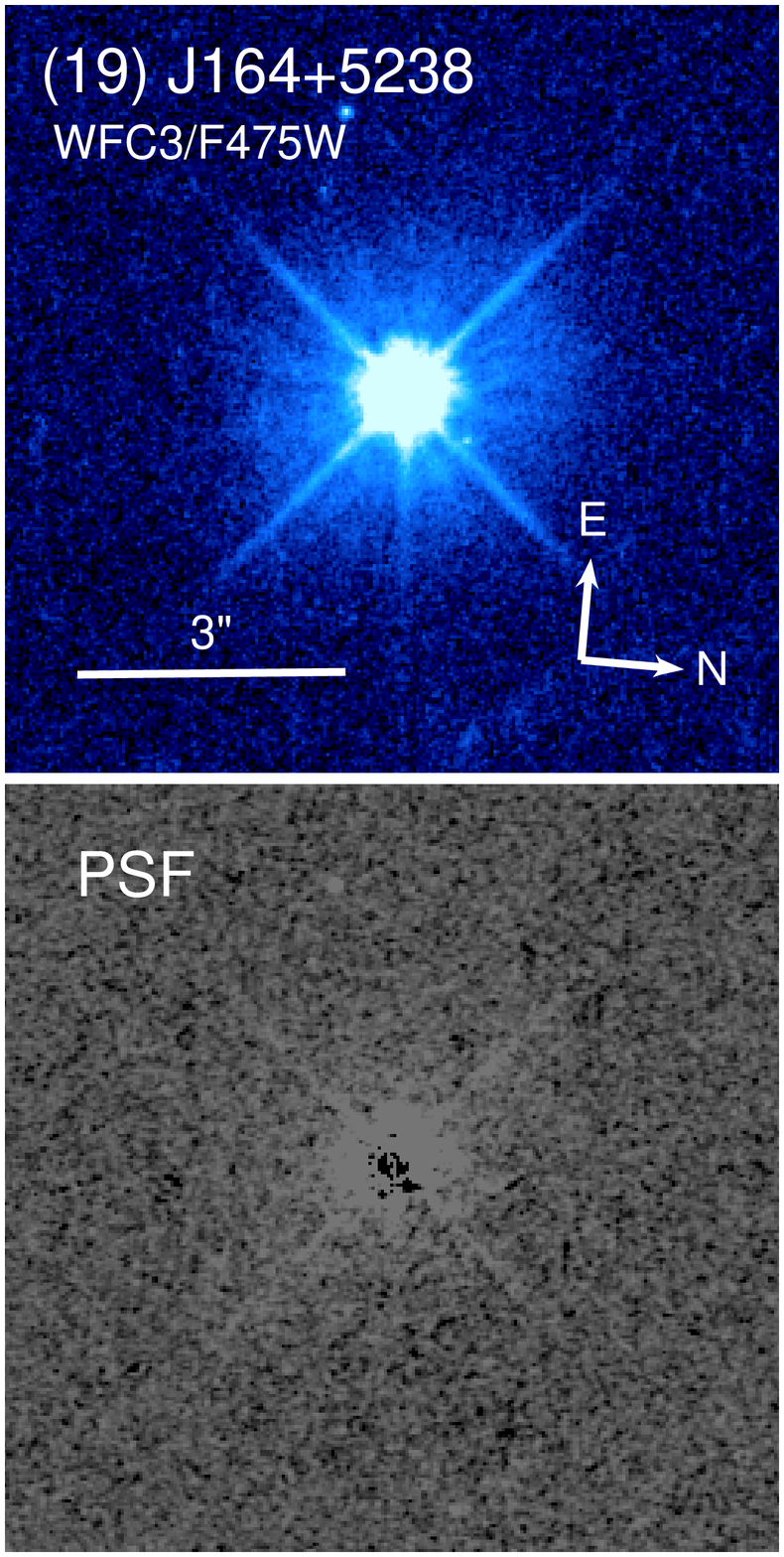,width=0.233\linewidth,clip=} & \epsfig{file=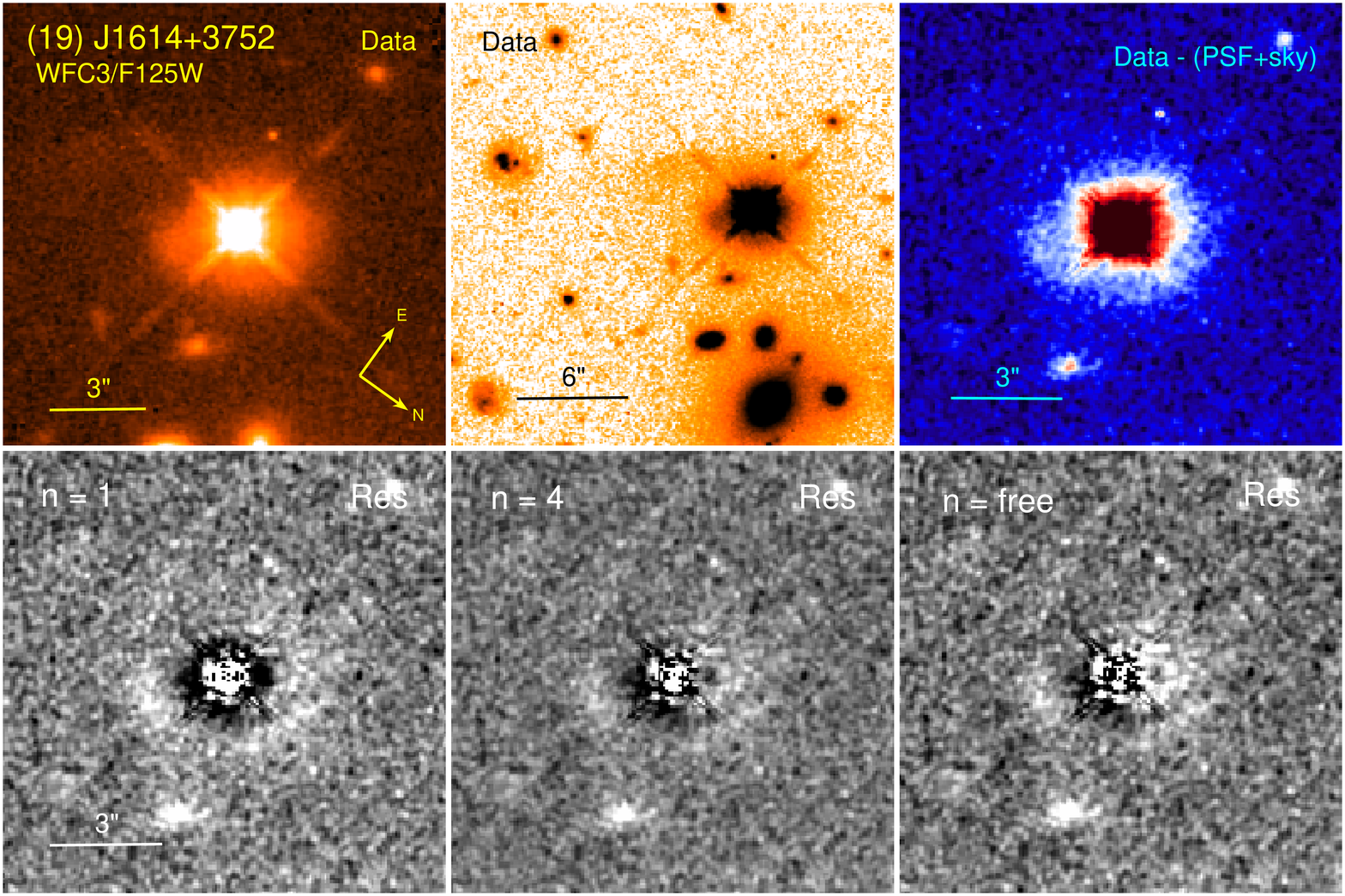,width=0.70\linewidth,clip=} \\
\epsfig{file=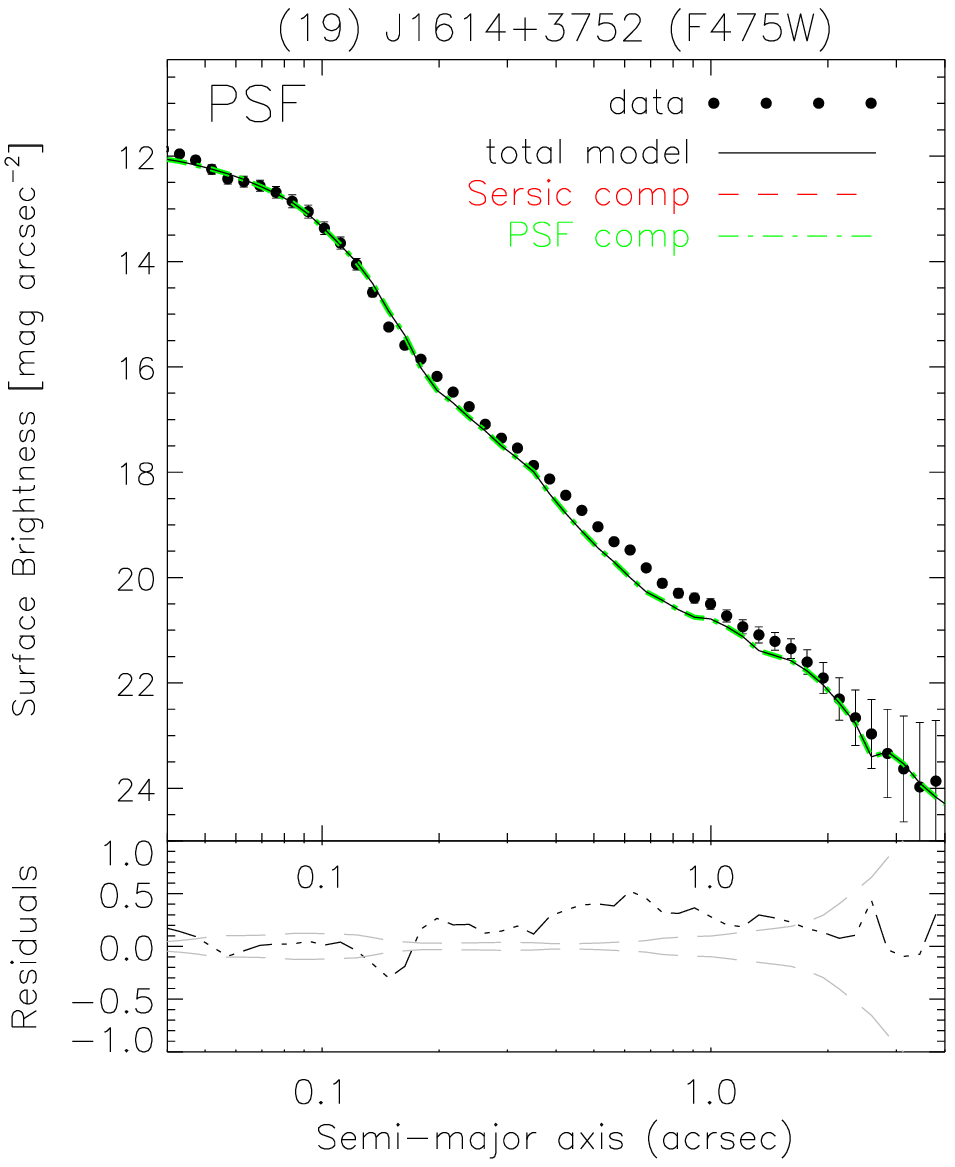,width=0.265\linewidth,clip=} &\epsfig{file=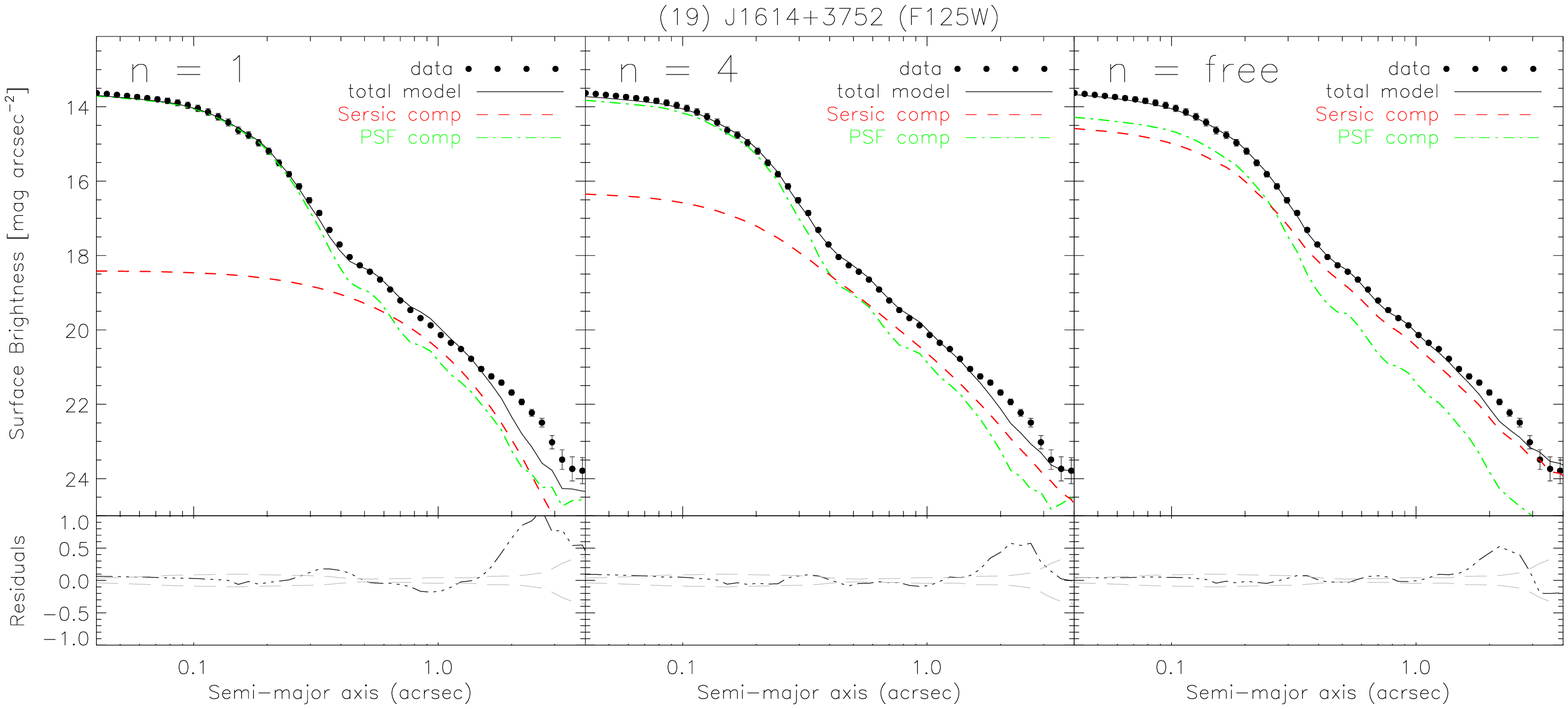,width=0.715\linewidth,clip=}
\end{tabular}
\caption{Object SDSS J1614+5238. Caption, as in Fig. \ref{fig:images1}.}
\label{fig:images19}
\end{figure}

\begin{figure}
\centering
\begin{tabular}{cc}
\epsfig{file=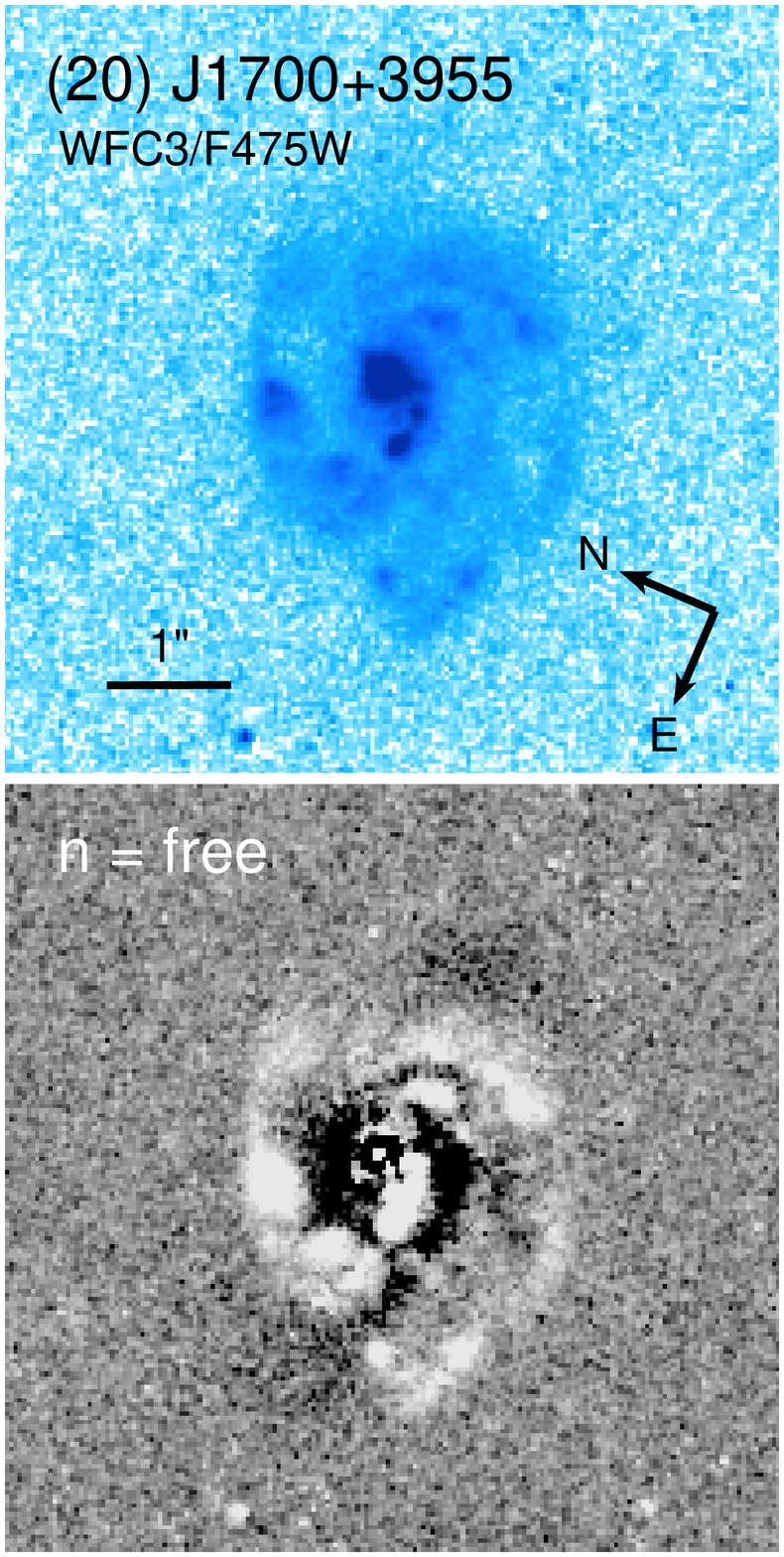,width=0.233\linewidth,clip=} & \epsfig{file=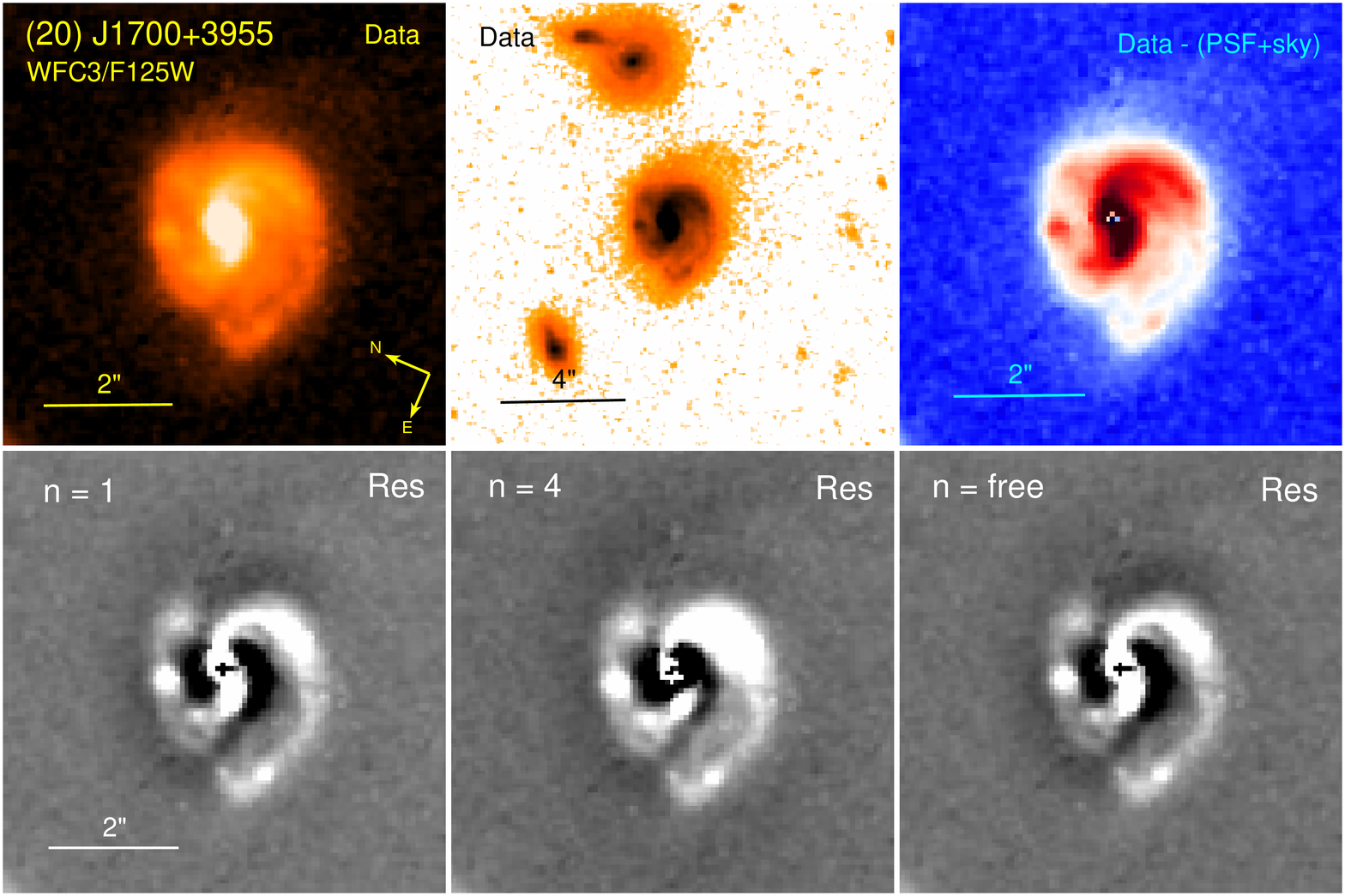,width=0.70\linewidth,clip=} \\
\epsfig{file=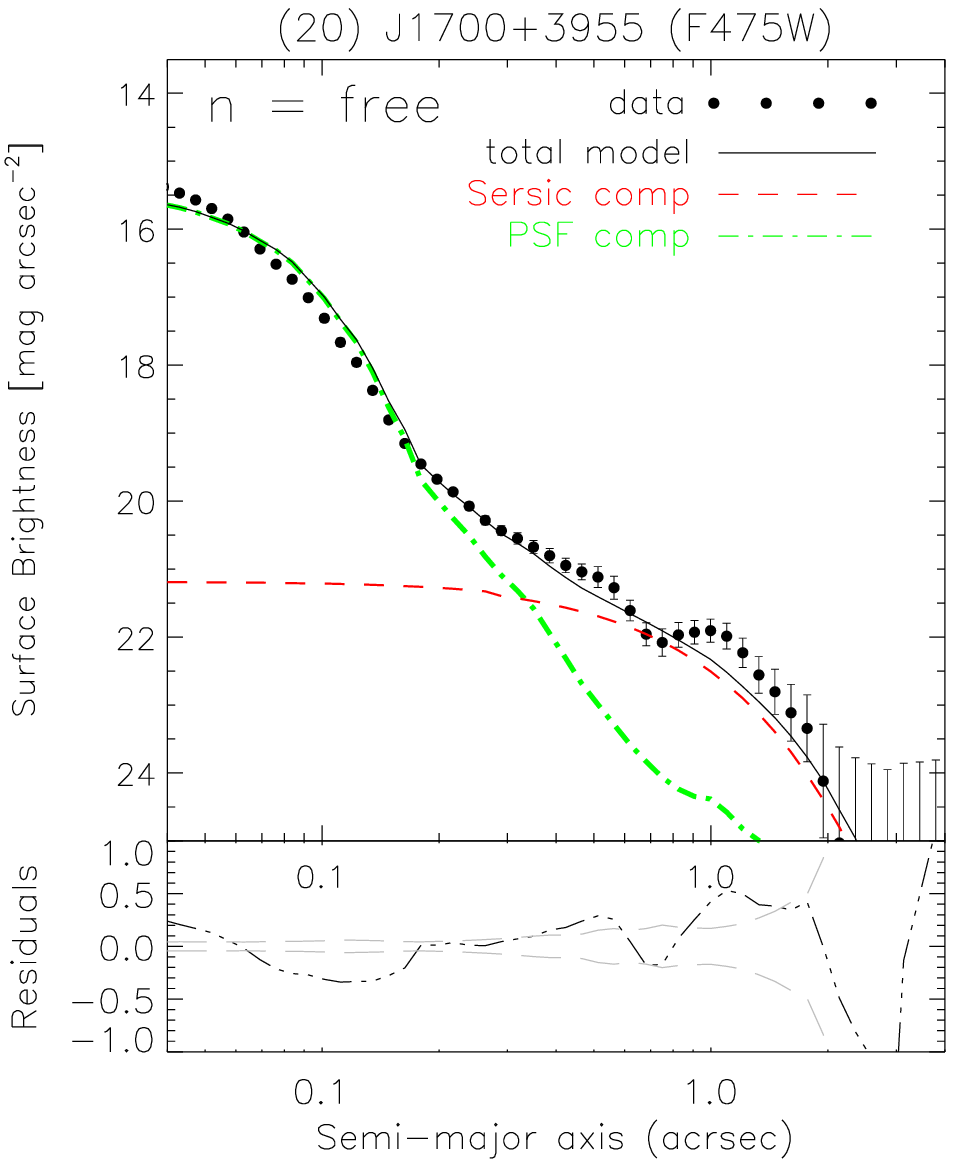,width=0.265\linewidth,clip=} &\epsfig{file=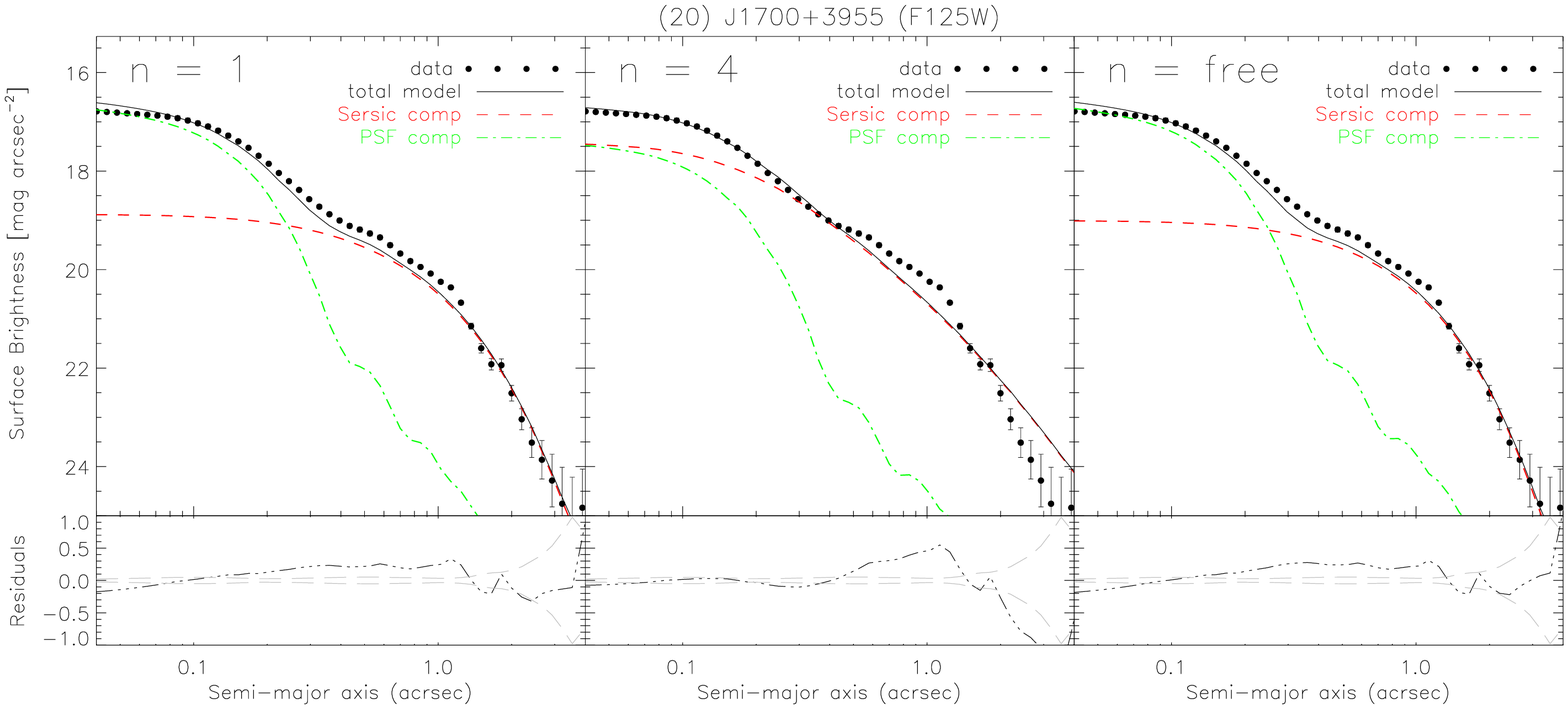,width=0.715\linewidth,clip=}
\end{tabular}
\caption{Object SDSS J1700+3955. Caption, as in Fig. \ref{fig:images1}.}
\label{fig:images20}
\end{figure}

\begin{figure}
\centering
\begin{tabular}{cc}
\epsfig{file=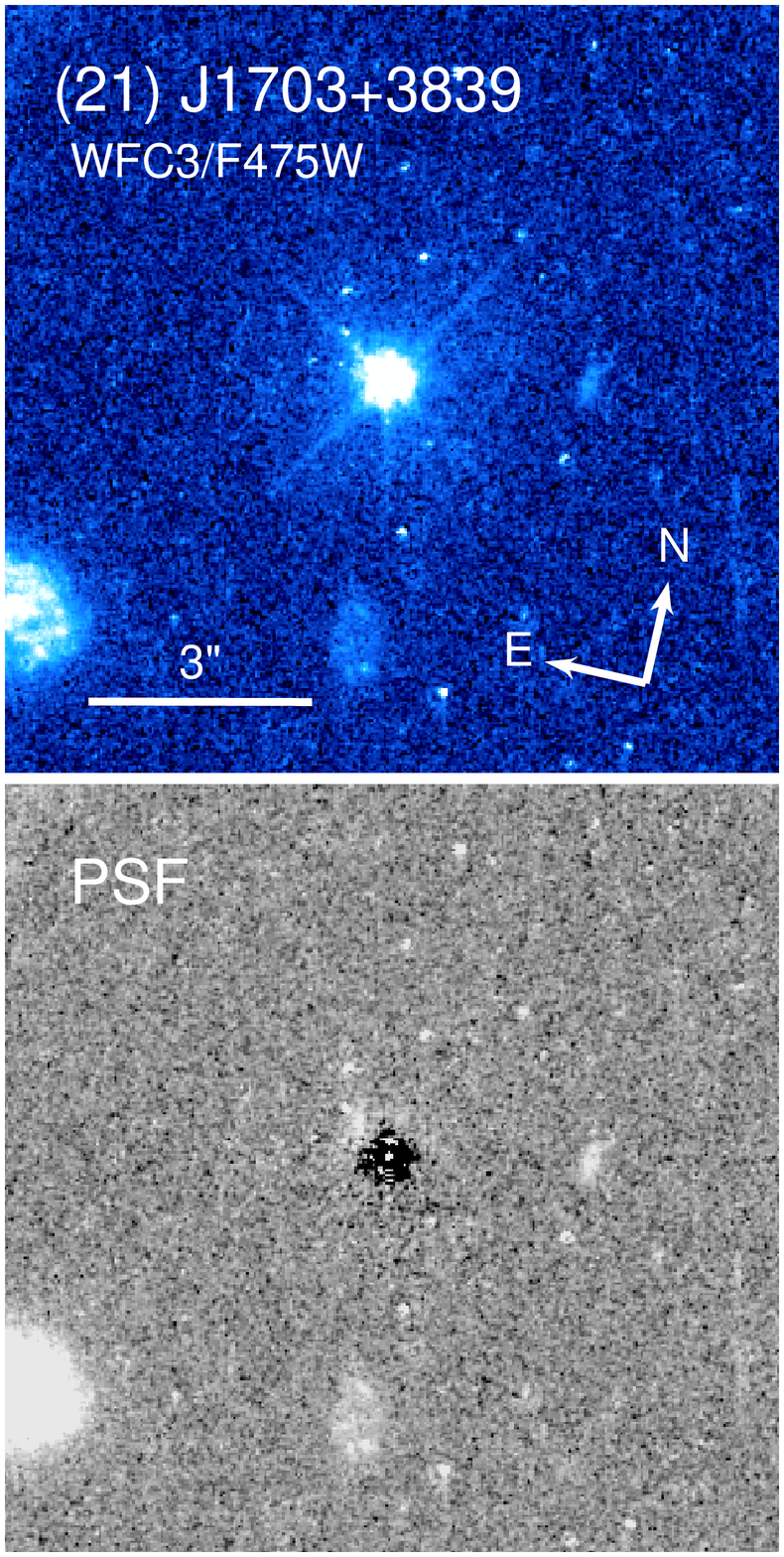,width=0.233\linewidth,clip=} & \epsfig{file=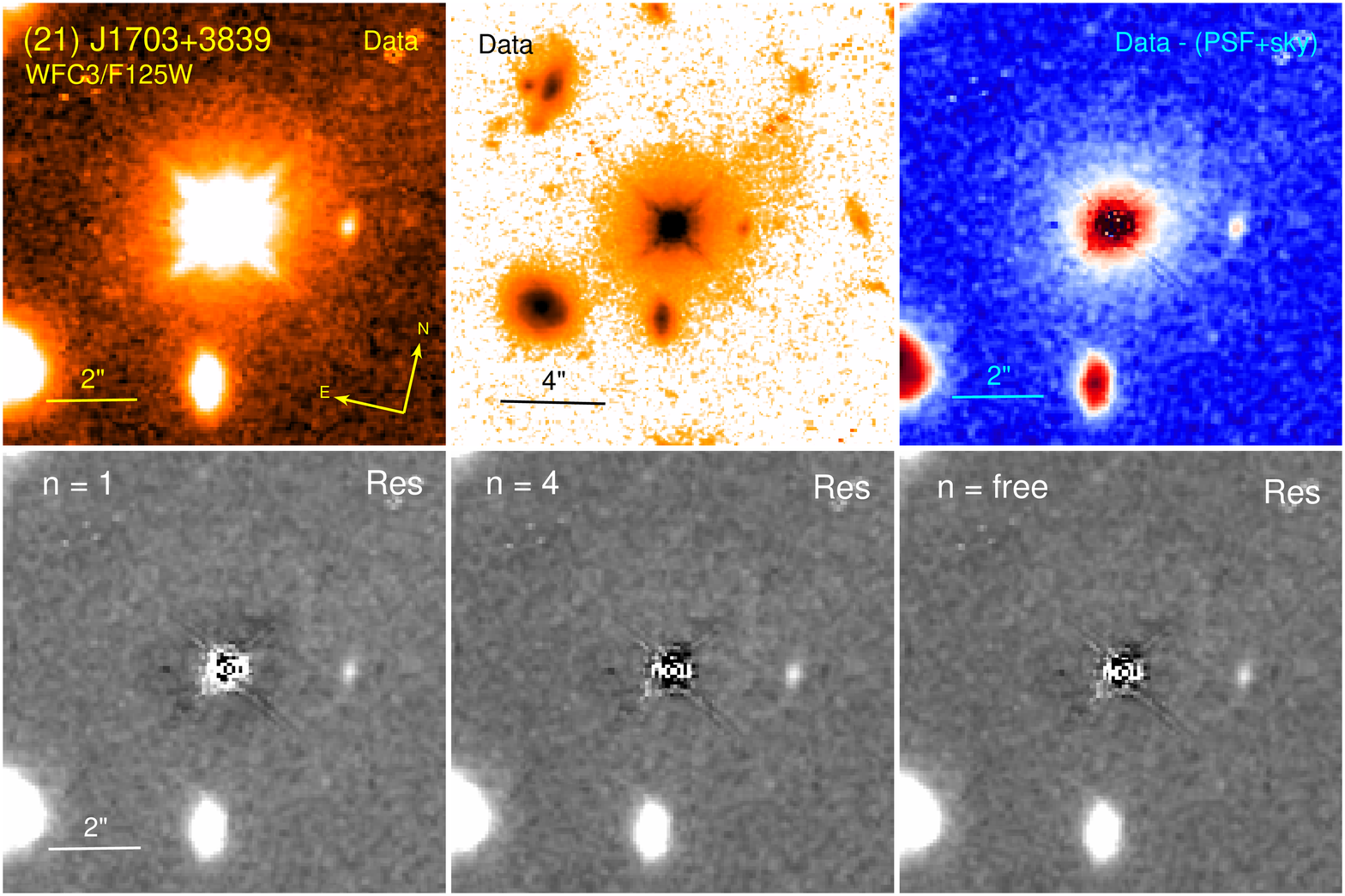,width=0.70\linewidth,clip=} \\
\epsfig{file=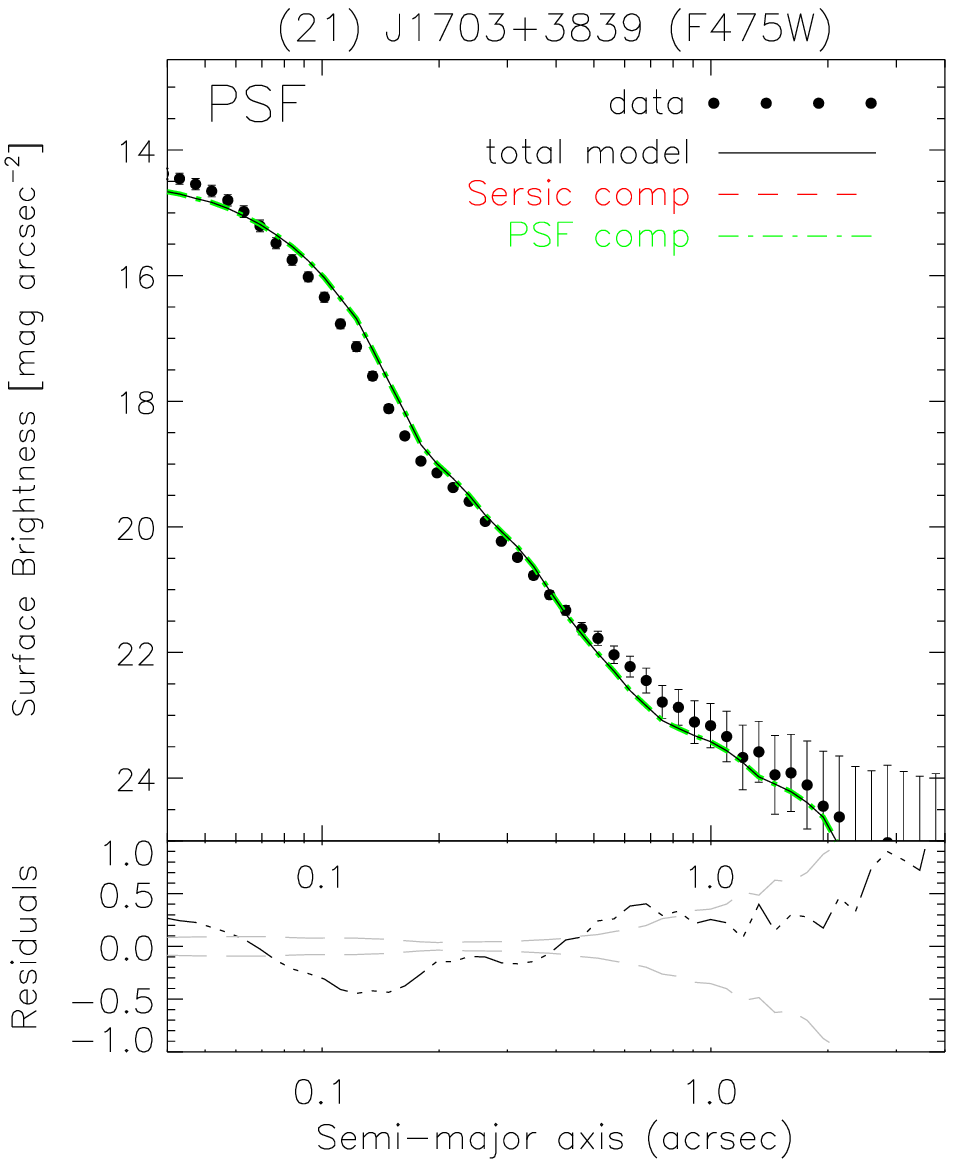,width=0.265\linewidth,clip=} &\epsfig{file=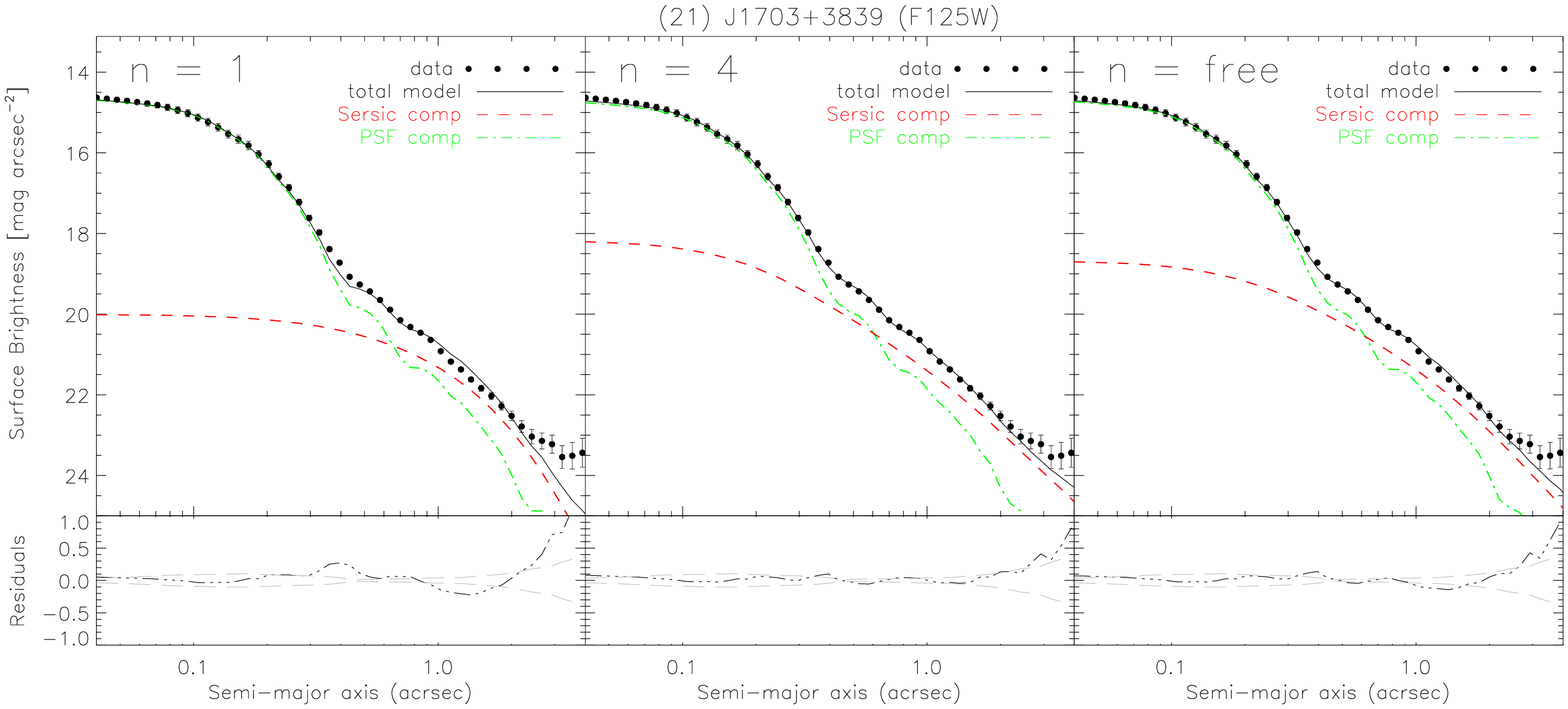,width=0.715\linewidth,clip=}
\end{tabular}
\caption{Object SDSS J1703+3839. Caption, as in Fig. \ref{fig:images1}.}
\label{fig:images21}
\end{figure}

\begin{figure}
\centering
\begin{tabular}{cc}
\epsfig{file=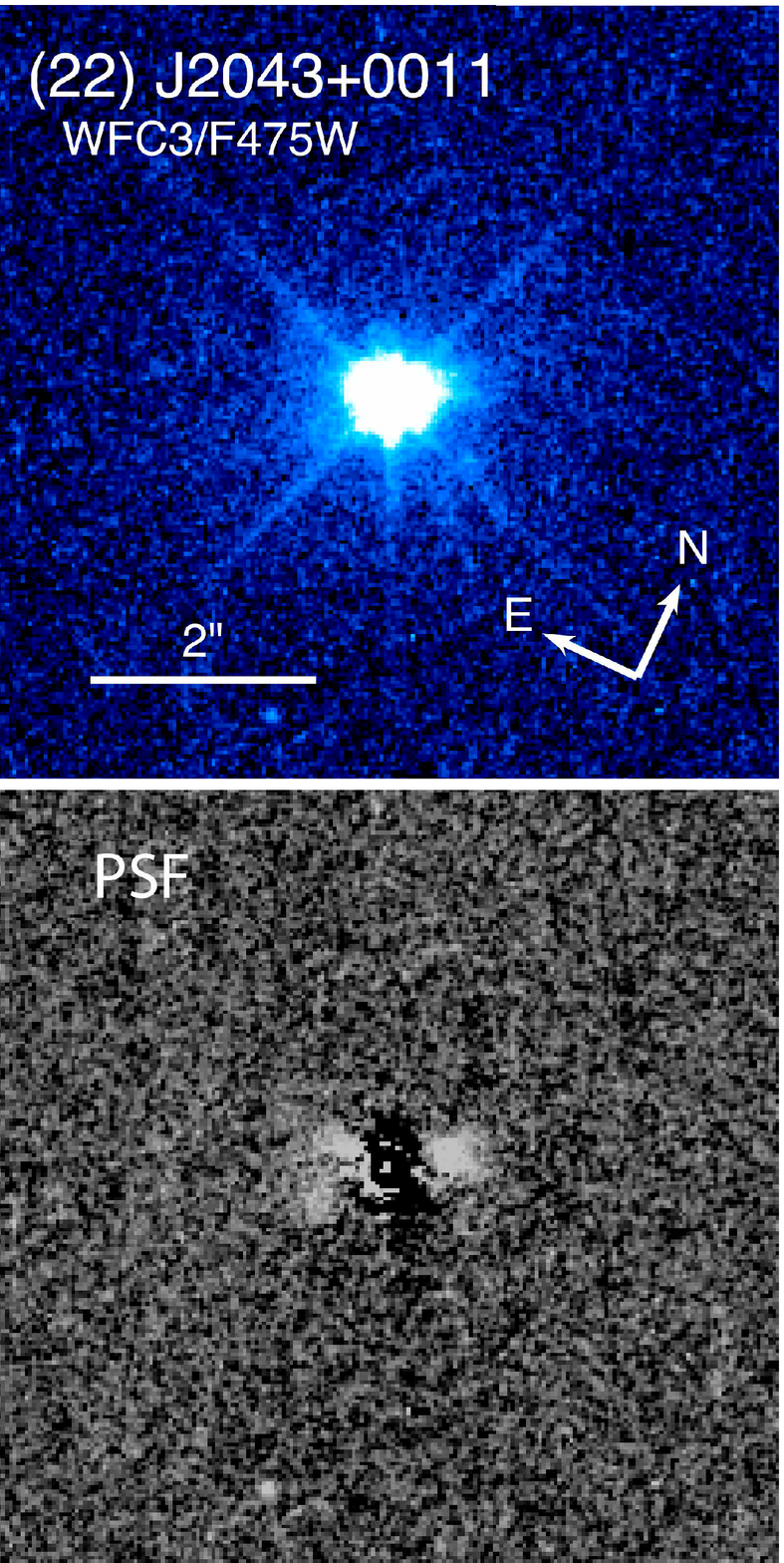,width=0.233\linewidth,clip=} & \epsfig{file=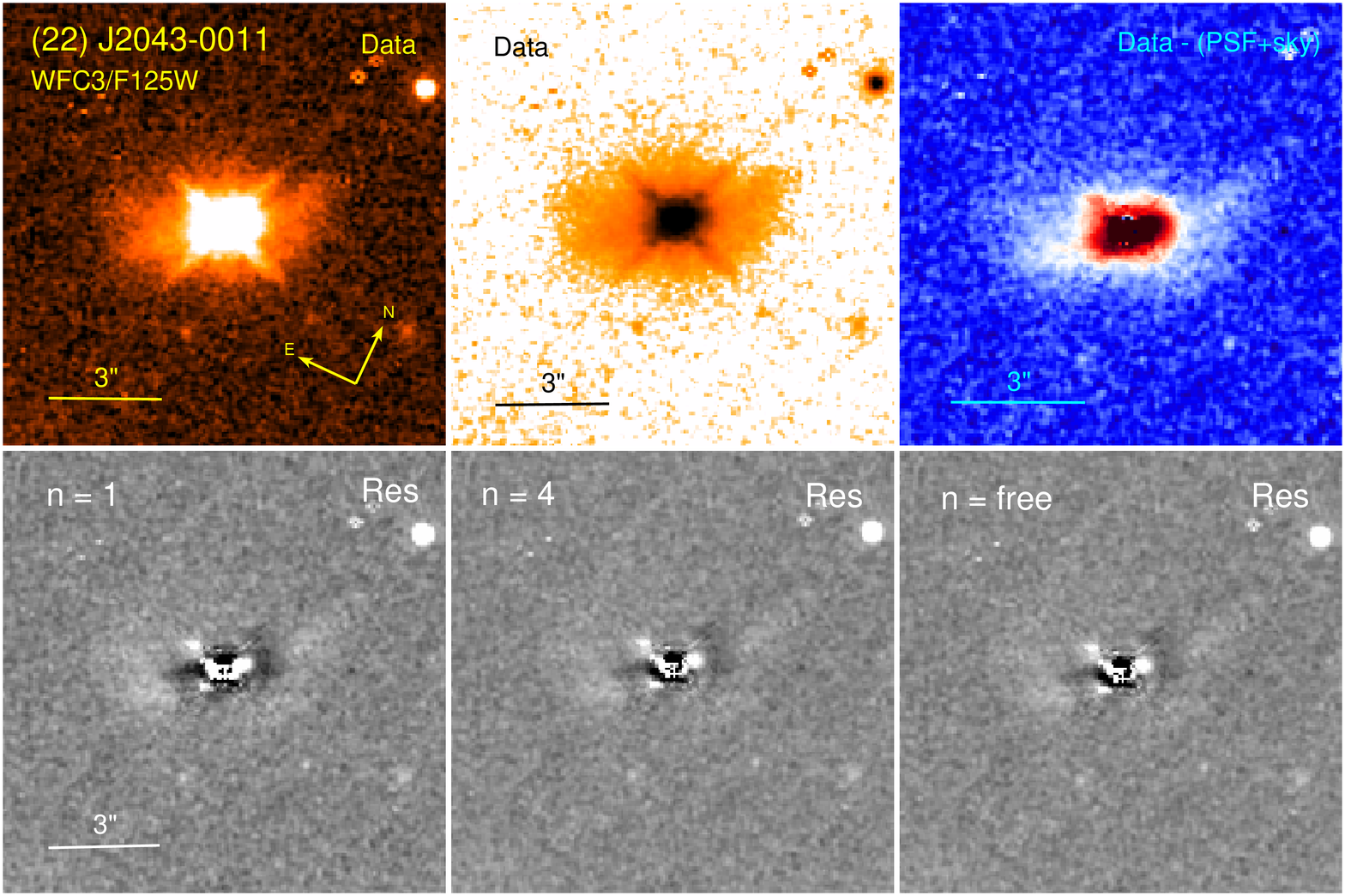,width=0.70\linewidth,clip=} \\
\epsfig{file=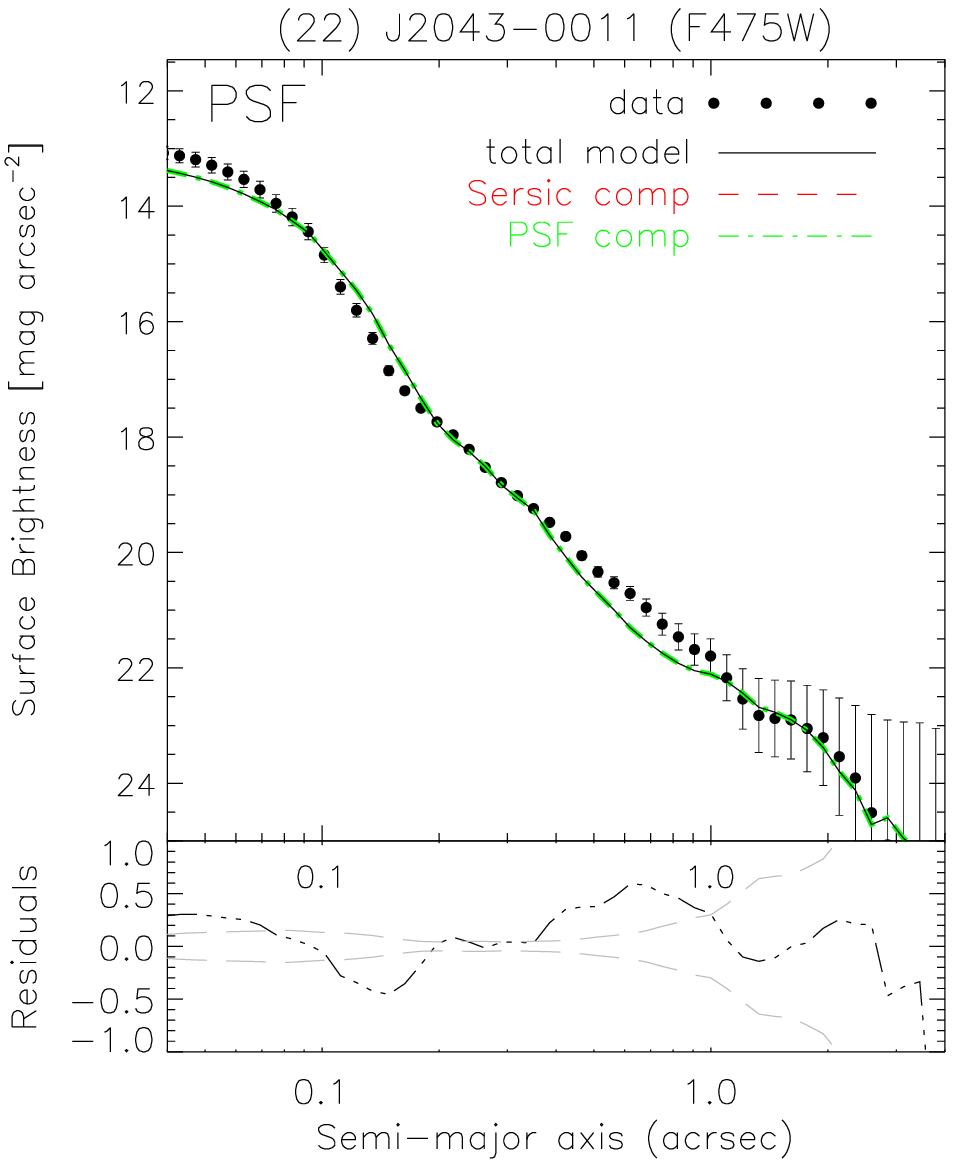,width=0.265\linewidth,clip=} &\epsfig{file=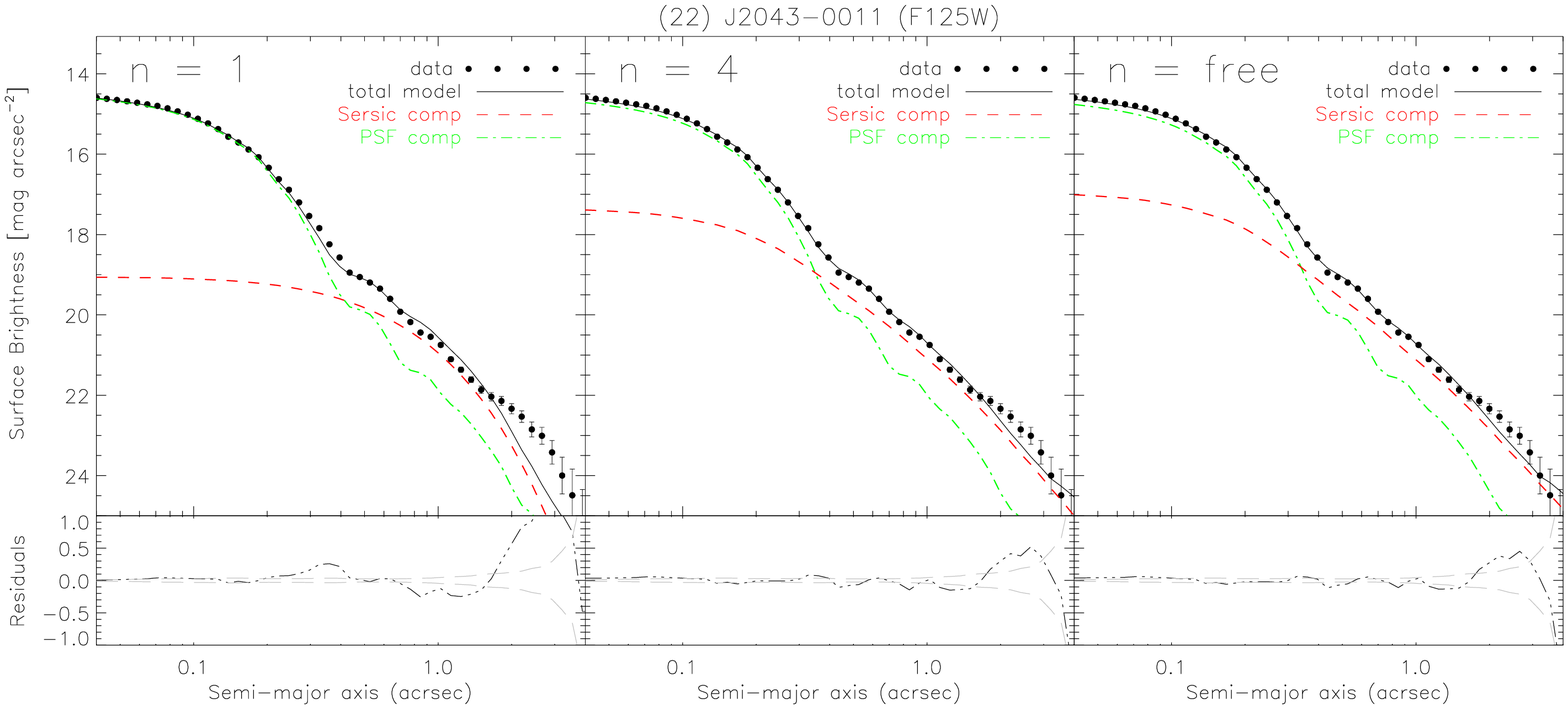,width=0.715\linewidth,clip=}
\end{tabular}
\caption{Object SDSS J2043+0011. Caption, as in Fig. \ref{fig:images1}.}
\label{fig:images22}
\end{figure}

The model parameters from the GALFIT analysis of the F475W and F125W images are listed in Tables~\ref{table:galfitf125w} and~\ref{table:galfitf475w}, respectively, and presented visually in Figures~\ref{fig:images1} through \ref{fig:images22}, one figure for each object. In the Figures, we show the F475W and F125W images (top row), the residuals after subtracting the GALFIT models (middle row), and radial surface brightness profiles of the data and of each component in the model, constructed with the $IRAF$ task $ellipse$ (bottom row). The task $ellipse$ performs ellipse fitting in 2D and averages the profiles to 1D. The residual images show how well the light profiles are fit by classical disk and bulge galaxy profiles. The radial intensity profiles give an idea of the relative contribution of each component as a function of distance from the center.

Table \ref{table:summary} offers a summary on the dominant morphology for each object. The best-fitting model for each object was determined based on the lowest reduced $\chi^2$ value and by visually inspecting the residuals. In cases where the reduced $\chi^2$ values for all models were close, a visual inspection of the residuals was used to decide on the best-fit morphology (see Figures~\ref{fig:images1}$-$\ref{fig:images22}).  In all cases but one, the morphological classification is based on the S\'{e}rsic index to which the n-free model converges; for J1051+5250, visual examination of the residuals qualitatively suggests the system is best-fit by a disk rather than a bulge, as implied by n-free = 8.0. Although most of the reduced $\chi^2$ values are close to unity, which indicates a good fit, the significant residuals in some cases warn us that these $\chi^2$ values are not to be trusted as an absolute measure of the goodness of the fit. A small value of  $\chi^2$ can be due to overestimating the errors, for instance. However, comparing $\chi^2$ values can be useful in discriminating the quality of different models for the same object.  We also note that some of the $\chi^2$ values for the F475W models are extremely high. Incomplete removal of cosmic rays due to having only two separate exposures for most F475W images could possibly account for some of the unrealistically high $\chi^2$ values. Visual inspection of the fits show that the residuals for those cases are minimal, regardless of the high $\chi^2$ values, and that those fits can be trusted.

 The high spatial resolution of the $HST$ images in F125W (FWHM$\lesssim$ 0{\farcs}16) and sufficiently long expose times allowed the host galaxy to be detected of 19/22 (86\%) of the objects; in three (14\%) of the targets, GALFIT did not converge to physically meaningful models and we consider those host unresolved. We use the F125W analysis to quantify the morphologies and interaction histories of the LoBAL, which are discussed in detail in \S~\ref{morphology}. The observations in F125W reveal the presence of a second source of emission close to the cores of nine (41\%) of the LoBALs in our sample.  Four of those become apparent only after model subtraction. Details on these possible interacting companions are listed in Table \ref{table:summary}. Seven of the nine companions are within 1.3$\arcsec$ ($\sim$8.3 kpc) projected separation from the QSO, and six of the nine are clearly detected in both F125W and F475W.  GALFIT companion models were successfully obtained for only three of these second nuclei: for objects J1054+0429 and J1419+4634 in F125W and for objects J0852+4920 and J1054+0429 in F475W; the model parameters for the companions are given in Tables~\ref{table:galfitf125w} and~\ref{table:galfitf475w}.

The F475W images (FWHM$\lesssim$0{\farcs}12) were modeled with a combination of a PSF and one S\'{e}rsic component, with an unconstrained S\'{e}rsic index (n = free). For 18 of the targets, this approach resulted in unrealistic physical parameters for the host component of the model (i.e., extremely small or large r$_e$, and/or high $n$), which was an indication that a S\'{e}rsic component is not needed. We consider the host galaxy unresolved in F475W for these 18 (82\%) objects and model them solely with a PSF + a sky components. The results of those fits are listed in Table~\ref{table:galfitf475w}. The minimal resulting residuals from the PSF subtractions show that a host was not detected. Either the host emission at these bluer wavelengths is very compact or longer exposure times are necessary to detect the host light. Four (18\%) of the targets were successfully fit with a PSF + sky + S\'{e}rsic component. Two of the four were best-fit by an exponential disk profile (with n$<$1) in F475W, matching the dominant morphology for those objects from the modeling of the F125W images.  In close agreement is also the merger J0250+0009 - which displays spectacular "S"-shaped tidal tails - and is best-fit by disk-dominated morphology in F475W (n=0.71), but by an intermediate S\'{e}rsic index of n=1.5 in F125W.  The ongoing  merger J1011+5429 - with an apparently interacting second nucleus - shows a bulge-dominated morphology in F475W (n=3.85), but is best-fit by an intermediate S\'{e}rsic index of n=2.50 in F125W.  We note that all of the four hosts detected in F475W show very high levels of star formation in the SED models presented in \citet{Lazarova2012}, with SFRs ranging from 100 to 310 $M_\odot$ $yr^{-1}$. The one exception is the apparent merger J1614+3752, which has SFR estimate of 326$_{-19} ^{+20}$ $M_\odot$ $yr^{-1}$, but the host is not resolved by GALFIT in F475W (see Fig.~\ref{fig:images19}), suggesting a centrally-concentrated starburst. Since the majority of the host galaxies were not detected in F475W, any further discussion of the hosts in this section refers to the findings from F125W images.

\subsection{Interaction Classification} \label{interaction}

The WFC3/F125W images and residuals (Figures~\ref{fig:images1}$-$\ref{fig:images22}) were visually examined to assess the level of disturbance in the hosts and any signs of merger activity. Table~\ref{table:summary} summarizes the findings, which are (somewhat subjectively) based on the majority opinion of the team members. 

Identifying galaxy mergers observationally is challenging due to the snapshot in time single observations offer and the limitation imposed by a unique viewing angle for an event that lasts Gyrs. In addition, these LoBALs are type-1 QSOs, in which there is high contrast between the central emission from the accretion disk and the host galaxy, quantified with the ratio of the PSF-to-host model intensities, $I_{QSO}/I_{host}$  (discussed in $\S$~\ref{nuclear}). 

Various signs of tidal interaction - double nuclei, distorted morphologies, tidal tails, plumes, bridges, shell-like structure, and debris - are visible in 14/22 (64\%) of the LoBALs, while five systems (23\%) show no distinguishable tidal features. Three of the objects (14\%) are PSF-dominated and for them a resolved host galaxy profile could not be fit by GALFIT; we consider those hosts unresolved. If we use a conservative merger classification - classifying as ongoing or recent mergers only objects showing clear tails, or clumps near the QSO, that appear to be connected via a bridge of emission to the host - we estimate that at least 45\% of the LoBALs show signs of tidal interaction while none of the trends discussed later in the paper changed significantly. The data shows that mostly bright QSOs (with larger ratios of $I_{QSO}/I_{host}$) were conservatively re-classified as undisturbed, likely due to the difficulty of seeing faint merger structures in the vicinity of a bright nucleus with certainty.

Qualitatively, we classify the observed interaction stages using the following rubric:

\begin{itemize}
\item
{\it Ongoing Mergers} are systems showing signs of interaction in which two distinct nuclei are detected.
\item
{\it Mergers} are systems which show signs of tidal interaction, such as plumes, tidal tails, bridges, shells, and/or excess of nearby debris.
\item
{\it Undisturbed} are considered systems that show no apparent signs of tidal interaction. 
\item
{\it Unresolved} are systems for which GALFIT could not converge on a physically-meaningful component fit. 
\end{itemize}

We summarize the consensus on the observed interactions and tidal features in Table~\ref{table:summary}. According to the classification above, this sample of LoBALs consists of nine (41\%) ongoing mergers, five (23\%) mergers, and five (23\% undisturbed host galaxies. A conservative classification of the observable signs on interaction places the lower limit on those fractions at six (27\%) ongoing mergers, four (18\%) mergers, and nine (41\%) undisturbed hosts. Interestingly, although the majority of the resolved hosts show signs of interaction, the sample as a whole represents objects at different stages of the merger process, from double nuclei to advanced mergers with extended, low-surface-brightness tidal tails and shells. 

The ongoing mergers represent mostly early-stage mergers with nuclear separations $<$~10 kpc \citep{Nevin2019} if we take the distances at face value and without corrections for orientation.  In all nine of those systems, although we cannot independently confirm that second sources are associated with the LoBALs, we assume that such small projected separations are an unlikely chance alignment concurrent with other signs of tidal interaction. In 4/9 of the cases, the possible second nucleus becomes visible only in the residual images, after GALFIT model subtraction. In 7/9 of them (hence, 32\% of the entire sample), the second source is within 1$\farcs$3 ($<$10 kpc) in projection. All suspected second nuclei lie within 23 kpc (projected distance) from the LoBAL. Three other objects could potentially belong to the category of ongoing mergers, suggesting as much as half of the sample represents early-stage mergers, but spectroscopic observations are needed to rule out a chance foreground or background galaxy: object J0853+4633 is a merger with strangely-elongated core (Fig.~\ref{fig:images7}), possibly an unresolved second nucleus; J1140+5324 does not show any obvious signs of disturbance and is, thus, classified as undisturbed, but also shows a strangely-elongated central point sources (Fig.~\ref{fig:images13}); both the F125W and F475W images of object 1700+3955 reveal a possible companion, or a large clump, located $\lesssim$1$\arcsec$E from the core (Fig.~\ref{fig:images20}).

We searched the SDSS DR16\footnote{Sloan Digital Sky Survey (SDSS) Data Release (DR) 16: http://skyserver.sdss.org/dr16/} online database for photometric redshift information on any nearby sources present on the HST images within 30 kpc projected separation from the LoBALs. In Table \ref{table:nearby}, we summarize the findings on the nearby galaxies that are at the same redshift as the LoBAL (within the photometric redshift uncertainties), or make notes on sources with unknown redshift but of potential interest.

At least some of the LoBALs are found in crowded environments (i.e., objects J0852+4920, J1128+4823, J1309+0119, J1416+4634, and J1700+3955 have at least three nearby sources, and are all ongoing or recent mergers), and most LoBALs have at at least one nearby object. But we do not find these LoBALs to be in particularly crowded fields, similar to the weak dependence of quasar activity at low redshifts on clustering of the galaxy environment \citep[\eg][]{Wethers2021}.

\subsection{Morphological Classification} \label{morphology}

The WFC3/F125W images and residual maps after the GALFIT model subtraction were visually examined by all team members to determine the best-fit morphology (discussed in this section) and any apparent tidal features (discussed in \S~\ref{interaction}). The majority opinion was adopted as the consensus classification for each object. We classify each object into one of three categories, based on the S\'{e}rsic index of the best-fitting model: a disk-dominated  morphology (n$<$1.5), an intermediate morphology (1.5$>$n$>$3), or a bulge-dominated morphology  (n$>$3). The morphological classification is summarized in Table~\ref{table:summary}, where we refer to the dominant morphologies as disk, intermediate and bulge.

The relative fraction of each morphological type is visualized in the right panel of Figure~\ref{fig:hst}(a). We find that the majority (9/22; 41\%) of the LoBALs have bulge-dominated morphologies. An exponential disk profile provided the best fit to the surface brightness of four (18\%) objects: J0852+4920, J1051+5250, J1400-0129, and J1700+3955.  Intermediate morphology describes the hosts of six (27\%) objects, five of which are ongoing or recent mergers (defined in \S~\ref{interaction}) and only one is undisturbed. Of the four disk systems, half are ongoing or recent mergers (J0852+4920 and J1700+3955), while the majority (5/6) of the intermediate hosts and the majority (7/9) of the bulges are ongoing or recent mergers. We also note that all nine bulges have best-fitting models with n$\gg$4. These very cuspy profiles indicates strong core emission and, thus, a possible PSF mismatch or low-surface-brightness structure attempted to be fit by the extended wings of the high-n profiles. The  likelihood of the latter possibility is high, given seven of the nine bulges were found in ongoing or recent mergers. Three objects in the sample - J0835+4352, J0850+4451 and J1429+5238 - have unresolved hosts (with r$_e$~$<$~2.27 pixels, the FWHM of the PSF). 

If we had adopted a binary classification of the morphologies into disks (n$<$2) or bulges (n$>$2), four of the objects categorized as intermediate would be bulges and the sample as whole would consist of 59\% bulges and 27\% disks. Interestingly, the two objects classified as intermediate which would add to the disks are J0250+0009, a merger with spectacular "S"-shaped extended tidal tails, and J1309+0119, an ongoing merger with shell-like structure (see Fig.~\ref{fig:images3} and \ref{fig:images14}), possibly making the majority (4/6) of the disks mergers as well.

We illustrate the intersectional subcategories between the morphological and the interaction classifications and quote the number of objects in each sub-class in Figure~\ref{fig:bubbleplot}. The bulge-dominated and intermediate hosts are mostly ongoing and recent mergers. Half of the disk-dominated hosts are undisturbed, but there is also an ongoing merger and a recent merger among them. The large fraction (7/10) of ongoing and recent mergers among the objects with disk-dominated and intermediate morphologies might point to the importance of minor mergers, where the disturbance to the morphology is less dramatic; however, morphology might not be a good predictor for the interaction history given simulations show that, in principle, even major mergers can form disks \citep[e.g.,][]{Barnes2002,Springel2005,Robertson2006,Lotz2008,Zeng2021}.

We conclude that this sample of LoBALs is dominated by early-type morphologies, but we caution against over-interpreting these results, given the prevalence of merges in this sample. This parametric approach to modeling the surface brightness of the galaxy images (i.e., S\'{e}rsic index, n) assumes symmetric profiles, which may fail at reliably determining the structure of irregular galaxies, including merging systems \citep{Lotz2004}.

\subsection{Nuclear Emission Strength}  \label{nuclear}

We subdivide the LoBAL sample into subsamples based on the dominant morphology (disks, intermediate, bulges; defined in \S~\ref{morphology}) and interaction class (ongoing mergers, mergers, undisturbed; defined in \S~\ref{interaction}), and look for trends in the parameters derived from this study, as well as from the investigation of their infrared luminosities, star formation rates, and mid-infrared spectral properties in \citet{Lazarova2012}, and in data from the literature. Boxplots of various parameters are shown in Figures~\ref{fig:hst}, \ref{fig:spitzer} and \ref{fig:edd}; the data is listed in Tables~\ref{table:lumir}, \ref{table:lumuv} and \ref{table:bol}.

 The QSO emission is modeled with the inclusion of a PSF component to the model. We quantify the relative strength of the QSO emission by calculating the flux ratio of the PSF-to-host, $I_{QSO}/I_{HOST}$, using the best-fit GALFIT model components (Table~\ref{table:lumir}). This PSF intensity, normalized to that of the host galaxy, quantifies the contrast between the QSO and host emission (important for judging our in/ability to detect low surface-brightness features) and the dominance of the AGN emission (important in judging the central black hole growth). We find a range of  $I_{QSO}/I_{HOST}$ ratios, from 0.15-5.0, with 50\% (11/22) of the targets having host galaxies brighter than their QSO (i.e., $I_{QSO}/I_{HOST}$  $<$ 1) within the F125W bandwidth ($\lambda_{pivot}$ = 1249 nm; $\Delta\lambda$=302 nm). 
 
 In Figure~\ref{fig:hst}(f), we show $I_{QSO}/I_{HOST}$ as a function of morphology and interaction class. The ongoing mergers have the largest range; disks and intermediate morphologies have larger ranges and higher medians than the bulges, which show mostly host-dominated emission with $I_{QSO}/I_{HOST}$$<$1.
 
 Figure~\ref{fig:qsotohost-mips} plots the MIPS 24-to-70 $\mu$m flux ratio, $f_{24}/f_{70}$ as a function of $I_{QSO}/I_{HOST}$ \citep[IR data from][]{Lazarova2012}. A significant fraction of the objects have $f_{24}/f_{70}$~$<$~0.3 and $I_{QSO}/I_{HOST}$ $<$ 1.5, with the caveat that the FIR fluxes are dominated by upper limits, indicated by arrows.  \citet{Veilleux2006} notes that the ratio $I_{QSO}/I_{HOST}$ increases as the various AGN in their sample become more AGN-like (i.e., as the $f_{24}/f_{70}$ increases because the AGN contributes more flux to the mid-infrared 24 $\mu$m band). Similar positive correlation of increasing $24-$to$-70$ $\mu$m color with increasing QSO strength is present in our sample only among the bulges and intermediate morphology objects, suggesting the AGN dominates the emission in the F125W band. However, this trend is not observed for the four disks (objects J0852+4920, J1051+5250, J1400-0129, and J1700+3955), which have some of the lowest FIR color, $f_{24}/f_{70}$ $<$ 0.2, and some of the the strongest nuclear emission normalized to the host light.  
 
 In Figure~\ref{fig:qsotohost}, the PSF-to-host intensity ratios are shown as a function of the total infrared luminosity, $L_{IR}$, star formation rate, $SFR$, and maximum velocity of the Mg {\sc II} broad absorption line, $V_{max}$. There is no notable trend with outflow velocity, meaning that stronger AGN activity is not associated with faster outflows; nor a trend with L$_{IR}$. The former finding is in contrast to \citet{Farrah2012}, who find an anti-correlation between outflow strength and the SF contribution to the total IR emission for 0.8~$<$~z~$<$~1.8 FeLoBALs. Notable is that the four objects in our sample with the highest $SFRs$ ($>$150 $M_{\odot}$ yr$^{-1}$) are exclusively found in systems with $I_{QSO}/I_{HOST}<$1.5 (middle panel of Fig~\ref{fig:qsotohost}).

\subsection{Host Sizes and Magnitudes}
 
\begin{figure*}
\centering
\begin{tabular}{c}
\epsfig{file=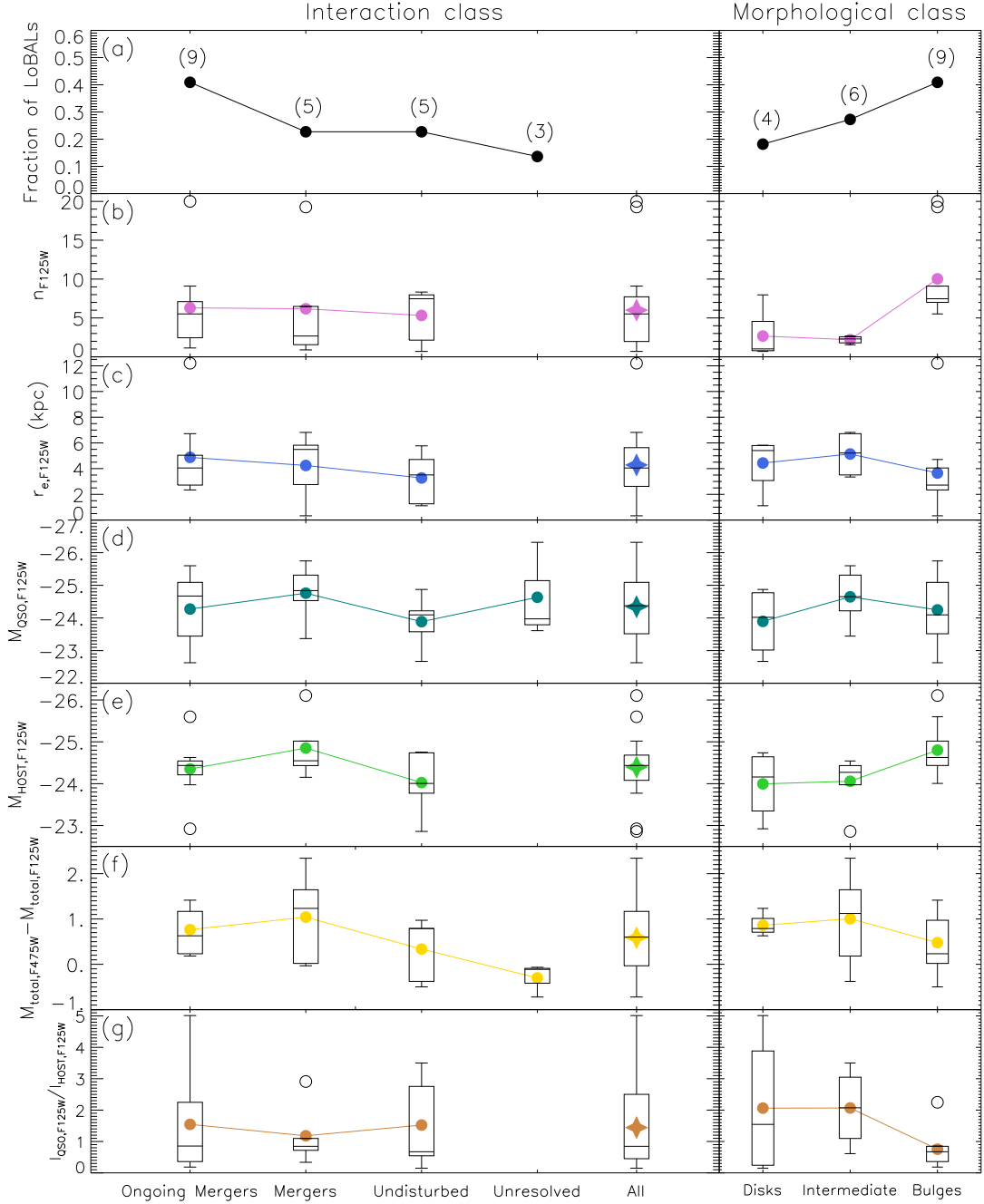,width=0.85\linewidth,clip=}  \\
\end{tabular}
\caption{HST data: Boxplots of some quantities from the GALFIT modeling of the HST/WFC3 images of this sample of LoBALs, divided into subsamples by interactions class (left column) and morphological class (right columns). The open box horizontal sides indicate the lower (25\%), median (50\%) and upper (75\%) quartiles, while the whiskers represent the range, with outliers farther above or below than 1.5 times the interquartile range plotted as open, black circles.  The average is plotted as filled color circles, connected with solid line across the subclasses to highlight any trends. Filled color star marks the average for the entire sample. For definitions of the classes, see Sections~\ref{interaction} and \ref{morphology}. (a) Fraction of LoBALs in each class, with numbers in parenthesis indicating the number of objects. (b) S\'{e}rsic index of the best-fitting GALFIT model for the F125W images, listed in Col.\ 3 of Table \ref{table:summary}. (c) Half-light radius (in kpc) of the surface brightness profile, listed in Col.\ (6) of Table \ref{table:galfitf125w}. (d) Absolute magnitude of the PSF in the F125W channel, corrected for Galactic extinction, listed in Col.\ (4) of Table \ref{table:lumir}. (e) Absolute magnitude of the host galaxy in the F125W channel, corrected for Galactic extinction, listed in Col.\ (5) of Table \ref{table:lumir}. (f) Colors M$_{F475W}$-M$_{F125W}$ for the total emission in F125W, listed in Col.\ (10) of Table \ref{table:lumuv}. (g) PSF to host component intensity ratio in F125W, listed in Col.\ (9) of Table \ref{table:lumir}. }
\label{fig:hst}
\end{figure*}

\begin{figure}
\centering
\begin{tabular}{c}
\epsfig{file=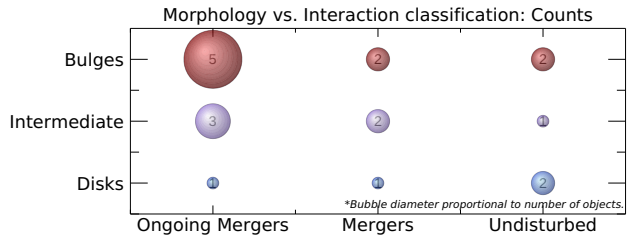,width=1.0\linewidth,clip=}  \\
\end{tabular}
\caption{Bubble plot visualizing the number of LoBALs in each intersectional categories between the morphological and interaction classifications. The diameter of each bubble scales with the number of objects, which is printed in the center. Ongoing and recent mergers have bulge-dominated hosts. However, note that disk-dominated morphologies are found among all interaction classes.}
\label{fig:bubbleplot}
\end{figure}

Figure~\ref{fig:hst} shows the effective radii and absolute magnitudes in F125W of the QSOs and the host galaxies (listed in Table~\ref{table:lumir}), grouped according to interaction class and morphology. The luminosities of the QSOs are comparable across all subclasses, with a range of the absolute F125W magnitudes from -22.6 to -26.3 (in AB mag; Fig.~\ref{fig:hst}(d)). The host absolute magnitudes in F125W range from -22.9 to -26.1 (in AB mag; Fig.~\ref{fig:hst}(e)), suggesting that the galaxies of these LoBALs are luminous and massive (Fig.~\ref{fig:bolhost}). The bulges and the undisturbed hosts have more compact surface brightness profiles (Fig.~\ref{fig:hst}(c)). Bulges also show more luminous hosts compared to the disks and the intermediate morphologies (right panel of Fig.~\ref{fig:hst}(e)).

\section{Discussion}\label{discussion}

In this section, we discuss the results of the interaction and morphology analysis of the HST/WFC3 images, presented in \S~\ref{interaction} and \ref{morphology}, in conjunction with results from their SEDs \citep{Lazarova2012}, archival SDSS spectra and the literature.

\subsection{The Hosts Not Seen in F475W}

The observations in F475W cover rest-frame $\sim$2700\AA ~to $\sim$3600\AA. This region is very sensitive to the age of the stellar populations, with young stars contributing significantly more than older ones.  Balmer emission lines and [O II] $\lambda$3727 will be excluded from this bandwidth even for the highest redshift objects in our sample.   Generally, for QSOs, one would have to worry about strong contribution from broad Mg\,II $\lambda$2800 emission in this filter; however, since these objects are LoBALs with broad, blue-shifted Mg\,II absorption lines, the nuclear Mg\,II emission is less intense due to the absorption than in normal QSOs. 

The host galaxies of only four (18\%) of the LoBALs are resolved in F475W and successfully modeled with GALFIT. Two of them (J1400+0129, and J1700+3955) are best-fit by a disk profile, consistent with the best-fit S\'{e}rsic indices in F125W; for the ongoing merger J1011+5429, the host is bulge-dominated in F475W, while the morphology is intermediate (n=2.5) in F125W; for the merger J0250+0009, which shows spectacular "S"-shaped extended tidal tails, the F125W emission is consistent with a close-to-disk intermediate morphology (n=1.6), while the F475W image is best modeled with a disk profile. 

Not detecting the majority (82\%) of the hosts in F475W indicates that there are not many O and B stars, i.e., not much current star formation. If dust obscuration was the reason the host light is not detected in F475W, one would expect that to be the case for the ULIRGs in this sample. However, all four detected hosts also show high star formation rates $\sim$100-300 $M_\odot$ $yr^{-1}$, as estimated by \citet{Lazarova2012} from the starburst contribution to the optical-to-FIR SEDs (see Table \ref{table:bol}). Three of the detected hosts (J0250+0009, J1011+5429, J1700+3955) are also three of the four objects in this sample classified as ULIRGs \citep{Lazarova2012}. The only host surprisingly not visible in F475W, which otherwise has SFR$\sim$326$_{-19} ^{+20}$ $M_\odot$ $yr^{-1}$ and ULIRG-level of IR emission, is J1614+3752, which also has the highest bolometric luminosity in the entire sample (Log($L_{bol}/L_{\odot})$$\sim$12.63) and the highest total infrared luminosity (Log($L_{IR}/L_{\odot})\sim$12.52; left panel of Fig.~\ref{fig:qsotohost}), yet very low dust extinction ($A_V$$\sim$0.10; see Table \ref{table:bol}). In F475W, J1614+3752 has the most luminous PSF in the entire sample, with $m_{PSF,F475W}\sim$ 15.77 (Table~\ref{table:galfitf475w}), which may point to a very luminous QSO emission overwhelming the stellar light, or the extinction to the star-forming regions may be greater than the extinction to the quasar.

Alternatively, any current star formation may be occurring in a compact region close to the nucleus and appear as part of the unresolved central source (i.e., within the central 0.77 kpc, corresponding to the 0$\farcs$12 spatial resolution of F475W at the median redshift, z = 0.55). Such circumnuclear starbursts \citep[on scales $<$ 100 pc;][]{Hickox2018}, expected to form naturally as a result of gas inflow to the central regions of the galaxy, have been predicted \citep[\eg][]{Thompson2005,Hopkins2016} and observed \citep[\eg][]{Davies2009}. In some starburst galaxies, they are also associated with fast ($\sim$1000$-$3000 km s$^{-1}$) outflows \citep{DiamondStanic2021}. It is also possible that the nuclear emission at these much shorter wavelengths swamps a much fainter host galaxy emission. The QSO-to-host intensity ratios, $I_{QSO}/I_{host}$, for the four objects for which comparison is possible, are much (3$-$8 times) higher in F475W than those in F125W (Tables \ref{table:lumir} and \ref{table:lumuv}). Obscuration by dust in the host galaxies can, in principle, absorb some of the short-wavelength light, but the SED modeling results for the dust extinction at the systemic redshift do not seem to support that for the majority of the sample (median $A_V$$\sim$0.43; Table \ref{table:bol}; also Fig.~\ref{fig:spitzer}).

The F475W observations presented here are not particularly deep, with surface brightness limit of 25.7$\pm$0.5 mag arcsec$^{-2}$ at the 1$\sigma$ level (or 24.4$\pm$0.5 mag arcsec$^{-2}$ at the 3$\sigma$ level). We dedicated one HST orbit per object, which allowed for an average total exposure time of $\sim$1550 s per target in F475W (Table~\ref{table:observations}). Previous imaging studies of quasar hosts at $z~\sim$ 0.2 by \citet{Canalizo2007} and \citet{Bennert2008} were able to demonstrate the presence of shells and tidal tails in galaxies first classified as ellipticals, but their observations with HST ACS/WFC in F606W dedicated five orbits per target, allowing for $\sim$11,000 seconds of total exposure time per object and the detection of fine structure with surface brightness of 28.3$-$29.6 mag arcsec$^{-2}$.  Hence, deeper observations might be needed to detect the host light and fine structure in our targets, as well. The lack of detections might be due to lack of young stars, and, at least for the objects identified as mergers in F125W, might suggests that the progenitor galaxies were not gas-rich disks, or that ongoing star formation has been suppressed in the last few hundred Myr (the lifetimes O-B stars, 1-200 Myr). 

In the absence of host magnitude estimates in F475W, we calculate the $M_{F475W}$-$M_{F125W}$ colors for the total emission and find that the undisturbed and unresolved hosts are much bluer than the ongoing and recent mergers (Table~\ref{table:lumuv}, Fig~\ref{fig:hst}(f)). That might indicate more luminous nuclear emission or possibly unresolved centrally-concentrated young stellar population, which can result from gas-rich major mergers, consistent with various spectroscopic studies of quasar stellar populations \citep[e.g.,][]{Canalizo2007,Jahnke2007}. Hypothetically, if we assume that 100\% of these LoBALs are mergers at various stages, hence, signs of tidal interaction are present also in the undisturbed and unresolved hosts, but could not be detected due to the surface brightness limit of these observations, then the $M_{F475W}$-$M_{F125W}$ color plot in Fig.~\ref{fig:hst}(f) shows an interesting trend of increasingly bluer colors from ongoing and recent mergers to undisturbed hosts to unresolved hosts. That picture would be consistent with a scenario in which the interaction stages (defined in \S~\ref{interaction}) represent progressive merger stages, with the undisturbed and unresolved objects representing more advanced mergers when the AGN dominates the emission. This possibility is also supported by the large $I_{QSO}/I_{HOST}$ ratios in the undisturbed objects (Fig.~\ref{fig:hst}(g)). But given the lack of merger signatures in the undisturbed (23\%) and unresolved hosts (14\%), it is also possible that the QSO emission in them is fueled instead by minor mergers or secular processes, yet that places a conservative upper limit of 37$-$41\%s on the optically-selected LoBALs not triggered by major mergers.

\subsection{Black Hole Activity}\label{mbh}

In the merger-driven scenario for the emergence of QSOs \citep[\eg][]{Silk1998,DiMatteo2005,Hopkins2005}, strong black hole accretion is predicted to increase when the distance between the black holes rapidly decreases \citep{Kawaguchi2020} and to peak shortly after nuclear coalescence \citep[\eg][]{Hopkins2008}. \citet{Farrah2022} show that observationally in ULIRGs and note that there is a significant super-Eddington phase. The QSO at this stage is predicted to emits at close to its Eddington limit, which gives rises to extreme outflows, at velocities similar to those observed in LoBALs and other BAL QSOs.  A dust-reddened, IR-luminous quasar phase is expected to be associated with this brief  "blowout" stage preceding the more typical optical quasar \citep[\eg][]{Hopkins2008,Blecha2018}. If LoBALs represent QSOs caught in this transient outflow stage, finding them at various stage of the merger process (\S~\ref{doublenuclei}) and in hosts with only moderate levels of obscuration \citep[see Fig.~\ref{fig:spitzer}; also][]{Lazarova2012} questions this model. The observations presented here support a picture in which ultrafast outflows arise at various stages of the merger, likely associated with episodes of rapid accretion onto the SMBHs and quasar-level emission \citep[\eg][]{Stickley2014}. If this picture is indeed correct, we would expect LoBALs to emit at the Eddington limit.

Figure~\ref{fig:edd}(c) shows boxplots of the black hole masses of the LoBALs grouped by interaction and morphology. We estimate the black hole masses from the SDSS DR7 spectra using the calibration of the single-epoch virial black hole mass relation by \citet{Park2012}, which assumes that the broad line region (BLR) clouds are virialized, and thus the black hole mass can be estimated as $M_{BH}$ =  $R_{BLR}v^2G^{-1}$ with the 5100\AA ~continuum luminosity, calibrated via the reverberations mapping technique, giving an estimate of the size of the BLR, $R_{BLR}$, and the FWHM of the broad H$_\beta$ line used as an approximation of the velocity of the clouds, $v$, around the black hole.  We note that the presence of extreme outflows in LoBALs may affect the width of the broad H$_\beta$ line. In fact, broad blue-shifted emission lines have been observed in BAL QSOs \citep[e.g.,][]{Hall2007,Ji2012}.  Yet another concern is that, in many of the systems here, the host galaxy contributes significant fraction of the flux in the optical, so the 5100\AA ~continuum needs to be corrected for host galaxy light. Hence, the absolute values of these masses are to be compared with caution, given the caveats noted here.
 
 The black hole masses of these LoBALs range from M$_{BH}\sim$3.2 $\times$ 10$^7$ to 1.0 $\times$ 10$^9$~M$_\odot$, with an average (median) of 3.3 $\times$ 10$^8$~M$_\odot$ (2.4 $\times$ 10$^8$~M$_\odot$) (Table~\ref{table:bol}, Fig.~\ref{fig:edd}(c)). The uncertainties in M$_{BH}$ from single-epoch optical spectra are dominated by an uncertainty in the virial factor and the scatter in the size-luminosity relationship rather than from propagated FWHM measurement errors; \citet{Park2012} suggest $\sim$0.35 dex as typical lower limit for the overall uncertainty.

Using the optical-to-FIR SED models in \cite{Lazarova2012}, we estimate the QSO bolometric luminosities, $L^{QSO}_{bol}$, corrected for galactic and host galaxy reddening and host galaxy contamination in the mid- and far-IR, by integrating only the AGN contributions to the SED, which include the AGN power law continuum plus the mid-IR power law from hot dust close to the center, heated by UV/optical photons from the accretion disk. We note that these are rough underestimates of the intrinsic $L^{QSO}_{bol}$, given the SEDs do not extend to the UV and X-ray bands and only light toward the observer is considered; however, some findings suggest that LoBALs might be intrinsically X-ray weak \citep[e.g.,][]{Teng2015, Luo2014} and at least some of the UV light is accounted for through the reprocessed mid-IR dust emission (for detailed discussion on double-counting AGN emission in estimating $L_{bol}$ from SEDs, see \citet{Krawczyk2013}).  

The values of $L^{QSO}_{bol}$ are listed in Table~\ref{table:bol} along with the corresponding Eddington ratios, $L_{bol}/L_{Edd}$, where the Eddington luminosity is estimated as $L_{Edd}=1.25\times10^{38}M_{BH}$($M_\odot$) erg s$^{-1}$. The LoBALs span an order of magnitude in $L_{bol}$ $\sim$ 1.7$\times10^{45}$ $-$ 1.6$\times10^{46}$ erg s$^{-1}$ (Fig.~\ref{fig:edd}(a)). The average (median) $L_{bol}$ is $\sim$6.7$\times10^{46}$ erg s$^{-1}$ (6.5$\times10^{46}$ erg s$^{-1}$). There is marginal trend of increasing median $L_{bol}$ from ongoing mergers (6.1$\times10^{45}$ erg s$^{-1}$) toward mergers (6.7$\times10^{45}$ erg s$^{-1}$) and unresolved host (7.7$\times10^{45}$ erg s$^{-1}$), with the undisturbed hosts having the lowest values (5.1$\times10^{45}$ erg s$^{-1}$).  

All LoBALs show sub-Eddington accretion, with an average (median) Eddington ratio of 24\% (15\%) and a wide range of 5$-$92\% (Fig.~\ref{fig:edd}(b)).  The recent mergers have the higher median (31\%) Eddington ratios, compared to the 10\%$-$13\% found for the ongoing mergers, the undisturbed and the unresolved hosts. As a check for the validity of our $L_{bol}/L_{Edd}$ estimates we note that one of the LoBALs in the sample with unresolved host, J0850+4451, was recently studied by \citet{Leighly2018}, who find accretion rate $\sim$ 0.06 $L_{Edd}$, consistent with 6\% estimated by our analysis. 

Similar or lower Eddington ratios are found for lower \citep[e.g.,][]{McLeod2001,Floyd2004,Veilleux2006,Veilleux2009} redshifts QSOs. The Eddington ratios for these LoBALs are on average higher than those found by \citet[RQQ $\sim7\pm1\%$]{Dunlop2003} for a sample of PG QSOs dominated by bulges, which were later shown to be advanced merger remnants displaying spectacular shell in deep $HST$ images \citep{Canalizo2007,Bennert2008}. They are also higher than the average 6\% found for large quasar sample 0.2 $<$ z $<$ 1 by \citet{Li2021}. This agrees with a scenario in which the occurrence of LoBALs is associated with higher levels of BH accretion.

The mean Eddington ratio of our LoBALs (24\%) and the wide range of values (5$-$92\%) is also consistent with the mean of 22\% and range ($\sim$3$-$100\%) found by \citet{Schulze2017} for a sub-sample of low-redshift LoBALs. In that study, they focus on comparing the black hole masses, Eddington ratios and SEDs of LoBALs at 1.3$<$z$<$2.5 to a control sample of non-BAL QSOs, but augment their high redshift sample with objects at z$<$1 from the literature. While their main conclusion is that there no statistically significant difference between LoBALs and non-BAL QSOs, they note that their z$\sim$0.6 LoBALs have slightly higher mean Eddington ratios (22\% vs.\ 17\%) and lower $M_{BH}$ (by $\sim$0.1 dex) than the control. Note that, unlike the LoBALs in our sample which were selected using the more inclusive Absorption Index (AI$>$0), their LoBALs are chosen via the traditional Balnicity Index (BI$>$0) criterion, the caveats of which we discuss in $\S$\ref{trends}. The \citet{Schulze2017} low-z LoBALs have slightly higher mean M$_{BH}$ than our LoBALs (5 $\times$ 10$^8$~M$_\odot$ vs.\ 3.3 $\times$ 10$^8$~M$_\odot$, respectively); and with the exception of one object, their entire high-redshift LoBAL sample has extremely massive black holes (M$_{BH}>10^9$ M$_\odot$) not seen among the low-redshift LoBALs.

\subsection{Double Nuclei Dominate in F125W}\label{doublenuclei}

The observations with the F125W filter cover rest-frame $\sim$6900 to $\sim$9300\AA, a region dominated by the continuum from older stellar populations and one that excludes H$\alpha$ for the redshift range of this sample.  The QSO power-law emission contributes significantly less to this bandpass, thus there is a lower contrast between the QSO and the host galaxy. 

Variety of tidal features are observed  and classified visually (for summary, see Table~\ref{table:summary}): double nuclei, bridges, shell-like structures, and various tails described as plume, straight, curved, "S"-shaped, "V"-shaped, after \citet{Ren2020}. The majority (14/22, 64\%) of the sample show merger signatures in this channel, with dominance of double nuclei (9/14) among them at close projected separations, $\Delta$x (7 of the 9 at $\Delta$x $<$ 10 kpc). A conservative classification suggests a merger fraction of at least 45\% (10/22), with at least 27\% (6/22) double nuclei. Excluding considerations for viewing angle, the close separations suggest that 32\% (7/22) of the sample are late-stage mergers, as defined by \citet{Nevin2019} (i.e., early-stage: $\Delta$x $>$ 10 kpc; late-stage: $\Delta$x $\approx$ 1$-$10 kpc; post-coalescence; $\Delta$x $<$ 1 kpc). Comparison to simulations of gas-rich major mergers of equal mass suggest that those potential double nuclei systems might represent a merger stage 0.5$-$1 Gyr prior to nuclear coalescence \citep{Ji2014,Lotz2008}. 

The dominance of merger signatures in these bona fide broad-line (i.e., type-1) QSOs strongly supports the evolutionary explanation for LoBALs as young QSOs resulting from galaxy interactions, but the wide range of tidal features is evidence they represent various stages of the merger process. For that reason, and the fact that their SMBHs are accreting at high rates (average L$_{bol}$/L$_{Edd}$ $\sim$ 24\%; \S~\ref{mbh}), LoBALs might be a phase in the life of type-1 QSOs that exhibit extreme winds in ionized gas as a result of a recent fueling episode \citep[for optical spectra, see the Appendix in][]{Lazarova2012}. Finding that 2/3 of these objects are in apparent mergers offers strong support for the evolutionary explanation for LoBALs as young QSOs resulting from mergers, and most likely major mergers. However, discovering that half of the mergers (or as much as 1/3 of the whole sample) represents a merger stage prior to nuclear coalescence and the other half showing a single nucleus swaddled in tidal features consistent with more advanced mergers suggests we are observing different snapshots of the merger process. This might be consistent with a model proposing that AGN activity happens during multiple short episodes of accretion on timescales of 10$^{4-5}$ yrs \citep{Schawinski2012} rather than the extended AGN lifetimes of 10$^{6-8}$ yrs needed to grow the black holes we observe in the centers of galaxies \citep[\eg][]{Martini2004,Soltan1982}. 

Simulations of binary disk mergers by \citet{Stickley2014} suggest that quasar-level of accretion occur only during four short phases in the entire merger process: (I) shortly after the first pass, (II) between the second and third passes, (III) during and immediately following nuclear coalescence, and (IV) long after nuclear coalescence. They note that the most luminous quasars (L$_{bol}$ $\sim$ 10$^{45}$ erg s$^{-1}$) would be observed during period II and III, and black hole masses measured during period I and II would reflect only one of the progenitor SBMHs, thus, be lower than M$_{BH}$ after coalescence. In this paradigm, the ongoing mergers with double nuclei in our sample might be objects caught during either episode I or II when the nuclear separations are still large; the mergers post coalescence would represent period III, while the undisturbed and unresolved objects, if merger remnants, might be caught during stage IV. We see marginal evidence in support of this pictures in Fig.~\ref{fig:edd}(c), which shows boxplots of the black hole estimates (\S~\ref{mbh}) across the interaction classes. There is a trend of increasing median M$_{BH}$ from ongoing mergers ($\sim$1.7 $\times$10$^8$~M$_\odot$) toward mergers ($\sim$2.4 $\times$10$^8$~M$_\odot$), undisturbed ($\sim$3.4 $\times$10$^8$~M$_\odot$) and unresolved hosts ($\sim$5.4 $\times$10$^8$~M$_\odot$), with the largest range and lowest mass black holes found among the ongoing and recent mergers.

However, it is likely that the fueling mechanisms at least in some of the five undisturbed, but resolved hosts, are not major mergers, but driven by bar instabilities and nuclear spirals \citep[\eg][]{Combes2019,Smethurst2021}. For instance, in the two undisturbed hosts best-fit by disk profiles, J1051+5250 and J1400-0129 (Fig.~\ref{fig:images10} and \ref{fig:images15}), we observe compact spiral-arm-like structures and significant residuals. Object J1028+5929, classified as undisturbed, is best-fit by an intermediate S\'{e}rsic index (n=2.2; Table~\ref{table:summary}) possibly  indicating the presence of an underlying disk component. While for object J1140+5324 we cannot rule out an ongoing merger given the cuspy S\'{e}rsic profile (n=7.5), significant residuals to all model fits, and the strangely elongated PSF emission (Fig.~\ref{fig:images13}) suggesting a potentially unresolved second nucleus. Similarly elongated central emission is observed in the ongoing merger J2043-0011 (Fig.~\ref{fig:images22}), where the residuals after the PSF subtraction revealed a highly-symmetrical, spherical object located $\sim$0$\farcs$7 NW from the nucleus (Table~\ref{table:summary}). Thus, two of the hosts classified as undisturbed might be ongoing or recent mergers, suggesting merger fraction as high as 77\% (17/22), and leaving only two objects (9\%) in the sample that are possibly experiencing fueling due to secular processes.

\subsection{Lessons from the Tidal Features}\label{tails}

Particularly interesting are the extended, low-surface-brightness tidal tails, bridges and plumes seen in twelve (12/22, $55\%$) of the objects (see Table~\ref{table:summary} for a summary) because those can be used to obtain very rough timescales for the merger event and reveal information about the progenitor galaxies. sTidal tails in mergers form from the spiral arms of disk galaxies \citep{Toomre1972}. In five ($23\%$) of the objects, the tails extend for more than 6$\arcsec$ ($\sim$38 kpc).  Assuming that the tail material travels at the typical rotation speed of 240 km s$^{-1}$ in the plane of the sky, the dynamical time for the formation of such tails would be on the order of 150 Myr.

The tails can also be used to reveal some characteristics of the progenitor galaxies. \citet{Barnes2016} note that a merger involving a disc would produce a bridge toward the galaxy’s companion and a tail stretching in the opposite direction. Bridges are seen in three of the objects (Table~\ref{table:summary}), and in J1054+0429, in particular, it extends toward a PSF-dominated source $\sim$9$\arcsec$NE (Fig.~\ref{fig:images11}), which we cannot confirm is at the same redshift at the LoBAL. In most of the cases only one extended tidal tail is visible, which is either a projection effect (when our line of sight is along one of the tails), or it can be a result of a spiral-elliptical merger, or a merger of counterrotating disk galaxies during which one of the disks can remain fairly undisturbed  \citep[i.e.,][]{Hibbard1999}. Object J0250+0009 is the most spectacular merger with two long, filamentary, "S"-shaped tidal tails extending in opposite directions (Figure~\ref{fig:images3}). Its north-west tail bifurcates closer to the central galaxy. The extended tail geometry suggests that this is a prograde major merger of two disk galaxies, i.e., the disk spin axes were aligned with the orbital axis of the merger system \citep[i.e.,][]{Mihos2004}. The nearly-circular arches we see in J0231-0831 (see Fig.~\ref{fig:images1}) resemble shell-like structure, similar to the ones shown in the time evolution simulation by \citet[their Fig. 2]{Quinn1984} of a minor merger with mass ratio 1:100 between a disk and a massive potential.  It is also possible we are seeing tidal tails extending in bipolar direction viewed in projection along an axis defined by the stretch of the tails, or tails curled in more tightly due to a massive halo \citep{Barnes2016}. A possible shell is also visible for object J1309+0119 (Fig.~\ref{fig:images13}). Shells are shown to form from nearly-radial minor or intermediate merger of galaxies \citep[\eg][]{Hendel2015}, which implies that minor merger might also be responsible for some of the distorted morphologies. The galaxy-scale, spiral-arms-like structures in J1700+3955 are possibly large, bright tidal tails (Figure~\ref{fig:images20}) due to their spatial extend (20-26 kpc), apparent lopsidedness and the presence of a low-surface brightness plume extending westward. This becomes more apparent in the residual images, and is further supported by the presence of a possible second nucleus within 1$\arcsec$ from the center, detected in both F125W and F475W. Extended, low surface brightness tidal tails are also visible in J1128+4823 (curved tail $\geq$8$\arcsec$S), J1419+4634 (straight tail $\geq$7$\arcsec$NNW), J1614+3752 (curved tail $\geq$7$\arcsec$SW), J1703+3839 (curved tail $\geq$6$\arcsec$NW), and in J0853+4633 (plume wrapping azimuthally around the source). 

The prevalence of tidal tails suggests that many of the mergers in this sample involved at least one gas-rich disk galaxy, which, in principle, can provide ample amount of gas to fuel the central black holes and ignite the AGN activity \citep[e.g.,][]{DiMatteo2005,Hopkins2006}. In addition, the majority of the sample has bulge-dominated hosts with centrally-concentrated light profiles (i.e., S\'{e}rsic indices n$\gg$4) and objects with intermediate (transitional) morphologies, all of which are dominated by ongoing and recent mergers (Fig.~\ref{fig:bubbleplot}). In light of the merger hypothesis \citep[e.g.,][]{Barnes1992,Hernquist1993,Cox2006}, this may suggest that gas-rich major mergers play a dominant role in the emergence of some of these low-z LoBALs. While there is a possibility that some are also minor mergers, \citet{Ji2014} note that the merger features in minor mergers show for longer time due to the longer time it takes the merger to proceed, but are more difficult or impossible to detect in shallow images with surface brightness $\sim$ 25 mag arcsec$^{-2}$, requiring $\mu_{limit}$ $>$ 28 mag arcsec$^{-2}$. We conclude that any tidal features detected in the observations presented here are most likely due to major mergers, given the low surface brightness limit of our images (i.e., $\mu_{F125W}$ $\sim$ 24.7$^{+0.2}_{-0.3}$ mag arcsec$^{-2}$).

\subsection{Trends with SFR, Outflow Velocity, Balnicity and Absorption Index, and Dust Extinctions }\label{trends}
 
Most of our LoBALs show relatively low levels of current star formation, in comparison to ULIRGs and other mergers. Could it be due to quenching by the fast outflows that characterize them as LoBALs?
 
\begin{figure*}
\centering
\begin{tabular}{c}
\epsfig{file=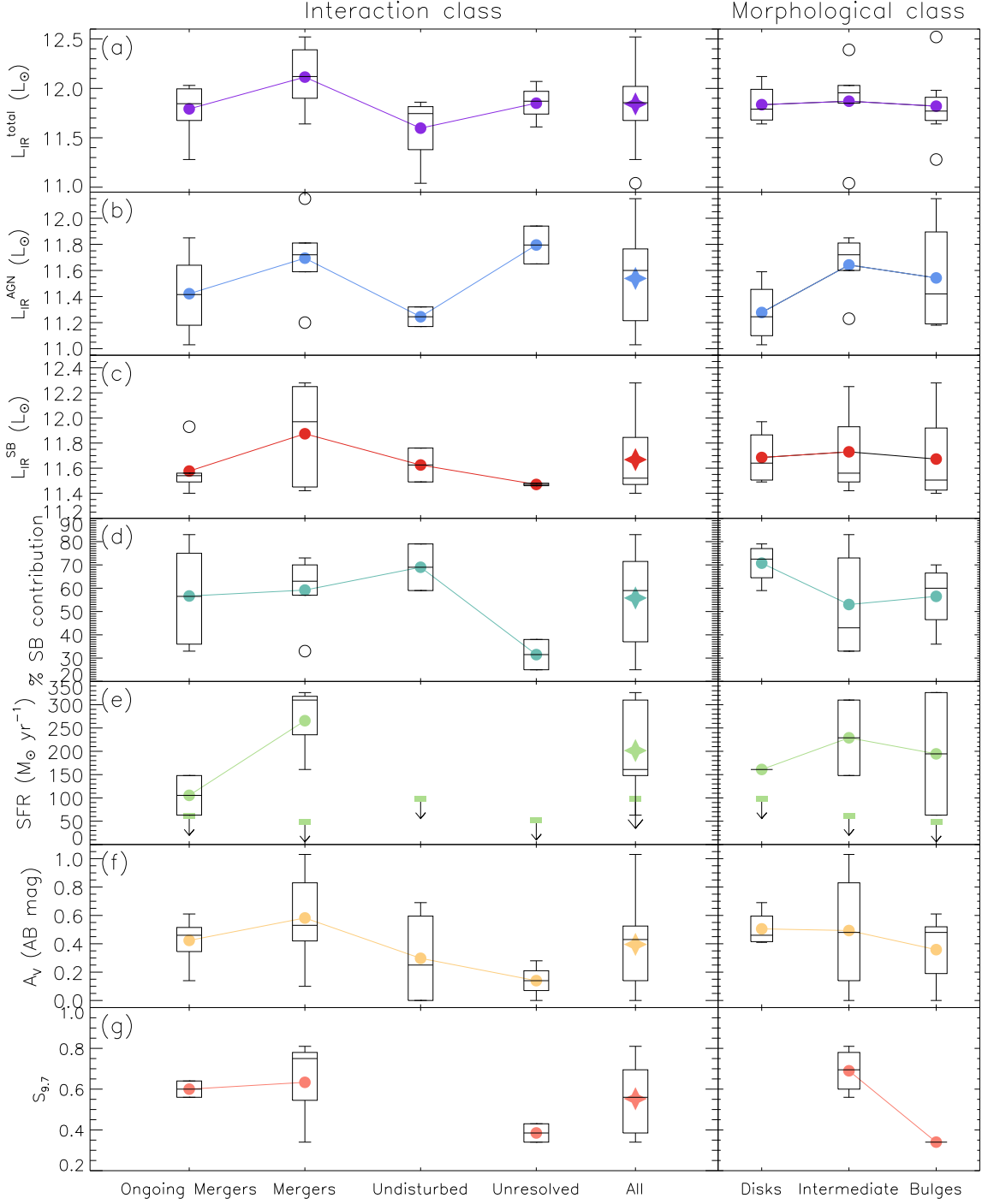,width=0.9\linewidth,clip=}  \\
\end{tabular}
\caption{SED data: Boxplots of some quantities derived from the optical-to-FIR SED modeling of this sample of LoBALs from \citet{Lazarova2012}, divided into subsamples by interactions class (left column) and morphological class (right columns) as classified from the HST imaging presented in this paper. For plot details, see the caption of Fig.~\ref{fig:hst}. (a) Total infrared luminosity, integrated from the SED between 8-1000$\mu$m. (b) Starburst infrared luminosity, corrected to AGN contribution. (c) AGN infrared luminosity, excluding contributions from star formation. (d) Percentage starburst contribution to the total infrared luminosity. (e) Star formation rates, estimated from the starburst contribution to the FIR. Since the FIR fluxes are dominated by upper limits, here we show boxplots and averages of the detections only and upper limits for the non-detections, indicated with downward arrows from the horizontal green bars. (f) Absolute extinction in $V$, assuming SMC extinction law for the QSO + host system. (g) Silicate 9.7$\mu$m emission strength. }
\label{fig:spitzer}
\end{figure*}

 In \citet{Lazarova2012}, we estimate the SFRs for this sample from the star formation contribution to the far-infrared from multi-component SED fitting that accounts for the significant AGN contribution to the 8-1000$\mu$m band. We found that the far-infrared MIPS 70 and 160 $\mu$m bands are dominated by upper limits, suggesting SFR upper limit of $<$98 $M_{\odot}$ yr$^{-1}$ for the sample (Table~\ref{table:bol}), with only four object having significant star formation from detections, with SFRs $\sim$ 150-330 $M_{\odot}$ yr$^{-1}$.  In Figures~\ref{fig:spitzer} (c) and (e), we see a trend in the starburst infrared luminosity, L$_{IR}^{SB}$, and, hence, the SFRs among the interaction classes: the average SFR in the ongoing mergers with FIR detections (106 M$_\odot$ yr$^{-1}$) is less than that in the mergers (266 M$_\odot$ yr$^{-1}$), while the undisturbed and the unresolved hosts have only upper limits ($<$98 M$_\odot$ yr$^{-1}$ and $<$52 M$_\odot$ yr$^{-1}$, respectively). Less pronounced, but similar, trend is seen in the median dust extinction, $A_V$, of the QSO, as estimated from the SED fitting assuming SMC-like extinction law (Fig.~\ref{fig:spitzer}(f)). We see similar reddening in the ongoing (0.48 mag) and recent mergers (0.53 mag)s, and a dramatic decrease in the undisturbed (0.25 mag) and unresolved hosts (0.14 mag). If we assume a natural time evolution from ongoing mergers to recent mergers toward undisturbed and unresolved hosts, it seems that SF is not yet triggered in the ongoing mergers, reaches a peak, dusty phase in the mergers and has already been quenched in the undisturbed and unresolved hosts.  We note that at least some of the undisturbed hosts might be triggered by secular processes (as discussed in \S~\ref{doublenuclei}).
 
\begin{figure*}
\centering
\begin{tabular}{c}
\epsfig{file=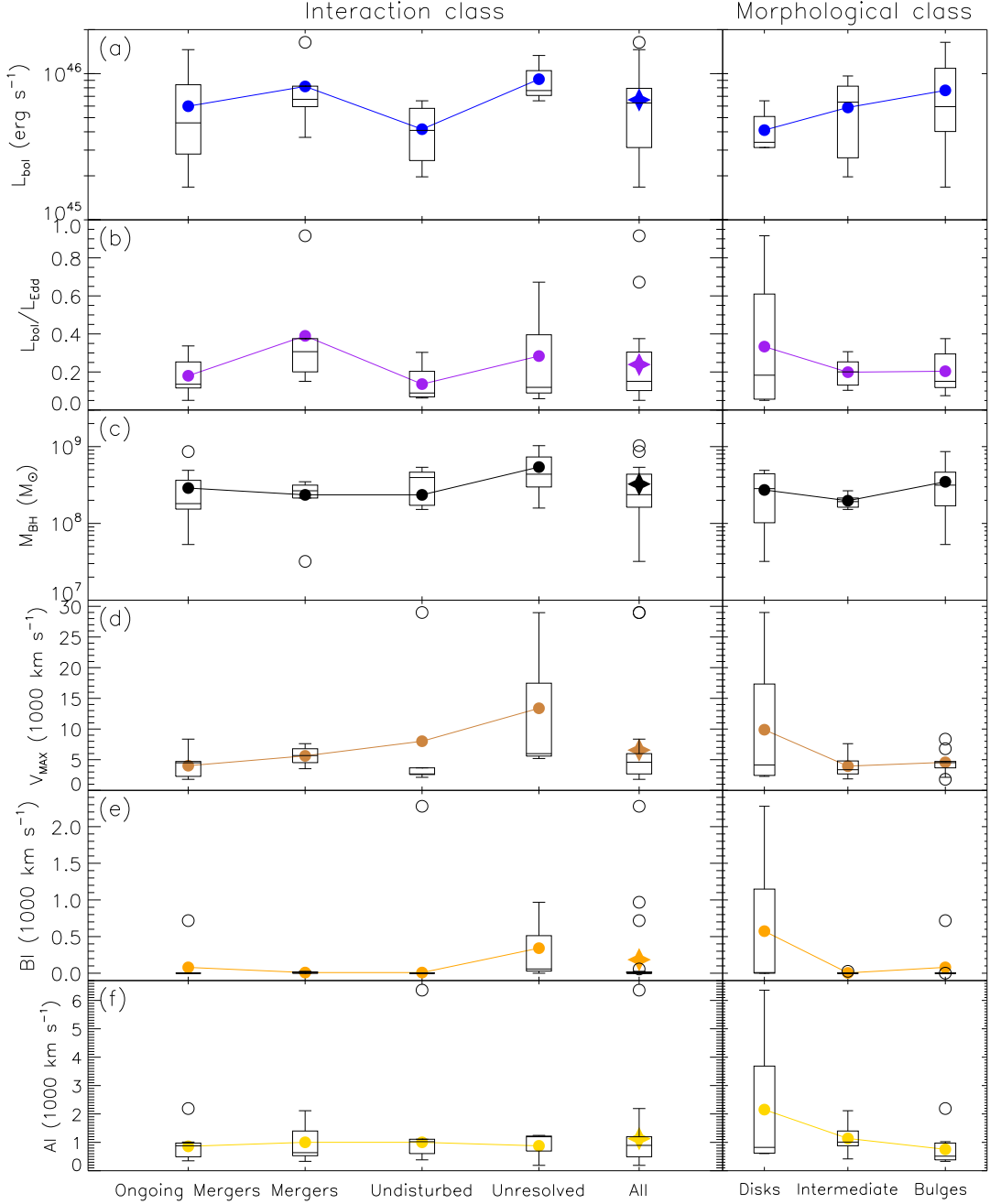,width=0.85\linewidth,clip=}  \\
\end{tabular}
\caption{SED and literature data: Boxplots of some quantities from the literature or derived from the SED modeling of this sample of LoBALs, divided into subsamples by interactions class (left column) and morphological class (right columns). For plot details, see the caption of Fig.~\ref{fig:hst}. (a) Bolometric luminosities of the QSOs, integrated from the optical through FIR SEDs from \citet{Lazarova2012}. (b) Eddington ratios. (c) Black hole masses estimated with the single epoch virial black hole mass relation by \citet{Park2012} using the SDSS spectra of the LoBALs.  (d) Maximum velocity of the Mg {\sc II} 2800\AA  ~broad absorption line from \citet{Trump2006}. (e) Balnicity Index (BI), taken from the BALQSOs catalog by \citet{Trump2006}, is the traditional BAL classification criterion by \citet{Weymann1991}; BI is a modified equivalent width of all continuous BAL troughs at least 10\% below the continuum and at least 2000 km s$^{-1}$ wide, integrated beyond the first 3000 km s$^{-1}$ to avoid host and intervening systems contamination. Traditionally classified BAL QSOs required BI$>$0. (f) Absorption Index (AI), taken from the BALQSO catalog of \citet{Trump2006}, is a true equivalent width measuring all absorption below 10\% of the continuum and at least 1000 km s$^{-1}$ wide, starting from zero velocity. \citet{Trump2006} classifies objects with AI$>$0 as BAL QSOs. }
\label{fig:edd}
\end{figure*}

\begin{figure}
\centering
\begin{tabular}{c}
\epsfig{file=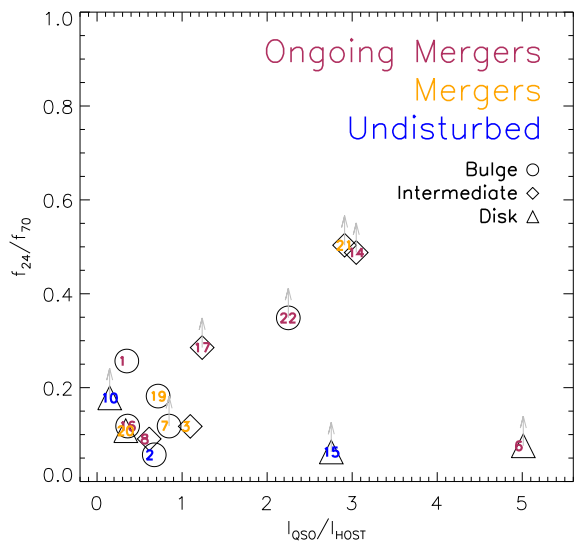,width=1.0\linewidth,clip=}  \\
\end{tabular}
\caption{MIPS 24-to-70 $\mu$m colors as a function of the HST/F125W PSF-to-host intensity ratios. The object number is inscribed inside the plotting symbol, color-coded according to interaction class: ongoing mergers in maroon, mergers in orange, and undisturbed in blue. Plotting symbols distinguish between the dominant morphology: circle for bulge-dominated, triangle for disk-dominated, and diamond for intermediate morphology. Gray arrows indicate upper limits on the FIR MIPS photometry. Seven objects in the sample are not shown as they do not have available FIR datas.}
\label{fig:qsotohost-mips}
\end{figure}

\begin{figure*}
\centering
\begin{tabular}{c}
\epsfig{file=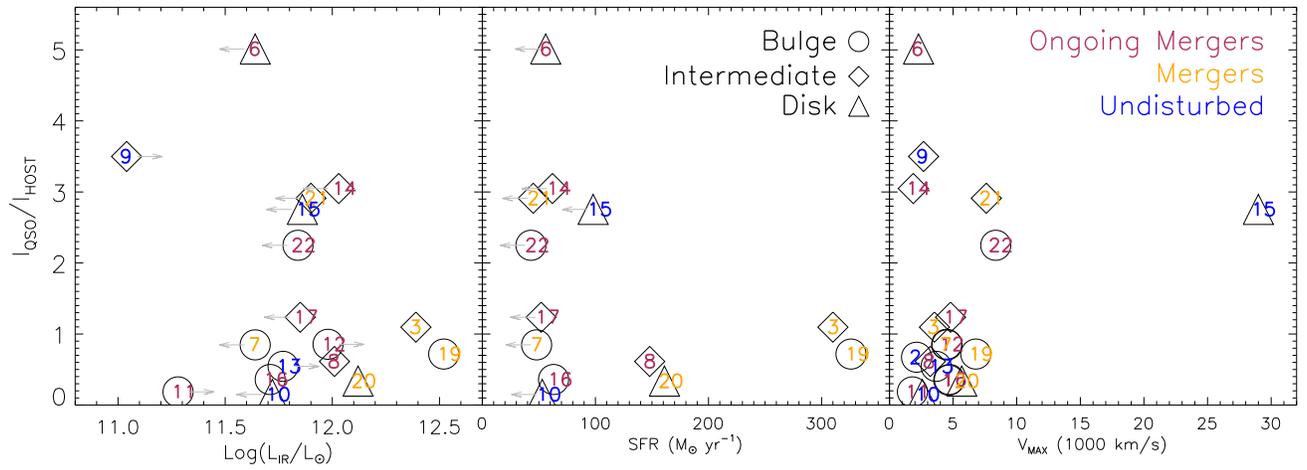,width=1.0\linewidth,clip=}  \\
\end{tabular}
\caption{WFC3/IR F125W PSF-to-host component intensity ratio as a function of total infrared luminosity (left panel), star formation rate (middle panel), and maximum outflow velocity (right panel). Plotting symbols distinguish between the dominant morphology: circle for bulge-dominated, triangle for disk-dominated, and diamond for intermediate morphology. The object number is inscribed inside the plotting symbol, color-coded according to interaction class: ongoing mergers in maroon, mergers in orange, and undisturbed in blue. Gray arrows indicate upper limits on the far-infrared MIPS photometry.}
\label{fig:qsotohost}
\end{figure*}

\begin{figure}
\centering
\begin{tabular}{c}
\epsfig{file=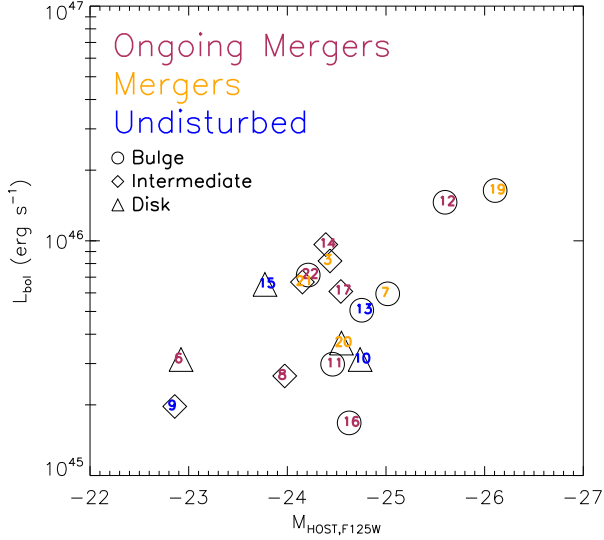,width=1.0\linewidth,clip=}
\end{tabular}
\caption{Bolometric AGN luminosity vs.\ host galaxy absolute magnitude in WFC3 IR/F125W for this sample of LoBALs. Plotting symbols as in the caption of Fig.~\ref{fig:qsotohost-mips}.}
\label{fig:bolhost}
\end{figure}

 Interestingly, these trends anti-correlate with an increase of the average maximum velocity of the Mg {\sc II} broad absorption line,  V$_{max}$ \citep[taken from][listed in Table~\ref{table:bol}]{Trump2006}. Among the interaction classes, shown in Figure~\ref{fig:edd}(d), the undisturbed ($\sim$8,000 km s$^{-1}$) and unresolved ($\sim$13,400 km s$^{-1}$) hosts have the fastest mean outflows compared to the ongoing ($\sim$4,000 km s$^{-1}$) and recent mergers ($\sim$5,600 km s$^{-1}$). The V$_{max}$ trend is not seen in the median values. Among the morphology classes, disks have the fastest average outflows and show the widest range of speeds. An outflow through lower gas density environment would be faster as it would experience less deceleration by intervening material. The outflows in the unresolved hosts might be faster if the ISM near the BH is less enshrouded in gas as a result of a dry merger, hence the lack of detectable low surface brightness features in those hosts, such as tails and bridges which form from disk spiral arms. It is also possible that an earlier outflow event - as might be the case if quasar-level accretion is episodic during a merger \citep[\eg][]{Stickley2014} and is associated with fast outflows when it occurs -  has cleared out much of the gas, thus allowing for faster winds in the later merger stages (see the discussion in \S~\ref{doublenuclei}). 

Moreover, faster outflows might be associated with more advanced merger stages, in hosts where merger signatures have already faded beyond observable levels. In this scenario, low surface brightness merger features would be more difficult to characterize in the objects with the fastest outflows, such as the ones seen in BAL QSOs by selection. The more inclusive selection of BALs in the catalog by \citet{Trump2006}, from which this sample is drawn, uses an Absorption Index (AI) and includes lower velocity outflows than the traditional Balnicity Index (BI) criterion by \citet{Weymann1991}, or the modified version of BI used by \citet{Gibson2009}. While AI is a true equivalent width measuring all blue-shifted absorption dipping below 10\% of the continuum with a minimum width of at least 1000 km s$^{-1}$, starting from zero velocity shift, the traditional BI is a modified equivalent width of all continuous BAL troughs at least 10\% below the continuum and at least 2000 km s$^{-1}$ wide, integrated beyond the first 3000 km s$^{-1}$ to avoid host and intervening systems contamination. Figures~\ref{fig:edd} (d), (e) and (f) show boxplots of V$_{max}$, BI and AI, respectively, for our LoBALs, grouped by interaction and morphology. As Figure~\ref{fig:edd}(e) clearly demonstrates, all but one of the ongoing and recent merger in this sample would have been missed with the traditional BI selection (BI$>$0) because those objects have the lowest velocity outflows (Fig.~\ref{fig:edd}(d)). The objects with the largest BI and AI values, as shown in Fig.~\ref{fig:edd}(d) and (f), are the undisturbed hosts and the disks in which secular processes might play a role. This is counterintuitive: since the BI and AI reflect both the width (i.e., velocity) and depth (i.e., mass driven out) of the trough, one would expect driving out more mass to be associated with a particularly luminous, merger-driven quasar phase. But in this sample, it appears that the more moderately windy LoBALs are the ones associated with apparent mergers. This raises strong concerns for potential biases in studies of BAL QSO populations selected in more traditional ways that find prevalence of disks and lack of mergers \citep[\eg][]{Villforth2019}.

We previously estimated the silicate dust emission feature at 9.7 $\mu$m, which is seen only in emission and is detected in the mid-infrared $Spitzer$/IRS spectra in only seven of the 20 objects in the sample \citep[see Table 7 in][]{Lazarova2012}; $MIR$ spectra of the ongoing merger J0231$-$0831 and the merger J0231$-$0933, both of which are bulge-dominated, were not obtained during that campaign due to the early start of the $Spitzer$ warm mission. In Figure~\ref{fig:spitzer}(g), we plot the strength of the feature for the various subclasses. Among the interaction classes, it follows the trend seen in the SFRs (Fig.~\ref{fig:spitzer}(e)): the median silicate emission is higher in the mergers (S$_{9.7}\sim$0.75) than in the ongoing mergers (S$_{9.7}\sim$0.64); it has lower value in the unresolved hosts (S$_{9.7}\sim$0.43), and is not detected in the undisturbed hosts. Among the morphological classes, it is not detected in the disks and is twice as high in the intermediate morphologies (S$_{9.7}\sim$0.75) compared to the bulges (S$_{9.7}\sim$0.34). This suggest that quenching of the star formation (i.e., low SFRs) may be linked to decrease in obscuration, which would be the case if the reservoir of cold gas was cleared by outflows that also removed the material providing the obscuration. It might also suggest that the outflows in the ongoing and recent mergers are more dusty, while mostly dust-free in the undisturbed and disk hosts. It is worth noting that V$_{max}$ is the highest among the undisturbed, the disk-dominated and the unresolved hosts, and simultaneously the silicate feature is not detected in the former two groups, while being the weakest in the latter category.

In summary, we see dusty, slower outflows in the mergers and dust-free, faster outflows in the non-mergers. This might be consistent with the radiation-dependent unification model proposed by \citet{Ricci2017}, which we discuss in the following section (\S~\ref{orientation}).

 \subsection{Evolution, Orientation, or Both}\label{orientation}
  
The observed emission in AGN at near- and mid-infrared wavelengths suggests the existence of a dusty, obscuring toroidal structure, the nature and geometry of which are still uncertain. AGN unification models \citep[e.g.,][]{Antonucci1993,Urry1995} invoke the so-called dusty torus to explain the difference between type-1 (broad-line) and type-2 (narrow-line) AGN as variations in viewing angle, face-on vs.\ edge-on, respectively \citep[for review, see][]{Netzer2015}. As our view of the obscuring structure changed from static, smooth, dusty doughnut \cite[e.g.,][]{Pier1992} to clumpy clouds \citep[e.g.,][]{Krolik1988,Nenkova2008}, recent MIR interferometric observations \citep[\eg][]{Asmus2016} suggest a more dynamic view of the obscuration as arising from dust in the polar regions located in the wall of the ionization cone, in which case the "torus" emission might be from a hollow, cone-shaped extension of dusty accretion disk winds \citep[\eg][]{Gallagher2015,Stalevski2017}.
 
 Particularly interesting is that the most prominent line in the optical spectra of AGN arising from the narrow-line region (NLR), the [O{III}] $\lambda$5007 line, is very weak or absent in these LoBALs, as can be seen in their SDSS spectra \citep[appendix in][]{Lazarova2012}. Strong [O{III}] emission is such a hallmark of AGN activity that it is used, in combination with other lines, in emission line diagrams distinguishing AGN and star forming galaxies \citep{Baldwin1981,Kewley2006}. The only four objects that show any notable emission (i.e., EW([O{III}])$>$20{\AA}; objects J0250+0009, J1011+5429, J1054+0429, J1419+4634) are all spectacular ongoing and recent mergers, three of which have the most prominent double nuclei (see Figures~\ref{fig:images3}, \ref{fig:images8}, \ref{fig:images11}, and \ref{fig:images16}). The NLR emission arises from lower density gas photoionized by the AGN at kpc-scales that has direct view of the unobscured radiation from the central source. Weak narrow lines might indicate higher gas density and/or obscuration of the ionizing radiation, or very few lines-of-sight through which the light can escape to reach farther into the galaxy and the NLR, invoking the "dusty cocoon" model \citep[][]{Hall2002}. If we interpret the lack of [O{III}] emission as a light travel time delay between the onset of AGN activity and the time the radiation reaches the host galaxy \citep{Schawinski2015}, then the objects with weaker emission should be earlier in the process of merging. Alternatively, it is possible that, as a result of a merger, the central source is still enshrouded in gas and gust, and hence there are no or few clear lines-of-sight reaching the NLR. Studies have shown that the strength and the shape of the [O{III}] line are affected by dust obscuration in AGN, leading to suppression of the emission and asymmetry in the profile as a result of greater extinction in the red wing \citep[e.g.,][]{Zakamska2016,DiPompeo2018}. The second nuclei in the objects with notable [O{III}] emission are the most prominent ones, suggesting these are possibly mergers of more equal mass ratio. The stronger [O{III}] emission may be coming from a second nucleus, something we can constrain with the long-slit spectroscopic observations obtained with the {Keck} telescope, results from which will be presented in a future paper.

 The low-ionization absorptions characterizing LoBALs suggest low ionization potential due to dust shielding.  \citet{Gallagher2015} propose that the dust is in the wind, but the winds observed have been previously interpreted as preferential line-of-sight to the observer skimming the edge of the obscuring torus \citep[\eg][]{Weymann1991,Gallagher2007}.  Recent high-resolution observations have resolved the MIR-emitting structure in AGN and support a cone geometry model with dusty polar winds in the ionization cones \citep[][and references therein]{Honig2012,Stalevski2019}. In this framework, LoBALs might be objects observed through lines-of-sight along the wall of the dusty cone or grazing its edge. That picture is consistent with detection of the silicate dust feature at 9.7 $\mu$m only in emission, as observations are larger inclinations (i.e., more edge-on) would give rise to Si absorptions. 
   
 \citet{Ricci2017} propose a radiation-regulated unification scheme for the obscuration in AGN that can be considered in the framework of an evolutionary scenario for episodic black hole growth. They suggest that radiation pressure acting on dusty gas is the main driver shaping the distribution of the dusty obscuring material, with significantly higher covering factors (i.e., more narrow opening angles) for lower Eddington ratios. If a merger event triggered a gradually increasing AGN accretion, when the Eddington ratios exceed 0.02$-$0.06, radiation-driven outflows would expel most of the obscuring gas and dust on very short timescales \citep{Ricci2017}. In this paradigm, LoBALs can be the brief dusty outflow phase - possibly observed mostly along polar lines-of-sight - when the AGN is transitioning from low to high accretion and as a result transforming the dusty, obscuring structure from small to large opening angles. In this picture, the slower, dusty outflows characterizing ongoing and recent merger are preferentially observed along polar sightlines, while the faster, dust-free outflows we find in the undisturbed and unresolved sources might be observed at larger inclinations in objects with larger opening angles. This is also consistent with some LoBALs, such as the disk-dominated hosts, arising due to secular processes and having large opening angles due to the higher Eddington ratios we observe (right panel in Fig.~\ref{fig:edd}(b)), thus having faster winds that escape through less dusty lines of sight (i.e., not along polar sightlines). On the other hand, if the ongoing mergers (i.e., those with double nuclei), recent mergers (single nuclei), and the unresolved hosts represent progressive stages of a merger sequence with gradually increasing accretion rates, as seen in the increasing range of their Eddington ratios toward higher values (see left panel of Fig.~\ref{fig:edd}(b)), the radiation-dependent unification would explain why the merging system are observed through more dusty polar lines of sight, while the larger opening angle (i.e., smaller covering fraction for the circumnuclear obscurer) would allow for a larger range of inclinations. The observational support for this model is based on an X-ray-selected sample of AGN, which might not apply to LoBALs until we solve the mystery of their intrinsic X-ray weakness \citep[e.g.,][]{Luo2013,Teng2014,Luo2014}. Additionally, it is not clear when and how long it would take for an orderly torus-like structure to form in the circumnuclear regions of merging systems, hence, a simple unification model might not be applicable to major mergers \citep{Netzer2015}, and possibly to these merger-dominated LoBALs (\S~\ref{interaction}).

\subsection{What we know about LoBALS and the Challenges for Comparison}\label{compare}
 
Currently, we do not have a control sample of QSOs matched in host galaxy magnitude, redshift and availability of observations in HST/WFC3 of matching depth, thus we resort to comparing the findings with other studies. However, given the unusual properties that LoBALs exhibit across the electromagnetic spectrum, it is not clear how to best select a control sample. In many studies, LoBALs stand out as outliers. A comprehensive effort to understand the nature of LoBAL QSOs needs to address all their observed peculiarities, but how characteristic of the entire class of objects those are, is uncertain due to the often anecdotal nature of the findings. LoBALs were first thought to be rare, making up only 1$-$3\% of optically-selected QSO samples \citep[e.g.,][]{Trump2006}, but were later found in much larger numbers ($>$40\%) when infrared selection was used \citep[e.g.,][]{Dai2008,Urrutia2009,Dai2012}. From surveys we know that LoBALs are type-1 QSOs, with broad emission lines characteristic of that classification, yet tend to have redder spectra and very weak or no narrow-line emission \citep[e.g.,][]{Weymann1991,Reichard2003,Trump2006,Gibson2009}. Particularly notable is the strong Fe{II}$-$weak [O{III}] correlation \citep[e.g.,][]{Weymann1991,Runnoe2013}, which is also the case for this sample \citep{Lazarova2012}. Most bizarre has been the discovery of LoBALs' intrinsic X-ray weakness in recent observations with $NuSTAR$, not due to obscuration and absorption, but true suppression of the "corona" responsible for the Compton up-scattering of UV photons from the accretion disk to X-ray frequencies \citep[e.g.,][]{Luo2013,Teng2014,Luo2014}. In the radio, LoBALs were first thought to be mostly radio-quiet or intermediate \citep[e.g.,][]{Francis1993}, thus limiting the comparison samples and warning of limitations with radio-selection, but were later observed in radio-loud systems as well \citep[e.g.,][]{Brotherton1998,Becker2000}. Of greatest interest have been the location and energetics of the observed outflows as a gauge of the wind's ability to affect the growth of the galaxy, either by shock-heating or by blowing out the available gas supply. Recent work suggests that BAL outflows are on parsec scale and have sufficient kinetic luminosity \citep[e.g.,][]{Leighly2018,Hamann2019} to provide the theoretically predicted feedback capable of impacting galaxy evolution \citep{Hopkins2010}. But detailed absorption line spectral analysis of FeLoBALs by \citet{Leighly2022} and \citet{Choi2022a,Choi2022b} find the outflows to be located at a wide range of distances - from parsec to kiloparsec scale - and discover two distinct accretion-state populations with possibility different acceleration mechanisms. Considering that dusty winds at nuclear scales are a potential explanation for the colors of red quasars \citep{CalistroRivera2021}, our optically-selected sample of LoBALs might not be representative of the wider population of LoBALs. For now, to this list of unusual properties, our study adds the fact that optically-selected LoBALs, at least at low redshifts, might be predominantly found in mergers, unlike even the highly-dust-reddened subclass of FeLoBALs (\S~\ref{comparison}).

 \subsection{Comparison to Other Studies}\label{comparison}

\begin{figure}
\centering
\begin{tabular}{c}
\epsfig{file=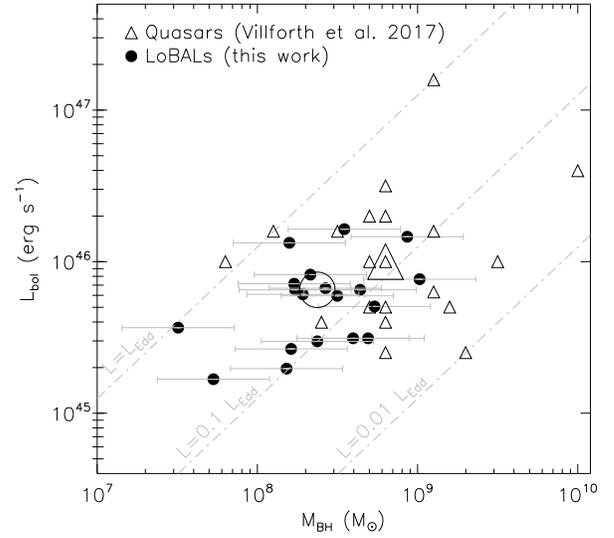,width=1.0\linewidth,clip=} \\
\end{tabular}
\caption{Bolometric AGN luminosity vs.\ black hole mass. For the LoBALs, plotted with filled black circles, L$_{bol}$ was integrated from the optical-to-FIR SEDs and the M$_{BH}$ was estimated from the FWHM of H$_\beta$ and the 5100\AA~luminosity using the single-epoch virial black hole mass relation by \citet{Park2012}. Black hole mass uncertainties are assumed to be large, dominated by the uncertainty in the virial factor, with lower limit $\pm$0.35 dex (see \citet{Park2012} for discussion). For comparison, over-plotted with open triangles are the values for the z$\sim$0.6 quasars of \citet{Villforth2017}, taken from \citet{Shen2011}, who estimate L$_{bol}$ from L$_{5100}$ using a bolometric correction and M$_{BH}$ from single-epoch spectra. With large open symbols we show the median value for each sample: circle for LoBALs and triangle for the \citet{Villforth2017} quasars. The gray, parallel dash-dot lines mark lines of constant Eddington ratios (L/L$_{bol}$) at 100\%, 10\%, and 1\%.}
\label{fig:bol}
\end{figure}

Overall, these LoBALs show higher fraction of mergers (14/22, 64\%) than the sample of 20 optically and X-ray selected 0.5$<$z$<$0.7 quasars of \citet{Villforth2017}. They perform similar morphological analysis on HST WFC3/F160W (H-band) images of comparable depth ($\mu\sim$25 mag arcsec$^{-2}$), and find only 25\% (5/20) with signs of disturbance. The two samples have objects of comparable bolometric luminosities and black hole masses, with only marginally higher median values for the Villforth \etal sample.  In Figure~\ref{fig:bol}, we show $L^{QSO}_{bol}$ as a function of M$_{BH}$ for both, with diagonals demarcating lines of constant Eddington ratios. \citet{Villforth2017} conclude that the mergers are not prevalent in the host galaxies of luminous quasars, with most objects in their sample having disk-like morphologies and only 4/20 bulges (20\%). In stark contrast, the LoBALs here are three times more likely to reside in mergers, and have hosts with intermediate (6/22, 27\%) and bulge-dominated morphologies (9/22, 41\%). Our sample has smaller number (3/22, 14\%) of unresolved hosts (vs.\ 5/20, 25\%), which may indicate their quasars, if product of a merger, would represent a more advanced stage when the AGN overwhelms the emission. We also note that X-ray selection was used by the Villforth \etal to identify their quasar sample, and at this point it is not clear how the X-ray-weak LoBALs are related to X-ray-luminous AGN.
 
At lower redshifts, NIR adaptive optics (H-band) observations of 32 PG QSOs at z~$<$~0.3 by \citet{Guyon2006} find morphologies somewhat similar to our LoBALs. They report $\sim$36\% ellipticals, $\sim$39\% disks, and $\sim$25\% undetermined, but significantly lower fraction ($\sim$30\%) of tidally disturbed systems. Comparable low fraction of mergers is found in a sample of 29 red QSOs at 0.14~$>$~z~$>$~0.6 observed with $HST$/WFC2 in I-band by \citet{Marble2003}, with less (14\%) ellipticals, more (31\%) disks, and mostly (55\%) undetermined morphologies.
 
Similarly, studies of FeLoBALs host galaxies do not support a merger connection. FeLoBALs are a sub-class of LoBALs that, in addition to the low-ionization outflows in Mg{II}, exhibit broad blue-shifted absorption in iron in their UV spectra, suggesting a much lower ionization potential and shielding of the iron-absorbing region, possibly by large column densities and/or dust. \citet{Farrah2007} find support for that in a sample of nine FeLoBALs 1.0~$<$~z~$<$~1.8, all of which are associated with ULIRGs, suggesting FeLoBALs may be a transition phase between a dust-obscured luminous starburst and an emerging quasar. They are heavily reddened \citep[\eg][]{Dunn2015} and suspected to be dust-enshrouded young QSOs as a result of a merger \citep{Glikman2012,Urrutia2009}.

Despite the above, which would be consistent with their host galaxies being mergers, HST WFC3/F160W observations of the host galaxies of 10 FeLoBALs at 0.6~$<$~z~$<$~1.1 by \citet{Villforth2019} do not show excess of merger signatures compared to luminous blue non-BAL quasars. Villforth \etal report only 10\% mergers and 30\% of the hosts showing signs of disturbance, and conclude that FeLoBALs are incompatible with the extreme mergers seen among the heavily reddened quasars of \citet{Urrutia2008}. Half of the objects are unresolved in their data, and many of the FeLoBALs have bolometric luminosities higher than the comparison sample of blue quasars, possibly due to the higher Eddington ratios \citep{Villforth2019}. A similar fraction of non-detections is reported by \cite{Lawther2018} for a small sample of four 0.89~$<$~z~$<$~2.04 FeLoBALs imaged with $HST$/ACS (rest UV) and NICMOS (rest optical). Their data reveals two host detections in the optical and none in the UV band, and they rule out any active star formation. Although they note that the NICMOS observations  (1-4 orbits per object) are not sensitive enough to detect faint morphological distortions, the conclusion is that the host galaxies of FeLoBALs have observed properties that are consistent with those of non-BAL quasars of comparable luminosity, i.e. quiescent or moderately star-forming ellipticals.

In contrast, \citet{Urrutia2008} find a high merger fraction (11/13, 85\%) in a sample of dust-reddened type-1 quasars 0.4~$<$~z~$<$~1, selected with a combination of the $FIRST$ radio and the $2MASS$ infrared surveys. These are a population of objects which, in the evolutionary paradigm for AGN fueling, would represent an earlier, dust-enshrouded stage in the life of quasars. Similar results are reported by \citet{Glikman2015} for dust-reddened quasars at z$\sim$2, objects spanning higher range of L$_{bol}$, 8/10 of which are found in merging galaxies. Interestingly, the one very strong FeLoBAL in the \citet{Glikman2015} study was the only object in their sample that was not resolved. While our optically-selected LoBALs share the similarly high fraction of mergers with those two studies, their quasars are selected in the radio and infrared and are at higher redshifts, representing a potentially biased samples toward mergers at a time closer to the peak of cosmic star formation and black hole activity \citep{Madau2014}. Although, we note that in a $WISE$-selected sample of AGN at z$\sim$2 targeting hot, dust-obscured objects, \citet{Farrah2017} do not find evidence for connection to mergers. 

 Another population of QSOs suspected to be caught in the short transition from a merger-induced starburst and quasar activity toward a quiescent remnant are the post-starburst quasars (PSQs) \citep[\eg][]{Brotherton1999,Canalizo2001,Canalizo2013,Cales2015}. Those are QSOs nested in post-starburst galaxies with a significant young stellar population ($\sim$ few 100 Myr) and evidence for the abrupt quenching of the star formation event that birth them. The AGN activity and the starburst are believed to have been triggered by a merger, and AGN feedback is the suspected culprit for the rapid termination of SF in the hosts. HST/ACS observations by \citet{Cales2011} show that significant fraction (17/29, 59\%) the hosts of PSQs at z~$<$~1 have morphological disturbances, but are equally present in disks and bulges. Host galaxy spectra of our LoBALs obtained with $Keck$/LRIS and ESI, suggest that at least some show Balmer absorption lines, telltale signature of post-starburst stellar populations, but those results will be presented in a future work.

 \subsection{The Radio Properties of LoBALs Might be Clue to Their Nature }\label{radio}

Contrary to conclusions drawn from their SEDs \citep[\eg][]{Gallagher2007,Lazarova2012}, LoBALs are emerging as fundamentally different, not only from typical type-1 QSOs, but also from FeLoBALs. The radio emission in LoBALs might be worth investigating in more detail and, indeed, two clues point to distinct radio properties for the LoBAL population. The first clue comes from stacking analysis of $FIRST$ radio survey data by \citet{White2007} who found that BAL quasars are brighter radio sources than non-BALs, and LoBALs, in particular, are much more likely to be radio-intermediate than either the HiBALs or non-BALs. In that study, they do not separate FeLoBALs from the general population of LoBALs, but, as we see from the optical morphologies of LoBALs presented here and FeLoBAL studies in the literature, they are different in their connection to mergers. The second clue relates to what distinguishes FeloBALs: it appears that $FIRST-2MASS$$-$selected samples of type-1 quasars identify an unusually high number of FeLoBALs \citep{Urrutia2009,Glikman2012}, suggesting an association with radio detection and dust-reddening, while previous studies have found that LoBALs are radio-quiet or intermediate \citep[e.g.,][]{Brotherton1998,Becker2000} and \citet{Lazarova2012} find only moderate levels of dust extinction in the hosts. And, unlike the LoBALs presented here, the optical and UV morphologies of FeLoBALs are mostly unresolved and do not seem to associate strongly with observable recent merger activity \citep{Lawther2018,Villforth2019}.  Hence, LoBALs are emerging as different from FeLoBALs in showing (i) more disturbed optical morphologies, (ii) less dust obscuration, and (iii) lower radio fluxes.

Is it possible that the outflows in LoBALs are driven by emerging radio jets? Hypothetically, if LoBALs are young, emerging radio sources triggered by a major merger, and FeLoBALs represent more advanced merger remnants, with more developed jets, and thus higher level of detectable radio emission, then it would be easier to detect faint merger signatures in the hosts of LoBALs, consistent with the limited observations of their morphologies. A preliminary look at the data from the VLA Sky Survey\footnote{Very Large Array Sky Survey: https://science.nrao.edu/vlass} for our sample of LoBALs shows incidence of detections (6/22, 27\%) that is higher than the typical $\sim$10\% for the general population of AGN. Most of the emission is very faint and compact, but notably 4 of the 6 detections have two or more distinct radio components. Some credibility to this idea comes from recently discovered compact radio emission in red quasars. \citet{Rosario2021} compare the radio properties of red and blue QSOs, and while confirming previous results that they have similar incidences of radio jets and lobes on larger scales ($>$10 kpc), find excess radio emission in red QSOs on smaller, galaxy scales ($<$10 kpc). They suggest that the primary mechanism that generates the enhanced radio emission in red QSOs is arising from AGN-driven compact jets or shocks produced by dusty AGN-driven winds \citep{Rosario2020,Zakamska2014}. However, this idea is yet to be supported by data, and might not be consistent with the curious transformation of an FeLoBAL to a LoBAL reported by \citet{Hall2011}. 

We conclude that mergers must play part of any effort to explain the emergence of LoBALs, and how they fit within any framework to unify the zoo of various AGN types. At least 45$-$64\% (and possibly as high as 77\%) of these LoBALs are involved in an ongoing or recent merger. We place an upper limit of 9$-$37\% on the LoBALs fueled by secular processes. Our data does not support the idea that the outflows observed in LoBALs are responsible for quenching any merger-induced star formation because only four objects have SFR $\geq$ 150 M$\odot$ yr$^{-1}$ and the majority of the mergers already have low levels of SFR measured only as upper limits of $<$98 M$\odot$ yr$^{-1}$.  If the outflow phase is indeed short, then outflows must be triggered during the various stages of the merger process since we observe LoBALs in ongoing mergers with double nuclei at small separations, tidally disturbed hosts, as well as in disks with no signs of tidal interaction. Since AGN activity, i.e., black hole accretion, is triggered with increasing intensity during each subsequence pericenter passage in a mergers \citep[e.g.,][]{VanWassenhove2012}, then it is possible that LoBALs are short outflow phases observed at various episodes when quasar-level accretion occurs \citep{Stickley2014}.

\section{Conclusions}\label{conclusion}

We present the first morphological analysis of the host galaxies of a complete, volume-limited sample of 22 optically-selected LoBAL QSOs at 0.5~$<$~z~$<$~0.6. Using high-resolution HST WFC3 IR/F125W and UVIS/F475W imaging observations, we visually classify the level of disturbance and perform two-dimensional GALFIT modeling to determine the past interaction history and the dominant morphology in each host. The results are interpreted in conjunction with  insights gained from their optical-to-FIR SEDs presented in \citet{Lazarova2012}. Our results can be summarized as follows:

\begin{itemize}[topsep=10pt]
\setlength\itemsep{0.5em}
\setlength{\parskip}{0pt}
\item Our imaging campaign was successful in resolving the host emission for the majority (19/22, 86\%) of the objects in F125W (rest-frame $\sim$6900\AA~to $\sim$9300\AA, sensitive to older stellar populations). The host galaxies are mostly undetected in F475W (rest-frame $\sim$2700\AA ~to $\sim$3600\AA, sensitive to young stars), with only four (18\%) detections among objects with notably high SFRs $\sim$ 150-330 $M_{\odot}$ yr$^{-1}$.
\item  Signs of an ongoing or recent merger are apparent in 64\% (14/22) of the hosts in the F125W images - ranging from double nuclei, extended low surface brightness tidal tails, plumes, bridges and debris - with a conservative merger fraction of at least 45\% (10/22). Given the fairly shallow observations presented here, the detected merger features are most likely due to major mergers as the more subtle morphological disturbances from minor mergers would not be detectable at the low surface brightness limit of the images, $\mu_{F125W}\sim$ 24.7$^{+0.2}_{-0.3}$ mag arcsec$^{-2}$, which were achieved with one $HST$ orbit per target.
\item  The disturbed systems are late-stage mergers, but represent various episodes of the merger process. Double nuclei are observed in 9/22 (41\%), with a conservative estimate of at least 27\%, and are dominated by systems with nuclear separations $<$10 kpc, which account for at least 1/4, and as much as 1/3, of the entire sample. Recent mergers showing a single nucleus and disturbed morphologies are observed in 5/22 (23\%) of the objects.  Five (23\%) of the resolved LoBALs do not show signs of tidal interaction, of which only two have disk-dominated morphologies and apparent spiral arms, possibly suggesting fueling via secular processes in those cases.
\item The high fraction of disturbed morphologies in LoBALs requires that any effort to explain their connection to the broader class of QSOs needs to account for their interaction history. Our results are consistent with an evolutionary model in which the occurrence of LoBAL outflows might be related to episodes of quasar-level accretion during different stages of a merger.
\item The morphologies of LoBALs are mostly bulge-dominated (9/22, 41\%), with those models converging to large S\'{e}rsic indices (n$\gg$4), indicating centrally concentrated profiles, PSF mismatch or extended low surface brightness features. Four (18\%) of the LoBALs have disk-dominated morphologies, and six (27\%) are best-fit by intermediate S\'{e}rsic indices (1.5$<$n$<$3). 
\item We compare various properties of this sample divided into groups by interaction stage and morphology. We find slower, dusty outflows among the mergers and faster, dust-free ones in the undisturbed and unresolved hosts, consistent with the latter representing more advanced AGN-dominated merger stages. 
\item Our results might be consistent with a radiation-dependent unification paradigm, applicable in the framework of an evolutionary scenario for episodic black hole growth, in which the covering fraction depends on the Eddington ratio. In this paradigm, LoBALs can be the brief dusty outflow phase - possibly observed mostly along polar lines-of-sight - when the AGN is transitioning from low to high accretion and as a result transforming the dusty obscuring structure from small to large opening angles. This framework and our data also support secular fueling for some systems.
\item Our data does not support the idea that the outflows observed in LoBALs are responsible for quenching any merger-induced star formation because only four objects have SFR $\geq$ 150 M$_\odot$ yr$^{-1}$ and the majority of the mergers already have relatively low levels of SFR measured only as upper limits of $<$98 M$_\odot$ yr$^{-1}$ and comparable to a control sample of type-1 non-BAL QSOs.
\item We find that these LoBALs have massive black holes (average M$_{BH}\sim$ 3.3 $\times$  10$^8$~M$_\odot$; range 3.2 $\times$ 10$^7$ to 1.0 $\times$ 10$^9$~M$_\odot$), high bolometric luminosities (average $L_{bol}$ $\sim$6.7$\times10^{46}$ erg s$^{-1}$; range 1.7$\times10^{45}$ $-$ 1.6$\times10^{46}$ erg s$^{-1}$), and emit at high Eddington rates, with an average Eddington ratio of $\sim$24\% in a wide range from 5\% to 92\%.
\item  Higher fraction of mergers is found in this sample of optically-selected LoBALs than among unobscured type-1 QSOs and FeLoBALs from the literature. While we do not have UV spectra of these targets to rule out any FeLoBAL interlopers, observations of the morphologies of FeLoBALs and LoBAL strongly suggest that our less-obscured LoBALs have stronger association with mergers.
\item We caution against biases in the selection of BAL QSOs, and demonstrate that all but one of the ongoing and recent mergers in this sample would have been excluded if a traditional BAL sample selection criterion (i.e., Balnicity Index~$>$~0) was used, which might explain why some studies find more disks and less mergers in their samples. {The moderately windy LoBALs in this sample are the ones associated with apparent mergers.}
\end{itemize}
\setlength{\parskip}{0pt}
In summary, to the list of strange and unusual properties of LoBALs, our study adds the fact that LoBALs, at least at low redshifts, are predominantly found in mergers, unlike even the highly-dust-reddened subclass of FeLoBALs.

\begin{acknowledgments}

We would like to thank the anonymous referee for their thoughtful suggestions to improve this article. Support for this work was provided by the National Science Foundation, under grant number AST 0507450 and AST 1817233, by NASA through a grant from the Space Telescope Science Institute (Program GO-11557), which is operated by the Association of Universities for Research in Astronomy, Incorporated, under NASA contract NAS5-26555, and by the CARES Act Stimulus Funds for Research, Scholarship, and Creative Works summer 2021 grant at the University of Northern Colorado. KR was funded in the summer by NASA's Colorado Space Grant Consortium grant, managed at the University of Northern Colorado by Dr. Matthew Semak. MSL would like to thank Dr. Chien Y. Peng for helpful discussions on using GALFIT. MSL would like to thank Maya Davidson, former undergraduate student at the University of Northern Colorado, for her assistance with the VLA Sky Survey. VNB gratefully acknowledge assistance from NSF Research at Undergraduate Institutions (RUI) grant AST-1909297. Note that findings and conclusions do not necessarily represent views of the NSF. This research has made use of the NASA/IPAC Extragalactic Database (NED) which is operated by the Jet Propulsion Laboratory, California Institute of Technology, under contract with the National Aeronautics and Space Administration. Funding for the Sloan Digital Sky Survey (SDSS) has been provided by the Alfred P. Sloan Foundation, the Participating Institutions, the National Aeronautics and Space Administration, the National Science Foundation, the U.S. Department of Energy, the Japanese Monbukagakusho, and the Max Planck Society. The SDSS Web site is http://www.sdss.org/. The SDSS is managed by the Astrophysical Research Consortium (ARC) for the Participating Institutions. The Participating Institutions are The University of Chicago, Fermilab, the Institute for Advanced Study, the Japan Participation Group, The Johns Hopkins University, Los Alamos National Laboratory, the Max-Planck-Institute for Astronomy (MPIA), the Max-Planck-Institute for Astrophysics (MPA), New Mexico State University, University of Pittsburgh, Princeton University, the United States Naval Observatory, and the University of Washington. The National Radio Astronomy Observatory is a facility of the National Science Foundation operated under cooperative agreement by Associated Universities, Inc.

\end{acknowledgments}

\facility{HST (WFC3)}


\begin{table*}
\scriptsize
\caption{Details of the HST WFC3 Observations}
\centering
\begin{tabular}{cccccccccc}
\hline \hline
\#	&	SDSS Object ID	&	RA	&	DEC	&	z	&	\multicolumn{2}{c}{Total exposure time (s)}	&\multicolumn{2}{c}{Number of frames}	& Scale \\ \cline{6-7} \cline{8-9}
	&		        &	(J2000)	&	(J2000)	&		&  F125W	&	F475W		&  F125W	&	F475W & (kpc arcsec$^{-1}$)\\
(1)	&	(2)	        &	(3)	&	(4)	&	(5)	&	(6)	&	(7)	                &	 (8) & (9) & (10)        \\
\hline\hline																					
1	&	J023102.49$-$083141.2	&	02 31 02.500	&	$-$08 31 41.28	&	0.587	&	1006	&	1536	&2 &2 &	6.617\\
2	&	J023153.63$-$093333.5	&	02 31 53.643	&	$-$09 33 33.57	&	0.555	&	1006	&	1536	&2 &2 & 6.440\\
3	&	J025026.66+000903.4	&	02 50 26.660	&	+00 09 03.40	&	0.597	&	1006	&	1530	&2 &2 &	6.670\\
4	&	J083525.98+435211.2	&	08 35 25.980	&	+43 52 11.30	&	0.569	&	 906	&	1486	&2 &4 &	6.519\\
5	&	J085053.12+445122.5	&	08 50 53.120	&	+44 51 22.50	&	0.541	&	 906	&	1486	&2 &4 &	6.359\\
6	&	J085215.66+492040.8	&	08 52 15.663	&	+49 20 40.88	&	0.567	&	1006	&	1692	&2 &2 &	6.508\\
7	&	J085357.87+463350.6	&	08 53 57.880	&	+46 33 50.60	&	0.549	&	1006	&	1436	&2 &4 &	6.405\\
8	&	J101151.95+542942.7	&	10 11 51.950	&	+54 29 42.70	&	0.536	&	1006	&	1748	&2 &2 &	6.329\\
9	&	J102802.32+592906.6	&	10 28 02.320	&	+59 29 06.70	&	0.535	&	1006	&	1556	&2 &4 &	6.323\\
10	&	J105102.77+525049.8	&	10 51 02.770	&	+52 50 49.90	&	0.543	&	1006	&	1748	&2 &2 &	6.370\\
11	&	J105404.73+042939.3	&	10 54 04.730	&	+04 29 39.30	&	0.579	&	1006	&	1530	&2 &2 &	6.574\\
12	&	J112822.41+482309.9	&	11 28 22.410	&	+48 23 10.00	&	0.543	&	 906	&	1536	&2 &4 &	6.359\\
13	&	J114043.62+532439.0	&	11 40 43.620	&	+53 24 39.00	&	0.530	&	1006	&	1492	&2 &4 &	6.293\\
14	&	J130952.89+011950.6	&	13 09 52.890	&	+01 19 50.60	&	0.547	&	 806	&	1472	&2 &4 &	6.394\\
15	&	J140025.53$-$012957.0	&	14 00 25.540	&	$-$01 29 57.00	&	0.584	&	1006	&	1530	&2 &2 &	6.601\\
16	&	J141946.36+463424.3	&	14 19 46.370	&	+46 34 24.30	&	0.547	&	1006	&	1692	&2 &2 &	6.394\\
17	&	J142649.24+032517.7	&	14 26 49.243	&	+03 25 17.71	&	0.529	&	1006	&	1530	&2 &2 &	6.287\\
18	&	J142927.28+523849.5	&	14 29 27.280	&	+52 38 49.50	&	0.595	&	1006	&	1492	&2 &4 &	6.659\\
19	&	J161425.17+375210.7	&	16 14 25.170	&	+37 52 10.70	&	0.553	&	 906	&	1448	&2 &4 &	6.429\\
20	&	J170010.83+395545.8	&	17 00 10.828	&	+39 55 45.82	&	0.577	&	1006	&	1604	&2 &2 &	6.563\\
21	&	J170341.82+383944.7	&	17 03 41.821	&	+38 39 44.77	&	0.554	&	1006	&	1604	&2 &2 &	6.434\\
22	&	J204333.20$-$001104.2	&	20 43 33.200	&	$-$00 11 04.30	&	0.547	&	 806	&	1472	&2 &4 &	6.394\\
\hline 
\end{tabular}
\label{table:observations}
\begin{flushleft}
\tablecomments{Col.\ (1): Object number in the sample. Col.\ (2): Official SDSS designation. In subsequent tables, the objects will be referred to by their truncated SDSS designation. Cols. (3) and (4): Optical positions taken from NED, where units of right ascension are hours, minutes, and seconds, and units of declination are degrees, arcminutes, and arcseconds. Col.\ (5):  Redshift as listed in NED. Col.\ (6) and (7):  Total integration time, in seconds, of the WFC3/IR-F125W and WFC3/UVIS-F475W observations. Col.\ (8) and (9): Number of frames combined with multidrizzle. Col.\ (10): Physical scale in kpc arcsec$^{-1}$.}
\end{flushleft}
\end{table*}

\begin{table*}
\scriptsize
\caption{WFC3/F125W: GALFIT results}
\centering
\begin{tabular}{ccccccccccc}
\hline\hline
\#  & Object ID & Model & n   & \multicolumn{2}{c}{r$_{e}$} & b/a & PA  & $m_{n}$ & $m_{PSF}$ & $\chi^2$ \\ \cline{5-6}
  &  &  &    & (pixels) & (kpc) &  & (deg)  & (AB mag) & (AB mag) & \\
(1) & (2)            & (3)   & (4) & (5)     & (6) & (7) & (8)     & (9)       & (10)&(11)\\
\hline\hline
 01 & J0231$-$0831 & S\'{e}rsic+PSF  &    9.10 &    5.41 &    2.52 &    0.63 &   64.19 &   18.21 &   19.35 & 
2.0 \\
    &                       & deVauc+PSF   &  [4.00] &    6.45 &    3.00 &  [0.85] &   63.55 &   18.53 &   18.97 & 
2.1 \\
    &                       & Exp+PSF     &  [1.00] &    9.60 &    4.47 &  [0.85] &   61.23 &   18.98 &   18.67 & 
2.3 \\
 02 & J0231$-$0933 & S\'{e}rsic+PSF  &    8.33 &   10.20 &    4.71 &    0.81 &  -20.04 &   18.50 &   18.93 & 
1.2 \\
    &                       & deVauc+PSF   &  [4.00] &    8.78 &    4.06 &    0.82 &  -18.41 &   18.81 &   18.79 & 
1.2 \\
    &                       & Exp+PSF     &  [1.00] &    9.22 &    4.26 &    0.82 &  -15.04 &   19.29 &   18.64 & 
1.4 \\
 03 &   J0250+0009 & S\'{e}rsic+PSF  &    1.56 &   12.22 &    5.49 &    0.58 &   47.34 &   18.23 &   18.13 & 
1.7 \\
    &                       & deVauc+PSF   &  [4.00] &   11.93 &    5.36 &    0.58 &   47.19 &   17.93 &   18.30 & 
1.9 \\
    &                       & Exp+PSF     &  [1.00] &   12.59 &    5.66 &    0.58 &   47.54 &   18.35 &   18.09 & 
1.7 \\
 04 &   J0835+4352 & S\'{e}rsic+PSF  &    5.95 &    0.87* &    0.40* &    0.87 &   73.15 &   17.50 &   18.60 & 
2.7 \\
    &                       & deVauc+PSF   &  [4.00] &    0.96* &    0.44* &    0.81 &   78.11 &   17.46 &   18.77 & 
2.8 \\
    &                       & Exp+PSF     &  [1.00] &    1.27* &    0.58* &    0.72 &   76.29 &   17.47 &   18.89 & 
3.2 \\
 05 &   J0850+4451 & S\'{e}rsic+PSF  &    9.53 &    1.09* &    0.48* &    0.87 &  -51.73 &   16.84 &   18.83 & 
3.6 \\
    &                       & deVauc+PSF   &  [4.00] &    1.72* &    0.76* &    0.95 &   52.27 &   17.09 &   18.24 & 
4.2 \\
    &                       & Exp+PSF     &  [1.00] &    5.45 &    2.42 &    0.98 &  -53.13 &   17.95 &   17.31 & 
6.1 \\
 06 &   J0852+4920 & S\'{e}rsic+PSF  &    1.14 &   11.10 &    5.04 &    0.62 &   54.35 &   19.65 &   17.90 & 
1.6 \\
    &                       & deVauc+PSF   &  [4.00] &   14.42 &    6.55 &    0.61 &   55.14 &   19.19 &   17.94 & 
1.6 \\
    &                       & Exp+PSF     &  [1.00] &   10.89 &    4.95 &    0.62 &   54.33 &   19.68 &   17.90 & 
1.6 \\
 07 &   J0853+4633 & S\'{e}rsic+PSF  &    6.48 &    6.17 &    2.76 &    0.57 &  -45.65 &   17.47 &   17.65 & 
2.4 \\
    &                       & deVauc+PSF   &  [4.00] &    7.01 &    3.14 &    0.58 &  -45.91 &   17.67 &   17.53 & 
2.5 \\
    &                       & Exp+PSF     &  [1.00] &    9.14 &    4.09 &    0.59 &  -46.70 &   18.16 &   17.35 & 
3.8 \\
 08 &   J1011+5429 & S\'{e}rsic+PSF  &    2.45 &    7.59 &    3.35 &    0.84 &  -44.29 &   18.46 &   18.99 & 
2.9 \\
    &                       & deVauc+PSF   &  [4.00] &    7.09 &    3.13 &    0.83 &  -41.26 &   18.28 &   19.18 & 
2.9 \\
    &                       & Exp+PSF     &  [1.00] &    7.69 &    3.40 &    0.84 &  -40.75 &   18.72 &   18.83 & 
3.1 \\
 09 &   J1028+5929 & S\'{e}rsic+PSF  &    2.15 &    7.95 &    3.51 &    0.85 &   56.45 &   19.57 &   18.21 & 
1.4 \\
    &                       & deVauc+PSF   &  [4.00] &    7.03 &    3.10 &    0.85 &   56.14 &   19.34 &   18.26 & 
1.4 \\
    &                       & Exp+PSF     &  [1.00] &    8.69 &    3.83 &    0.86 &   57.00 &   19.82 &   18.18 & 
1.4 \\
 10 &   J1051+5250 & S\'{e}rsic+PSF  &    7.97 &    2.50 &    1.11 &    0.60 &  -41.94 &   17.73 &   19.80 & 
1.7 \\
    &                       & deVauc+PSF   &  [4.00] &    4.29 &    1.91 &    0.62 &  -42.14 &   18.06 &   18.87 & 
1.8 \\
    &                       & Exp+PSF     &  [1.00] &    8.15 &    3.62 &    0.60 &  -41.15 &   18.60 &   18.30 & 
2.1 \\
 11 &   J1054+0429 & S\'{e}rsic+PSF  &    7.01 &    5.09 &    2.34 &    0.74 &  -56.50 &   18.15 &   19.98 & 1.5 \\
     &  +Companion                     & deVauc   & [4.00]  & 10.15    &  4.66  & 0.66    & -3.69  & 20.61   &    &  \\

    &   QSO                    & deVauc+PSF   &  [4.00] &    2.57 &    1.18 &    0.70 &  -49.53 &   18.28 &   20.83 & 1.6 \\
    &  +Companion  & deVauc   &  [4.00] & 28.03   & 12.96   & 0.89    & -30.19  & 19.40   &    &  \\

    &   QSO                   & Exp+PSF     &  [1.00] &    4.50 &    2.06 &    0.71 &  -48.84 &   18.89 &   19.22 & 2.0 \\
    &  +Companion    & deVauc   &  [4.00] & 37.44    &  17.18  & 0.87    & -29.72  & 19.02   &    &  \\

 Continues   &                       &   &   &    &    &    &   &   &    &  \\
\hline 
\end{tabular}
\label{table:galfitf125w}
\begin{flushleft}
\tablecomments{Col.\ (1): Object number in the sample. Col.\ (2): Truncated SDSS designation. Col.\ (3): Components of the GALFIT model, each also includes a sky component: S\'{e}rsic refers to a S\'{e}rsic profile with S\'{e}rsic index free to vary; deVauc refers to a de Vaucouleurs profile with fixed $\it n$=4; Exp refers to an Exponential profile with fixed $\it n$=1.  Col.\ (4): S\'{e}rsic index, $\it n$. Cols. (5) and (6): Half-light radius of the S\'{e}rsic component in pixels and in kpc, respectively. A star (*) indicates $r_{e}<$1 kpc, the resolution of the F125W images at these redshifts. Col.\ (7): Axis ratio (minor/major) of the S\'{e}rsic component.  Col.\ (8): Position angle of major axis of the S\'{e}rsic component, in degrees East of North. (9): Apparent F125W (J) magnitude of the S\'{e}rsic component, in AB magnitudes. Average error in magnitude is 0.3 mag. Col.\ (10): Apparent F125W (J) magnitude of the PSF component, in AB magnitudes, not corrected for Galactic or host extinction. Col.\ (11): Reduced $\chi^2$. Square brackets indicate that the parameter was kept fixed in the fitting process. For details, see $\S$~\ref{onemodel}. }
\end{flushleft}
\end{table*}

\addtocounter{table}{-1}
\begin{table*}
\scriptsize
\caption{$-$Continued}
\centering
\begin{tabular}{ccccccccccc}
\hline\hline
\#  & Object ID & Model & n   & \multicolumn{2}{c}{r$_{e}$} & b/a & PA  & $m_{n}$ & $m_{PSF}$ & $\chi^2$ \\ \cline{5-6}
  &  &  &    & (pixels) & (kpc) &  & (deg)  & (AB mag) & (AB mag) & \\
(1) & (2)            & (3)   & (4) & (5)     & (6) & (7) & (8)     & (9)       & (10)&(11)\\
\hline\hline
 12 &   J1128+4823 & S\'{e}rsic+PSF  & 20.00 &    9.18 &    4.05 &    0.61 &  -52.56 &   16.85 &   17.02 & 
5.6 \\
    &                       & deVauc+PSF   &  [4.00] &    8.91 &    3.93 &    0.64 &  -52.28 &   17.50 &   16.83 & 
6.7 \\
    &                       & Exp+PSF     &  [1.00] &   16.66 &    7.35 &    0.64 &  -53.49 &   17.96 &   16.67 & 
8.0 \\
 13 &   J1140+5324 & S\'{e}rsic+PSF  &    7.47 &    2.87 &    1.26 &    0.64 &   86.68 &   17.65 &   18.31 & 
2.2 \\
    &                       & deVauc+PSF   &  [4.00] &    3.79 &    1.66 &    0.66 &   86.40 &   17.88 &   18.07 & 
2.4 \\
    &                       & Exp+PSF     &  [1.00] &    7.27 &    3.19 &    0.66 &   85.02 &   18.49 &   17.73 & 
3.6 \\
 14 &   J1309+0119 & S\'{e}rsic+PSF  &    1.80 &   15.05 &    6.72 &    0.99 &   13.92 &   18.07 &   16.86 & 
3.1 \\
    &                       & deVauc+PSF   &  [4.00] &   17.57 &    7.84 &    1.00 &  -78.23 &   17.77 &   16.89 & 
3.2 \\
    &                       & Exp+PSF     &  [1.00] &   14.59 &    6.51 &    0.99 &    7.25 &   18.25 &   16.84 & 
3.2 \\
 15 & J1400$-$0129 & S\'{e}rsic+PSF  &    0.69 &   12.51 &    5.77 &    0.79 &   22.91 &   18.85 &   17.75 & 
1.9 \\
    &                       & deVauc+PSF   &  [4.00] &   11.22 &    5.17 &    0.88 &   31.43 &   18.39 &   17.84 & 
2.2 \\
    &                       & Exp+PSF     &  [1.00] &   12.11 &    5.58 &    0.81 &   23.58 &   18.77 &   17.76 & 
2.0 \\
 16 &   J1419+4634 & S\'{e}rsic+PSF  &    7.10 &   27.35 &   12.20 &    0.83 &  -41.67 &   17.86 &   18.97 & 1.1 \\
      &  +Companion                     & S\'{e}rsic  & 9.45  & 0.67   &  0.30  & 0.52   & -33.28  & 20.25   &    &  \\
   &        QSO               & deVauc+PSF   &  [4.00] &   19.57 &    8.73 &    0.84 &  -37.90 &   18.11 &   18.82 & 1.1 \\
      &  +Companion                     & S\'{e}rsic  &  3.64 &  0.67  &   0.30 & 0.60   & -36.24  & 20.42   &    &  \\
    &     QSO                  & Exp+PSF     &  [1.00] &   13.13 &    5.85 &    0.86 &  -36.37 &   18.71 &   18.66 & 1.5 \\
      &  +Companion                     & S\'{e}rsic  & 18.04   & 0.81   &  0.36  & 0.53   & -20.56  & 20.06   &    &  \\
    
 17 &   J1426+0325 & S\'{e}rsic+PSF  &    2.58 &   11.25 &    4.94 &    0.90 &  -62.87 &   17.83 &   17.60 & 
1.8 \\
    &                       & deVauc+PSF   &  [4.00] &   10.98 &    4.82 &    0.90 &  -64.63 &   17.68 &   17.66 & 
1.8 \\
    &                       & Exp+PSF     &  [1.00] &   11.67 &    5.12 &    0.89 &  -57.72 &   18.14 &   17.51 & 
2.2 \\
 18 &   J1429+5238 & S\'{e}rsic+PSF  &    0.70 &  0.21* &    0.10* &  0.05 &   19.91 &   17.63 &   16.81 & 
7.5 \\
    &                       & deVauc+PSF   &  [4.00] &   10.48 &    4.87 &    0.15 &  -44.78 &   21.95 &   16.39 & 
7.8 \\
    &                       & Exp+PSF     &  [1.00] &  0.29* &    0.13* &    0.83 &    1.71 &   17.49 &   16.87 & 
7.5 \\
 19 &   J1614+3752 & S\'{e}rsic+PSF  & 19.27 &    0.74 &    0.33 &    0.87 &  -87.13 &   16.40 &   16.76 & 
6.0 \\
    &                       & deVauc+PSF   &  [4.00] &    5.34 &    2.40 &    0.86 &  -85.77 &   17.36 &   16.26 & 
6.3 \\
    &                       & Exp+PSF     &  [1.00] &    9.66 &    4.34 &    0.84 &  -88.54 &   17.95 &   16.17 & 
7.0 \\
 20 &   J1700+3955 & S\'{e}rsic+PSF  &    0.89 &   12.70 &    5.82 &    0.76 &   -9.38 &   18.07 &   19.25 & 
3.3 \\
    &                       & deVauc+PSF   &  [4.00] &   13.34 &    6.12 &    0.67 &   -0.64 &   17.72 &   19.99 & 
4.2 \\
    &                       & Exp+PSF     &  [1.00] &   12.65 &    5.80 &    0.75 &   -8.53 &   18.05 &   19.28 & 
3.3 \\
 21 &   J1703+3839 & S\'{e}rsic+PSF  &    2.68 &   15.19 &    6.82 &    0.86 &  -58.79 &   18.34 &   17.18 & 
2.5 \\
    &                       & deVauc+PSF   &  [4.00] &   15.37 &    6.90 &    0.85 &  -60.40 &   18.20 &   17.21 & 
2.6 \\
    &                       & Exp+PSF     &  [1.00] &   15.61 &    7.01 &    0.88 &  -54.93 &   18.66 &   17.14 & 
2.7 \\
 22 & J2043$-$0011 & S\'{e}rsic+PSF  &    5.52 &    6.10 &    2.72 &    0.45 &  -82.25 &   18.23 &   17.35 & 
2.2 \\
    &                       & deVauc+PSF   &  [4.00] &    7.25 &    3.23 &    0.46 &  -82.05 &   18.37 &   17.31 & 
2.2 \\
    &                       & Exp+PSF     &  [1.00] &   10.56 &    4.70 &    0.48 &  -81.86 &   18.90 &   17.21 & 
2.4 \\
\hline 
\end{tabular}
\begin{flushleft}
\scriptsize{}
\end{flushleft}
\end{table*}

\begin{table*}
\scriptsize
\caption{WFC3/F475W: GALFIT results.}
\centering
\begin{tabular}{ccccccccccc}
\hline\hline
\#  & Object ID & Model & n   & \multicolumn{2}{c}{r$_{e}$} & b/a & PA  & $m_{n}$ & $m_{PSF}$ & $\chi^2$ \\ \cline{5-6}
  &  &  &    & (pixels) & (kpc) &  & (deg)  & (AB mag) & (AB mag) & \\
(1) & (2)            & (3)   & (4) & (5)     & (6) & (7) & (8)     & (9)       & (10)&(11)\\
\hline\hline
 01 & J0231$-$0831 & PSF  & $\cdots$& $\cdots$ &$\cdots$ & $\cdots$        &  $\cdots$      &  $\cdots$      &  18.33 & 2.1 \\
 02 & J0231$-$0933 & PSF  & $\cdots$& $\cdots$ &$\cdots$ & $\cdots$       & $\cdots$       &$\cdots$        &  18.82 & 1.8 \\
 03 &   J0250+0009 & S\'{e}rsic+PSF  &    0.71 &   30.95 &    6.95 &   0.48 &  45.08 &  21.19$\pm$0.01 &  19.82 & 1.7 \\
 04 &   J0835+4352 & PSF  & $\cdots$        & $\cdots$&   $\cdots$ &  $\cdots$       &  $\cdots$       & $\cdots$        &  16.35 & 2.9 \\
 05 &   J0850+4451 & PSF  & $\cdots$        & $\cdots$&  $\cdots$  & $\cdots$       & $\cdots$       & $\cdots$       &  16.52 & * \\
 06 &   J0852+4920 & PSF  & $\cdots$ & $\cdots$ &    $\cdots$  & $\cdots$ & $\cdots$ & $\cdots$ &  18.27 & * \\
  &   +Companion & S\'{e}rsic  &  [1.05] &  [3.13] &  0.71   & [0.18] & [70.54] & [23.20] &   &  \\
 07 &   J0853+4633 & PSF  &  $\cdots$       & $\cdots$&    $\cdots$ & $\cdots$       & $\cdots$       & $\cdots$       &  16.71 & 3.5 \\
 08 &   J1011+5429 & S\'{e}rsic+PSF  &    3.85 &    8.57 &    1.89 &   0.66 & -28.99 &  20.94$\pm$6.73 &  19.18 & * \\
 09 &   J1028+5929 & PSF  & $\cdots$        & $\cdots$&    $\cdots$ &$\cdots$        & $\cdots$       & $\cdots$        &  17.53 & 2.2 \\
 10 &   J1051+5250 & PSF  & $\cdots$        & $\cdots$&    $\cdots$ & $\cdots$       & $\cdots$       & $\cdots$       &  18.34 & * \\
 11 &   J1054+0429 & PSF  &  $\cdots$&  $\cdots$& $\cdots$    &$\cdots$ & $\cdots$ &$\cdots$ &  19.22 & 1.7 \\
  &   +Companion & S\'{e}rsic  &  [1.77] &  [2.78] &  0.64   & [0.84] & [-18.45] & [23.78] &  & \\
 12 &   J1128+4823 & PSF  & $\cdots$        &  $\cdots$& $\cdots$    & $\cdots$       & $\cdots$       & $\cdots$       &  16.35 & * \\
 13 &   J1140+5324 & PSF  & $\cdots$        &  $\cdots$&    $\cdots$ & $\cdots$       & $\cdots$       & $\cdots$       &  16.65 & * \\
 14 &   J1309+0119 & PSF  & $\cdots$        &  $\cdots$&    $\cdots$ & $\cdots$       &  $\cdots$      & $\cdots$       &  16.61 & 3.9 \\
 15 & J1400$-$0129 & S\'{e}rsic+PSF  &    0.31 &   27.43 &    6.32 &   0.71 &  22.48 &  21.01$\pm$0.01 &  18.12 & 1.7 \\
 16 &   J1419+4634 & PSF  & $\cdots$        &  $\cdots$&    $\cdots$ & $\cdots$       & $\cdots$       & $\cdots$       &  18.91 & 1.8 \\
 17 &   J1426+0325 & PSF  & $\cdots$        &  $\cdots$&    $\cdots$ &  $\cdots$      & $\cdots$       & $\cdots$       &  18.00 & 1.9 \\
 18 &   J1429+5238 & PSF  &  $\cdots$ &  $\cdots$ &   $\cdots$ &   $\cdots$ & $\cdots$ &  $\cdots$ &  16.25 & * \\
 19 &   J1614+3752 & PSF  &  $\cdots$       &  $\cdots$&    $\cdots$ & $\cdots$       & $\cdots$       &  $\cdots$      &  15.77 & * \\
 20 &   J1700+3955 & S\'{e}rsic+PSF  &    0.74 &   30.15 &    6.91 &   0.60 & -15.30 &  20.13$\pm$0.01 &  19.36 & 1.9 \\
 21 &   J1703+3839 & PSF  &  $\cdots$       & $\cdots$ &    $\cdots$ & $\cdots$       & $\cdots$       & $\cdots$       &  18.38 & * \\
 22 & J2043$-$0011 & PSF  &    $\cdots$ &   $\cdots$ &    $\cdots$ &   $\cdots$ & $\cdots$ &  $\cdots$ &  17.07 & * \\
\hline 
\end{tabular}
\label{table:galfitf475w}
\begin{flushleft}
\tablecomments{Col.\ (1): Object number in the sample. Col.\ (2): Truncated SDSS designation. Col.\ (3): Components of the GALFIT model: PSF = only PSF + sky; S\'{e}rsic+PSF = PSF + sky + S\'{e}rsic profile, with $\it n$ free to vary. Col.\ (4): S\'{e}rsic index, $\it n$. Cols.\ (5) and (6): Half-light radius of the S\'{e}rsic component in pixels and in kpc, respectively. Col.\ (7): Axis ratio (minor/major) of the S\'{e}rsic component.  Col.\ (8): Position angle of major axis of the S\'{e}rsic component, in degrees East of North. Col.\ (9): Apparent F475W (SDSS g') magnitude of the S\'{e}rsic component, in AB magnitudes, not corrected for Galactic or host extinction; errors are from GALFIT. Col.\ (10): Apparent F475W (SDSS g') magnitude of the PSF component, in AB magnitudes. Col.\ (11): Reduced $\chi^2$. Square brackets indicate that the parameter was kept fixed. Square brackets ([ ]) in Cols. (4)-(9) indicate that the parameter was kept fixed after the initial fit converged. A star (*) indicates reduced $\chi^2 >100$.}
\end{flushleft}
\end{table*}

\begin{table*}
\caption{Summary of the HST WFC3/F125W Morphologies and Tidal Interactions.}
\scriptsize
\centering
\begin{tabular}{cccccll}
\hline \hline
\#	&	Object ID	& S\'{e}rsic     & Morphological & Interaction &  \multicolumn{2}{l}{Visual features}                                    \\ \cline{6-7} 
&                       &   index    & class   & class & Second nucleus    & Tidal features                  \\
(1)   &(2)                    &(3)                            &(4)                    &(5)            &(6)    &(7)              \\
\hline\hline																					
1	&	J0231$-$0831	& 9.1	 & Bulge & Ongoing Merger & Yes, $\sim$1$\arcsec$N (6.6 kpc)        & $\frac{1}{2}$-arc shell at r$\sim$1$\farcs$3E; $\frac{1}{4}$-arc shell at r$\sim$0$\farcs$9SW  \\
2	&	J0231$-$0933	& 8.3	 & Bulge & Undisturbed & $\cdots$ 	& Possible curved tail $\sim$4$\arcsec$N; "V"-shape $\sim$3$\farcs$5 SE	   \\
3	&	J0250+0009	& 1.6	 & Intermediate   & Merger  &  $\cdots$ 	& "S"-shaped tails extending $\sim$13$\arcsec$N and $\sim$7$\arcsec$S      \\
4	&	J0835+4352	& * & $\cdots$ & $\cdots$ & $\cdots$ & $\cdots$  \\
5	&	J0850+4451	& * & $\cdots$ & $\cdots$ & $\cdots$ & $\cdots$  \\
6	&	J0852+4920	& 1.1	 & Disk   & Ongoing Merger & *Yes, $\sim$0$\farcs$8S (5.2 kpc)        &  Bridge $\sim$4$\farcs$5S 	 \\
7	&	J0853+4633	& 6.5	 & Bulge & Merger  & Elongated PSF	& Plume $\sim$SE    \\
8	&	J1011+5429	& 2.5	 & Intermediate  & Ongoing Merger & Yes,$\sim$1$\farcs$2W (7.6 kpc)         & Plume $\sim$W	 	 \\
9	&	J1028+5929	& 2.2	 & Intermediate  & Undisturbed  & $\cdots$ 			& Possible faint tail $\sim$N 	\\
10	&	J1051+5250	& 8.0	 & Disk & Undisturbed & $\cdots$ 			& $\cdots$ 	 \\
11	&	J1054+0429	& 7.0	 & Bulge & Ongoing Merger & Yes,$\sim$ 1$\arcsec$SW (6.6 kpc)       & Plume $\sim$W; Bridge $\sim$9$\arcsec$NE toward PSF source  \\
12	&	J1128+4823	& 20.0	 & Bulge & Ongoing Merger & *Yes, $\sim$1$\farcs$3SE (8.3 kpc)	& Plume $\sim$E; Curved tail $\sim$8$\arcsec$S 	\\
13	&	J1140+5324	& 7.5	 & Bulge & Undisturbed & Elongated PSF 			& $\cdots$ 	\\
14	&	J1309+0119	& 1.8	 & Intermediate  & Ongoing Merger & Gal$\sim$3$\farcs$5NNE (22.4 kpc)	& $\frac{1}{4}$-arc shell at r$\sim$2$\arcsec$; Bridge to debris $\sim$6$\arcsec$SE    \\
15	&	J1400$-$0129	& 0.7	 & Disk   & Undisturbed & $\cdots$ 			& $\cdots$  \\
16	&	J1419+4634	& 7.1	 & Bulge & Ongoing Merger & Yes,$\sim$0$\farcs$9S (5.8 kpc)       & Straight tail $\sim$7$\arcsec$NNW; Curved tail $\sim$1$\farcs$2SE    \\
17	&	J1426+0325	& 2.6	 & Intermediate  & Ongoing Merger & *Yes,$\sim$2$\arcsec$N (12.6 kpc)	& Plume $\sim$SE 	 \\
18	&	J1429+5238	& * & $\cdots$  & $\cdots$ & $\cdots$ 			& $\cdots$ 	 \\
19	&	J1614+3752	& 19.3	 & Bulge & Merger & $\cdots$ 			& Curved tail $\sim$7$\arcsec$SW; Plume $\sim$NW   \\
20	&	J1700+3955	& 0.9	 & Disk   & Merger &  Clump $\sim$0.6$\arcsec$ (3.9 kpc)	& Plume $\sim$W; "S"-shaped clumpy tails (see residuals)	 \\
21	&	J1703+3839	& 2.7	 & Intermediate  & Merger & $\cdots$ 			& Curved tail $\sim$6$\arcsec$NW    \\
22	&	J2043$-$0011	& 5.5	 & Bulge & Ongoing Merger  & *Yes, $\sim$0$\farcs$7NW (4.5 kpc)			& Plume-like emission extending along SE$-$NW axis   \\
\hline 
\end{tabular}   
\label{table:summary}
\begin{flushleft}
\tablecomments{Col.\ (1): Object number in the sample. Col.\ (2): Truncated SDSS designation.  Col.\ (3): S\'{e}rsic index of the best-fit model based with the lowest reduced $\chi^2$; a star (*) indicates PSF-dominated source. Col.\ (4): Morphological classification based on the S\'{e}rsic index of the the best-fit model and visual examination of the residuals; in all cases but object (10) J1051+5250, the classification is based on n-free; Disk = disk-dominated morphology (n-free$<$1.5); Bulge = bulge-dominated morphology (n-free$>$3); Intermediate = intermediate morphology (1.5$<$n-free$<$3); $\cdots$ indicates PSF-dominated source that could not be fit by a resolved S\'{e}rsic model. Col.\ (5): Interaction classification: Ongoing Merger = presence of two nuclei and signs of tidal disturbance; Merger = single nucleus with tidal features indicative of a merger; Undisturbed = no visible signs of tidal interaction. Col (6): Indicating the presence of a possible second nucleus, at the listed projected separation from the central source.  A preceding star (*) indicates that the second nucleus becomes apparent after the PSF subtraction, in the residuals. Objects 7 and 13 show an elongated central emission from the LoBAL, suggesting the presence of a possibly unresolved second nucleus. Col.\ (7): Indicating if tidal features are visible, in what direction and the approximate projected length for tails and radius for shells: shell = shell-like feature, either $\frac{1}{2}$- or $\frac{1}{4}$-circle, at the projected radius, r; tails are described as plume, straight, or curved after \citet{Ren2020}.}
\end{flushleft}
\end{table*}

\begin{table*}
\scriptsize
\caption{Sources in the immediate neighborhood of the LoBALs.}
\centering
\begin{tabular}{cccl}
\hline \hline
\#	&	Object ID	&   \multicolumn{2}{c}{Nearby sources}      \\   \cline{3-4} 
  &                  &   \#           & Description and Approximate Location                \\
(1)   &(2)                    &(3)              & (4)                 \\
\hline\hline																					
1	&	J0231$-$0831	&1  & faint, debris-like $\sim$10$\arcsec$SW \\
2	&	J0231$-$0933	&2 & compact source $\sim$4$\arcsec$E; distorted galaxy $\sim$8$\arcsec$W  \\
3	&	J0250+0009	&3 & 3 compact objects at $\sim$16$\farcs$7NNE, $\sim$10$\farcs$3NW, and $\sim$7$\farcs$8WNW  \\
4	&	J0835+4352	&2 & compact objects at $\sim$6$\arcsec$SW and $\sim$17$\farcs$4SW  \\
5	&	J0850+4451	&7 & distorted galaxy $\sim$8$\farcs$4SE (z$\sim$0.55$\pm$0.15) with tidal tail toward LoBAL; \\
	&		& & 6 compact sources at $\sim$5$\arcsec$W, $\sim$7$\arcsec$E, $\sim$13$\arcsec$SSW, $\sim$14$\arcsec$WSW, $\sim$12$\arcsec$W, and $\sim$13$\arcsec$W  \\
6	&	J0852+4920	&4 &  red galaxy $\sim$4$\farcs$4S (z$\sim$0.58$\pm$0.04) with a bridge toward LoBAL; compact source $\sim$6$\farcs$2SSW; \\
	&		& &  distorted galaxy $\sim$7$\arcsec$E; distorted galaxy $\sim$16$\farcs$3SE with tail extending $\sim$E \\
7	&	J0853+4633	&4 & compact objects at $\sim$8$\farcs$6ESE and $\sim$4$\farcs$4SSW;	 \\
	&		& & distorted galaxies at $\sim$27$\farcs$2W (z$\sim$0.43$\pm$0.11) and $\sim$24$\arcsec$W  \\
8	&	J1011+5429	&Many & Many debris-like sources in projection\\ 
9	&	J1028+5929	&8 & compact sources at $\sim$10$\arcsec$W, $\sim$13$\arcsec$ESE, $\sim$14$\arcsec$NE, and $\sim$6$\arcsec$NNW; \\
	&		                 & & debris-like at $\sim$5$\farcs$3N and $\sim$7$\arcsec$N;  \\
	&		& &  galaxies at $\sim$23$\arcsec$S (z$\sim$0.54$\pm$0.04) and $\sim$29$\arcsec$S (z$\sim$0.54$\pm$0.04) \\
10	&	J1051+5250	&4 & distorted galaxy $\sim$7$\farcs$4SSW; debris-like at $\sim$8$\arcsec$SW and $\sim$8$\farcs$5SW;	  \\
	&		& & distorted galaxy at  $\sim$23$\farcs$2NW (z$\sim$0.42$\pm$0.14)  \\
11	&	J1054+0429	&3 &  compact object at $\sim$6$\arcsec$WNW; \\
	&		                 & & bridge to PSF-dominated object at $\sim$9$\farcs$5ENE and galaxy $\sim$7$\farcs$8E	 \\
12	&	J1128+4823	&6 & galaxies at $\sim$4$\farcs$8E (z$\sim$0.52$\pm$0.04), $\sim$15$\farcs$2NNW (z$\sim$0.57$\pm$0.04), and $\sim$22$\farcs$4ENE (z$\sim$0.45$\pm$0.12); \\
	&		& & compact sources at $\sim$5$\arcsec$NW and $\sim$7$\arcsec$N; \\
	&		& & PSF-dominated source $\sim$14$\farcs$4S in direction of LoBAL's tidal tail \\
13	&	J1140+5324	&3 & compact source $\sim$5$\farcs$1SE with tail extending $\sim$northward toward the LoBAL; \\
	&		& & two distorted compact sources at $\sim$8$\farcs$9N and $\sim$9$\arcsec$N \\
14	&	J1309+0119	&6 &  compact distorted galaxies at $\sim$3$\farcs$5NNE, $\sim$3$\farcs$5NE, $\sim$6$\farcs$4E, and $\sim$8$\farcs$5NE;  \\
	&		& &  galaxy at $\sim$11$\farcs$3W (z$\sim$0.60$\pm$0.12); compact source $\sim$16$\arcsec$WNW; many debris \\
15	&	J1400$-$0129	&3 & compact object at $\sim$9$\farcs$4NNW; distorted sources at $\sim$12$\farcs$5S and $\sim$16$\farcs$6W 	 \\
16	&	J1419+4634	&4 & pair of merging galaxies at $\sim$6$\farcs$8S (z$\sim$0.55$\pm$0.06); \\
	&		                 & & distorted spiral galaxy at $\sim$8$\farcs$3W (z$\sim$0.38$\pm$0.16);\\
	&	&	&    compact galaxy at $\sim$19$\farcs$3NNW (z$\sim$0.49$\pm$0.06); galaxy at $\sim$14$\arcsec$N  (z$\sim$0.60$\pm$0.13)\\
17	&	J1426+0325	&4 & galaxy $\sim$14$\farcs$7S (z$\sim$0.40$\pm$0.17); distorted compact sources $\sim$7$\farcs$1E and $\sim$6$\farcs$7N; \\
	&		& & distorted galaxy $\sim$15$\arcsec$N (z$\sim$0.40$\pm$0.13); many debris \\
18	&	J1429+5238	&3 & compact sources at $\sim$3$\farcs$8SW and $\sim$6$\farcs$3N; \\
	&		                 & & compact red galaxy at $\sim$14$\arcsec$N (z$\sim$0.50$\pm$0.08); debris	 \\
19	&	J1614+3752	&1 & distorted compact merger $\sim$14$\arcsec$SW (z$\sim$0.53$\pm$0.18) at end of of LoBAL's tidal tail; \\
	&		& & many debris and projected compact sources \\
20	&	J1700+3955	&6 & distorted compact sources $\sim$5$\farcs$4ENE and $\sim$7$\farcs$8N; many compact sources nearby;\\
	&		& & distorted spiral galaxy $\sim$27$\farcs$4SW (z$\sim$0.57$\pm$0.13); merger of galaxies $\sim$23$\arcsec$NE (z$\sim$0.61$\pm$0.16); \\
	&		& &  compact galaxies at $\sim$23$\farcs$3E (z$\sim$0.46$\pm$0.14) and $\sim$21$\farcs$3NW (z$\sim$0.47$\pm$0.22) \\
21	&	J1703+3839	&3 & compact source at $\sim$2$\farcs$7W; many compact sources and debris;  \\
	&		& & galaxy $\sim$18$\farcs$6N (z$\sim$0.44$\pm$0.09) in direction of the LoBAL's extended tidal tail;  \\
	&		& & galaxy merger at $\sim$7$\farcs$1NE misclassified as a "star" by SDSS  \\
22	&	J2043$-$0011	&1 & compact source at  $\sim$6$\farcs$3NNW; many compact sources nearby \\
\hline 
\end{tabular}   
\label{table:nearby}
\begin{flushleft}
\tablecomments{List of sources in the immediate neighborhood and nearby galaxies within 30 kpcs projected separation. Here we do not include the possible second nuclei listed in Col.\ 4 of Table \ref{table:summary}. Col.\ (1): Object number in the sample. Col.\ (2): Truncated SDSS designation. Col.\ (3): Total number of nearby sources. Note that most objects have more sources than listed here, but we only note those of possible interaction. Col.\ (4): Approximate location of the nearby sources. Redshifts are listed where data was available; all redshifts are photometric redshifts taken from SDSS DR16 (https://www.sdss.org/dr16/). Nearby galaxies, in projection, are listed only if they are at the same redshift of the LoBAL within the uncertainties.}
\end{flushleft}
\end{table*}

\begin{table*}
\scriptsize
\caption{WFC3/F125W Absolute magnitudes and luminosities from the GALFIT models.}
\centering
\begin{tabular}{ccccccccc}
\hline\hline
\# & Object ID &  $M_{total}^{F125W}$ & $M_{QSO}$ & $M_{HOST}$ & $L_{total}$ & $L_{QSO}$ & $L_{HOST}$ &  I$_{QSO}$/I$_{HOST}$  \\  \cline{3-5} \cline{6-8}
 &    & \multicolumn{3}{c}{(AB mag)} & \multicolumn{3}{c}{(Log($L/L_\odot$))} & \\
(1) & (2) & (3) & (4) & (5) & (6) & (7) & (8) & (9) \\
\hline\hline
   1 &   J0231$-$0831&  -24.76 & -23.30 & -24.44 &  11.34 &  10.75 &  11.21 &   0.35 \\
  2 &   J0231$-$0933&  -24.57 & -23.58 & -24.01 &  11.26 &  10.87 &  11.04 &   0.67 \\
  3 &   J0250+0009 &   -25.24 & -24.53 & -24.43 &  11.53 &  11.25 &  11.21 &   1.10 \\
  4 &   J0835+4352 &   -25.41 & -23.97 & $\cdots$ &  11.60 &  11.02 & $\cdots$ & $\cdots$ \\
  5 &   J0850+4451 &   -25.76 & -23.61 & $\cdots$ &  11.74 &  10.88 & $\cdots$ & $\cdots$ \\
  6 &   J0852+4920 &   -24.87 & -24.67 & -22.92 &  11.38 &  11.30 &  10.60 &   5.01 \\
  7 &   J0853+4633 &   -25.68 & -24.84 & -25.02 &  11.71 &  11.37 &  11.44 &   0.85 \\
  8 &   J1011+5429 &   -24.49 & -23.44 & -23.97 &  11.23 &  10.81 &  11.02 &   0.61 \\
  9 &   J1028+5929 &   -24.49 & -24.22 & -22.86 &  11.23 &  11.12 &  10.58 &   3.50 \\
 10 &   J1051+5250 &   -24.89 & -22.67 & -24.74 &  11.39 &  10.50 &  11.33 &   0.15 \\
 11 &   J1054+0429 &   -24.64 & -22.63 & -24.46 &  11.29 &  10.49 &  11.22 &   0.19 \\
 12 &   J1128+4823 &   -26.27 & -25.43 & -25.60 &  11.94 &  11.61 &  11.67 &   0.86 \\
 13 &   J1140+5324 &   -25.23 & -24.09 & -24.75 &  11.53 &  11.07 &  11.34 &   0.54 \\
 14 &   J1309+0119 &   -25.91 & -25.60 & -24.39 &  11.80 &  11.67 &  11.19 &   3.05 \\
 15 &   J1400$-$0129&  -25.21 & -24.87 & -23.77 &  11.52 &  11.38 &  10.94 &   2.75 \\
 16 &   J1419+4634 &   -24.96 & -23.52 & -24.63 &  11.42 &  10.84 &  11.29 &   0.36 \\
 17 &   J1426+0325 &   -25.42 & -24.77 & -24.54 &  11.60 &  11.34 &  11.25 &   1.24 \\
 18 &   J1429+5238 &   -26.32 & -26.32 & $\cdots$ &  11.96 &  11.96 & $\cdots$ &   $\cdots$ \\
 19 &   J1614+3752 &   -26.69 & -25.75 & -26.11 &  12.11 &  11.73 &  11.88 &   0.72 \\
 20 &   J1700+3955 &   -24.86 & -23.37 & -24.55 &  11.38 &  10.78 &  11.25 &   0.34 \\
 21 &   J1703+3839 &   -25.63 & -25.31 & -24.15 &  11.69 &  11.56 &  11.10 &   2.91 \\
 22 &   J2043$-$0011&  -25.49 & -25.09 & -24.21 &  11.63 &  11.47 &  11.12 &   2.25 \\
\hline 
\end{tabular}
\label{table:lumir}
\begin{flushleft}
\tablecomments{Col.\ (1): Object number in the sample. Col.\ (2) SDSS designation. Col.\ (3): Absolute total magnitudes in F125W, from the PSF + S\'{e}rsic models. Col.\ (4): Absolute QSO magnitudes in F125W, from the PSF model. Col.\ (5): Absolute host galaxy magnitudes in F125W, from the S\'{e}rsic model. Col.\ (6)-(8); Total, QSO, and host galaxy luminosities in F125W. Col.\ (9): QSO-to-host galaxy intensity ratios in F125W. }
\end{flushleft}
\end{table*}

\begin{table*}
\scriptsize
\caption{WFC3 F475W Absolute magnitudes and luminosities from the GALFIT models.}
\centering
\begin{tabular}{cccccccccc}
\hline\hline
\# & Object ID & $M_{total}^{F475W}$ & $M_{QSO}$ & $M_{HOST}$ & $L_{total}$ & $L_{QSO}$ & $L_{HOST}$ & I$_{QSO}/I_{HOST}$ & $M_{total}^{F475W}$-$M_{total}^{F125W}$ \\ \cline{3-5} \cline{6-8}
 &   & \multicolumn{3}{c}{(AB mag)} & \multicolumn{3}{c}{(Log($L/L_\odot$))}& &  \\ 
(1) & (2) & (3) & (4) & (5) & (6) & (7) & (8) & (9) & (10)  \\
\hline\hline
   1 &   J0231$-$0831&  -24.19 & -24.19 &   $\cdots$ &  11.53 &  11.53 &  $\cdots$ &  $\cdots$ &   0.57 \\
  2 &   J0231$-$0933&  -23.59 & -23.59 &   $\cdots$ &  11.29 &  11.29 &  $\cdots$ &  $\cdots$ &   0.97 \\
  3 &   J0250+0009 &   -22.90 & -22.62 & -21.25 &  11.01 &  10.90 &  10.35 &   3.53 &   2.34 \\
  4 &   J0835+4352 &   -26.13 & -26.13 &   $\cdots$ &  12.30 &  12.30 &  $\cdots$ &  $\cdots$ &  -0.72 \\
  5 &   J0850+4451 &   -25.83 & -25.83 &   $\cdots$ &  12.18 &  12.18 &  $\cdots$ &  $\cdots$ &  -0.07 \\
  6 &   J0852+4920 &   -24.24 & -24.24 &   $\cdots$ &  11.55 &  11.55 &  $\cdots$ &  $\cdots$ &   0.63 \\
  7 &   J0853+4633 &   -25.72 & -25.72 &   $\cdots$ &  12.14 &  12.14 &  $\cdots$ &  $\cdots$ &  -0.03 \\
  8 &   J1011+5429 &   -23.42 & -23.22 & -21.46 &  11.22 &  11.14 &  10.44 &   5.06 &   1.07 \\
  9 &   J1028+5929 &   -24.87 & -24.87 &   $\cdots$ &  11.80 &  11.80 &  $\cdots$ &  $\cdots$ &  -0.38 \\
 10 &   J1051+5250 &   -24.10 & -24.10 &   $\cdots$ &  11.49 &  11.49 &  $\cdots$ &  $\cdots$ &   0.79 \\
 11 &   J1054+0429 &   -23.26 & -23.26 &   $\cdots$ &  11.16 &  11.16 &  $\cdots$ &  $\cdots$ &   1.38 \\
 12 &   J1128+4823 &   -26.04 & -26.04 &   $\cdots$ &  12.27 &  12.27 &  $\cdots$ &  $\cdots$ &   0.23 \\
 13 &   J1140+5324 &   -25.72 & -25.72 &   $\cdots$ &  12.14 &  12.14 &  $\cdots$ &  $\cdots$ &  -0.50 \\
 14 &   J1309+0119 &   -25.73 & -25.73 &   $\cdots$ &  12.14 &  12.14 &  $\cdots$ &  $\cdots$ &   0.18 \\
 15 &   J1400$-$0129&  -24.42 & -24.35 & -21.46 &  11.62 &  11.59 &  10.43 &  14.3 &   0.79 \\ 
 16 &   J1419+4634 &   -23.55 & -23.55 &   $\cdots$ &  11.27 &  11.27 &  $\cdots$ &  $\cdots$ &   1.41 \\
 17 &   J1426+0325 &   -24.25 & -24.25 &   $\cdots$ &  11.55 &  11.55 &  $\cdots$ &  $\cdots$ &   1.17 \\
 18 &   J1429+5238 &   -26.44 & -26.43 & $\cdots$  &  12.43 &  12.42 &  $\cdots$ & $\cdots$  &  -0.11 \\
 19 &   J1614+3752 &   -26.67 & -26.67 &   $\cdots$ &  12.52 &  12.52 &  $\cdots$ &  $\cdots$ &   0.02 \\
 20 &   J1700+3955 &   -23.63 & -23.20 & -22.43 &  11.30 &  11.13 &  10.82 &   2.03 &   1.23 \\
 21 &   J1703+3839 &   -23.99 & -23.99 &   $\cdots$ &  11.45 &  11.45 &  $\cdots$ &  $\cdots$ &   1.65 \\
 22 &   J2043$-$0011&  -25.26 & -25.25 & $\cdots$  &  11.95 &  11.95 &   $\cdots$  & $\cdots$  &   0.23 \\
\hline 
\end{tabular}
\label{table:lumuv}
\begin{flushleft}
\tablecomments{Col.\ (1): Object number in the sample. Col.\ (2) SDSS designation. Col.\ (3): Absolute total magnitudes in F475W, from the PSF + S\'{e}rsic models. Col.\ (4): Absolute QSO magnitudes in F475W, from the PSF model. Col.\ (5): Absolute host galaxy magnitudes in F475W, from the S\'{e}rsic model. Col.\ (6)-(8); Total, QSO, and host galaxy luminosities in F475W. Col.\ (9): QSO-to-host galaxy intensity ratios in F475W. Col.\ (10): F475W-F125W colors, from the total absolute magnitudes (PSF+S\'{e}rsic models). }
\end{flushleft}
\end{table*}

\begin{table*}
\scriptsize
\caption{Other quantities for this sample of LoBALs.  }
\centering
\begin{tabular}{ccccccccccc}
\hline\hline
\# & Object ID & $D_L$ & E(B-V) & $A_V$ & $f_{24}/f_{70}$&V$_{max}$ & SFR & $M_{BH}$ & $L_{bol}^{QSO}$ & $L_{bol}$/$L_{Edd}$\\ 
 &   & (Mpc) & (Galactic) & (QSO) && (km s$^{-1}$) & ($M_\odot$ $yr^{-1}$) & (Log(M/$M_\odot$))  & (Log($L/L_\odot$)) &  \\
(1) & (2) & (3) & (4) & (5) & (6) & (7) & (8) & (9) & (10) & (11) \\
\hline\hline
  1 &   J0231$-$0831 &  3494.8 &    0.04 &     $\cdots$ & 0.26 &   4662. & $\cdots$ &    8.16 &     $\cdots$  &     $\cdots$     \\
  2 &   J0231$-$0933 &  3430.4 &    0.03 &     $\cdots$ & 0.06 &   2129.  & $\cdots$ &    8.67 & $\cdots$ &     $\cdots$   \\
  3 &   J0250+0009 &  3126.5 &    0.07 &    1.03  & 0.12 &   3533. & 310$_{-30} ^{+25}$ &   8.33 &  12.33 &    0.31 \\
  4 &   J0835+4352 &  3295.3 &    0.03 &    0.00 & $\cdots$ &  28968. & $\cdots$ &     8.64  &   12.23 &    0.12 \\
  5 &   J0850+4451 &  3105.6 &    0.03 &    0.14 & 0.32 &   5191.  &  $<$49&    9.01 &   12.30 &    0.06\\
  6 &   J0852+4920 &  3281.2 &    0.02 &    0.41 & 0.08 &   2283. & $<$56 &   8.69 &   11.91 &    0.05  \\
  7 &   J0853+4633 &  3168.5 &    0.02 &    0.53 & 0.12 &   4523. & $<$48 &    8.50 &   12.19 &    0.15  \\
  8 &   J1011+5429 &  3070.7 &    0.01 &    0.52 & 0.09 &   3203. & 148$_{-23} ^{+16}$ &  8.21 &  11.84 &    0.13 \\
  9 &   J1028+5929 &  3063.8 &    0.01 &    0.00 & $\cdots$ &   2682. & $\cdots$ &   8.18 &   11.71 &    0.10 \\
 10 &   J1051+5250 &  3119.5 &    0.01 &    0.50 & 0.18 &   2583. & $<$53 &    8.60 &   11.91 &    0.06  \\
 11 &   J1054+0429 &  3366.3 &    0.04 &    0.51 & $\cdots$ &   1803. & $\cdots$ &  8.37 &   11.89 &    0.10\\
 12 &   J1128+4823 &  3119.5 &    0.02 &    0.61 & $\cdots$ &   4496. & $\cdots$ &     8.94 &   12.58 &    0.14 \\
 13 &   J1140+5324 &  3029.0 &    0.01 &    0.00 & $\cdots$ &   3665. & $\cdots$ &    8.73 &   12.12 &    0.08 \\
 14 &   J1309+0119 &  3147.5 &    0.04 &    0.14 & 0.49 &   1869. & $<$62 &   $\cdots$  &   12.40 &  $\cdots$ \\
 15 &   J1400$-$0129 &  3409.0 &    0.05 &    0.69 & 0.06 &  28993. & $<$98 &   8.24 &   12.23 &    0.30 \\
 16 &   J1419+4634 &  3140.5 &    0.01 &    0.48 & 0.12 &   4712. & 63 & 7.73 &   11.64 &    0.25 \\
 17 &   J1426+0325 &  3029.0 &    0.04 &    0.44 & 0.29 &   4800. & $<$52 &   8.28  &   12.20 &    0.25\\
 18 &   J1429+5238 &  3480.5 &    0.01 &    0.28 & 0.62 &   5981. & $<$52 &      8.20  &   12.54 &    0.67\\
 19 &   J1614+3752 &  3189.6 &    0.02 &    0.10 & 0.18 &   6789. & 326$_{-19} ^{+20}$ &  8.54 &   12.63 &    0.38\\
 20 &   J1700+3955 &  3359.2 &    0.02 &    0.42 & 0.10 &   5677. & 161$_{-18} ^{+20}$ &    7.51 &   11.98 &    0.92  \\
 21 &   J1703+3839 &  3196.6 &    0.04 &    0.83 & 0.50 &   7611. & $<$45 &  8.43 &   12.24 &    0.20 \\
 22 &   J2043$-$0011 &  3133.5 &    0.06 &    0.28 & 0.35 &   8354. & $<$43 &    8.23 &   12.27 &    0.34\\
\hline 
\end{tabular}
\label{table:bol}
\begin{flushleft}
\tablecomments{Col.\ (1): Object number in the sample. Col.\ (2) SDSS designation. Col.\ (3): Distance luminosity from NED. Col.\ (4): Galactic dust extinction. Col.\ (5): Dust extinction at the systemic redshift estimated from SED modeling by \cite{Lazarova2012}. Col.\ (6): Sprizer MIPS 24-to-70 $\mu m$ flux ratios from \cite{Lazarova2012}. Col.\ (7): Maximum outflow speed of the Mg II broad-absorption tough from \cite{Trump2006}. Col.\ (8): Star formation rates estimated from the starburst contribution to the optical-to-FIR SEDs by \citet{Lazarova2012}; $<$ indicates an upper limit due to non-detection the FIR $Spitzer$/MIPS bands. Col.\ (9): Black hole masses estimated using the calibration of the single-epoch virial black hole mass relation by \citet{Park2012}. Col.\ (10): Bolometric luminosities of the QSO estimated from integrating only the AGN components in the optical-to-FIR SED models by \cite{Lazarova2012}. Col.\ (11): Eddington ratios.}
\end{flushleft}
\end{table*}

\clearpage

\clearpage

\bibliographystyle{aasjournal}
\bibliography{REF}

\end{document}